\documentclass{rspublic}
\usepackage{amsmath}
\usepackage{amsfonts}
\usepackage{amssymb}
\usepackage[Symbol]{upgreek}
\usepackage[nointegrals]{wasysym}
\usepackage{mathrsfs}
\usepackage{graphicx}
\newcommand{\real}{{\mathbb R}}
\begin{document}
\title[Global Well-posedness of Navier-Stokes Equations]{\large Integral Invariance and Non-linearity Reduction for Proliferating Vorticity Scales in Fluid Dynamics}
\author[F. Lam]{F. Lam}
%
%
\label{firstpage}
\maketitle
\begin{abstract}{Navier-Stokes and Euler Equations; Vorticity; Laminar Flow; Transition; Turbulence; Diffusion; Kinetic Theory of Gases; Randomness}

An effort has been made to solve the Cauchy problem of the Navier-Stokes equations of motion in the whole space $\real^3$. It is shown that vorticity associated with any fluid motion is a direct consequence of conservation of angular momentum, and hence our effort has been concentrated on solving the vorticity equation. It is proved that the sum of the three vorticity components is a time-invariant in fluid motion. Two separate methods of solution have been used. The first one is based on an interpolation theory in Sobolev spaces. It has been proved that, given smooth, localized initial data with finite energy and enstrophy, the vorticity equation admits a global, unique and smooth solution. In our second approach, the vorticity equation has been converted into a non-linear integral equation by means of similarity reduction. The solution of the integral equation has been constructed in a series expansion. The series is shown to converge for initial data of finite size and its analytic properties are extremely intricate. 
Nevertheless, it has been found that the complete vorticity field is characterized, as an instantaneous description, by a multitude of vorticity constituents. The flow field is composed of vortical elements of broad spatio-temporal scales. Every individual element has its own distinct strength apportioned according to viscous diffusion and influence of the Biot-Savart induction. Specifically, the vorticity constituents are vast in quantity. 

The mathematical solutions assert that the Navier-Stokes dynamics is deterministic in nature. The law of energy conservation holds over the entire course of flow evolution. Inference of the solutions leads itself to a satisfactory account for the observed dynamic characteristics of transition process, and of turbulent motion without recourse to instability theory or bifurcation mechanism. In essence, the non-linearity in the equations of motion dictates that every fluid motion ultimately evolves into turbulence as long as the initial data are sufficiently large. The flow evolution is a strong function of the specification of the initial vorticity. In particular, any fluid motion eventually decays in time and restores to its stationary thermodynamic equilibrium state. It is shown that, in the limit of vanishing viscosity, the equations of motion cannot develop flow-field singularities in finite time.

By revisiting the Maxwell-Boltzmann kinetic theory for dilute gases, we consider Maxwellian molecules with cut-off as a generic model. It is found that the density function possesses a phase-space distribution resembling the continuum turbulence provided that molecules' initial conditions are appropriate. In a qualitative sense, the apparent macroscopic randomness of turbulence can be attributed to a ramification of molecular fluctuations, since the viscous dissipation of the mechanical energy is non-uniform and irregular among the multitudinous vortices of small scales. Consequently, the equations of Navier-Stokes dynamics do not contain a separate entity which has been contrived to account for the stochasticity nature of turbulence.

\end{abstract}
\tableofcontents
\section{Introduction}
In the Eulerian description of the motion of an incompressible, homogeneous Newtonian fluid, the principle of mass conservation leads to the continuity equation
\begin{equation} \label{eq:continuity}
	\nabla . u =0.
\end{equation}
The velocity vector $u=u(x,t)$ has the components $u_i,i{=}1,2,3$. The space variable is denoted by $x=(x_1,x_2,x_3)$. The momentum equation for fluid dynamics is derived from Newton's second law of motion. In the absence of a prescribed force, it reads
\begin{equation} \label{eq:ns}
	{\partial u}/{\partial t} + (u . \nabla) u = \nu \Delta u   - {\rho}^{-1} \nabla p, 
\end{equation}
where the scalar quantity $p$ is the pressure, and $\Delta$ is the Laplacian. The density and the viscosity of the fluid are denoted by $\rho$ and $\mu$ respectively. The kinematic viscosity is $\nu =\mu/\rho$. The system of the equations (\ref{eq:continuity}) and (\ref{eq:ns}) is known as the Navier-Stokes equations (Navier 1823; Stokes 1845). They are derived on the basis of the continuum hypothesis (see, for example, Lamb 1975; Prandtl 1952; Serrin 1959; Ladyzhenskaya 1969; Schlichting 1979; Batchelor 1973). For ideal or inviscid flows ($\mu=0$), the system is called the Euler equations (Euler 1755).

We are interested in the global well-posedness of the initial value problem, or the Cauchy problem,  for the system in the whole space $\real^3$. 
The initial condition is given by
\begin{equation} \label{eq:ns-ic}
 u(x,t=0) = u_0(x),  \;\;\; x \in \real^3. 
\end{equation} 
In addition, we assume that $\nabla.u_0 = 0$. To simplify our analysis, we mainly consider regular, localized initial data:
\begin{equation} \label{eq:ic-localization}
	\Big \| \Big( 1 + | \: x \: | \Big )^{k_0} \: \frac{\partial^{\alpha_0} u_0 } {\partial x^{\alpha_0}} \Big \|_{L^{\infty}(\real^3)} \;{<}\; \infty 
\end{equation}
for any values of $k_0$ and $\alpha_0$. The smoothness requirement, $u_0 \in C^{\infty}$, is a strong restriction. In practice, we frequently encounter initial flows that are not necessarily infinitely differentiable; the localization may be limited to a few values of $k_0$ and $\alpha_0$. Instead of specifying the localization, we may rely on the concept of compactness. For instance, we require $u_0 \in C_c^1(\real^3)$ for $C^1$ initial velocity with compact supports. 

Taking divergence of (\ref{eq:ns}) and making use of the continuity, we obtain a Poisson equation for the pressure 
\begin{equation} \label{eq:p-poisson}
	\Delta p(x;t)= - \rho \sum^{3}_{i,j=1} \Big( \frac{\partial u_j}{\partial x_i} \: \frac{\partial u_i}{\partial x_j}\Big)(x;t) = -\rho \Uppi(x;t).
\end{equation}
We have written the independent variables as $(x;t)$ in order to emphasize the fact that, at every instant of time, equation (\ref{eq:p-poisson}) is a kinematic equation as opposed to a dynamic one. Because the differential equation does not contain a term like $\partial p/\partial t$, it does not describe any time-evolution of the pressure. Extra care must be taken in evaluating quantities like $\partial_t \big({\nabla} p(\cdot,t)\big)$, $\int_0^t {\nabla} p(\cdot,\tau) \rd \tau$. Due to the incompressibility hypothesis, the pressure is non-local; any variation in the velocity gradients will instantaneously affect the pressure at any other space locations. This apparent deficiency in physics is in parallel to the infinitely fast propagation speed in the pure initial value problem of heat equation (see, for example, Courant \& Hilbert 1966; Sobolev 1964; John 1982; Evans 2008). 

The vorticity is the curl of the velocity,
\begin{equation*} 
\omega = \nabla {\times} u. 
\end{equation*}
The concept of vorticity is of great assistance as it introduces essential simplifications in the mathematical theory (Helmholtz 1858; Thomson 1869; Lamb 1975). We will demonstrate that the equation governing the vorticity dynamics can be derived from first principles of physics, so that vorticity is no longer an abstraction in mathematics. Compared with velocity, vorticity is more amendable to local and global analyses. Flow development in space-time can best be viewed as the evolution of vorticity field which is characterized by interaction of the shears arising from velocity differences.  

To evaluate flow quantities during the evolution, it may be convenient, in certain circumstances, to trace fluid particles by the Lagrangian description. A fluid particle is a fluid material point that moves with the local velocity. Let $a(x,t)$ be particle's position at time $t$ which is at $a$ with respect to the reference time $t_0$. The initial reference position at $t_0$ is usually taken as $a=a(x,t_0)$. By integrating the following relation in time,
\begin{equation} \label{eq:lagrangian-frame}
	\frac{\partial a(x,t)}{\partial t} = u\big(a(x,t),t\big),
\end{equation}
we can calculate, at least in principle, particle's position $a(x,t)$ for all time once we have the full knowledge of the Eulerian velocity field. In practice it is generally much involved to integrate the deceptively straightforward relation defined in (\ref{eq:lagrangian-frame}) as it largely implies strongly non-linear functions of time, even for simple flows. Furthermore, the Eulerian velocity must be a dynamic quantity which must come from the Navier-Stokes equations or the Euler equations. We accentuate the fact that one cannot simply make use of the velocity derived from the Biot-Savart relation (see (\ref{eq:biot-savart}) below) for substituting the dynamics. The velocity field seconded from the elliptic equation is a result of an infinite-range instantaneous interaction (albeit an anomalous causality) and hence contains no time-wise information.

Introducing a scaling parameter $\lambda{>}0$, one can undertake algebraic manipulations of (\ref{eq:continuity}), (\ref{eq:ns}) and (\ref{eq:vorticity}) by means of transformations $x{\rightarrow}(x/\lambda)$ and $t {\rightarrow} (t/ \lambda^2)$. This procedure of dimensional analysis shows that if the triplet $(u,p, \omega)$ solves the Navier-Stokes system, so does $(u_{\lambda},p_{\lambda}, \omega_{\lambda})$. For fixed fluid properties $\rho$ and $\mu$, the scaled solutions are
\begin{equation} \label{eq:ns-scaling}
  \begin{split}
	u_{\lambda}( x,t) & = {\lambda}^{-1} u\: ( {\lambda}^{-1} x, {\lambda}^{-2} t ), \\
	p_{\lambda}( x,t) & = {\lambda}^{-2} p\:( {\lambda}^{-1} x, {\lambda}^{-2} t), \\
	\omega_{\lambda} (x,t) & = {\lambda}^{-2} \omega \: ( {\lambda}^{-1} x, {\lambda}^{-2} t). \\
	\end{split}
\end{equation}
These scaling properties are derived on the basis of a {\itshape formal} procedure as the numerical value of the scale parameter has no specific upper bounds. Hence it may be assigned to an arbitrarily large value. We have a paradoxical situation; we are free to zone in on every small-scale motion, possibly beyond the fundamental dimensions of matters, {\itshape without} solving the equations of fluid dynamics! The suggestion is that, for some values of $\lambda$, the aggregate variation due to fine scale motions would always invalidate the continuum hypothesis. 

Over a time interval $t \in [0,T],T{>}0$, the kinetic energy of flow motion is 
\begin{equation*} 
\frac{\rho}{2}\int_{\real^3} u^2(x,t) \rd x. 
\end{equation*} 
Since we are working in $\real^3$, the energy may appear to be infinite for an observer moving with a constant finite speed along any straight path which can be arbitrarily far away from the origin. It is defensible that the energy appears to be unbounded. Such an anomaly is due to our choice of frame of reference. As a remedy, we reformulate our fluid dynamics problem by choosing a Galilean transform so that the observer becomes stationary relative the fluid motion. Thus the energy remains finite unless the velocity is out of bounds during the flow development. In subsequent analysis, we take it for granted that a Galilean transform is effected. 

As in many applications, we deal with the Navier-Stokes equations for a fluid enclosed by a smooth, impermeable boundary of finite size, the initial condition must be supplemented by the no-slip Dirichlet boundary condition. For fluid motions in $\real^3$, the ``boundary condition'' takes the form of decay:
\begin{equation*}
u \rightarrow 0 \;\;\; \mbox{as} \;\;\; |x| \rightarrow {\infty}.
\end{equation*}
The decay specification is natural and its physical explanation is evident. However, it is of importance to appreciate that the decay is merely a qualitative statement. In rigorous mathematical analyses, what is crucial are quantitative decay rates, such as $u \sim |x|^{-k},\; k>0$. In general, they are not known {\itshape a priori}. 
  
The Navier-Stokes equations are a system of non-linear parabolic partial differential equations. In $\real^3$, the absence of a solid boundary simplifies our problem since there are no external sources for generating vorticity. Once the initial data are specified, there do not exist a characteristic velocity and a reference length scale during the flow evolution. Thence the equations are in a canonical form where the only parameter of dynamic similarity is related to the property of the fluid $\nu^{-1}$. It has long been conjectured that a singular behaviour may develop during flow evolution (Oseen 1927; Leray 1934$b$). Allowance must be made for solutions of the equations to be rough functions which have limited regularity in space or in time. The concept of weak solutions was introduced by Leray. In practice, we have to deal with distribution solutions; no {\itshape a priori} assumptions should be made on the integrability and differentiability on the triplet $(u,p,\omega)(x,t)$. In three space dimensions, some {\itshape a priori} bounds are known to exist but they do not lend themselves to the solution of the global regularity. Nevertheless, we observe that an equation governing the vorticity dynamics enjoys certain symmetry which in turn imposes a cancellation condition at infinity so that an invariance principle holds over the entire flow evolution. 

To attack the Cauchy problem, we introduce two separate courses of analysis. The first one focuses on the method of interpolation for certain Sobolev spaces. A substantial analysis is devoted to deriving coercive {\itshape a priori} bounds. The well-posedness for the smooth initial data follows without difficulty. We then address the global regularity of the Navier-Stokes equations by extending the properties of the smooth solution to initial data $u_0 \in H_0^{1}(\real^3)$. 

Since our ultimate aim of solving the Navier-Stokes equations is to elucidate on the nature of turbulence, it is of importance to evoke our second method which solves a non-linear integral equation for the vorticity by construction. The analytical structure of the solution represents a complex vorticity field from which many ingredients for turbulence can be found. On the basis of our solution of the Navier-Stokes equations, we make an attempt to account for the transition phenomenon in Reynolds' pipe flow experiments (Reynolds 1883). Furthermore, it is a well-established experimental fact that turbulent fluid motions exhibit intensive, irregular fluctuations at high Reynolds numbers. The kinetic energy is supplied on the continuum and dissipated at the microscopic scales due to molecular friction. It is intuitively clear that we ought not be able to adequately understand many observed characters of turbulence without any knowledge of fluid's microscopic properties. 
Consequently, an effort is made to inquire the connection between the continuum fluid dynamics and the kinetic theory of gases (Maxwell 1867; Boltzmann 1905). 

The study of the Navier-Stokes equations and turbulence is an immense subject; there exists a large collection of literature. It must be admitted that it is almost impractical for one to comprehend all the technical details. So we would like to stress our priority for solving the problems of fluid dynamics; it is not our main concern here to conduct an overview of the past work. To grasp basic developments in the mathematical theory, one may consult recent reviews and monographs (see, for example, Rosenhead 1963; Serrin 1963; Ladyzhenskaya 1969; Temam 1977; Constantin \& Foias 1988; Stuart 1991; Doering \& Gibbon 1995; Lions 1996; Foias {\em et al} 2001; Temam 2001; Lemarie-Rieusset 2002; Cannone 2003; Ladyzhenskaya 2003; Germain 2006; Heywood 2007; Doering 2009). 
For the development of the equations from a historical prospective, consult Darrigol (2002).

The following list on turbulence is by no means comprehensive. These works and the references cited therein contain detailed accounts on theoretical and practical issues relating to turbulence, notably on the work by Reynolds, Richardson, Prandtl, Taylor, Kolmogorov, Kraichnan and others. Dedicated monographs on theory of turbulence are available (see, for example, Batchelor 1953; Bradshaw 1971; Tennekes \& Lumley 1972; Leslie 1973; Hinze 1975; Monin \& Yaglom 1975; Townsend 1976; McComb 1990; ; Saffman 1992; Chorin 1994; Pope 2000; Tsinober 2001; Davidson 2004; Lesieur 2008). We recommend the review articles by Prandtl 1925; von K\'{a}rm\'{a}n 1948; von Neumann 1949; Lin 1959; Corrsin 1961; Saffman 1978; Liepmann 1979; Frisch \& Orszag 1990; Bradshaw 1994; Hunt 2000; Moffatt 2000; Lumley \& Yaglom 2001.

Most mathematical symbols and notations used in the present work are standard. The parabolic cylinder in space-time $(x,t)$ is written as $\real^3 {\times} [0,t]$ for given fixed $0{<}t{\leq}T$, where time $T$ is finite and given. We use multi-index notation for the spatial partial derivatives of order $\alpha$,
	$\partial_x^{\alpha} f(x) = {\partial ^{\alpha} f(x) } / ({\partial x_1^{\alpha_1} {\cdots} \partial x_m^{\alpha_m}  } )$,
for all multi-indexes $\alpha{=}(\alpha_1, \cdots, \alpha_m)$ with $|\alpha|{=}\alpha_1 + \cdots + \alpha_m$.
All $\alpha$'s are non-negative integers. For integrals over $\real^3$, we write $\int(\cdot)\rd x$ for $\int_{\real^3}(\cdot)\rd x$. It is convenient to work in the space of
\begin{equation*}
	Y = L^{2} (\real^3) \cap C^0_c (\real^3),
\end{equation*}
where $C^0_c (\real^3)$ denotes the set of continuous functions with compact supports, normed by $\| \cdot \|_{L^{\infty}}$. The norm $\|\omega\|_{L^{2}}$ has the dimensions of  $\mbox{Length}^{3/2}\: \mbox{Time}^{-1}$ and the same dimensions hold for $(\nu \: t)  \|\omega\|_{L^{\infty}}$. When we refer to space $Y$, we mean the intersection of two Banach spaces of functions as $\|\cdot\|_Y{=}\|\cdot\|_{L^{2}} {+}  \|\cdot\|_{L^{\infty}}$, assuming that the spaces have been normalized so that equality's dimensions are consistent.
We use shorthand notations $L^p_x$ and $L^p_t$ for space $L^p$ and time $L^p$ respectively.
  
Many mathematical symbols have multiple meanings in different sections but extra care has been taken to ensure that their uses do not cause confusion.
\section{Vorticity as a physical characteristic}
\begin{figure}[ht] \centering
  {\includegraphics[width=5cm]{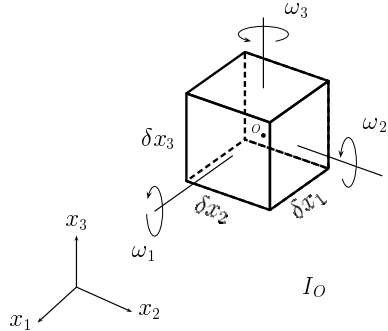}}
  \vspace{3mm}
  \caption{A fluid element is represented here as a cube with its mass centre at $O$ having vorticity vector $\omega$ and moment of inertia $I_O$. The origin of the local co-ordinates system $(x_1,x_2,x_3)$ is at the mass centre. }\label{fig:vemt} 
\end{figure}

Consider an infinitesimal fluid element whose sides are $\delta x_1, \delta x_2$ and $\delta x_3$. Its mass centre is at $O$ which is moving with velocity $u$ and rotating with angular velocity $\omega$/2, see figure~\ref{fig:vemt}. 
The instantaneous rate of deformation of the element can be written as a sum of a symmetric tensor for strains and an antisymmetric tensor for rotations:
\begin{equation} \label{eq:strain-tensors}
	\frac{\partial{u_i}}{\partial{x_j}}= \frac{1}{2} \Big( \frac{\partial{u_i}}{\partial{x_j}}  + \frac{ \partial{u_j}}{\partial{x_i}} \Big)
	+ \frac{1}{2} \Big( \frac{\partial{u_i}}{\partial{x_j}}  - \frac{\partial{u_j}}{\partial{x_i}} \Big) = S_{ij}+R_{ij}.
\end{equation}
The deviatoric stress tensor $\tau_{ij}$ is related to the strain tensor by 
\begin{equation} \label{eq:shear-stresses}
	\tau_{ij} = 2 \mu (S_{ij} - \delta_{ij} \nabla. u /3)=2 \mu S_{ij}.
\end{equation}
The symbol $\delta_{ij}$ is the Kronecker delta. The antisymmetric tensor $R_{ij}$ equals to the angular velocity. The vorticity is twice of the instantaneous local angular velocity (Stokes 1845; Truesdell 1954). The total stress in the $x_j$-direction is given by
\begin{equation*} 
	\sigma_{ij} = - p \:\delta_{ij} + \tau_{ij}.
\end{equation*}

\begin{figure}[ht] \centering
  {\includegraphics[width=5.7413cm]{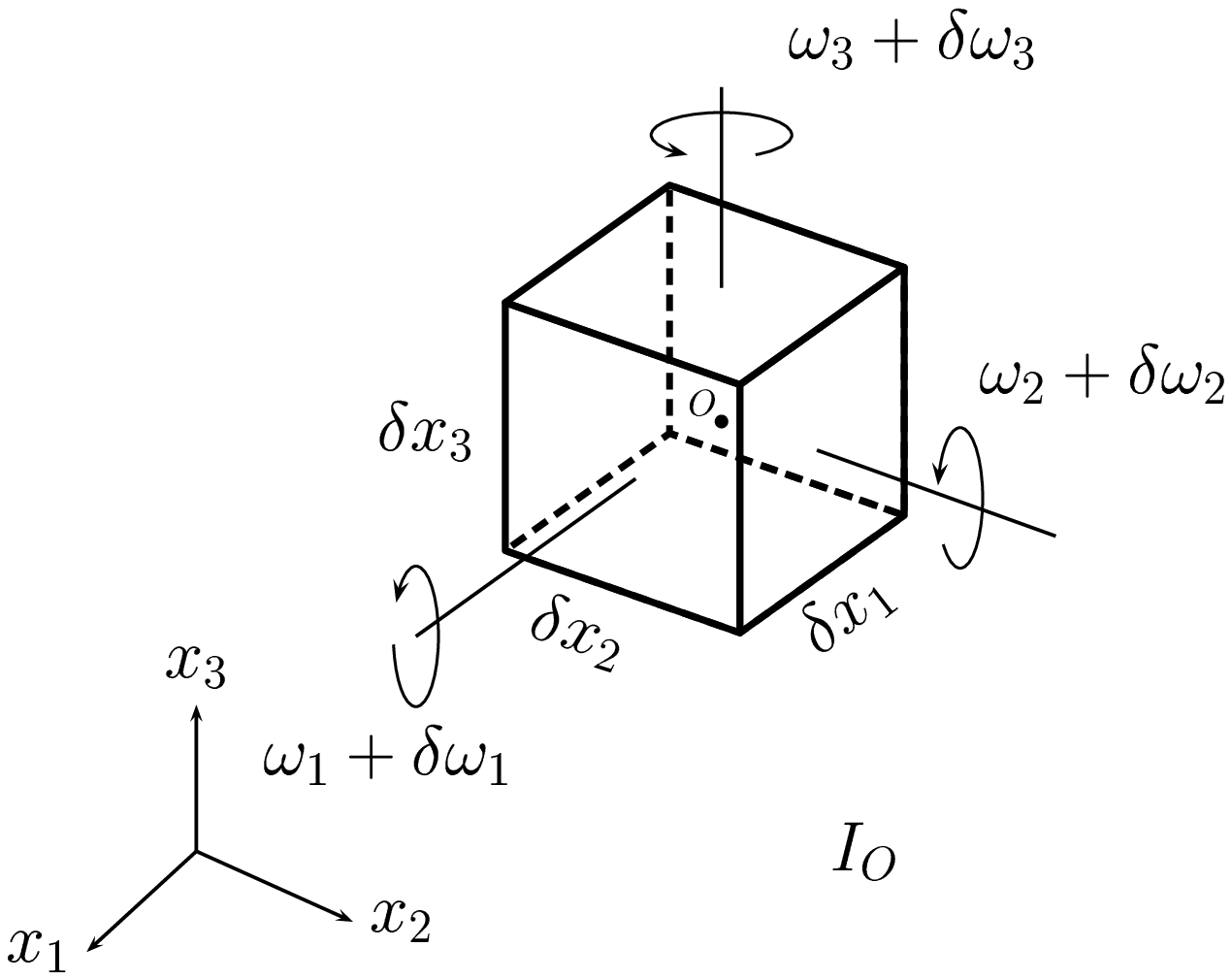}}
  {\includegraphics[width=4.9563cm]{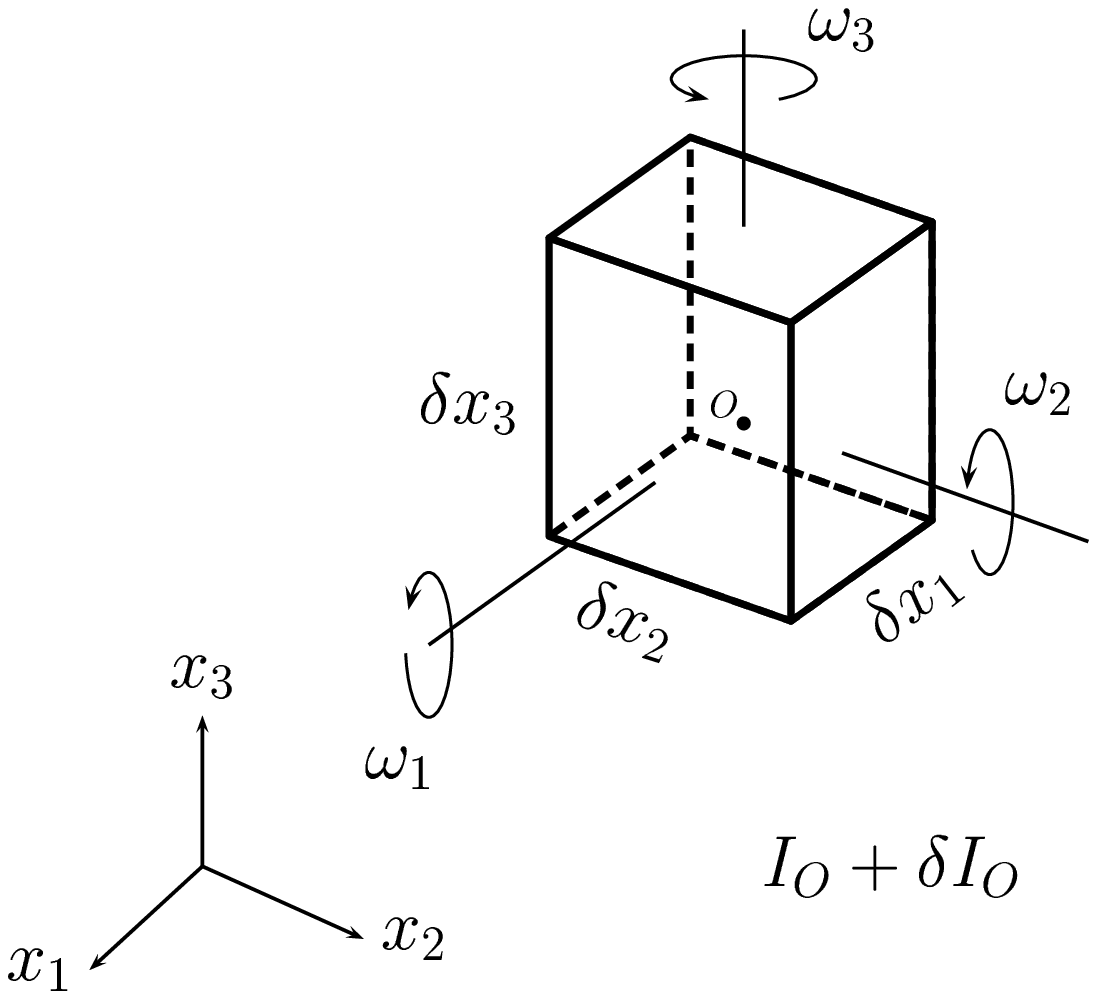}}
  \vspace{3mm}
  \caption{Illustration of the rate of change of angular momentum on the fluid element: the change of angular velocity at fixed moment of inertia (left sketch), and the variation in moment of inertia $\delta I_O$ at fixed angular velocity due to displaced mass centre.  }\label{fig:deft} 
\end{figure}

Over an infinitesimal time interval $\delta t$, the velocity changes to $u{+}\delta u$ and the vorticity to $\omega {+} \delta \omega$. 
The changes in the angular momentum can be considered to consist of two parts. The first one is associated with the change of angular velocity as if the element behaves like a rigid body rotating about the axes, see the left sketch in figure~\ref{fig:deft}. Because of symmetry, the angular momentum is about the axes passing through the mass centre of the element. The axes are in fact the instantaneous principal axes. The moment of inertia of the element about one (say $x_1$) of its principle axes is given by $m(\delta x_2)^2$ or $m( \delta x_3)^2$, where the element is treated as a particle, and hence its actual geometry is immaterial. Denote the mass of the fluid element by $m({=} \rho \delta x_1 \delta x_2 \delta x_3 {=} \rho \delta V)$.
The moment of inertia is simply given by the diagonal matrix ${\mathbf I}$ whose entries are
\begin{equation*}
	I_{ii}= m (\delta x_i)^2.
\end{equation*}
The second part of the angular momentum is related to the change of the geometric shape because the element is being deformed during the motion. The result of the deformation leads to the loss of the geometric symmetry so that the mass centre no longer coincides with its geometric centre, see the right sketch in figure~\ref{fig:deft}. To determine the instantaneous angular momentum, we make use of the inertia tensor about point $O$. The tensor has the form
\begin{equation*}
{\mathbf J} = m \left( \begin{array}{ccc}
{(\delta x_2)^2 {+} (\delta x_3)^2} & - \delta x_1 \delta x_2  & - \delta x_1 \delta x_3    \\
 & & \\
- \delta x_2 \delta x_1       & (\delta x_3)^2 {+} (\delta x_1)^2 & - \delta x_2 \delta x_3  \\
 & & \\
- \delta x_3 \delta x_1 & - \delta x_3 \delta x_2 & (\delta x_1)^2 {+} (\delta x_2)^2 \\
\end{array} 
\right).
\end{equation*}
This is the inertia tensor for rigid body: $J_{ij}{=}m[r^2 \delta_{ij} {-} r_i r_j]$, where $r$ is the position vector. This formula is well-known in mathematical physics (see, for example, Byron \& Fuller 1969; Landau \& Lifshitz 1976). The off-diagonal entries are the products of inertia. The entries, ${-}\delta x_i \delta x_j$, stand for the moment of inertia around the $i$-axis when the element rotates about the $j$-axis with angular velocity $\omega_j/2$. 

Thus the rate of change of angular momentum is found from the following two relations:
\begin{equation*}
	\frac{m}{2} {\mathbf I} \frac{\rd {\omega} } {\rd t} \; \Big|_{\mbox{\small{Rigid body}}} + \frac{m}{2}   \frac{\rd {\mathbf J}}{ \rd t} \omega \; \Big|_{\mbox{\small{Deformed body}}} = \frac{m}{2}\:\big(L_{R} + L_{D}\big), 
\end{equation*}
and
\begin{equation*}
	\frac{\rd m} {\rd t} \; {\mathbf J} {\omega},
\end{equation*}
where ${\omega}{=}(\omega_1 \; \omega_2 \; \omega_3)$ is vorticity column vector. The rate of change in mass is
\begin{equation*}
	\frac{\rd m}{\rd t} = \frac{\partial \rho}{\partial t} \delta V + 
	\frac{\partial (\rho V)}{\partial t} = \frac{\partial \rho}{\partial t} \delta V + 
 \frac{\partial (\rho u_i)}{\partial x_i} \delta V.
\end{equation*}
The differential equality reduces to, after simplification,
\begin{equation} \label{eq:comp-continuity}
{\partial \rho}/{\partial t}  + \nabla.\big(\rho u \big)={D \rho}/{D t}  + \rho (\nabla.u)=0
\end{equation}
in view of the principle of mass conservation. The continuity equation for incompressible flows is recovered according to hypothesis $D \rho/D t{=}0$.

\begin{figure}[ht] \centering
  \includegraphics[width=7cm]{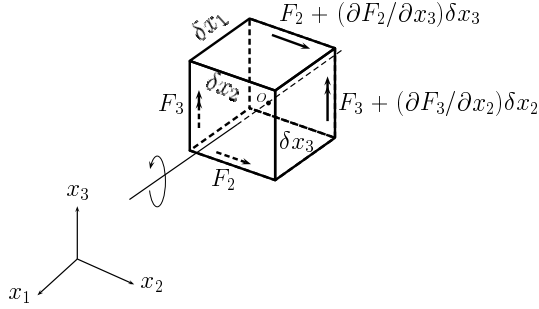}
  \vspace{3mm}
  \caption{Forces $F_2$ and $F_3$ are the stresses on the planes ${x_1{-}x_2}$ and ${x_1{-}x_3}$ of the fluid element respectively. On each of the planes, the stresses consist of one normal component and two shears.  The resultant torque due to the forces is calculated about the local axis through the centre $O$ being parallel to the ${x_1}$-axis.  The net torque is only half of the total stress infinitesimals in the absence of body force.}\label{fig:sfs} 
\end{figure}

The total force per unit volume in the $x_j$-direction can be calculated from the normal and the shear stresses,
\begin{equation*}
	{\partial \sigma_{ij}}/{\partial x_i} = -\partial p /\partial x_j + \partial \tau_{ij} / \partial x_i.
\end{equation*}
Now we derive the rate of change of angular momentum in the $x_1$-direction. The contribution relating to the rigid body rotation per unit mass per unit element area has the form  
\begin{equation*}
 \begin{split}
	L_{R} & = \frac{\partial \omega_1} {\partial t} + \frac{\partial \omega_1} {\partial u_1} \frac{\partial u_1} {\partial x_1} \frac{ \delta x_1} {\delta t} + \frac{\partial \omega_1} {\partial u_2} \frac{\partial u_2} {\partial x_2} \frac{ \delta x_2} {\delta t} + \frac{\partial \omega_1} {\partial u_3} \frac{\partial u_3} {\partial x_3} \frac{\delta x_3} {\delta t} \\
	& = \frac{\partial \omega_1} {\partial t} + u_1 \frac{\partial \omega_1} {\partial x_1} + u_2 \frac{\partial \omega_1} {\partial x_2} + u_3 \frac{\partial \omega_1} {\partial x_3}.
 \end{split}
\end{equation*}
Since $\omega_i{=}\omega_i(u_i)$ and $u_i{=}u_i(x_i,t)$, from the first row of the tensor ${\mathbf J}$, the contribution due to deformation is given by
\begin{equation*}
 \begin{split}
  L_{D} & =   \omega_1 \Big( \delta x_2 \frac{ \delta x_2}{\delta t} + \delta x_3 \frac{\delta x_3}{\delta t} \Big) - \omega_2 \delta x_2 \frac{\delta x_1}{\delta t} - \omega_3 \delta x_3 \frac{\delta x_1}{\delta t}  \\
  & =  \omega_1  \Big( \frac{\partial u_2}{\partial x_2} (\delta x_2)^2 + \frac{\partial u_3}{\partial x_3} (\delta x_3)^2 \Big) - \omega_2 \frac{\partial u_1}{\partial x_2} (\delta x_2)^2 
  -\omega_3 \frac{\partial u_1}{\partial x_3} (\delta x_3)^2.
 \end{split}	
\end{equation*}
Let the net shear force on the $x_1{-}x_2$ plane be $F_2$, see figure~\ref{fig:sfs}. 
The infinitesimal change in torque due to shear $F_2$ per unit mass between the two opposite planes is
\begin{equation*}
	\delta x_3 \frac{\partial(F_2 \delta x_1 \delta x_2)} {\partial x_3}.
\end{equation*}
As the pressure acts in the normal direction to the surfaces of the element, the resultant force in the $x_2$ direction is a pure shear force
\begin{equation*}
	F_2 = \frac{\partial \tau_{12}}{\partial x_1} + \frac{\partial \tau_{22}}{\partial x_2} + \frac{\partial \tau_{32} }{\partial x_3}.
\end{equation*}
Similarly, the shear in the $x_3$ direction is
\begin{equation*}
	F_3 = \frac{\partial \tau_{13}}{\partial x_1} + \frac{\partial \tau_{23}}{\partial x_2} + \frac{\partial \tau_{33} }{\partial x_3}.
\end{equation*}
Hence the net torque per volume about the $x_1$-axis is given by
\begin{equation*}
	\frac{1}{2}\Big(\; \frac{\partial F_3} {\partial x_2} - \frac{\partial F_2} {\partial x_3} \; \Big) = \frac{1}{2} \: \mu \Delta \omega_1,
\end{equation*}
where the last quantity has been obtained by direct calculation using (\ref{eq:strain-tensors}), (\ref{eq:shear-stresses}) and (\ref{eq:comp-continuity}). If an external force is present, the additional torque due to this force must be included. In the limit $\delta x_i{\rightarrow}0$, the sides of the fluid element are indistinguishable so that $(\delta x_1)^2{=}(\delta x_2)^2{=}(\delta x_3)^2$. Newton's second law for the rate of change of angular momentum gives rise to an identity: $\big(L_{R}{+}L_{D}\big)/m=\mu \Delta \omega_1$.
In view of the continuity, this relation yields the equation for the $x_1$-component of vorticity:
\begin{equation*}
	(\partial/\partial t + u. \nabla)\omega_1 - (\omega . \nabla) u_1 = \nu \Delta \omega_1.
\end{equation*}
Two similar equations for the other two vorticity components can be derived by cyclic permutations. Thus the vorticity equation is written as
\begin{equation} \label{eq:vorticity}
	{\partial \omega}/{\partial t} - \nu \Delta \omega = (\omega . \nabla) u  - (u . \nabla )\omega.
\end{equation}
The vorticity field is solenoidal
\begin{equation} \label{eq:vorticity-continuity}
\nabla. \omega=0. 
\end{equation}
For the Cauchy problem, the initial vorticity is specified as 
\begin{equation} \label{eq:vort-ic}
 \omega (x,0) = \nabla {\times} u_0 (x) = \omega_0 (x).
\end{equation} 
We assume that the localization requirements in (\ref{eq:ic-localization}) apply.

{\itshape The vorticity equation for incompressible flows can be derived from the first principle of conservation of angular momentum. Thus the vorticity must be regarded as a physical characteristic in fluid motions governed by the Navier-Stokes dynamics. }
\section{The vorticity Cauchy problem and {\itshape a priori} bounds}
With the vector identity, 
\begin{equation*}
	\nabla (F.G) = F {\times} (\nabla {\times} G) + G {\times} (\nabla {\times} F) + (F. \nabla) G + (G. \nabla)F,
\end{equation*}
for two vectors $F$ and $G$, the Navier-Stokes momentum equation can be rewritten as
\begin{equation} \label{eq:ns-mod}
	{\partial u}/{\partial t} - \nu \Delta u = u {\times} (\nabla {\times} u)  - \nabla \chi ,
\end{equation}
where $\chi{=}p/\rho+u^2/2$ denotes the Bernoulli-Euler pressure. This form of the equation takes advantage of the symmetry in quantity $\varepsilon_{ijk} u_j \omega_k$ that is a spatial invariant in $\real^3$. We take divergence of (\ref{eq:ns-mod}) to get an analogous equation to (\ref{eq:p-poisson}):
\begin{equation} \label{eq:chi-poisson}
	\Delta \chi (x;t) = \big( \omega . \omega - u. (\nabla{\times}\omega) \big)(x;t).
\end{equation}

As allowance must be made for the presence of a possible singularity in the velocity-pressure field, we begin our analysis by using a standard mathematical device. We introduce a sequence of regularizations on the solutions of the momentum equation. For fixed $N{>}0$, we put $\delta{=}T/N$. The symbol $\delta$ is referred as the mollification parameter, and we wish to examine the behaviour of the flow field when the parameter becomes arbitrarily small. We use notation $\phi_{\delta}(u)$ to denote the mollification of the velocity $u$. In essence, the mollification is a convolution in space and in time
\begin{equation} \label{eq:mollifiers}
	 \phi_{\delta}(u) (x,t) = \delta^{-4} \int_{\real}  \int_{\real^3} {\widetilde \phi} \Big( \frac{y}{\delta}, \frac{s}{\delta} \Big) \: u (x{-}y,t{-}s) \rd y \rd s.
\end{equation}
The kernel function, $ {\widetilde \phi}(x,t) {\in} C^{\infty}_c(1{<}t{<}2,|x|{<}1)$, is a mollifier which is a smooth function in space and in time with compact supports. In addition, it has the properties of ${\widetilde \phi}{\geq}0$, and 
\begin{equation*}
	\int_{\real} \int_{\real^3}  {\widetilde \phi}(x,t) \rd x \rd t =  1.
\end{equation*}
In view of the well-known properties of mollifiers (see, for example, Adams \& Fournier 2003; Majda \& Bertozzi 2002; Brezis 2011), the mollification preserves the solenoidal property of $u$:
\begin{equation} \label{eq:mollified-nabla-u}
 \nabla.\phi_{\delta}(u)=\phi_{\delta}(\nabla.u)=0.
\end{equation}
Similarly, the mollification of the vorticity implies that
\begin{equation} \label{eq:mollified-nabla-omega}
 \nabla.\phi_{\delta}(\omega)=\phi_{\delta}(\nabla.\omega)=0.
\end{equation}
When it is necessary, the initial data may also be mollified:
\begin{equation}
	{\psi}_{\delta}(u_0) (x) = \delta^{-3} \int_{\real^3}  {\widetilde \psi} \Big( \frac{y}{\delta}\Big) \: u_0 (x{-}y) \rd y ,
\end{equation}
where the mollifier function ${\widetilde \psi}$ is non-negative, and its space integral equals to unity. Let $u{=}u_N$ and $\chi{=}\chi_N$. The mollified momentum equation reads
\begin{equation} \label{eq:mollified-momentum}
	{\partial u} / {\partial t} - \nu \Delta u = \phi_{\delta}(u) {\times} (\nabla {\times} u) - \nabla \chi.
\end{equation}
The continuity constraint (\ref{eq:continuity}) remains unchanged. The values of $\phi_{\delta}(u)$ at time $t$ depend solely on the values of $u$ at positive time in the interval $(t{-}2\delta, t{-}\delta)$.
This procedure of mollification has been used to construct weak or generalized solutions of the Navier-Stokes equations in the work of Leray 1934$b$; Scheffer 1976; Caffarelli {\em et al} 1982; Constantin 1990. The orthodox Leray mollification refers to smoothing the momentum equation in the form
\begin{equation*} 
	{\partial u} / {\partial t} - \nu \Delta u = - (\phi_{\delta}(u).\nabla) u - \rho^{-1} \nabla p.
\end{equation*}
\subsection*{Bounds derived from vorticity equation}
Taking curl of equation (\ref{eq:mollified-momentum}) and taking the solenoidal constraints (\ref{eq:mollified-nabla-u}) and~(\ref{eq:mollified-nabla-omega}) into account, we obtain 
\begin{equation} \label{eq:mollified-vort}
	{\partial \omega} / {\partial t} - \nu \Delta \omega = (\omega.\nabla)\phi_{\delta}(u) -(\phi_{\delta}(u).\nabla)\omega,
\end{equation}
where the vorticity $\omega{=}\omega_N$. This dynamic equation has an important property of being symmetric with respect to the velocity mollification.

We consider the components of (\ref{eq:mollified-vort}) and let 
\begin{equation} \label{eq:vort-rhs}
	{\mathscr R}_i(x,t, u,\omega) = \Big( (\omega.\nabla)\phi_{\delta}(u_i) -(\phi_{\delta}(u).\nabla)\omega_i \Big)(x,t).
\end{equation}
Carrying out an integration over space $\real^3$ on the sum of three components, integrating by parts and taking the continuity constraints (\ref{eq:vorticity-continuity}) and (\ref{eq:mollified-nabla-u}) into account, we deduce the summability condition,
\begin{equation} \label{eq:vort-rhs-integrability}
	\sum_{i=1}^{3} \int {\mathscr R}_i(x,t,u,\omega) \rd x = 0.
\end{equation}
We notice that this identity is independent of the mollification parameter $\delta$. This observation suggests that it is advantageous to work in terms of the total vorticity,
\begin{equation} \label{eq:total-vort}
	\upomega = \omega_1 + \omega_2 + \omega_3.
\end{equation}
By Duhamel's principle, the total vorticity satisfies the following scalar integral equation $(t{>}s)$:
\begin{equation} \label{eq:duhamel}
	\upomega(x,t) = \int Z(x{-}y,t) \upomega_0(y) \rd y +  \int_0^{t} \! \int Z(x{-}y,t{-}s) \sum_{i=1}^{3}  {\mathscr R}_i(y,s,u,\omega)  \rd y \rd s,
\end{equation}
where $\upomega_0$ is the initial total vorticity. The integral kernel $Z$ is the fundamental solution of the heat equation in $\real^3$,
\begin{equation} \label{eq:heat-kernel}
	Z(x{-}y,t) = (4 \pi \nu t)^{-{3}/{2}} \:  \exp \Big( - \frac{|x-y|^2}{4 \nu t } \Big), \;\;\; t>0.
\end{equation}
A further integration followed by a trivial application of Fubini's theorem enables us to infer the invariance of the total vorticity:
\begin{equation} \label{eq:vorticity-invariance}
\frac{\rd }{\rd t} \int \upomega(x,t) \rd x = 0
\end{equation}
thanks to the well-known properties of the heat kernel. 

{\itshape The total vorticity is shown to be Lebesgue integrable in space and in time provided that the initial total vorticity is an integrable function on $\real^3$. The invariance principle (\ref{eq:vorticity-invariance}) states that, in any fluid motion, the total vorticity is conserved during the entire course of flow evolution. In particular, if the initial total vorticity is zero, the total vorticity remains zero at all subsequent time.}
 
Consider any subset $A_{T} {\subset} \real^3 {\times} [0,T]$. Since there exists a non-zero subset $A_0 {\subset} A_{T}$ such that $\upomega {=} \upomega_0$, it implies that $ \upomega $ does not vanish altogether. It follows that the invariance principle (\ref{eq:vorticity-invariance}) implies 
\begin{equation*} 
	\int \upomega(x,t) \rd x = \int \upomega_0(x) \rd x \;{\leq} \;  \|\upomega_0\|_{L^1 (\real^3)}.
\end{equation*}
(If the initial velocity has compact support, then the initial energy is finite. Thus the invariance principle renders the left-hand side to zero.) 
Consider $\upomega$ as a measurable, real-valued function, we write $\upomega = \upomega^{+}  - \upomega^{-}$, where both $\upomega^{+}{=}\mbox{max}(\upomega,0)$ and $\upomega^{-} {=}{-}\mbox{min}(\upomega,0)$ are measurable, non-negative and {\itshape{finite}}. Hence the set $\{\upomega {=} {+} {\infty}\} \cup\{\upomega {=} {-} {\infty}\}$ has measure zero. By the Archimedean property of the real numbers (see, for example, Royden \& Fitzpatrick 2010), we deduce that every component of vorticity satisfies the bound,
\begin{equation*} 
	\big\|\omega_i(x)\big\|_{L^1 (\real^3)} \;{\leq}\; {\mathbb N} \; \big\|\upomega_0(x)\big\|_{L^1 (\real^3)},
\end{equation*}
except possibly on a set of measure zero. The symbol ${\mathbb N}{=}{\mathbb N}(T)$ denotes a natural number and it can be suitably chosen for any $t {\in} [0,T]$. Making use of the elementary identity
\begin{equation*}
	\omega_1^2 + \omega_2^2 + \omega_3^2 \: {\leq} \: \big( |\omega_1^2| + |\omega_2^2| + |\omega_3^2| \big)^2,
\end{equation*}
and of the fact that $|(\omega_0)_i|{\leq}|\omega_0|$, the integrability bound can be upgraded to
\begin{equation} \label{eq:vort-l1-norm}
	\big\|\omega(x)\big\|_{L^1 (\real^3)} \;{\leq}\; {\mathbb N} \; \big\|\omega_0(x)\big\|_{L^1 (\real^3)}
\end{equation}
for some natural number ${\mathbb N}$. The vorticity invariance is well-known for the Euler equations (see, for example, Batchelor 1969; Majda \& Bertozzi 2002). The existence of vorticity $L^1$-bound in terms of initial energy norm was shown by Constantin (1990) in a periodic domain and by Qian (2009) on a 3D torus.

Differentiating the mollified vorticity components in (\ref{eq:mollified-vort}) yields
\begin{equation*}
	{\partial (\partial_x \omega_i) } / {\partial t} - \nu \Delta (\partial_x \omega_i) = \partial_x {\mathscr R}_i(x,t,u,\omega).
\end{equation*}
We validate the following integral condition analogous to (\ref{eq:vort-rhs-integrability}):
\begin{equation*}
\sum_{i=1}^3 \int \partial_x {\mathscr R}_i (x,t,u,\omega)\rd x = 0.
\end{equation*}
Thus
\begin{equation} \label{eq:invariance-d-omega}
\frac{\rd }{\rd t} \int \partial_x \upomega(x,t) \rd x = 0,
\end{equation}
because the operators ${\partial_x}$ and the gradient operator ${\nabla}$ commute. By analogy, we see that the invariance  (\ref{eq:invariance-d-omega}) implies the summability of $\partial_x \omega$:
\begin{equation} \label{eq:d-vort-l1-norm}
	\big\| \partial_x \omega \big\|_{L^1 (\real^3)} \;{\leq}\; {\mathbb N} \; \big\|\partial_x \omega_0 \big\|_{L^1 (\real^3)}
\end{equation}
for some natural number ${\mathbb N}$. Since the vorticity and its gradient are integrable in $\real^3$, we establish the {\itshape a priori} bound
\begin{equation} \label{eq:vort-crit-norm}
	\omega \in L^{3/2}(\real^3)
\end{equation}
by virtue of the Sobolev embedding theorem (see, for example, Adams \& Fournier 2003; Brezis 2011). The norm number $3/2$ is just the Sobolev conjugate of unity. In view of the scaling properties given in (\ref{eq:ns-scaling}), this bound is scale-invariant and hence it is a critical bound. 
In fact, we can generalize the integrability procedure as follows. For any integer $\alpha{>}0$, we derive the following equation for the vorticity derivatives:
\begin{equation*} 
	{\partial (\partial_x^{\alpha} \omega_i)} / {\partial t} - \nu \Delta (\partial_x^{\alpha} \omega_i) = \partial_x^{\alpha} {\mathscr R}_i(x,t,u,\omega).
\end{equation*}
We confirm that the space-wise integral renders the sum ($\sum_{i=1}^3 \partial_x^{\alpha} {\mathscr R}_i $) to zero. Thus we have the invariance,
\begin{equation} \label{eq:space-vort-int} 
\frac{\rd }{\rd t} \int \partial_x^{\alpha} \upomega(x,t) \rd x = 0,\;\;\;\alpha>0,
\end{equation}
and the bound,
\begin{equation*}
\big \| \partial_x^{\alpha} \omega(x) \big\|_{L^1 (\real^3)} \:{\leq}\: {\mathbb N}_{\alpha} \big\|\partial_x^{\alpha} \omega_0(x) \big\|_{L^1 (\real^3)},
\end{equation*}
for some natural number ${\mathbb N}_{\alpha}$. Although the invariance of the total vorticity holds for the space derivative of arbitrary order, it is sufficient to restrict our derivations to cases $\alpha{=}2,3,4$. Now the vorticity belongs to the Sobolev space $W^{3,1}$ or $L^{\infty}$. We conclude that the vorticity is {\itshape a priori} bounded for fixed $t \in [0,T]$
\begin{equation} \label{eq:vort-infty-bound}
	\big\|\omega(x)\big\|_{L^{\infty}(\real^3)}\:{\leq}\: {\mathbb N}\:\big\|\omega_0(x)\big\|_{L^{\infty}(\real^3)}\;\;\;a.e.
\end{equation}
Evidently, this bound is independent of the mollification parameter $\delta$. 
Interpolation of Lebesgue spaces $L^1$ and $L^{\infty}$ furnishes the estimate
\begin{equation} \label{eq:vort-lp-bound}
	\big\|\omega(x)\big\|_{L^p(\real^3)}\:{\leq}\: C \:\big\|\omega_0(x)\big\|_{L^p(\real^3)},\;\;\;1 \leq p \leq \infty,
\end{equation}
where constant $C{=}C(T,p)$. One of the important consequences is that the enstrophy, $\|\omega\|_{L^2}$, which controls the dissipation of kinetic energy, is bounded a.e. in space -- a presumption stipulating the related dynamic bound $|\omega(\cdot,t)|$. 

In general, the invariance (\ref{eq:space-vort-int}) implies the spatial smoothness,
\begin{equation*}
	\omega(x) \in C^{\alpha}_B(\real^3),
\end{equation*}
because the procedures for integrability consistently enable us to verify $ \omega \in W^{3{+}\alpha,1}$.

The following equation of Poisson type defines a {\itshape kinematic} relationship between the vorticity and the solenoidal velocity: 
\begin{equation} \label{eq:stream-func-poisson}
	\Delta \psi(x;t) = - \omega(x;t),
\end{equation}
where $\psi$ denotes a solenoidal stream-function vector (see, for example, Lamb 1975).
Since $\omega {\in} L^1(\real^3)$, there exists a distribution solution $\psi {\in} L^1_{loc}(\real^3)$. Consequently, the Laplacian has an inverse (see, for example, Folland 1995; Gilbarg \& Trudinger 1998). The velocity is recovered by computing $\nabla{\times}\psi$ at every instant of time, where the curl operation is interpreted as a distributional derivative of the stream-function. Explicitly, the velocity is given by the Biot-Savart law
\begin{equation} \label{eq:biot-savart}
u(x;t)= \frac{1}{4 \pi} \int \frac{(x-y)}{|x-y|^3} {\times} \omega(y;t) \rd y = {\cal K}{*}\omega(x;t).	
\end{equation}
We put special emphasis on our use of notation $(x;t)$ (cf. (\ref{eq:p-poisson})) in order to remind ourselves of the fact that {\itshape velocity-vorticity relation (\ref{eq:biot-savart}) alone does not define a function of time $t$}.
The gradient in the Biot-Savart law is related to $\nabla {\cal K}$ which defines a singular operator of classical Calder\'{o}n-Zygmund type.  Thus a variant of the Calder\'{o}n-Zygmund theorem for solenoidal velocity can be readily derived:
\begin{equation} \label{eq:cz}
	\big\|\nabla u(x) \big\|_{L^p(\real^3)}\:{\leq}\: C \: \big\|\omega(x) \big\|_{L^p(\real^3)},\;\;\; 1 < p < \infty.
\end{equation}
Applying the Hardy-Littlewood-Sobolev inequality, we have
\begin{equation*}
	\big\| u(x)\big\|_{L^q(\real^3)} \: {\leq} \: C \: \big\| \nabla u(x) \big\|_{L^p(\real^3)}, \;\;\; 1/q = 1/p-1/3,
\end{equation*}
where $C$ denotes a constant and can be evaluated sharply (see, for example, \S4 of Lieb \& Loss 1997). 
Hence a relation between the vorticity and the velocity can be deduced:
\begin{equation} \label{eq:sobolev-vel}
	\big\| u(x)\big\|_{L^q(\real^3)} \: {\leq} \: C \: \big\| \omega(x) \big\|_{L^p(\real^3)}, \;\;\; 1/q = 1/p-1/3
\end{equation}
for any $1 {<} p {<} 3$ and $C=C(p,q)$. 

Next we introduce a device so as to partition domain $\real^3$ into two parts: a ball centred at the origin with finite radius $r_0$, $\mathscr{B}(0,r_0)$ and its complement $\mathscr{B}^c{=}\real^3 {\setminus} \mathscr{B}$. In view of Young's inequality  for convolution, the velocity represented in (\ref{eq:biot-savart}) can be estimated according to
\begin{equation} \label{eq:domain-partition}
	\big\|u(x) \big\|_{L^r(\real^3)}\:{\leq}\:\Big( \big\| {\cal K} \big\|_{L^p(\mathscr{B})} \big\|\omega(x)\big\|_{L^q(\real^3)} + \big\| {\cal K} \big\|_{L^{p'}(\mathscr{B}^c)} \big\|\omega(x)\big\|_{L^{q'}(\real^3)} \Big),
\end{equation}
where $1{+}1/r{=}1/p{+}1/q{=}1/{p'}{+}1/{q'}$, $3/2{<}r{\leq}\infty$, and $1{\leq}p,q,p',q'{\leq}\infty$.
We readily find that the energy must be bounded, {\itshape viz}
\begin{equation*} 
	\big\|u(x)\big\|_{L^2(\real^3)}\:{\leq}\:  C_1 \: \big\|\omega_0(x)\big\|_{L^1(\real^3)}+C_2 \: \big\|\omega_0\big\|_{L^{2}(\real^3)},
\end{equation*}
where $C_1$ and $C_2$ are constants. By the same token, we establish
\begin{equation} \label{eq:u-space-bound}
	\big\|u(x)\big\|_{L^{\infty}(\real^3)}\:{\leq}\:  C_1 \: \big\|\omega_0(x)\big\|_{L^{\infty}(\real^3)}+C_2 \: \big\|\omega_0\big\|_{L^{2}(\real^3)}.
\end{equation}

So far we have not yet evoked the {\itshape incompressibility} hypothesis, $\nabla.u{=}0$, which in fact defines a finite sum or a zero sum of three (extended) real numbers. Intuition suggests that every member, $\partial u_i/\partial x_i$ $(i{=}1,2,3)$, must be {\itshape finite} at every time $t$, or $|\partial u_i/\partial x_i| {<} \infty$. The finiteness may be inferred from the properties of the reals (see, for instance, \S 3F of Beals 2004). Instructively the expressions, $(+ \infty)+(- \infty)$ and $(- \infty)+(+ \infty)$, are {\itshape undefined}. If one member becomes infinite while the other two are bounded, their sum violates the hypothesis in one direction to infinity (either $- \infty$ or $+ \infty$). Similar arguments holds if two members are unbounded. Should the trace of the velocity Jacobian be out of bounds, the immeasurable manifold would insinuate either a vacuum or an infinity plenum. 

By the differential form of the conservation, we have
\begin{equation*}
0=\nabla.u =\nabla.u_0 \leq {\mathbb N}_1 \: \|\partial_x u_0(x) \|_{L^{\infty}(\real^3)}, \;\;\; \forall x \in \real^3,
\end{equation*}
where ${\mathbb N}_1$ is a natural number. Conceptually, it is impossible for a gradient on the left to be unbounded. We now show that it is indeed the case.

Applying Gauss' divergence theorem, we obtain that, from the integral form of the mass conservation,
\begin{equation*}
	\int (\nabla.u)(x) \rd x = \int_{\partial S} (u \cdot \vec{n})(x) \rd x = 0,
\end{equation*}
where $\partial S$ denotes a spherical surface whose radius $r{\rightarrow}\infty$, and $\vec{n}$ the outward normal on the surface. This apparently simple integral relation enables us to derive two important properties of the velocity. The first one is the rate of the velocity decay
\begin{equation*}
	|u| \:{\sim}\: O\big(\:|x|^{-2}\:\big)\;\;\;\mbox{as}\;\;\;|x| \rightarrow \infty. 
\end{equation*}
This condition is consistent with the bound $\omega {\in} L^1(\real^3)$.
The second is the summability of the velocity gradient because
\begin{equation*}
	\int (\nabla.u)(x) \rd x \:{\leq}\: \|\nabla u_0\|_{L^1(\real^3)}.
\end{equation*}
It follows that every component, $\|\partial u_i/\partial x_i \|_{L^1} < {\mathbb N}_1 \|\nabla u_0\|_{L^{1}}$, where ${\mathbb N}_1$ is a natural number. Combining this result with bound (\ref{eq:vort-lp-bound}), we assert that
\begin{equation} \label{eq:du-space-l1-bound}
	\big\| \nabla u(x) \big\|_{L^1(\real^3)} \:{\leq}\: C \: \big\|\nabla u_0(x)\big \|_{L^{1}(\real^3)},
\end{equation}
where $C$ is constant. The Archimedean principle for the reals suggests that $|\partial^{\alpha}_x u_0(x) |$ $< {\mathbb N}_{\alpha} \|\omega_0(x) \|_Y \; \forall x \in \real^3$, where ${\mathbb N}_{\alpha}$ denotes some natural numbers. To be specific, we may write the bound as $C(\|\omega_0\|_Y)$, where $C$ is a constant. For convenience, we also denote
\begin{equation} \label{eq:vort-lq-y}
	\| \omega (x)\|_q \:{\leq}\: C \: (\| \omega_0 (x)\|_Y),\;\;\; 1 \leq q \leq \infty.
\end{equation}
Similarly, $\nabla \omega$ can be specifically bounded in this way, as implied in (\ref{eq:space-vort-int}). 

The identity, $\nabla{\times}(\nabla{\times}A)=\nabla(\nabla.A)-\Delta A$, shows that
the velocity Laplacian, 
\begin{equation*}
	\Delta u = -\nabla{\times}\omega,
\end{equation*}
is just a kind of vorticity. This relation is a direct consequence of the continuity. The solenoidal quantity, $\nabla{\times}\omega$, is known as divorticity and satisfies the following dynamic equation:
\begin{equation*}
	{\partial (\nabla{\times}\omega)_i } / {\partial t} - \nu \Delta (\nabla{\times}\omega)_i = \nabla{\times}\big({\mathscr R}_i(x,t,u,\omega)\big).
\end{equation*}
As the curl is a differential operator, by analogy, the invariance and integrability properties established for vorticity apply equally to the divorticity. The strength of the double curl is measured by palinstrophy, $(\nabla{\times}\omega)^2/2$. It is clear that
\begin{equation*}
	\big\|\nabla {\times}\omega(x) \big\|_{L^{q}(\real^3)} \leq \big\|\nabla \omega(x) \big\|_{L^{q}(\real^3)}, \;\;\; 1 \leq q \leq \infty,
\end{equation*}
as all the $12$ derivatives of $\nabla {\times}\omega$ are in $\nabla \omega$ which has a total of $18$ velocity derivatives. This simple rule can be applied to higher differentiations of the divorticity.
Essentially, the velocity Jacobian can be found from the expression
\begin{equation*}
	\nabla u(x;t) = - \int \nabla_x {\cal N}(x,y) (\nabla{\times}\omega)(y;t) \rd y = {\cal M}*(\nabla{\times}\omega)(x;t),
\end{equation*}
where ${\cal N}(x,y)=1/(4 \pi |x{-}y|)$ is the Newtonian potential.
Following the idea of $\real^3$ partition (cf. (\ref{eq:domain-partition})), we readily establish that
\begin{equation*} 
\begin{split}
	\big\|\nabla u(x)\big\|_{L^{\infty}(\real^3)} & \leq  C_1 \: \big\|(\nabla{\times}\omega)(x)\big\|_{L^{\infty}(\real^3)}+C_2 \: \big\|(\nabla{\times}\omega)(x)\big\|_{L^{2}(\real^3)} \\
	& \leq  C_1 \: \big\|\nabla\omega(x)\big\|_{L^{\infty}(\real^3)}+C_2 \: \big\|\nabla\omega(x)\big\|_{L^{2}(\real^3)} < \infty
	\end{split}
\end{equation*}
By interpolation, we conclude that
\begin{equation} \label{eq:du-space-lq-bound}
	\big\|\nabla  u(x)\big\|_{L^q(\real^3)} \leq C \:(\|\omega_0\|_Y), \;\;\; 1\leq q \leq \infty. 
\end{equation}
Evidently, bounds (\ref{eq:du-space-l1-bound}) and (\ref{eq:du-space-lq-bound}) supplement (\ref{eq:cz}). 

To determine whether solutions of the Navier-Stokes equations are smooth, we need {\itshape a priori} bounds for the time variable. We observe that
\begin{equation} \label{eq:vort-beta-time-deriv}
	\partial_t (\partial_t^{\beta} \upomega) - \nu \Delta (\partial_t^{\beta} \upomega) = \partial_t^{\beta} \Big( \sum_{i=1}^3 {\mathscr R}_i(x,t,u,\omega) \Big)
\end{equation}
for any $\beta{\geq}0$. It is clear that the right-hand side is still in $L^1_x$. Thus
\begin{equation} \label{eq:time-vort-int}
\frac{\rd }{\rd t} \int \partial_t^{\beta} \upomega(x,t) \rd x = 0,\;\;\; t \in [T_L,T],
\end{equation}
where $T_L>0$. Smooth and unique solutions of the Navier-Stokes equations over time interval $[0,T_L]$ are known to exist, as a result of {\itshape local} in-time analyses (see, for example, Leray 1934$b$; Hopf 1951; Ladyzhenskaya 1969; Temam 1977; Heywood 1980). Specifically, the local time $T_L$ depends on the norm size of initial data (\ref{eq:ns-ic}). The delicate issue of how the smooth solutions assume the initial data as $t \rightarrow 0$ has been fully vindicated by Heywood (1980, 2007). Consequently, invariance (\ref{eq:time-vort-int}) must be valid from $t=0$ or 
\begin{equation} \label{eq:time-vort-int-bound}
	\frac{\rd }{\rd t} \int \partial_t^{\beta} \upomega(x,t) \rd x = 0,\;\;\; t \in [0,T].
\end{equation}
Clearly, $\omega=\omega_N$ for any value of $N$; this bound is independent of the mollification parameter $\delta$. Once again, we emphasize that bound (\ref{eq:time-vort-int-bound}) {\itshape alone does not} imply any time-wise {\itshape a priori} bound for $u$ and $p$ with exception of the classical time interval $0 {\leq} t {\leq} T_L$, because neither the pressure relation (\ref{eq:p-poisson}) nor the Biot-Savart law (\ref{eq:biot-savart}) contains temporal information. 

Let $\beta=1,2,3$ in (\ref{eq:time-vort-int-bound}). In view of the local in-time solutions, we deduce that $\omega(\cdot,t) \in W^{3,1}\: \forall t \in [T_{\varepsilon},T]$, where $0 {<} T_{\varepsilon} {\leq} T_L$. (This choice of the lower time bound is to avoid unnecessary complications in specifying the initial data.) Obviously we may continue this process of upgrading regularity. In conclusion, we assert that
\begin{equation} \label{eq:vort-space-time-bound}
	\omega(x,t) \in C^{\alpha}_B(\real^3)\:C^{\beta}_B([T_{\varepsilon},T]),\;\;\; \alpha,\beta \geq 0.
\end{equation}

Bounds (\ref{eq:vort-infty-bound}) and (\ref{eq:time-vort-int-bound}) represent a stronger version of the vorticity integral theorem proved by Beale {\itshape et al} (1984).
\subsection*{Bounds derived from momentum equation}
To derive time-wise bounds on velocity, such as $\|\nabla u(\cdot,t)\|_{\infty}$, we must make use of the dynamic equation in the Navier-Stokes system. Since the equation governing the pressure is elliptic in nature, it turns out that there are no direct methods to derive {\itshape a priori time-wise} bounds for the pressure, or more relevantly, for the pressure gradient. The best we can do is to bound the gradient in space at every instant of time as soon as we have a full knowledge of the vorticity. The lack of {\itshape a priori} bound may well reflect the fact that pressure must be a passive variable in fluid motions except for its role in the generating initial motions. It is vorticity which drives flow development. As soon as the velocity and its gradients have been calculated from the initial data, we can tabulate the pressure at every spatio-temporal location $(x,t)$ and from the tabulated data we can evaluate any rate of change in the pressure.
 
As a starting point to solve the Navier-Stokes system, equation (\ref{eq:ns}) has often been rewritten as
\begin{equation} \label{eq:leray-projection}
	{\partial u} / {\partial t} - \nu \Delta u = - {\cal P}\big((u.\nabla) u \big),
\end{equation}
where ${\cal P}$ stands for a pseudo-differential operator (see, for example, Kato 1984; Lemarie-Rieusset 2002; Cannone 2003; Germain 2006). 
For every $q$ in $1<q<\infty$, ${\cal P}$ is a bounded Helmholtz-Leray orthogonal projection from the space $L^q$ of the velocity vector fields to the subspace of $L^q$ consisting of all solenoidal vector fields. We do not need the explicit form of the operator but we notice some of its important properties:
\begin{equation*}
	{\cal P} (u) = u,\;\;\; {\cal P} (\partial_t u)=\partial_t u,\;\;\; {\cal P} (\nabla p) = 0.
\end{equation*}
Applying the Helmholtz-Leray projection to the Navier-Stokes momentum equation, we arrive at an evolution equation for the velocity in Banach space
\begin{equation*}
	\rd u / \rd t  + \nu {\mathbf A} u = - {\cal P} \big( (u. \nabla)u \big),
\end{equation*}
where ${\mathbf A}{=}{-}{\cal P} \Delta$ is the Stokes operator. In $\real^3$, we have ${\mathbf A}(u){=}{-}{\cal P} \Delta(u) {=} {-} \Delta u$. The differential equation can also be expressed as
\begin{equation} \label{eq:mom-semi-group}
	u(x,t) = \re^{-t {\mathbf A}} u_0(x) - \int_0^t \re^{-(t-\tau) {\mathbf A}} {\cal P}\big((u.\nabla)u \big) \rd \tau.
\end{equation}
In no reference to vorticity, we have assumed that it would be legitimate to transform (\ref{eq:continuity}) and (\ref{eq:ns}) into (\ref{eq:leray-projection}) or (\ref{eq:mom-semi-group}), without justification. Strictly, the driving function $\Pi$ in (\ref{eq:p-poisson}) must be shown to be in $L^1_{loc}(\real^3)$, at least, for the pressure Poisson equation to have an inverse. This bound in turn demands that the Jacobian elements, $\partial_{x_i}u_j$, have to be in $L^2(\real^3)$ or better in view of H\"{o}lder's inequality. The anticipated $L^2$-bound is just the enstrophy and is unequivocal in understanding of the physics involved. In fact, it has never been clear how to circumvent the difficulty due to the lack of a stronger {\itshape a priori} bound on velocity. A particular case of analysis in a periodic domain in conjunction with periodic boundary conditions may be justified (Foias {\itshape et al} 2001). In many analyses, this delicate issue has simply been brushed aside; it has taken for granted that the expression (\ref{eq:mom-semi-group}) is an equivalent form of the Navier-Stokes system beyond the classical time $T_L$ without {\itshape a fortiori} consideration. Would a flow remain regular beyond $T_L$? It is precisely in this sense that a complete knowledge of the driving function is crucial. On the ground of rigorousness, we may appreciate the reason why Leray's formulation of weak solutions (Leray 1934$b$) has a classic status as far as irregular motions are concerned. The related issues and implications of writing the Navier-Stokes dynamics as the Banach-spaced evolution equation has been discussed at length by Cannone (2003).

In present analysis, we are mainly interested in establishing bounds $\|u(\cdot,t)\|_{\infty}$ and $\|\nabla u(\cdot,t)\|_{\infty}$.  The limitation of the semi-groups approach is apparent. The well-known Gagliardo-Nirenberg-Sobolev inequality takes the form
\begin{equation*}
	\big\|u(x)\big\|_q\:{\leq}\: C \:\big\|\nabla u(x)\big\|^{\lambda}_s \:\big\| u(x)\big\|^{1-\lambda}_r, 
\end{equation*}
where $1 {\leq} s,r {\leq} \infty$, and $1/q{=}\lambda(1/s{-}1/3){+}(1{-}\lambda)/r$. We would like to stress the fact that this is an elliptic bound. Hence it {\itshape does not} apply to interpolation for any time-wise bounds, such as $\|u(\cdot,t)\|_{\infty}$.

The pressure is governed by (\ref{eq:p-poisson}). It is only the particular solution that is relevant in the dynamic equation. Specifically, the pressure gradient is given by
\begin{equation} \label{eq:nabla-p}
	\nabla p (x) = \rho \: {\cal M}* \Uppi (x).
\end{equation}
In view of Young's inequality for convolution, we deduce that
\begin{equation} \label{eq:nabla-p-lp-bound}
	\big\|\nabla p (x) \big \|_{L^q(\real^3)} \leq C \Big( \; \big\| \omega_0(x)  \big\|^2_{2q} + \big\| \omega_0(x)  \big\|^2_{4q/(q+2)} \Big) \leq C(\|\omega_0\|_Y),
\end{equation}
where	$3/2 < q < \infty$. In particular, we establish that
\begin{equation} \label{eq:nabla-p-linf-bound}
	\big\|\nabla p (x) \big \|_{L^{\infty}(\real^3)} \leq C \Big( \; \big\| \nabla u (x)  \big\|^2_{\infty} + \big\| \nabla u (x)  \big\|^2_4 \:\Big)\leq C(\|\omega_0\|_Y).
\end{equation}
Recall the fact that the actual pressure, 
\begin{equation*}
p(x)=-\rho \int {\cal N}(x{-}y) \:\Uppi(y) \rd y + \const,
\end{equation*}
appears implicitly in the equations of motion. As it is specifically gauged relative to a finite constant ($p_0(t)$ which cannot be determined by the dynamic equations), we shall not write out any explicit pressure bounds.

As a consequence of our analyses conducted so far, the driving term in the momentum (\ref{eq:ns}), $(u.\nabla)u + \nabla p/\rho$,
is bounded in space in its $L^{q}$ norm $(q{>}3/2)$, and hence it is integrable in time. Thus it is justified to convert the equation into a non-linear integro-differential equation for $u$:
\begin{equation} \label{eq:mom-ie}
\begin{split}
	u(x,t) & = \int {\mathbf Z}(x{-}y,t) u_0(y) \rd y \\
	\quad & {-} \int_0^t \! \int {\mathbf Z}(x{-}y,t{-}\tau) \Big( (u.\nabla) u  {+} \nabla p/\rho \Big)  (y,\tau) \rd y \rd \tau = {\mathscr F}(x,t) {-} {\mathscr G}(x,t).
	\end{split}
\end{equation}
The matrix $\mathbf Z$ is diagonal with elements $Z$. Integral equation (\ref{eq:mom-ie}) and Poisson's equation (\ref{eq:p-poisson}) form a system defining distributional solutions for ($u,p$).

Now the principal interest lies in the fact that the integral equation contains a scheme which, in conjunction with bounds on vorticity, enables us to derive certain time-wise bounds for the velocity. By virtue of H\"{o}lder's inequality, it is straightforward to verify that, for $t>0$,
\begin{equation} \label{eq:mom-ie-u}
	\big\|{\mathscr F}(\cdot,t) \big\|_q \:{\leq}\:C \: (\nu t)^{-3/2\:(1/s-1/q)} \: \big\| u_0 \big\|_s,
\end{equation}
and
\begin{equation} \label{eq:mom-ie-du}
	\big\|\nabla {\mathscr F}(\cdot,t)\big\|_q \:{\leq}\:C \: (\nu t)^{-3/2\:(1/s-1/q)-1/2} \: \big\| u_0 \big\|_s.
\end{equation}
In these formulas, $C{=}C(q,s)$, $1{\leq}s{\leq}q{\leq}{\infty}$ except the case $s{=}1,q{=}\infty$.

By applying Minkowski's inequality, we establish 
\begin{equation} \label{eq:mom-ie-p}
	\big\|{\mathscr G}(\cdot,t)\big\|_q \:{\leq}\:C \int_0^t (\nu t{-} \nu \tau)^{-3/2\:(1/s-1/q)} \: \Big( \big\| u \nabla u \big \|_s {+} \big\| \nabla p \big\|_s \Big) (\cdot,\tau) \rd \tau,
\end{equation}
and
\begin{equation} \label{eq:mom-ie-dp}
	\big\|\nabla {\mathscr G}(\cdot,t)\big\|_q \:{\leq}\:C \int_0^t  (\nu t{-} \nu \tau)^{-3/2\:(1/s-1/q)-1/2} \: \Big( \big\| u \nabla u \big \|_s {+} \big\| \nabla p \big\|_s \Big) (\cdot,\tau) \rd \tau.
\end{equation}
From the estimates (\ref{eq:mom-ie-u}) and (\ref{eq:mom-ie-p}), we get 
\begin{equation*} 
\begin{split}
	\|u(\cdot,t)\|_{\infty}  & \leq  C_1 (\nu t)^{-1/4} \|u_0\|_6 + C_2 \int_0^t (\nu t {-} \nu \tau)^{-3/4} (\|u\|_6 \: \|\nabla u\|_3) (\cdot,\tau) \rd \tau  \\
	\quad & {\hspace{3cm}}+ C_3 \int_0^t (\nu t {-} \nu \tau)^{-3/4}\|\nabla p \|_2 (\cdot,\tau) \rd \tau.
	\end{split}
\end{equation*}
By virtue of (\ref{eq:nabla-p-lp-bound}), (\ref{eq:cz}) and (\ref{eq:sobolev-vel}), or (\ref{eq:u-space-bound}) and (\ref{eq:du-space-lq-bound}), we obtain a time-wise bound for the velocity,
\begin{equation} \label{eq:u-time-bound-y}
	\big\|u(\cdot,t)\big\|_{\infty}  \leq C \: (\nu t)^{-1/4},\;\;\;t>0,
\end{equation}
where $C{=}C( T, \|\omega_0\|_Y)$. 
In view of the estimates (\ref{eq:mom-ie-du}) and (\ref{eq:mom-ie-dp}), we have
\begin{equation*} 
\begin{split}
	\|\nabla u(\cdot,t)\|_{\infty}  & \leq  C_1 (\nu t)^{-3/4} \|u_0\|_6 + C_2 \int_0^t (\nu t {-} \nu \tau)^{-3/4} (\|u\|_{12} \: \|\nabla u\|_{12}) (\cdot,\tau) \rd \tau  \\
	\quad & {\hspace{3cm}}+ C_3 \int_0^t (\nu t {-} \nu \tau)^{-3/4}\|\nabla p \|_6 (\cdot,\tau) \rd \tau.
	\end{split}
\end{equation*}
Hence the velocity gradient is essentially bounded in time:
\begin{equation} \label{eq:du-time-bound-y}
	\big\|\nabla u(\cdot,t)\big\|_{\infty}  \leq C \: (\nu t)^{-3/4},\;\;\;t>0,
\end{equation}
where $C{=}C( T, \|\omega_0\|_Y)$. Estimates (\ref{eq:u-time-bound-y}) and (\ref{eq:du-time-bound-y}) substantiate the bounds in (\ref{eq:u-space-bound}) and (\ref{eq:du-space-lq-bound}). As anticipated, these estimates depend on initial data $\|\omega_0\|_Y$ in a non-linear fashion.
\section{Adjoint equations of motion}
The Navier-Stokes momentum equation can be viewed as a quasi-linear partial differential equation if we treat the gradient in the inertia term $(u.\nabla)u$ as an unknown. Equally, the mathematical structure of vorticity equation (\ref{eq:vorticity}) allows an interpretation that the solenoidal velocity and its gradient can be regarded as coefficient functions.  In this respect, the vorticity equation is nothing more than a linear partial differential equation. By analogy, the non-linear velocity term, $(u.\nabla)u \equiv (f(\omega).\nabla)u$ in (\ref{eq:ns}), can also be treated as a coefficient function of vorticity in view of the Biot-Savart law. In fact, equation (\ref{eq:ns-mod}) explicitly contains this idea in a different form. For the present purposes, it is useful to derive corresponding adjoints for the momentum and vorticity equations. These adjoints are not necessarily linked to each other. For the theory of adjoints for differential equations and their applications, consult Lanczos (1961), Courant \& Hilbert (1966) and Sobolev (1964).
\subsection{Adjoint vorticity equation}
To simplify our discussion, let us rewrite vorticity equation (\ref{eq:vorticity}) as
\begin{equation}
	L {\omega} = A_0 {\omega} + A_1 \frac{\partial{{\omega}}}{\partial{x_1}} + A_2 \frac{\partial{{\omega}}}{\partial{x_2}} + A_3 \frac{\partial{{\omega}}}{\partial{x_3}} - B {\omega} = 0.
\end{equation}
The coefficients matrices, $A_i,\: i{=}1,2,3$, are diagonal. Each $A_i$ contains identical elements $u_i$. The matrix $A_0$ is also diagonal and its elements are $\partial_t{-}\nu \Delta$. The matrix $B$ is the velocity Jacobian
\begin{equation*}
	B = \left( \begin{array}{ccc}
{\partial u_1}/{\partial x_1} & {\partial u_1}/{\partial x_2} & {\partial u_1}/{\partial x_3}    \\
 & & \\
{\partial u_2}/{\partial x_1} & {\partial u_2}/{\partial x_2} & {\partial u_2}/{\partial x_3}    \\
 & & \\
{\partial u_3}/{\partial x_1} & {\partial u_3}/{\partial x_2} & {\partial u_3}/{\partial x_3} 
\end{array} 
\right).
\end{equation*}
As stated in the derivation of the bound (\ref{eq:vort-infty-bound}), we have established {\itshape a priori} bound $\omega {\in} C_B^k,k{\geq}2$. Hence it is justified to derive adjoint vorticity in space $C^2$. Both the velocity and the vorticity decay at infinity ($|x|\rightarrow \infty$) in view of the {\itshape a priori} bounds of the preceding section. In general, an adjoint vorticity is calculated from the bilinear Lagrange-Green identity:
\begin{equation*}
	\int_0^t \! \int \Big( {{\omega}^{\dagger}}' L {\omega} - {\omega}' L^{\dagger} {{\omega}}^{\dagger} \Big) \rd x \rd t =0.
\end{equation*}
By direct computation, we obtain
\begin{equation*}
	L^{\dagger} {\omega}^{\dagger} = A^{\dagger}_0 {\omega}^{\dagger} - A'_1 \frac{\partial{{\omega}^{\dagger}}}{\partial{x_1}} - A'_2 \frac{\partial{{\omega}^{\dagger}}}{\partial{x_2}} - A'_3 \frac{\partial{{\omega}^{\dagger}}}{\partial{x_3}} - B' {\omega}^{\dagger} =0,
\end{equation*}
where the matrix $A_0^{\dagger}$ is diagonal, and all its elements have identical entry $\partial_t{+}\nu \Delta$.
Hence the adjoint vorticity in its component-wise form is found to be
\begin{equation} \label{eq:vort-adj}
	\frac {\partial \omega^{\dagger}_i} {\partial t} + \nu \Delta \omega^{\dagger}_i = u_j \frac{\partial \omega^{\dagger}_i}{\partial x_j} + \frac{\partial u_j }{\partial x_i} \omega^{\dagger}_j.
\end{equation}
The adjoint starting condition is specified as
\begin{equation*}
	\omega^{\dagger}(x,T)=\omega_0(x).
\end{equation*}
The solenoidal symmetry in $u$ has not been preserved in the adjoint Jacobian as shown in the last term. 

Equations (\ref{eq:vort-adj}) are equivalent to a system of linear integral equations as the adjoint vorticity can be expressed as
\begin{equation*} 
	\omega^{\dagger}_i(x,t) = \int_T^t Z^{\dagger}(x{-}y,t{-}T) (\omega_0)_i(y) \rd y + \int_T^t \! \int \sum_{j=1}^3 Q_{ij}(x,t,y,s) \omega^{\dagger}_j(y,s) \rd y \rd s. 
\end{equation*}
The kernel $Z^{\dagger}$ is the adjoint of the heat kernel $Z$. The elements of the kernel $Q_{ij}$ can all be expressed in terms of $u$, $\nabla u$ and $Z^{\dagger}$. Thus the existence and regularity of the vorticity $\omega^{\dagger}(x,t)$ can be established for $x \in \real^3,\; 0\leq t \leq T$ according to the well-known theories of linear integral equations. Moreover, $\omega^{\dagger} \rightarrow \omega_0$ as $t \rightarrow T$.  
\subsection{Adjoint momentum equation}
It is not so obvious how to derive adjoint momentum because of the incompressibility hypothesis giving rise to a Poisson-type equation (cf. (\ref{eq:p-poisson})). Nevertheless, we have to deal with a mixed problem of parabolic-elliptic type. Let the solution column vector be
\begin{equation*}
	\phi = (u_1 \;\;\; u_2 \;\;\; u_3 \;\;\; \chi )',
\end{equation*}
where the prime denotes the matrix transpose.
We express the Navier-Stokes equation (\ref{eq:ns-mod}) and the pressure relation (\ref{eq:chi-poisson}) in matrix form 
\begin{equation}
	M \phi = D_0 \phi + D \phi + D_1 \frac{\partial{\phi}}{\partial{x_1}} + D_2 \frac{\partial{\phi}}{\partial{x_2}} + D_3 \frac{\partial{\phi}}{\partial{x_3}} =0.
\end{equation}
The leading matrix $D_0$ is given by
\begin{equation*}
	D_0 = \left( \begin{array}{cccc}
\partial_t {-} \nu \Delta & 0 & 0 & 0 \\
0 & \partial_t {-} \nu \Delta & 0  & 0 \\
0 & 0 & \partial_t {-} \nu \Delta  & 0 \\
0 & 0 & 0 & \Delta \\ 
\end{array} 
\right).
\end{equation*}
The vorticity matrix $D$ is
\begin{equation*}
	D = \left( \begin{array}{cccc}
0 & -\omega_3 & \omega_2 & 0 \\
\omega_3 & 0 & -\omega_1 & 0 \\
-\omega_2 & \omega_1 & 0 & 0 \\
(\nabla{\times} \omega)_1 & (\nabla{\times} \omega)_2 & (\nabla{\times} \omega)_3 & 0 \\ 
\end{array} 
\right).
\end{equation*}
The other coefficient matrices are given by
\begin{equation*} 
	D_1 {=} \left( \begin{array}{cccc}
0 & 0 & 0 & 0 \\
0 & 0 & 0 & 0 \\
0 & 0 & 0 & 0 \\
0 & -\omega_3 & \omega_2 & 0 \\ 
\end{array} 
\right),
\end{equation*}
\begin{equation*}
	D_2 {=} \left( \begin{array}{cccc}
0 & 0 & 0 & 0 \\
0 & 0 & 0 & 0 \\
0 & 0 & 0 & 0 \\
\omega_3 & 0 & -\omega_1 & 0  \\ 
\end{array} 
\right),
\;\;\;\;\;\;
	D_3 {=} \left( \begin{array}{cccc}
0 & 0 & 0 & 0 \\
0 & 0 & 0 & 0 \\
0 & 0 & 0 & 0 \\
-\omega_2 & \omega_1 & 0 & 0 \\ 
\end{array} 
\right).
\end{equation*}
The corresponding bilinear identity yields the expression for an adjoint,
\begin{equation*}
	M^{\dagger} {\phi}^{\dagger} = D_0^{\dagger} {\phi}^{\dagger} + D' {\phi}^{\dagger} - \Big( D'_1 {\phi}^{\dagger}\Big)_{x_1} - \Big( D'_2 {\phi}^{\dagger}\Big)_{x_2} - \Big( D'_3 {\phi}^{\dagger}\Big)_{x_3} =0,
\end{equation*}
where the matrix subscripts denote differentiation.
The presence of the elliptic part due to the pressure introduces no technical difficulties as Poisson's equation holds at every moment of time.  The matrix $D_0^{\dagger}$ is diagonal with identical element $\partial_t{+}\nu \Delta$. It follows that an adjoint system for the equations of motion can be derived:
\begin{equation} \label{eq:adj-ns}
	\begin{split}
	\partial_t u^{\dagger} + \nu \Delta u^{\dagger} + u^{\dagger} {\times}\omega + \omega {\times}(\nabla \chi^{\dagger}) &= 0, \\
	\Delta \chi^{\dagger} & =0, \\
	u^{\dagger}(x,T) &= u_0(x).
	\end{split}
\end{equation}
The key feature here is that the adjoint pressure is harmonic.
As expected, the velocity is no longer solenoidal. Bound (\ref{eq:vort-lp-bound}) implies that a solution of (\ref{eq:adj-ns}) exists. In particular, the adjoint pressure $\chi^{\dagger}$ must be bounded and hence it is a constant according to Liouville's theorem. In other words, the adjoint reference pressure $p_0^{\dagger}$ must be constant. Consequently, system (\ref{eq:adj-ns}) reduces to
\begin{equation*} 
	\partial_t u^{\dagger} + \nu \Delta u^{\dagger} + u^{\dagger} {\times}\omega = 0,\;\;\;
	u^{\dagger}(x,T) = u_0(x).
\end{equation*}
This is a classical linear diffusion equation and its solution is well-known. 
\section{Global regularity of smooth solution}
By virtue of {\itshape a priori} bounds (\ref{eq:u-time-bound-y}) and (\ref{eq:du-time-bound-y}), we may take  
\begin{equation} \label{eq:vel-vort-pde}
	\nabla u \in L^{\infty}((0,T],L^3(\real^3)) \;\;\; \mbox{and} \;\;\ u \in  L^{\infty}((0,T],L^4(\real^3)).
\end{equation}
Thus we are considering vorticity equation (\ref{eq:vorticity}) as a system of second order {\itshape linear} parabolic equations with measurable coefficients as we interpret $u \equiv u(x,t)$ and $\nabla u \equiv \nabla u(x,t)$. For parabolic systems with H\"{o}lder continuous coefficients, regularity theories have been well established (see, for example, Nash 1958; Aronson 1968; Ladyzhenskaya {\itshape et al} 1968; Friedman 1964; Edel'man 1969).  
The theory of parabolic equations with coefficients in general Lebesgue class has been developed by Aronson (1968) based on the work of Aronson \& Serrin (1967). The criteria for regular solutions are that the coefficient functions belong to
\begin{equation} \label{eq:pde-coeff-criteria}
	\begin{split}
	u \in L^r((0,T],L^q(\real^3)),\;\;\; & 3/q+2/r \; {<} \; 1,\;\;\; 2 < q,r \leq \infty,\\
	\nabla u \in L^{r'}((0,T],L^{q'}(\real^3)),\;\;\;& 3/q'+2/r' \;{<} \; 2,\;1 < q',r' \leq \infty,
	\end{split}
\end{equation}
where the functions $u$ and $\nabla u$ are supposed to be compactly supported. Similar theories have been developed for the coefficient functions which are not required to be compact supported (see, for example, Zhang 1995; Liskevich \&  Semenov 2000). 
\subsection*{Vorticity by fundamental solution}
Since the coefficients of the vorticity equation belong to the admissible class (\ref{eq:pde-coeff-criteria}), we have to deal with the fundamental solution in generalized sense. It turns out, by mollifying the coefficients, the generalized fundamental solution has been shown to exist. More importantly, it inherits the lower and upper bounds of its classical counterpart.
In particular, the generalized solution of the Cauchy problem for (\ref{eq:vorticity}) has an integral representation,
\begin{equation} \label{eq:parabolic-pde-vort-sol}
	\omega(x,t) = \int {\mathbf \Gamma}(x,t,y,s) \; \omega_0(y) \rd y ,
\end{equation}
where ${\mathbf \Gamma} (x,t,y,s)$ is the weak fundamental solution. It is a $3 {\times} 3$ matrix of functions defined for $(x,t), (y,s) \in {\real^3{\times}[0,T]}$ with $t{>}s$. For fixed $(y,s)$, ${\mathbf \Gamma}$ satisfies the vorticity equation. If the initial data are continuous, then
\begin{equation*}
	\lim_{t \rightarrow s} \int {\mathbf \Gamma}(x,t,y,s) \; \omega_0(y) \rd y = \omega_0(x).
\end{equation*}
The fundamental solution can be found by the parametrix method which consists in a scheme of approximations (Courant \& Hilbert 1966; Friedman 1964). The successive approximations are then obtained by solving a system of integral equations of Volterra-Fredholm type. The most important property of the fundamental solution relevant to our analysis is that every element of ${\mathbf \Gamma}$ possesses a Gaussian upper bound as well as a lower bound:
\begin{equation} \label{eq:gaussian-bounds}
	\frac{C_1} {(\nu(t{-}s))^{3/2}} \exp \Big( {-} C_2 \frac { |x{-}y|^2}{ \nu (t{-}s)} \Big) \;{\leq}\; \Gamma_{ij} \;{\leq} \;\frac{C_3} {(\nu(t{-}s))^{3/2}} \exp \Big( {-} C_4 \frac { |x{-}y|^2}{ \nu (t{-}s)} \Big),
\end{equation}
where $\Gamma_{ij}{=}\Gamma_{ij}(x,t,y,s)$ for every $i,j{=}1,2,3$. The constants, $C_k{=}C_k(T)$, are all positive. To ascertain the uniqueness of the fundamental solution, we make use of the fact that the coefficients in vorticity adjoint (\ref{eq:vort-adj}) have the same properties as those given in (\ref{eq:vel-vort-pde}). The existence of the adjoint $\omega^{\dagger}$ follows. Hence ${\mathbf \Gamma}^{\dagger}(y,s,x,t)$ exists and ${\mathbf \Gamma}^{\dagger}(y,s,x,t)=( {\mathbf \Gamma} (x,t,y,s))'$, where the prime denotes the matrix transpose.
\subsection{Existence and uniqueness of smooth solution}
We intend to establish a sequence of approximations to solutions of the vorticity equation. The approximations are denoted by $\omega^{(k)}{=}\omega^{(k)}(x,t)$ and they are defined in the following iterative relations. The associated velocity is denoted by $u^{(k)}{=}u^{(k)}(x,t)$. Let $u^{(0)}{\equiv}0$ and $\nabla u^{(0)}{\equiv}0$. For $k=1,2,3,\cdots$, the approximations are found by solving
\begin{equation} \label{eq:iterative-vorticity}
 \begin{split}
	{\partial \omega^{(k)}}/{\partial t} - \nu \Delta \omega^{(k)} & = (\omega^{(k)}.\nabla) u^{(k-1)} - (u^{(k-1)} . \nabla ) \omega^{(k)}, \\
	u^{(k)} (x;t)& = {\cal K} {*} \omega^{(k)}(x;t), \\ 
	\omega^{(k)}(x,0)& =\nabla{\times}u_0(x). 
 \end{split}
\end{equation}
First we notice that $\nabla.u^{(k)}{=}0$ and $\nabla.\omega^{(k)}{=}0$. 
Following the analysis established in preceding sections, the invariance properties of vorticity suggest that {\itshape a priori} bounds,  
\begin{equation} \label{eq:vel-k-apriori-bounds}
	u^{(k)} \in L^{\infty}((0,T],L^4(\real^3)), \;\;\; \nabla u^{(k)} \in L^{\infty}((0,T],L^3(\real^3)), 
\end{equation}
hold for $k=1,2,3,\dots$. For $k{=}1$, $\omega^{(1)}$ is just the solution of the pure initial value problem for heat equation $\partial \omega^{(1)} / {\partial t} {-} \nu \Delta \omega^{(1)}{=}0$. It can be readily solved by the method of Fourier transform (John 1982; Evans 2008). Explicitly, 
\begin{equation*} 
	\omega^{(1)}(x,t) = \int   \mathbf\Gamma^{(0)} (x,t,y) \; \omega_0(y) \rd y,
\end{equation*}
where the kernel matrix $\mathbf\Gamma^{(0)} \equiv \mathbf Z $. 
In view of the bounds (\ref{eq:vel-k-apriori-bounds}), the vorticity 
for $k>1$ is given by  
\begin{equation} \label{eq:vort-k-lp-sol}
	\omega^{(k)}(x,t) = \int \mathbf\Gamma^{(k-1)} (x,t,y) \; \omega_0(y) \rd y,
\end{equation}
where $\mathbf\Gamma^{(k-1)}$ denotes the fundamental solution whose elements have bounds satisfying (\ref{eq:gaussian-bounds}), and they are functions of the iterates $u^{(k-1)}, \nabla u^{(k-1)}$. More specifically, every component of the vorticity is bounded. Hence we find
\begin{equation} \label{eq:vort-k-lq-lp-bound}
 \begin{split}
	\big\| \omega^{(k)}(\cdot,t)\big\|_{q} & \: {\leq} \: C \: (\nu t)^{-3/2\:(1/r-1/q)} \: \| \omega_0 \|_{r}, \\
		\big\| \nabla \omega^{(k)}(\cdot,t)\big\|_{q} & \: {\leq} \: C \: (\nu t)^{-3/2\:(1/r-1/q)-1/2} \: \| \omega_0 \|_{r},
 \end{split}
\end{equation}
where $1{\leq}r{\leq}q{\leq}{\infty}$ except the case $r{=}1, q{=}\infty$, and $C{=}C(r,q,T,\omega_0)$. 
For $1{\leq}s{\leq}\infty$, any multi-index $\alpha>0$ and $t>0$, we have 
\begin{equation} \label{eq:vel-vort-k-lp}
	\big\| \partial_x^{\alpha}  \omega^{(k)} (x)\big\|_{L^s(\real^3)} \: {\leq} \: C \: \big\| \partial_x^{\alpha} \omega_0 (x)\big\|_{L^s(\real^3)}.	
\end{equation}
This relation is a direct consequence of H\"{o}lder's inequality.

To see that these inequalities imply regularity, we first infer that $|\partial_t \omega^{(k)}|$ must be bounded in view of the first equation of (\ref{eq:iterative-vorticity}). Evidently, the velocity functions, $u^{(k)}$ and $\nabla u^{(k)}$, are bounded. A solution for Poisson's equation (\ref{eq:p-poisson}) can be expressed as (cf. \ref{eq:nabla-p}))
\begin{equation} \label{eq:dp-bs}
	\nabla p^{(k)} (x;t) = \rho \: {\cal M}* \Uppi^{(k)}(x;t).
\end{equation}
By virtue of formula (\ref{eq:nabla-p-lp-bound}), we deduce that
\begin{equation*}
	\big\| \nabla p^{(k)} (x) \big\|_2 < \infty,
\;\;\;
	\big\| \nabla p^{(k)} (x) \big\|_6 < \infty.
\end{equation*}
In parallel to estimates (\ref{eq:mom-ie-u}) and (\ref{eq:mom-ie-p}), we obtain a time-wise bound for the velocity iterates,
\begin{equation} \label{eq:u-k-time-bound}
	\big\|u^{(k)}(\cdot,t)\big\|_{\infty}  \leq C \: (\nu t)^{-1/4},\;\;\;t>0,
\end{equation}
where $C{=}C( T, \|\omega_0\|_Y)$. 
Similarly, in view of the estimates (\ref{eq:mom-ie-du}) and (\ref{eq:mom-ie-dp}), we find the bound for the velocity gradient,
\begin{equation} \label{eq:du-k-time-bound}
	\big\|\nabla u^{(k)}(\cdot,t)\big\|_{\infty}  \leq C \: (\nu t)^{-3/4},\;\;\;t>0,
\end{equation}
where $C{=}C( T, \|\omega_0\|_Y)$. To show that $|\nabla p^{(k)}|$ is bounded, we make use of the procedures leading to (\ref{eq:domain-partition}) in conjunction with interpolation theory. Thus it is straightforward to estimate
\begin{equation} \label{eq:p-k-bound}
	\big\|\nabla p^{(k)}(x)\big\|_{\infty}\:{\leq}\: C \big( \|\omega_0(x)\|_Y \big).
\end{equation}

Since it is illegitimate to take time-differentiation on the pressure gradient in (\ref{eq:dp-bs}) or on the Biot-Savart relation in (\ref{eq:iterative-vorticity}), we must restore to the governing dynamics equation to get $\partial_t u^{(k)}$.
Corresponding to the vorticity iterates, the momentum equation for $u^{(k)}$ reads
\begin{equation} \label{eq:iterative-momentum}
	\partial_t u^{(k)} = \nu \Delta u^{(k)} - (u^{(k)} . \nabla) u^{(k)}  - {\rho}^{-1} \nabla p^{(k)}. 
\end{equation}
As all the terms on the right-hand side are bounded in view of (\ref{eq:vort-k-lq-lp-bound}), (\ref{eq:u-k-time-bound}) and (\ref{eq:p-k-bound}), the boundedness of $| \partial_t u^{(k)} |$ is established. From the vorticity equation, both $| \partial_t \partial^{\alpha}_x \omega^{(k)}|$ and $\| \partial_t \partial^{\alpha}_x \omega^{(k)}\|_{L^2}$ are bounded. The solution of Poisson's equation shows that $|\partial_x^{\alpha} (\nabla p^{(k)})|$ is finite at every instant of time. It follows that, from the momentum equation, $| \partial_t \partial^{\alpha}_x u^{(k)}|$ is bounded. It is clear that we can continue this process as many times as we wish. The uniform boundedness of $\| \partial_t \partial^{\alpha}_x u^{(k)} \|_{L^{\infty}(\real^3)}$ for any values of $\alpha\:(>0)$ follows. 
Up to now, we are able to establish that the triplet $(u^{(k)},p^{(k)},\omega^{(k)})$ satisfy the iterations (\ref{eq:iterative-vorticity}) and (\ref{eq:iterative-momentum}) {\itshape classically}.

To assert that the solutions are smooth, i.e., $u, p \in C^{\infty}$, we must take {\itshape a priori} bound (\ref{eq:vort-space-time-bound}) into account. In particular, it is sufficient to deduce that the vorticity are essentially bounded in space and in time:
\begin{equation} \label{eq:vort-k-space-time-bound}
	\omega^{(k)} \in L_t^{\infty}L_x^{\infty}.
\end{equation}
Differentiating the first equation in (\ref{eq:iterative-vorticity}) with respect to time, we see that $\partial_t^2 \omega^{(k)}$ is bounded and this is a free gift. Then differentiating once more, we obtain
\begin{equation*}
	 \partial_t^2 \Big( (\omega^{(k)}.\nabla) u^{(k-1)} - (u^{(k-1)} . \nabla ) \omega^{(k)} \Big)= \partial_t^3 \omega^{(k)} - \nu \Delta(\partial_t^2 \omega^{(k)}).
\end{equation*}
The left-hand side can be expanded according to Leibnitz's rule and thus we assert that $\partial_t^2 u^{(k-1)}$ is finite by virtue of the vorticity in (\ref{eq:vort-k-space-time-bound}). By analogy, we can show that $|\partial_t^2 u^{(k)}| {<} \infty$.  From (\ref{eq:iterative-momentum}), we obtain the time variation in the gradient,
\begin{equation*}
	\partial_t (\nabla p^{(k)}/\rho) = \nu \Delta (\partial_t u^{(k)}) - \partial_t^2 u^{(k)} - \partial_t\big( (u^{(k)} . \nabla) u^{(k)} \big).
\end{equation*}
Every term on the right-hand side has been shown to be bounded in space and in time and thus $|\partial_t (\nabla p^{(k)})| {<} \infty$.
Evidently, these procedures can be repeated for any $\beta{>}0$. 
The pressure can be recovered from any one of the momentum components within an arbitrary constant of integration ($p_0(t)$ say). It is obvious that the pressure solution $p^{(k)} \in C^{\infty}$ for $x \in (\real^3)$ and $t \geq 0$. 

{\itshape In summary, given any initial velocity $u_0(x) \in C^{\infty} (\real^3)$ satisfying the localization (\ref{eq:ic-localization}), and let $T > 0$, iterative system (\ref{eq:iterative-vorticity}) gives rise to a sequence of smooth solutions $C^{\infty}(\real^3{\times}[0,T])$ with continuous dependence on the initial data. We denote the solutions by
\begin{equation*} 
\Big\{\omega^{(k)}(x,t),\;u^{(k)}(x,t) \Big\}^{\infty}_{k=0},\;\;\; k=1,2,3,\dots,
\end{equation*}
and $\nabla.u^{(k)}{=}0$, $\nabla.\omega^{(k)}{=}0$. For every multi-index $\alpha$ and every integer $\beta \geq 0$,
\begin{equation} \label{eq:vv-k-st-bound}
	\sup_{(x,t) {\in} \real^3{\times}[0,T]} \big| \partial_t^{\beta} \partial^{\alpha}_x \omega^{(k)}  \big|  \;{<}\; \infty,  \;\;\;
	\sup_{(x,t) {\in} \real^3{\times}[0,T]} \big| \partial_t^{\beta} \partial^{\alpha}_x u^{(k)} \big|  \;{<}\; \infty.
\end{equation}
Let $p^{(k)}(x,t)$ be the pressure associated with $u^{(k)}(x,t)$ and $\omega^{(k)}(x,t)$, then 
\begin{equation} \label{eq:p-k-st-bound}
	\sup_{(x,t) {\in} \real^3{\times}[0,T]} \big|  \partial_t^{\beta} \partial^{\alpha}_x (\nabla p^{(k)}) \big|  \;{<}\; \infty.
\end{equation}
Finally, we have the decays $u^{(k)} \rightarrow 0$,$\; \omega^{(k)} \rightarrow 0$, $\; \nabla p^{(k)} \rightarrow 0$ as $|x| \rightarrow \infty$}.

For fixed $T > 0$, we denote the solution difference of (\ref{eq:iterative-vorticity}), $k{=}0,1,2,3,\dots$, by
\begin{equation}
	R_k(t) = \sup_{0\:{<}\tau{\leq}\:t} \big\| \omega^{(k+1)}(\cdot,\tau) - \omega^{(k)}(\cdot,\tau) \big\|_Y.
\end{equation}
We set $\omega^{(0)}{=}0$. In view of (\ref{eq:u-k-time-bound}), we have, for $k{=}1,2,3,\dots$,
\begin{equation} \label{eq:u-infty-omega-y}
	\big\| u^{(k)}(\cdot,\tau)\big\|_{L^{\infty}(\real^3)} \:{\leq}\: C \: (\nu \tau)^{-1/4},
\end{equation}
where $C{=}C(T, \| \omega_0 \|_Y)$.

By Duhamel's principle and integrating by parts, vorticity (\ref{eq:iterative-vorticity}) reduces to
\begin{equation} \label{eq:omega-ieqn}
	\omega^{(k+1)}(x,t) = \int {\mathbf Z}(x{-}y,t) \omega_0(x) \rd x 
   + \int_0^t \!\! \int  V^{(k)}(x,t,y,\tau) \omega^{(k+1)} (y,\tau) \rd y \rd \tau.
\end{equation}
The coefficient matrix $V^{(k)}$ has the explicit expression
\begin{equation*} 
	V^{(k)}= \left( \begin{array}{ccc}
{-}Z_2 u^{(k)}_2 {-} Z_3 u^{(k)}_3 \!&\! Z_2 u^{(k)}_1  \!&\! Z_3  u^{(k)}_1 \\
 & & \\
Z_1 u^{(k)}_2 \!&\! {-}Z_1 u^{(k)}_1 {-} Z_3 u^{(k)}_3 \!&\! Z_3  u^{(k)}_2 \\
& & \\
Z_1 u^{(k)}_3 \!&\! Z_2  u^{(k)}_3 \!&\! {-}Z_1  u^{(k)}_1 {-} Z_2 u^{(k)}_2 
\end{array} 
\right),
\end{equation*} 
where short-hand notation $Z_i$ stands for $\nabla_{y_i} Z(x,t,y,\tau)$. We then take $L^{2}$ norm of the difference
\begin{equation*} 
	\omega^{(k+1)}(x,t)- \omega^{(k)}(x,t) = \int_0^t \!\! \int \Big( V^{(k)} \omega^{(k+1)} - V^{(k-1)} \omega^{(k)} \Big) (x,t,y,\tau) \rd y \rd \tau.
\end{equation*}
Similarly, we take $L^{\infty}$ norm. The results are used to form $\| \omega^{(k+1)}(\cdot,t) - \omega^{(k)}(\cdot,t) \|_Y$. Applying the bounds
\begin{equation*}
	\big\|\omega^{(k)}\big\|_{L^{2}(\real^3)}\;{\leq}\; C \: \big\| \omega_0 \big\|_Y, \;\;\; \big\|\omega^{(k)}\big\|_{L^{\infty}(\real^3)} \;{\leq}\; C \: \big\| \omega_0 \big\|_Y,
\end{equation*}
we deduce that
\begin{equation*} 
\begin{split}
	\big\|\omega^{(k+1)}(\cdot,t) - \omega^{(k)}(\cdot,t) \big\|_{Y} \:{\leq}\: \int_0^t  (\nu t {-} \nu \tau)^{-1/2} & \Big( C_1 \big\|\omega^{(k+1)}(\cdot,\tau) - \omega^{(k)}(\cdot,\tau) \big\|_{Y}  \\  
	 & + C_2 \big\| u^{(k)}(\cdot,\tau) - u^{(k-1)}(\cdot,\tau) \big\|_{L^{\infty}} \Big) \rd \tau.
\end{split}
\end{equation*}
The second integral term can be bounded by vorticity in virtue of (\ref{eq:u-infty-omega-y}). Thus
\begin{equation} \label{eq:qk-inequality}
	R_k(t) \;{\leq}\; C_1 \int_0^t (\nu t {-} \nu \tau)^{-1/2} R_k(\tau) \rd \tau + C_2 \int_0^t (\nu t {-} \nu \tau)^{-1/2} (\nu \tau)^{-1/4} R_{k-1}(\tau) \rd \tau,
\end{equation}
where $C_1$ and $C_2$ are constant, $C_1{=}C_1(\|\omega_0\|_Y)$, and $C_2{=}C_2(\|\omega_0\|_Y, T)$. Clearly $R_0$ is the solution of heat equation and is bounded by
\begin{equation*}
	R_0(t) = \sup_{0{<}\tau{\leq}t} \big\| \omega^{(1)}(\cdot,\tau) \big\|_Y \: {\leq} \: \: \big\| \omega_0 \big\|_Y.
\end{equation*}
By Gronwall's lemma, integral inequality (\ref{eq:qk-inequality}) can be reduced to
\begin{equation} \label{eq:qk-iteration-inequality}
	R_{k}( t) \: {\leq} \: A \: \int_0^t ( \nu t {-} \nu \tau)^{-1/2} (\nu \tau)^{-1/4} R_{k-1}( \tau) \rd (\nu \tau),
\end{equation}
where constant $A$ is given by $ \nu A {=} C_2 \exp (2 C_1 \sqrt{T/\nu} )$.
The first few values of $R_k$ are found to be
\begin{equation*} 
 \begin{split}
	R_{1}(t) & \: {\leq} \: A R_0 \: B(1/2,3/4) \; (\nu t)^{1/4}, \\ 
	R_{2}(t) & \: {\leq} \: A^2 R_0 \: B(1/2,3/4) \; B(1/2,4/4)  (\nu t)^{1/2}, \\ 
	R_{3}(t) & \: {\leq} \: A^3 R_0 \: B(1/2,3/4) B(1/2,4/4)  B(1/2,5/4) \; (\nu t)^{3/4},\\ 
	R_{4}(t) & \: {\leq} \: A^4 R_0 \: B(1/2,3/4) B(1/2,4/4)  B(1/2,5/4) \; B(1/2,6/4) (\nu t).
 \end{split}
\end{equation*}
Suppose that 
\begin{equation} \label{eq:q-k}
	R_{k}(t) \: {\leq} \: A^{k}  R_0 \: \Big( \prod_{n=1}^{k} B(1/2,(n{+}2)/4) \Big) \: (\nu t)^{{k}/{4}}.  
\end{equation}
Making use of the identity
\begin{equation*}
	\int_0^t (t{-}\tau)^{-1/2}\tau^{(k-1)/4} \rd \tau = B(1/2,(k{+}3)/{4}) \: t^{(k+1)/4},
\end{equation*}
we obtain
\begin{equation*} 
\begin{split}
  R_{k+1}(t)  \: &  {\leq} \: A \: R_{k}(t) \int_0^t ( \nu t {-} \nu \tau)^{-1/2} (\nu \tau)^{-1/4} (\nu \tau)^{k/4}\rd (\nu \tau) \\
  \quad \: &  {\leq} \:  A^{k+1}  R_0 \: \Big( \prod_{n=1}^{k+1} B(1/2,(n{+}3)/4) \Big) \: (\nu t)^{(k+1)/4}.
  	\end{split}
\end{equation*}
By induction, inequality (\ref{eq:q-k}) holds for $k{\geq} 1$.
In view of the property of the Euler Beta functions, $B(p,q){=}\Gamma(p)\Gamma(q)/\Gamma(p{+}q)$, we find that the series $\sum_{k=1}^{\infty}R_k(t)$ is majorized by
\begin{equation} \label{eq:q-series}
	R_0 \: \Gamma(3/4) \: \sum_{k=1}^{\infty} A^{k} \: \big(\nu T \big) ^{k/4} \: \Big(\Gamma\big({1}/{2}\big) \Big)^{k} \: / \: \Gamma \big( k/4{+}1 \big).
\end{equation}
Note that $\Gamma(3/4){\approx}1.225417$ and $\Gamma(1/2){=}\sqrt{\pi}$.
By D'Alembert ratio test, the radius convergence of the series in (\ref{eq:q-series}) is infinity for any fixed $t {\in} (0,T]$ because
\begin{equation*}
	R_{k+1}/{R_k} \:{\sim} \: \Big( C \sqrt[4]{T/\nu} \: \exp \big( 2 C_1 \sqrt{T/\nu} \big) \: \:/ \: k \Big) \: {\rightarrow} \: 0 \;\;\;\mbox{as}\;\;\; k{\rightarrow}\infty.
\end{equation*}

In view of the uniform convergence of series $\sum_{k=1}^{\infty}Q_k(t)$, we establish the existence of vorticity ($0 < t \leq T$):
\begin{equation} \label{eq:vort-time-existence}
 \omega(\cdot,t)=\lim_{k{\rightarrow}\infty} \omega^{(k)}(\cdot,t) \;\;\; \in C^1\big( \:(0,T],L^2(\real^3) \cap C^0(\real^3\:\big).
\end{equation}
By virtue of bounds (\ref{eq:vv-k-st-bound}) and (\ref{eq:p-k-st-bound}), in view of the Arzel{\`{a}}-Ascoli theorem and Cantor's diagonal process over the expanding compact subsets of $\real^3{\times}(0,T]$, we have established the stronger existence claim: The limits,
\begin{equation} \label{eq:vel-vort-spacetime-existence}
 \omega(x,t) = \lim_{k{\rightarrow}\infty} \omega^{(k)}(x,t)\;\;\;\mbox{and}\;\;\; u(x,t) = \lim_{k{\rightarrow}\infty} u^{(k)}(x,t),
\end{equation}
exist. In view of the classical local in-time solutions, we can extend the time interval to $[0,T]$. Specifically, $\omega \in C^{\infty}(\real^3{\times}[0,T])$, evidently $\omega$ depends continuously on the initial data. The uniform convergence of the vorticity and the velocity together with their derivatives of all orders in every compact subset of $\real^3{\times}[0,T]$ is asserted. 
The dynamic evolution of the pressure can be calculated according to
\begin{equation} \label{eq:dt-ddp}
	\partial_t (\nabla p/\rho) = \nu \Delta (\partial_t u) - \partial_t^2 u - \partial_t\big( (u . \nabla) u \big).
\end{equation}
This equation can be generalized to higher derivatives in space and in time.
In addition, all derivatives of $\omega(x,t)$, $u(x,t)$ are essentially bounded and decay at infinity. The equations of continuity, the vorticity, and the momentum are satisfied. 

Let $\tilde{\omega}(x,t)$ be another vorticity satisfying (\ref{eq:vorticity}) with $\tilde{\omega}(x,0){=}\omega_0$. Let the associated solenoidal velocity be $\tilde{u}{=}{\cal K} {*} \tilde{\omega}$, and denote the associated pressure by $\tilde{p}$. Evidently, all the {\itshape a priori} bounds for $u$ and $\omega$ are also valid for $\tilde{u}$ and $\tilde{\omega}$.
Hence it is straightforward to verify that
\begin{equation*}
 \begin{split}
	\partial_t(\omega-\tilde{\omega})- \nu \Delta (\omega - \tilde{\omega}) & 
= ((\omega- \tilde{\omega}). \nabla )u + (\tilde{\omega}.\nabla)(u-\tilde{u}) \\
	\quad & - ((u-\tilde{u}).\nabla)\omega - (\tilde{u}.\nabla)(\omega-\tilde{\omega}). 
 \end{split}
\end{equation*}
As every function on the right-hand side is integrable in space and in time, the vorticity difference is expressed as
\begin{equation*} 
 \begin{split}
	\omega(x,t)-\tilde{\omega}(x,t) & = \int_0^t {\mathbf Z}(x{-}y,t) (\omega_0-\tilde{\omega}_0) \rd y \\
	& + \int_0^t \! \int  \Big( (\nabla {\mathbf Z}.u) \:(\omega-\tilde{\omega})  - 
	(\nabla {\mathbf Z}.(\omega-\tilde{\omega})) \: u \Big) \rd y \rd \tau  \\
	& + \int_0^t \! \int  \Big( (\nabla {\mathbf Z}.(u-\tilde{u})) \:\tilde{\omega}  - 
	(\nabla {\mathbf Z}.\tilde{\omega}) \: (u-\tilde{u}) \Big) \rd y \rd \tau. 
 \end{split}
\end{equation*}
It follows that the difference in space $Y$ can be determined from inequality
\begin{equation*}
	\big\| \omega(\cdot, t)-\tilde{\omega}(\cdot, t) \big\|_Y  \:{\leq}\: \big\| \omega_0 - \tilde{\omega}_0 \big\|_Y + C \int_0^t (\nu t {-} \nu \tau)^{-1/2} \big\| \omega(\cdot, \tau)-\tilde{\omega}(\cdot, \tau) \big\|_Y \rd \tau. 
\end{equation*}
Hence we establish 
\begin{equation} \label{eq:vort-uniqueness}
	\big\| \omega(\cdot, t)-\tilde{\omega}(\cdot, t) \big\|_Y \: {\leq} \:  \big\| \omega_0 - \tilde{\omega}_0 \big\|_Y  \: \exp \Big( C(T) \sqrt{T/\nu} \Big),
\end{equation}
where the continuous dependence of $\| \omega{-}\tilde{\omega}\|_Y$ on the initial conditions is evident. We have seen that the vorticity difference remains zero $\forall t \in [0,T]$ if the initial data of vorticity coincide.

Alternatively, we ascertain the uniqueness from the consideration of energy by virtue of (\ref{eq:u-time-bound-y}) and (\ref{eq:du-time-bound-y}).
Denote $U(x,\tau){=}\tilde{u}(x,\tau){-}u(x,\tau)$ and note that $\nabla.U{=}0$. The momentum equations are combined to give
\begin{equation*}
	\partial_t U - \nu \Delta U = -U.\nabla U - U.\nabla u - u.\nabla U - \nabla(\tilde{p}-p)/\rho.
\end{equation*}
Taking the inner product with $U$, integrating over $\real^3$ and integrating by parts, we find that
\begin{equation*}
	\frac{1}{2}\frac{\rd }{\rd t} \big\| U \big\|_{L^2}^2 + \nu \big\| \nabla U \big\|_{L^2}^2 = -\int  U(U.\nabla )u \:\rd x \:{\leq}\: \sup_{x \in \real^3} \big\|\nabla u \big\|_{L^{\infty}}\: \big\|U \big\|_{L^2}^2. 
\end{equation*}
Because the term $\|\nabla U\|_{L^2}^2$ remains finite and is always non-negative, it is sufficient to write the final result in the form
\begin{equation}
	\big\|\tilde{u}(\cdot,t)-u(\cdot,t) \big\|_{L^2(\real^3)}^2  \:{\leq}\: C \: \big\|{\tilde{u}_0 - u_0 }\big\|_{L^2(\real^3)}^2,
\end{equation}
where constant
\begin{equation*}
	C = C \big(\: T, \big\| \omega_0 \big\|_Y \:\big) < \infty 
\end{equation*}
in view of bound (\ref{eq:du-time-bound-y}).

We conclude:
{\itshape Given $u_0(x) \in C^{\infty}$ having the localization property (\ref{eq:ic-localization}), there exists a unique, bounded solution of vorticity equation (\ref{eq:vorticity}) $\omega \in C^{\infty}(\real^3{\times}[0,T])$, which depends continuously on the initial data. The vorticity $\omega$ attains its initial value $\omega_0(x)=\nabla{\times}u_0(x)$ at $t=0$ and possesses the essential properties 
\begin{equation*} 
 \begin{split}
  \big\| \omega(\cdot,t) \big\|_{L^p(\real^3)} & \; {\leq} \; C(T) \: \big\| \omega_0 \big\|_{L^p(\real^3)}, \\
  \big\| \nabla \omega(\cdot,t) \big\|_{L^p(\real^3)} & \; {\leq} \; C(T) \: (\nu t)^{-1/2} \: \big\| \omega_0 \big\|_{L^p(\real^3)}, \;\;\; \;t>0,\\
 \end{split}
\end{equation*}
$\forall p \in [1,\infty]$. For every multi-index $\alpha$ and for every positive integer $\beta$,
\begin{equation*}
	\sup_{(x,t) {\in} \real^3{\times[0,T]}} \; \big| \partial^{\beta}_t \partial^{\alpha}_x \omega \big| \: {<} \: \infty.
\end{equation*}
The global regularity of the vorticity solution implies $u \in C^{\infty}(\real^3{\times}[0,T])$. The most important properties for the velocity are
\begin{equation*} 
 \begin{split}
	\big\| u(\cdot,t) \big\|_{L^{\infty}(\real^3)} & \:{\leq}\:C\:(\nu t)^{-1/4}, \\
	\big\| \nabla u(\cdot,t) \big\|_{L^{\infty}(\real^3)} & \:{\leq}\:C\:(\nu t)^{-3/4},\;\;\;t > 0,
 \end{split}
\end{equation*}
where the constants depend on $\|\omega_0\|_Y$ non-linearly.
Moreover, $\nabla p \;(\mbox{and}\; p) \in C^{\infty}$.
In addition, the following decay conditions hold at each time $t$:
\begin{equation*} 
	\big| \partial^{\beta}_t \partial^{\alpha}_x \omega(x,t) \big| \: {\rightarrow} \: 0,  \;\;\; \big| \partial^{\beta}_t \partial^{\alpha}_x u(x,t) \big| \: {\rightarrow} \: 0, \;\;\; \big| \partial^{\beta}_t \partial^{\alpha}_x (\nabla p)(x,t) \big| \: {\rightarrow} \: 0 \;\; \mbox{as} \;\; |x| \rightarrow \infty.
\end{equation*}
Furthermore, the law of energy conservation holds at all time.}
\subsection{Quantities of special interest in flow evolution}
First we show how helicity density $h(x,t)=\omega.u(x,t)$ evolves in viscous, incompressible flows. Taking dot product on (\ref{eq:ns-mod}) by $\omega$ and on (\ref{eq:vorticity}) by $u$, adding the products and simplifying, we obtain a dynamic equation governing the density,
\begin{equation} \label{eq:dyn-helicity}
	\partial_t h - \nu \Delta h = -2 \nu \nabla \omega. \nabla u + \nabla.\big( \omega (u^2 -\chi) - u h \big) = D.
\end{equation}
By Duhamel's principle, the density satisfies the scalar integral equation
\begin{equation*}
	h(x,t) = \int Z(x{-}y,t) h(x,0) \rd x + \int_0^t \!\int Z(x{-}y,t{-}s) D(y,s) \rd y \rd s,\;\;\;t>s.
\end{equation*}
Given the {\itshape a priori} bounds derived and the decay properties, the space integral over $\real^3$ renders the divergence term to zero. Thus helicity's evolution satisfies
\begin{equation} \label{eq:helicity-evo}
	\frac{1}{2} \frac{\rd }{\rd t} \int (\omega.u)(x,t) \rd x = - \nu \int (\nabla\omega . \nabla u)(x,t)\rd x.
\end{equation}
The density function does not appear to play a direct role in the equations of motion. For ideal fluids, the helicity is a measure of the degree and the strength of tangled vorticity filaments.

The function, $M{=}\omega{\times}u$, is known as the Lamb vector (see \S VII of Lamb 1975) which does not define a solenoidal field but is closely connected with the law of energy conservation. Taking the divergence of (\ref{eq:ns-mod}), we obtain
\begin{equation*}
	\Delta \chi(x;t) = - \nabla.(\omega{\times}u)(x;t).
\end{equation*}
The Poisson equation exemplifies the vector as the ``source'' of the Bernoulli-Euler pressure function, which, in particular, defines the stream-surfaces of constant energy in steady inviscid flows. In view of the identity,
$\Delta(\omega{\times}u) = \Delta \omega  {\times} u + \omega{\times}\Delta u + 2\nabla\omega {\times}\nabla u$,
and the momentum equation (\ref{eq:ns-mod}), we find that the Lamb vector satisfies the following dynamic equation:
\begin{equation} \label{eq:dyn-lamb}
	\partial_t M - \nu \Delta M = -2 \nu \nabla\omega{\times}\nabla u  -  (\nabla {\times}M){\times}u - (\nabla{\times}u) {\times} ( M + \nabla \chi) = R.
\end{equation}
The sum of the middle two terms involving $M$ is zero owning to the vector identity
\begin{equation} \label{eq:three-vectors}
	A{\times}(B{\times}C)=B(A.C)-C(A.B).
\end{equation}
The solution of the dynamic equation is given by
\begin{equation*}
	M(x,t) = \int {\mathbf Z}(x{-}y,t) M(x,0) \rd x + \int_0^t \!\int {\mathbf Z}(x{-}y,t{-}s) R(y,s) \rd y \rd s,\;\;\;t>s.
\end{equation*}
By the vector identity $\nabla{\times}(\nabla \phi)=0$ and the decays at infinity, the integral over the term containing the pressure gradient reduces to zero. Thus we establish that the integral of the Lamb vector evolves according to 
\begin{equation} \label{eq:lamb-evo}
	\frac{1}{2}\frac{\rd }{\rd t} \int (\omega{\times}u)(x,t) \rd x = - \nu \int (\nabla\omega {\times} \nabla u)(x,t)\rd x.
\end{equation}
In inviscid flows, this evolution relation becomes an invariance of fluid motion which is a special case of Noether's theorem for the principle of energy conservation.

In the dynamic constraints (\ref{eq:helicity-evo}) and (\ref{eq:lamb-evo}), each has a symmetry in the dissipation term which may be viewed as the counterpart of the enstrophy. As expected, all the dissipations are due to processes in three space dimensions. As a well-known fact in vector analysis, the velocity and the vorticity vectors in fluid motions can be orthogonally decomposed into one another in terms of the helicity density and the Lamb vector:
\begin{equation*}
|u|^2\:\omega = u(\omega.u) + u{\times}(\omega{\times}u),\;\;\;	|\omega|^2 \: u = \omega(\omega.u) - \omega{\times}(\omega{\times}u).
\end{equation*}
These relations in turn define a Pythagorean identity: $|u|^2	\:|\omega|^2 = |\omega.u|^2 + |\omega{\times}u|^2$.

The integrals (vector quantities),
\begin{equation*}
	I(t)=\frac{\rho}{2} \int x {\times} \omega \rd x \;\;\; \mbox{and} \;\;\; L(t)=\frac{\rho}{3} \int x {\times} (x {\times} \omega) \rd x,
\end{equation*}
are known as the fluid impulse and the fluid angular impulse respectively. 

In view of the identity $\Delta(x{\times}\omega) = x {\times} \Delta \omega  + 2\nabla{\times}\omega$,
we readily derive the following equation for the density of the impulse, $i=x{\times}\omega$:
\begin{equation} \label{eq:dyn-impulse}
	\partial_t i - \nu \Delta i =   (\nabla {\times}M){\times}x - M  - 2\nu \nabla{\times}\omega = Q_1 + Q_2 + Q_3=Q.
\end{equation}
The solution of this diffusion-type equation can be expressed as
\begin{equation*}
	i(x,t) = \int {\mathbf Z}(x{-}y,t) i(x,0) \rd x + \int_0^t \!\int {\mathbf Z}(x{-}y,t{-}s) Q(y,s) \rd y \rd s,\;\;\;t>s.
\end{equation*}
By virtue of a vector cancellation, the integral of $Q_1$ over $\real^3$ reduces to products of one of the components $\nabla{\times}M$ and a linear length scale at infinity. As the velocity decays sufficiently fast at infinity, the integral vanishes. By the continuity constraint, the space integral of $Q_2$ decays to zero. We assert that the impulse is an invariance in viscous flows:
\begin{equation} \label{eq:impulse-invariance}
	{\rd I(t)}/{\rd t} = 0 ,  \;\;\;t>0,
\end{equation}
provided that the initial fluid impulse, $I(0)$, is a known finite quantity.

Similarly, making use of equation (\ref{eq:dyn-impulse}), a dynamic equation for the impulse integrand $l = x {\times} (x {\times} \omega)$ can be derived:
\begin{equation} \label{eq:dyn-angular-impulse}
	\partial_t l - \nu \Delta l = 2\nu \big( (\nabla{\times}\omega){\times}x- \nabla{\times}(x{\times}\omega) \big) + u{\times}(x{\times}\omega)-(Q_1+Q_2){\times}x.
\end{equation}
Consider the space integral of $Q_1{\times}x$, integration by parts produces a zero boundary term, $| (M{\times}x){\times}x| \rightarrow 0$ as ${|x|\rightarrow \infty}$,
if the necessary decays at infinity are effected. Using the vector identity (\ref{eq:three-vectors}), the resulting two integrals from the integration neutralize the space integrals of the last two functions. Effectively, we are left with
\begin{equation*}
	l(x,t) = \int {\mathbf Z}(x{-}y,t) l(x,0) \rd x + 2 \nu \int_0^t \!\int {\mathbf Z}(x{-}y,t{-}s) \Big((\nabla{\times}\omega){\times}x - \nabla{\times}(x{\times}\omega) \Big) \rd y \rd s.
\end{equation*}
Integrating over $\real^3$, we verify that 
\begin{equation} \label{eq:angular-impulse-invariance}
	{\rd L(t)}/{\rd t} = 0 ,  \;\;\;t>0,
\end{equation}
assuming that the initial angular impulse, $L(0)$, is a given bounded quantity.

In viscous, incompressible fluid motions in $\real^3$, the impulse and the angular impulse coincide with the total momentum and the total angular momentum respectively; the physics of the invariant relations in (\ref{eq:impulse-invariance}) and (\ref{eq:angular-impulse-invariance}) is evident.
\section{Solution of finite initial energy and enstrophy}
If the initial velocity is in $L^2(\real^3)$, it is well-known that Leray-Hopf weak solutions exist for all time $t{>}0$  but the uniqueness is open (Leray 1934$b$; Hopf 1951). However given any smooth initial data, it is known that a unique smooth solution exists up to some time $T_L$, depending on the size of the data (Ladyzhenskaya 1969; Temam 1977; Heywood 1980).
This local regularity property suggests that the diffusive nature of viscosity {\itshape is} strong enough to smooth out any initial vorticity over a small time interval.
Numerous attempts have been made to study the global regularity of the Navier-Stokes equations (see, for example, Oseen 1927; Odqvist 1930; Kiselev \& Ladyzhenskaya 1957; Prodi 1959; Serrin 1962; Scheffer 1976; Caffarelli {\em et al} 1982; Kato 1984; Constantin \& Fefferman 1993; Lin 1998; Escauriaza {\em et al} 2003). 

In what follows, we address the well-posedness of the weak solutions. Our aim is to extend the properties of the smooth solutions to the initial data in $H_0^1(\real^3)$. Once again, we work in terms of the vorticity and we rely on a theory of non-linear integral equation. In two space dimensions, many attempts have been made to establish the well-posedness of the Navier-Stokes equations (see, for example, Leray 1934$a$; Ladyzhenskaya 1959). In the space of continuous functions, McGrath (1968) showed that there exist classical solutions to the vorticity equation in $\real^2$. Specifically, Brezis (1994) and Ben-Artzi (1994) proved the global regularity for vorticity in the scale-invariant space $L^1(\real^2)$, though, in the latter paper, the mathematical arguments leading to the time-wise bounds on velocity are open to debate. Nevertheless, a part of their approach is relevant to our overall strategy for the problem in three space dimensions.
\subsection{Properties of $\omega(x,t) {\in} C^{\infty}$}
In the preceding section, for smooth initial data, we have shown how to solve the vorticity integral equation to obtain the smooth vorticity $\omega {\in} C^{\infty}(\real^3{\times}[0,T])$ by a sequence of approximate solutions. 
Since the solutions of vorticity approximations (\ref{eq:iterative-vorticity}) have been completely determined, we assert that the solution of vorticity equation (\ref{eq:vorticity}) has the form
\begin{equation*}
	\omega(x,t) = \int \mathbf \Gamma (x{-}y,t) \omega_0(y) \rd y,
\end{equation*}
where $\mathbf \Gamma(x{-}y,t)$ is the fundamental solution. By an analogy to the bounds in (\ref{eq:vort-k-lq-lp-bound}), we calculate, for any $t{>}0$, that
\begin{equation} \label{eq:vort-lq-lp-bound}
 \begin{split}
	\big\| \omega(\cdot,t) \big\|_{L^q(\real^3)} & \: {\leq} \: C \: (\nu t)^{-3/2\:(1/p-1/q)} \; \big\| \omega_0 \big\|_{L^p(\real^3)}, \\
	\big\| \nabla \omega(\cdot,t) \big\|_{L^q(\real^3)} & \: {\leq} \: C \: (\nu t)^{-3/2\:(1/p-1/q)-1/2} \; \big\| \omega_0 \big\|_{L^p(\real^3)},
 \end{split}
\end{equation}
where $1{\leq}p{\leq}q{\leq}{\infty}$ except the case $p{=}1, q{=}\infty$, and the constant $C{=}C(p,q,T)$.
In particular, we have a time-wise bound for the vorticity in terms of the initial enstrophy
\begin{equation} \label{eq:vort-linfty-bound}
	\big\| \omega(\cdot,t) \big\|_{L^{\infty}(\real^3)}  \: {\leq} \: C \: (\nu t)^{-3/4} \; \big\| \omega_0 \big\|_{L^{2}(\real^3)}.
\end{equation}
\subsection*{Revised bounds on velocity}
To derive time-wise bounds for the velocity, we must start from momentum equation (\ref{eq:mom-ie}). First of all, we need some knowledge of the pressure.
From the pressure bound in (\ref{eq:nabla-p-lp-bound}), we find
\begin{equation*}
	\big\| \nabla p (x) \big\|_2 \:{\leq}\: C\big(\|\omega_0(x)\|_2\big),
\;\;\; 
	\big\| \nabla p (x)\big\|_6 \:{\leq}\: C\big(\|\omega_0(x)\|_2\big),
\end{equation*}
in view of the Archimedean principle for the real numbers.
Specifically, making use of (\ref{eq:mom-ie}), (\ref{eq:mom-ie-u}) and (\ref{eq:mom-ie-p}), $\|u\|_{\infty}$ can be found from  the integral inequality
\begin{equation} \label{eq:u-time-bound} 
\begin{split}
	\big\|u(\cdot,t) \big\|_{\infty}  \leq \: & C_1 (\nu t)^{-1/4} \big\|u_0 \big\|_6 + C_2 \int_0^t (\nu t {-} \nu s)^{-3/4} \big\|u\big\|_6 \big\| \nabla u\big\|_3 (\cdot,s) \rd s  \\
	\quad & {\hspace{3cm}}+ C_3 \int_0^t (\nu t {-} \nu s)^{-3/4} \big\|\nabla p \big\|_2 (\cdot,s) \rd s \\
	\quad \leq \: & C_1  (\nu t)^{-1/4} \big\|\omega_0 \big\|_2 + C_2 \int_0^t (\nu t {-} \nu s)^{-3/4} (\nu s)^{-1/4} (\cdot,s) \rd s  \\
	\quad \leq \: & \Big( C_1(T) \big\|\omega_0 \big\|_2 + C_2(T) \big\|\omega_0 \big\|^2_2 \: \Big) \; (\nu t)^{-1/4}, \;\;\; t>0.
	\end{split}
\end{equation}
We have made use of the time-wise bounds on vorticity in (\ref{eq:vort-lq-lp-bound}). 
Note that the velocity norm in (\ref{eq:u-time-bound}) depends on the initial data $\|\omega_0\|_{2}$ quadratically. The first factor is due to the linear diffusion and the second to the non-linearity.

Similarly, the gradient can be found by solving an inequality derived from (\ref{eq:mom-ie-du}) and (\ref{eq:mom-ie-dp}), namely, 
\begin{equation*} 
\begin{split}
	\big\|\nabla u(\cdot,t) \big\|_{\infty}  & \leq  C_1 (\nu t)^{-3/4} \big\|u_0 \big\|_6 + C_2 \int_0^t (\nu t {-} \nu s)^{-3/4} \big\|u \big\|_6 \big\|\nabla u \big\|_{\infty} (\cdot,s) \rd s  \\
	\quad & {\hspace{3cm}}+ C_3 \int_0^t (\nu t {-} \nu s)^{-3/4} \big\|\nabla p \big\|_6 (\cdot,s) \rd s \\
	\quad & \leq C_1 (\nu t)^{-3/4} \big\|\omega_0 \big\|_2 + C_2 \big\|\omega_0 \big\|_2 \int_0^t (\nu t {-} \nu s)^{-3/4} \big\|\nabla u \big\|_{\infty} (\cdot,s) \rd s \\
	\quad & {\hspace{2cm}} + C_3 \big\|\omega_0 \big\|^2_2 \int_0^t (\nu t {-} \nu s)^{-3/4} (\nu s)^{-3/4} (\cdot,s) \rd s.
	\end{split}
\end{equation*}
By Gronwall's lemma, we obtain an important bound on the gradient in terms of the initial enstrophy,
\begin{equation} \label{eq:du-time-bound}
	\big\|\nabla u(\cdot,t) \big\|_{\infty}  \leq C \big(\: T, \big\|\omega_0 \big\|_2 \:\big) \; (\nu t)^{-3/4}, \;\;\; t > 0.
\end{equation}

By considering $L^2$-bound of equation (\ref{eq:mom-ie}), we have
\begin{equation*} 
\begin{split}
	\big\|u(\cdot,t) \big\|_{2}  & \leq  C_1 \big\|u_0 \big\|_2 + C_2 \int_0^t (\nu t {-} \nu s)^{-3/4} \big\|u \big\|_2 \big\|\nabla u \big\|_{2} (\cdot,s) \rd s  \\
	\quad & {\hspace{3cm}}+ C_3 \int_0^t \big\|\nabla p \big\|_2 (\cdot,s) \rd s \\
	\quad & \leq C_1 \big\|u_0 \big\|_2 + C_2 \big\|\omega_0 \big\|_2 \int_0^t (\nu t {-} \nu s)^{-3/4} \big\|u \big\|_{2} (\cdot,s) \rd s \\
	\quad & {\hspace{2cm}} + C_3 \big\|\omega_0 \big\|^2_2 \int_0^t (\nu s)^{-1/4} (\cdot,s) \rd s.
	\end{split}
\end{equation*}
Thus  
\begin{equation} \label{eq:u2-by-mom-ie}
	\big\| u (\cdot,t) \big\|_{L^2(\real^3)} \: {\leq} \: \Big( C_1 \big\| u_0 \big\|_{L^2(\real^3)} + C_3 \big\| \omega_0 \big\|^2_{L^2(\real^3)} \Big) \exp\Big( C_2 \big\| \omega_0 \big\|_{L^2(\real^3)} \Big),
\end{equation} 
where constants, $C_1$ to $C_3$, depend on $T$. This is an alternative energy bound. 
\subsection*{Uniform bounds on vorticity perturbation}
Recall that the vorticity in (\ref{eq:vorticity}) satisfies the integral equation
\begin{equation} \label{eq:vort-ieqn}
	\omega(x,t) = \int {\mathbf  Z}(x{-}y,t) \omega_0(y) \rd y + 
	\int_0^t \! \int   \nabla {\times} {\mathbf Z}(x{-}y,t{-}s)  \: ( \omega {\times} u) (y,s) \rd y \rd s.
\end{equation}
Here the non-linearity may be viewed as a ``perturbation'' from the linear diffusion. To bound the perturbation by initial enstrophy, we first notice that the gradient of the diffusion kernel can be computed:
\begin{equation} \label{eq:del-heat-kernel-bound}
	\big\| \nabla Z (\cdot,t) \big\|_r = C(r) \: (\nu t)^{-{5}/{4}+{3}/{(2r)}} \big\|\omega_0 \big\|_{2},\;\;\; 2 \leq r \leq \infty.
\end{equation}

Leray's idea of weak solution for the Navier-Stokes equations (Leray 1934$b$) can be applied to the vorticity equation. Equation (\ref{eq:vort-ieqn}) defines the vorticity as a distribution solution. In fact, for all the smooth vector test functions $\psi$ which are compactly supported in $(\real^3 {\times} \real)$, we have
\begin{equation*} 
	\int_{\real} \int_{\real^3} \Big(  \omega \partial_t \psi + \nu \omega \Delta \psi - (\omega. \nabla) \psi u + (u. \nabla) \psi \omega \Big) \rd x \rd t=0.
\end{equation*}
Similarly we interpret the velocity solenoidal condition as 
\begin{equation*} 
	\int_{\real} \int_{\real^3} u. \nabla \phi \rd x \rd t=0,
\end{equation*}
where $\phi$ is a smooth test function with compact supports in $(\real^3 {\times} \real)$.
In the sense of distribution, we have a similar interpretation for the vorticity continuity.

The first term on the right of (\ref{eq:vort-ieqn}) is just the solution of the pure initial value problem for linear diffusion. If $\omega_0 \in C_c^{\infty}(\real^3)$, then 
\begin{equation*}
	(\nu t)^{3/4-3/(2p)} \big\| Z{*} (\omega_0)_i \big\|_{p} \rightarrow 0 \;\;\; \mbox{as} \;\;\; t \rightarrow 0
\end{equation*}
in $L^p(\real^3)$ for $2{<}p{\leq}\infty$ by virtue of (\ref{eq:vort-lq-lp-bound}). Hence if we can find a time $T^*$ such that the following bound is valid, namely,
\begin{equation*}
	\sup_{0{<}\tau{\leq}T^*} (\nu \tau)^{3/4-3/(2p)} \big\| Z{*} (\omega_0)_i \big\|_p \: < \: A^*,
\end{equation*}
where constant $A^*{=}A^*(p)$, then strong solutions up to $T^*$ for the vorticity equation can be constructed by techniques of successive approximations (e.g. Faedo-Galerkin method). The bound defines $T^*$ accordingly. However, when the initial vorticity  $\omega_0 \in L^2(\real^3)$, the limit,
\begin{equation*}
	\lim_{\tau \rightarrow 0^+} (\nu \tau)^{3/4-3/(2p)} \big\| Z{*} (\omega_0)_i \big\|_{p},
\end{equation*}
may not necessarily be uniform in some fixed bounded subset of $L^2$. This implies that a compactness proof may encounter problems in $\real^3$. Nevertheless, instead of considering the space $L^2$, we consider some relatively compact (or precompact) set ${\mathscr K}$ of $L^2(\real^3)$ as the closure of ${\mathscr K}$ is indeed compact. We attempt to show that
there exists a function $\vartheta(\nu t;{\mathscr K})\geq0$ such that 
\begin{equation} \label{eq:vort-delta-def}
	(\nu t)^{3/4-{3}/{(2p)})} \big\| {\mathbf Z}{*} \omega_0 \big\|_{p} \; {\leq} \; \vartheta(\nu t;{\mathscr K}),\;\;\;\omega_0 \in {\mathscr K} ,
\end{equation}
if ${\mathscr K} \subseteq L^{2}$ is precompact and $\forall p {\in} (2,\infty]$. We define  $\vartheta(\nu t;{\mathscr K})$ as
\begin{equation*}
	\sup_{0{<}\tau{\leq}t,\; \omega_0 \in {\mathscr K} } \: \Big\{ (\nu \tau)^{\:{3}/{(4p)}-{3}/{(2q)}} \: \big\|  {\mathbf Z} {*} \omega_0 \big\|_{L^q(\real^3)}\; \Big\}
\end{equation*}
for $2<q< \infty$. It is clear that $\vartheta(\nu t;{\mathscr K})$ is a monotonic non-decreasing function defined for $t{\geq}0$, and $\lim_{t{\rightarrow}0} \vartheta(\nu t;{\mathscr K}) =0$.
Since $\|\omega_0\|_{L^p}$ is bounded by a constant for all $\omega_0 {\in} {\mathscr K}$, then for all $t{>}0,\;\vartheta(\nu t;{\mathscr K})< C$ for some constant $C$.
This implies that $\vartheta(\nu t;{\mathscr K})$ is uniformly bounded. We write explicitly $\vartheta(\nu t;{\mathscr K})$ for $\vartheta(\nu t)$ to indicate its dependence on certain subsets of the initial data.

The first term on the right-hand side of (\ref{eq:vort-ieqn}) is bounded by (\ref{eq:vort-delta-def}).  Making use of (\ref{eq:del-heat-kernel-bound}) and Sobolev formula (\ref{eq:sobolev-vel}) in conjunction with Young's and H\"{o}lder's inequalities, the second term can be bounded:
\begin{equation} \label{eq:vort-lp-int}
	\big\| \omega(\cdot,t) \big\|_p \leq \vartheta(\nu t;{\mathscr K}) (\nu t)^{-3/4+{3}/{(2p)}} + C \int_0^t \! (\nu t{-} \nu s)^{-{5}/{4}+{3}/{(2r)}} \big\| \omega(\cdot,t)\big\|^2_p \: \rd s,
\end{equation}
where $\omega_0 {\in} {\mathscr K}$, and $1/p{+}1/r{=}4/3$.
For any $t{\in}(0,T]$, we define 
\begin{equation*}
	W_p(t)= \sup_{0{<}\tau{\leq}t} (\nu \tau)^{3/4-{3}/{(2p)}} \big\| \omega(\cdot,\tau) \big\|_p.
\end{equation*}
Since the initial vorticity $\omega_0$ is smooth, $\|\omega\|_p$ is bounded and continuous, the limit, $\lim_{t {\rightarrow} 0} W_p(t) \rightarrow 0$, holds. Thus $W_p(t) \in C([0,t{>}0))$. Inequality (\ref{eq:vort-lp-int}) can be rewritten as
\begin{equation} \label{eq:wp-quadratic}
	W_p(t) \; {\leq} \;  \vartheta(\nu t;{\mathscr K}) + C_* \:(\nu t) \: (W_p(t))^2.
\end{equation}
The constant is independent of ${\mathscr K}$ and $C_*{=}C(p) B(p) {>}0$, where  
\begin{equation*}
	B(p)= B \Big( {7}/{4}{-}{3}/{(2p)},{3}/{p}{-}1/2 \Big).
\end{equation*}
Note that $1{<}B(p){<}1.748$ for $2{<}p{<}3$. Because both $W_p(t)$ and $\vartheta(\nu t;{\mathscr K})$ are non-decreasing functions of $t({>}0)$, there exists a time $T^*{=}T^*({\mathscr K})({\leq}T)$ for all $\omega_0 {\in} {\mathscr K}$ such that  
\begin{equation} \label{eq:def-t-star}
	(\nu T^*) \vartheta(\nu T^*;{\mathscr K}) \; {\leq} \; \big( 4 C_* \big)^{-1}. 
\end{equation}
The quadratic inequality in (\ref{eq:wp-quadratic}) implies two possible bounds on $W_p$. The first one is related to the root (with the positive sign) 
\begin{equation*}
	(\nu t) W_p(t) \:{\geq}\: \big( 1 + \sqrt{1 - 4 C_* (\nu t) \: \vartheta \: } \: \big) \big( 2 C_* \big)^{-1}.
\end{equation*}
This root must be discarded because it does not conform with the limit condition 
\begin{equation*}
 \lim_{t{\rightarrow}0^+}(\nu t)W_p(t) \rightarrow 0.	
\end{equation*}
The other bound is defined by
\begin{equation*}
	(\nu t) W_p(t) \:{\leq}\: \big( 1 - \sqrt{1 - 4 C_* (\nu t) \: \vartheta \: }\:\big) \big( 2C_* \big)^{-1}.
\end{equation*}
It follows that
\begin{equation} \label{eq:wt-lp-bound}
	W_p(t) \; {\leq} \:  2\: \vartheta(\nu t;{\mathscr K}), \;\;\; \forall t \in (0,T^*{\leq}T).
\end{equation}

Interpolating between (\ref{eq:wt-lp-bound}) and (\ref{eq:vort-linfty-bound}), we obtain the following time-wise bound for vorticity,
\begin{equation} \label{eq:vort-l2-linfty-bound}
	\big\| \omega(\cdot,t) \big\|_{L^p(\real^3)} \: {\leq} \: C(T) \: \vartheta(\nu t;{\mathscr K}) \: (\nu t)^{-3/4+{3}/{(2p)}}, \;\;\; \forall p \in (2,\infty],
\end{equation}
where $\omega_0 \in {\mathscr K} \subseteq C_c^{\infty}(\real^3)$, being precompact in $L^{2}(\real^3)$.
In view of (\ref{eq:vort-l2-linfty-bound}) and (\ref{eq:u-time-bound}), we deduce a bound for velocity,
\begin{equation} \label{eq:vel-linfty-bound-strong}
\big\| u(\cdot,t) \big\|_{L^{\infty}(\real^3)} \:{\leq}\: (\nu t)^{-1/4} \: \vartheta(\nu t;{\mathscr K}) \big( C_1(T) + C_2(T) \: \vartheta(\nu t;{\mathscr K}) \big).
\end{equation}
As expected, this temporal bound is a non-linear function of $\vartheta$.
\subsection{Vorticity solution in space $L^2(\real^3){\cap}C^0_c(\real^3)$}
Our objective in this subsection is to show that the vorticity equation is well-posed in space $Y$.
A vorticity solution obtained from integral equation (\ref{eq:vort-ieqn}) is viewed as a result of an operator over the initial data
\begin{equation*}
	{\mathbf{S}}: \: C_c^{\infty} (\real^3)  \:  {\rightarrow} \: C^{\infty}(\real^3{\times}(0,T]), \;\;\; \omega(\cdot,t) \:{=}\: {\mathbf{S}} \omega_0,
\end{equation*}
where $\omega_0$ is the initial condition. First of all, we attempt to establish the well-posedness of the solution operator ${\mathbf{S}}$.

Let $\tilde{\omega}(x,t)$ denote the solution of (\ref{eq:vort-ieqn}) with the initial data $\tilde{\omega}_0 = \tilde{\omega}(x,0) \in C_c^{\infty}(\real^3)$. Denote the associated velocity by $\tilde{u}(x) = {\cal K} {*}\tilde{\omega}(x)$ and the pressure by $\tilde{p}(x,t)$. From the integral equations governing $\omega$ and $\tilde{\omega}$, we deduce that
\begin{equation} \label{eq:omega-phi-diff}
 \begin{split}
	\omega(x,t)-\tilde{\omega}(x,t)& =  \int {\mathbf Z}(x{-}y,t) \big(\omega_0(y) - \tilde{\omega}_0(y)\big) \rd y  \\
	\quad & + \int_0^t \!\int  \Big( (\nabla {\mathbf Z}.u) \:\big(\omega- \tilde{\omega}\big) + (\nabla {\mathbf Z}.\omega) \:\big(u- \tilde{u}\big) \Big) (y,s)\rd y \rd s \\
	\quad &-  \int_0^t \!\int \Big( (\nabla {\mathbf Z}.\tilde{\omega}) \:\big(u- \tilde{u}\big) 
	+  (\nabla {\mathbf Z}.\tilde{u}) \:\big(\omega- \tilde{\omega}\big) \Big) (y,s)\rd y \rd s,
 \end{split}
\end{equation}
where $\nabla{\mathbf Z}=\nabla{\mathbf Z}(x{-}y,t{-}s)$.
Recall the $L_p$-bound for vorticity in (\ref{eq:vort-lq-y}). In view of the momentum equations for $u$ and $\tilde{u}$, we calculate velocity difference, $u-\tilde{u}$.
Then we follow the derivation steps leading to estimate (\ref{eq:u-time-bound-y}) for the difference. We readily obtain
\begin{equation} \label{eq:u-diff-time-bound-y}
	\big\| u(\cdot,t)- \tilde{u}(\cdot,t) \big\|_{\infty} \; {\leq} \; C \; \big\| \omega_0 - \tilde{\omega}_0 \big\|_Y \; (\nu t)^{-1/4},\;\;\;t>0, 
\end{equation}
where $C$ depends on both $\|\omega_0\|_Y$ and $\|\tilde{\omega}_0\|_Y$. This result is expected as the pressure difference is linear. The two non-linear terms, $\|u \nabla u\|_2$ and $\|\tilde{u} \nabla \tilde{u}\|_2$, can be bounded separately.

Thus we infer from (\ref{eq:omega-phi-diff}) that
\begin{equation*}
	\big\| \omega(\cdot,t){-}\tilde{\omega}(\cdot,t) \big\|_Y \: {\leq} \: C_1 \: \big\| \omega_0 {-} \tilde{\omega}_0 \big\|_Y + C_2 \int_0^t  (\nu t {-} \nu s)^{-1/2} (\nu s)^{-1/4} \big\| \omega(\cdot,s){-}\tilde{\omega}(\cdot,s) \big\|_Y  \rd s.
\end{equation*}
By Gronwall's lemma, we deduce that
\begin{equation} \label{eq:vort-diff}
	\big\| \omega(\cdot,t)-\tilde{\omega}(\cdot,t) \big\|_Y \; {\leq} \; C \: \big\| \omega_0 - \tilde{\omega}_0 \big\|_Y, \;\;\; \forall t \in [0,T],
\end{equation}
where the constant $C{=}C(T,\|\omega_0\|_Y,\|\tilde{\omega}_0\|_Y)$.

Differentiating (\ref{eq:vort-ieqn}) and writing the result in full, we get
\begin{equation} \label{eq:nabla-vort-ieqn}
 \begin{split}
	\nabla \omega(x,t) = & \int \nabla \mathbf Z(x{-}y,t) \omega_0(y) \rd y \\ 
	&+ \int_0^t \! \int  \nabla {\mathbf Z} (x{-}y,t{-}s) \; \Big( \:(\omega . \nabla) u - (u . \nabla) \omega \: \Big)(y,s) \rd y \rd s.
	 \end{split}
\end{equation}
By virtue of bounds (\ref{eq:u-time-bound-y}) and (\ref{eq:du-time-bound-y}), we are able to estimate the contribution from the second integral. Direct calculations show that a bound for the vorticity gradient can be found by solving
\begin{equation*} 
\begin{split}
	\big\| \nabla \omega(\cdot,t) \big\|_{Y} \: {\leq} \:  C_1(\nu t)^{-1/2} & + C_2 \int_0^t (\nu t{-} \nu s)^{-1/2} (\nu s)^{-3/4} \rd s \\
	\quad & + C_3 \int_0^t (\nu t{-} \nu s)^{-1/2} (\nu s)^{-1/4}  \big\| \nabla \omega(\cdot,s) \big\|_{Y} \rd s,
\end{split}
\end{equation*}
where $C$'s are positive constants and depend on $\|\omega_0\|_Y$. Thus we obtain
\begin{equation} \label{eq:nabla-vort-norm-Y}
	\big\| \nabla \omega(\cdot,t) \big\|_Y  \: {\leq} \: C \big( \:T, \big\|\omega_0\big\|_Y \:\big) \: (\nu t)^{-1/2},\;\;\;0 < t \leq T.  
\end{equation}
Similarly, we find that
\begin{equation*}
 \begin{split}
 	\nabla\omega(x,t) - \nabla\tilde{\omega}(x,t) &=  \int \nabla{\mathbf Z}(x{-}y,t) \big(\omega_0(y) - \tilde{\omega}_0(y)\big) \rd y \\
	\quad & - \int_0^t \!\int \Big( \nabla {\mathbf Z} (u.\nabla) \:(\omega - \tilde{\omega}) + \nabla {\mathbf Z} ((u - \tilde{u}).\nabla)\:\tilde{\omega} \Big) (y,s) \rd y \rd s \\
	\quad & + \int_0^t \!\int \Big( \nabla {\mathbf Z} (\omega.\nabla) \:(u - \tilde{u} )  + \nabla {\mathbf Z} ((\omega - \tilde{\omega}).\nabla) \:\tilde{u} \Big) (y,s) \rd y \rd s \\
	& = I_1 + I_2 + I_3 + I_4 + I_5.
 \end{split}
\end{equation*}
First of all, we calculate
\begin{equation*}
	\big\|I_1 \big\|_Y \: {\leq} \: C \: (\nu t)^{-1/2} \: \big\| \omega_0-\tilde{\omega}_0 \big\|_Y. 
\end{equation*}
Second, in view of bounds (\ref{eq:u-time-bound-y}) and (\ref{eq:vort-diff}), we deduce that
\begin{equation*}
	\big\|I_2 \big\|_Y \: {\leq} \: C \:  \int_0^t (\nu t{-} \nu s)^{-1/2} (\nu s)^{-1/4} \big\| \nabla \omega(\cdot,s)- \nabla \tilde{\omega}(\cdot,s) \big\|_Y \rd s.
\end{equation*}
Applying bound (\ref{eq:u-diff-time-bound-y}), it is easy to establish that
\begin{equation*}
	\big\|I_3 \big\|_Y \: {\leq} \: C \:  B(1/2,1/2) \: \big\|\omega_0 - \tilde{\omega}_0 \big\|_Y.
\end{equation*}	
Furthermore, it is obvious that
\begin{equation*}
	\big\|I_5 \big\|_Y \: {\leq} \: C \: B(1/2,1/4) \: \big\|\omega_0 - \tilde{\omega}_0 \big\|_Y \: (\nu t)^{-1/4}.
\end{equation*}	
In evaluation of $I_4$, we find an analogous bound to (\ref{eq:u-diff-time-bound-y}),
\begin{equation} \label{eq:du-diff-time-bound-y}
	\big\| \nabla u(\cdot,t)- \nabla \tilde{u}(\cdot,t) \big\|_{\infty} \; {\leq} \; C \; \big\| \omega_0 - \tilde{\omega}_0 \big\|_Y \; (\nu t)^{-3/4},\;\;\;t>0. 
\end{equation}
In fact, the gradient difference can be computed by the derivation steps leading to estimate  (\ref{eq:du-time-bound-y}). We notice that the estimates of the non-linear terms, $\|u \nabla u\|_6$ and $\|\tilde{u} \nabla \tilde{u}\|_6$, and of the pressure gradients, $\|\nabla p\|_6$ and $\|\nabla \tilde{p}\|_6$, can be asserted without difficulty by virtue of (\ref{eq:mom-ie-u}) to (\ref{eq:mom-ie-dp}). Once we have bound (\ref{eq:du-diff-time-bound-y}) in place, we derive 
\begin{equation*}
	\big\|I_4 \big\|_Y \: {\leq} \: C \:  \big\|\omega_0 - \tilde{\omega}_0 \big\|_Y \: (\nu t)^{-1/4}.
\end{equation*}	
In summary, we obtain the integral inequality,
\begin{equation*}
 \begin{split}
 	\big\|\nabla\omega(\cdot,t) - \nabla\tilde{\omega}(\cdot,t) \big\|_Y \: {\leq} \: & \big( C_1 (\nu t)^{-1/2}  + C_2 (\nu t)^{-1/4} + C_3 \big) \: \big\|\omega_0 - \tilde{\omega}_0 \big\|_Y\\
 	 \quad & +  \int_0^t (\nu t{-} \nu s)^{-1/2} (\nu s)^{-1/4} \big\| \nabla \omega(\cdot,s) - \nabla \tilde{\omega}(\cdot,s) \big\|_Y \rd s.  
 \end{split}
\end{equation*}
Thus we conclude that
\begin{equation} \label{eq:nabla-vort-diff}
	\big\| \nabla \omega(\cdot,t) - \nabla \tilde{\omega}(\cdot,t) \big\|_Y \; {\leq} \; C \; \big\| \omega_0 - \tilde{\omega}_0 \big\|_Y \; (\nu t)^{-1/2}, \;\;\; \forall t \in (0,T],
\end{equation}
where $C{=}C\big( T,\|\omega_0\|_Y, \| \tilde{\omega}_0 \|_Y \big)$. 

We summarize the well-posedness of the vorticity solution operator, defined for $\omega_0 {\in} C_c^{\infty}(\real^3)$. 
Bounds (\ref{eq:vort-diff}) and (\ref{eq:nabla-vort-diff}) show that 
the operator ${\mathbf{S}}$ can be extended as a continuous operator as follows: ${\mathbf{S}}:\; Y \rightarrow C((0,T], Y)$.
The map ${\nabla} {\mathbf{S}}$ can be extended continuously to
\begin{equation*} 
\nabla {\mathbf{S}}:\; Y \rightarrow C((0,T], Y) \cap L^p_{loc}((0,T],Y), 
\end{equation*}
for any $1{\leq}p{<}2$. In addition, the function $\omega(t){=}({\mathbf{S}} \omega_0)(t)$ satisfies the vorticity equation weakly in $\real^3{\times}(0,T]$.

Let $\tilde{\omega}$ (and $\tilde{u}(x){=}{\cal K} {*} \tilde{\omega}(x)$) be a weak solution in $\real^3{\times}(0,T]$ of
\begin{equation*}
	{\partial \tilde{\omega}}/{\partial t} - \nu \Delta \tilde{\omega} = (\tilde{\omega} . \nabla )\tilde{u}  - (\tilde{u} . \nabla ) \tilde{\omega} = \nabla {\times} (\tilde{u} {\times} \tilde{\omega} )
\end{equation*}
with the initial data $\tilde{\omega}(x,0) = \omega_0(x) \in Y$. But equation (\ref{eq:vort-ieqn}) is also satisfied by $\tilde{\omega}$ with $\omega_0, \omega, u$ replaced by $\tilde{\omega}_0, \tilde{\omega}, \tilde{u}$. It is routine to show the vorticity difference
remains zero at all time if the initial data coincide.

The regularity of the solution ($\omega = {\mathbf{S}} \omega_0$) can be established by the well-known theories for parabolic equations (see, for example, Aronson \& Serrin 1967; Friedman 1964; Aronson 1968; Zhang 1995; Liskevich \& Semenov 2000). For every initial vorticity $\omega_0 {\in} Y$, the function $\omega(t){=}({\mathbf{S}} \omega_0)(t)$ belongs to $C^{\infty}((0,T]{\times}\real^3)$ and satisfies the vorticity equation. Furthermore, for every multi-index $\alpha$ and every integer $\beta{\geq}0$, the map, $\partial_t^{\beta} \partial_x^{\alpha} {\mathbf{S}}: Y \rightarrow C((0,T], Y)$,
is continuous in time (analogous to the uniqueness proof of the smooth solutions). Since the pressure gradient is known (cf. (\ref{eq:nabla-p-lp-bound})), the velocity can be evaluated from (\ref{eq:mom-ie}) for almost every $t \in (0,T]$. 
\subsection{Well-posedness of vorticity solution for $\omega_0 {\in} L^{2}(\real^3)$}
Let ${\mathscr K}{\subseteq} C^{\infty}(\real^3)$ be precompact in the $L^{2}$ topology.
We first show that, making use of the bounds (\ref{eq:vort-l2-linfty-bound}) and (\ref{eq:vel-linfty-bound-strong}), the family of mappings from $(0,T]$ into $L^{2}(\real^3)$, 
\begin{equation} \label{eq:t-mapping}
	t \rightarrow ({\mathbf{S}} \omega_0)(t),\;\;\; t\:{\geq}\:0, \; \omega_0 \in {\mathscr K},
\end{equation}
is equicontinuous and equibounded. By the properties of heat kernel (\ref{eq:del-heat-kernel-bound}) and the integral presentation (\ref{eq:vort-ieqn}), we have 
\begin{equation}
	\big\|{\mathbf{S}} \omega_0 - \omega_0 \big\|_{L^{2}(\real^3)} \; {\leq} \; \big\| Z*\omega_0 - \omega_0 \big\|_{L^{2}(\real^3)} + C \: \vartheta(\nu t; {\mathscr K}).
\end{equation}
This bound converges to $0$ as $t {\rightarrow} 0$ uniformly in $\omega_0 {\in} {\mathscr K}$.
In view of (\ref{eq:vort-l2-linfty-bound}) and the regularity of the operator ${\mathbf{S}}$, for any $\varepsilon {>} 0$ and any multi-index $\alpha$, we also have
\begin{equation} \label{eq:vort-alpha-Y-bound}
	\sup_{ \substack{ 0{<}\varepsilon {\leq} t {<}  {\infty} \\ \omega_0 \in {\mathscr K} }}  \Big\{ \big\| \partial_x^{\alpha} ({\mathbf{S}} \omega_0)(\cdot,t) \big\|_Y \Big \}\;{<}\; {\infty}.
\end{equation}
By the estimates (\ref{eq:u-time-bound}) and (\ref{eq:du-time-bound}), bound (\ref{eq:vort-alpha-Y-bound}) implies
\begin{equation} \label{eq:vel-alpha-Y-bound}
	\sup_{ \substack{ 0{<}\varepsilon {\leq} t {<}  {\infty} \\ \omega_0 \in {\mathscr K} }} \Big\{ (\nu t)^{1/2}  \big\| \partial_t \omega(\cdot,t) \big\|_{L^{2}(\real^3)} + \big\| u(\cdot,t) \big\|^2_{L^{\infty}(\real^3)} + (\nu t) \big\| \nabla u(\cdot,t) \big\|^2_{L^{\infty}(\real^3)} \Big \} \:{<}\: {\infty}. 
\end{equation}
The quantity on the left has the dimensions of energy per unit mass.
Thus these bounds imply the equicontinuity of (\ref{eq:t-mapping}) in $L^{2}(\real^3)$. Bound (\ref{eq:vort-l2-linfty-bound}) shows that the mappings are equibounded. 
Let $\{ \omega_0^{(n)}(x)\}_{n=1}^\infty \subseteq C_c^{\infty} (\real^3) $ be a Cauchy sequence in $L^{2}(\real^3)$ converging to $\omega_0(x) {\in} L^{2}(\real^3)$. If we take ${\mathscr K}{=}\{ \omega_0^{(n)}(x)\}_{n=1}^{\infty}$, the images $\{ {\mathbf{S}} (t) \omega_0^{(n)}\}_{n=1}^{\infty}$ are equicontinuous in $L^{2}(\real^3)$. 

We now establish the uniform convergence of the images. 
Let $\omega^{(n)}(x,t) = {\mathbf{S}} \omega_0^{(n)}$ and $u^{(n)}(x) =  {\cal K}{*}\omega^{(n)}(x)$. In parallel to (\ref{eq:omega-phi-diff}), we see that
\begin{equation} \label{eq:vort-n-m}
 \begin{split}
	(& \omega^{(n)} - \omega^{(m)})(x,t) =  \int \! {\mathbf Z}(x{-}y,t) ( \omega^{(n)}_0 - \omega^{(m)}_0 )(y) \rd y \\
	\quad & + \int_0^t \!\int \! \Big( (\nabla{\mathbf Z}.u^{(n)})(\omega^{(n)} - \omega^{(m)}) + (\nabla{\mathbf Z}.\omega^{(m)})(u^{(n)} - u^{(m)}) \Big) (y,s) \rd y \rd s  \\
	\quad & - \int_0^t \!\int \Big( (\nabla{\mathbf Z}.\omega^{(n)})(u^{(n)} - u^{(m)}) +  (\nabla{\mathbf Z}.u^{(m)})(\omega^{(n)} {-} \omega^{(m)}) \Big) (y,s) \rd y \rd s  \\
	\quad &  = J_1+J_2+J_3+J_4+J_5.
 \end{split}
\end{equation}
In view of the properties of the heat kernel, we find that, for $2{<}p{<}3$, 
\begin{equation*}
	\big\|J_1(\cdot,t) \big\|_p \; {\leq} \; C_0(p) \:(\nu t)^{-3/4+{3}/{(2p)}} \: \big\| \omega_0^{(n)}- \omega_0^{(m)} \big\|_{2}. 
\end{equation*}
By H\"older's inequality, the velocity-vorticity product in the second integrand becomes
\begin{equation*}
	\big\| u^{(n)}(\omega^{(n)}{-}\omega^{(m)}) \big\|_p \:{\leq}\: \big\| u^{(n)} \big\|_{\infty} \big\| \omega^{(n)}{-}\omega^{(m)} \big\|_p.
\end{equation*}
By virtue of bound (\ref{eq:vel-linfty-bound-strong}), integrals $J_2$ and $J_5$ can be estimated as
\begin{equation*}
\begin{split}
	\big\|J_2(\cdot,t) \big\|_p \: + \big\|J_5(\cdot,t) \big\|_p \:  {\leq} & \:  
	\vartheta(\nu t;{\mathscr K})\big( C_1(T)+C_2(T) \vartheta(\nu t;{\mathscr K}) \big) \times \\
	\quad & \int_0^t (\nu t {-} \nu s)^{-1/2} (\nu s)^{-1/4}  \big\| \omega^{(n)}(\cdot,s){-} \omega^{(m)}(\cdot,s) \big\|_p \: \rd s.
	\end{split}
\end{equation*}
By (\ref{eq:sobolev-vel}) and (\ref{eq:vort-l2-linfty-bound}), the integrand relating to $J_3$ is simplified as
\begin{equation*}
 \begin{split}
	\big\| \omega^{(m)}(u^{(n)}{-}u^{(m)}) \big\|_p \:{\leq}\: & \big\| \omega^{(m)}\|_3 \big\| u^{(n)}{-}u^{(m)}\|_q \\
	\quad \:{\leq}\: & C_3(T,p) \: \vartheta(\nu t;{\mathscr K})  (\nu t)^{-1/4} \big\| \omega^{(n)}{-}\omega^{(m)} \big\|_p.
 \end{split}
\end{equation*}
Hence we obtain the estimate,
\begin{equation*}
 \begin{split}
	\big\|J_3(\cdot,t) \big\|_p \:+ & \big\|J_4(\cdot,t) \big\|_p \\
	\quad \: {\leq} \: & C_3(T,p) \: \vartheta(\nu t;{\mathscr K})  \int_0^t (\nu t {-} \nu s)^{-1/2} (\nu s)^{-1/4}  \big\| \omega^{(n)}(\cdot,s){-} \omega^{(m)}(\cdot,s) \big\|_p \: \rd s.
 \end{split} 
\end{equation*}
Inserting the bounds on $J_1$ to $J_5$ into (\ref{eq:vort-n-m}) yields
\begin{equation} \label{eq:vort-n-m-lp}
\begin{split}
\big\|\omega^{(n)}(\cdot,t){-}\omega^{(m)} & (\cdot,t) \big\|_p \:{\leq}\: C_0(p)  \: (\nu s)^{-3/4+{3}/{(2p)}} \: \big\| \omega_0^{(n)}{-} \omega_0^{(m)} \big\|_{2} + \\
\quad & C({\vartheta}) \int_0^t (\nu t {-} \nu s)^{-1/2} (\nu s)^{-1/4}  \big\| \omega^{(n)}(\cdot,s){-} \omega^{(m)}(\cdot,s) \big\|_p \: \rd s,
\end{split}
\end{equation}
where
\begin{equation*}
	C(\vartheta)\:{\leq}\: \vartheta(\nu t;{\mathscr K}) \Big( C_1(T) + C_3(T,p) + C_2(T) \: \vartheta(\nu t;{\mathscr K}) \Big).
\end{equation*}
Recall that the function $\vartheta(\nu t;{\mathscr K})$ is non-decreasing. In view of Gronwall's lemma, we deduce that inequality (\ref{eq:vort-n-m-lp}) gives rise to 
\begin{equation} \label{eq:vort-n-m-lp-bound}
	\big\| \omega^{(n)}(\cdot,t) {-} \omega^{(m)}(\cdot,t) \big\|_{L^p(\real^3)} \; {\leq} \; C(p) \: (\nu t)^{-3/4+{3}/{(2p)}} \: \big\| \omega_0^{(n)} {-} \omega_0^{(m)} \big\|_{L^{2}(\real^3)}  
\end{equation}
for $0 < t \leq T^*$, where time $T^*$ is defined in (\ref{eq:def-t-star}). 

Taking $L^{2}$-norm of (\ref{eq:vort-n-m}), we find that
\begin{equation} \label{eq:vort-n-m-l2}
 \begin{split}
	 \big\| \omega^{(n)} &(\cdot,t)-\omega^{(m)}(\cdot,t) \big\|_{2} \;  {\leq} \;  \big\| \omega_0^{(n)}- \omega_0^{(m)} \big\|_{2} \\
	 \quad & {+} \int_0^t  ( \nu t{-} \nu s)^{-1/2} 
\Big( \big( \big\|u^{(n)} \big\|_q + \big\|u^{(m)} \big\|_q \big) \: \big( \big\|\omega^{(n)}-\omega^{(m)} \big\|_r \big) \\
	\quad & \hspace{2.5cm}{+} \big( \big\|\omega^{(m)} \big\|_r + \big\|\omega^{(n)} \big\|_r \big) \: \big( \big\|u^{(n)}-u^{(m)} \big\|_q \big)  \Big) (\cdot,s) \rd s ,
 \end{split} 
\end{equation} 
where $1/q + 1/r = 1/2$. 

Consider two motions, $\omega(x,t)$ and $\tilde{\omega}(x,t)$, with the respective velocities $u(x,t)$ and $\tilde{u}(x,t)$. We first notice that $\nabla{\times}(u{-}\tilde{u})=\omega{-}\tilde{\omega}$. Thus there exists a stream function $\bar{\psi}$ such that $\Delta{\bar{\psi}}={-}(\omega{-}\tilde{\omega})$. Hence a modified Biot-Savart relation follows:
\begin{equation*}
	(u-\tilde{u})(x;t) = {\cal K}*(\omega-\tilde{\omega})(x;t).
\end{equation*}
This equation implies that (\ref{eq:sobolev-vel}) also holds for the velocity difference. Choose 
\begin{equation*}
	q=15,\;\;\;r=30/13
\end{equation*} 
in (\ref{eq:vort-n-m-l2}). We find that
\begin{equation*}
	\big\| u \big\|_{15}\:{\leq}\:C\: \big\|\omega \big\|_{5/2},\;\;\; \big\| u-\tilde{u} \big\|_{15}\:{\leq}\:C\: \big\|\omega-\tilde{\omega} \big\|_{5/2}.
\end{equation*}
Now the vorticity norms can be bounded by (\ref{eq:vort-n-m-lp-bound}) and (\ref{eq:vort-lq-lp-bound}).
Consequently, integral inequality (\ref{eq:vort-n-m-l2}) can be simplified as
\begin{equation*}
 \begin{split}
	 \big\| \omega^{(n)} (\cdot,t)-\omega^{(m)}(\cdot,t) \big\|_{2} \; & {\leq} \;  \big\| \omega_0^{(n)}- \omega_0^{(m)} \big\|_{2} \\
	 \quad & + C \: \big\|\omega_0^{(n)}-\omega_0^{(m)} \big\|_{2} \: \int_0^t  ( \nu t{-} \nu s)^{-1/2} (\nu s)^{-1/4} \rd s. 
 \end{split} 
\end{equation*} 
For $0<t \leq T^*$, we assert that 
\begin{equation} \label{eq:vort-n-m-l2-bound}
	\big\|\omega^{(n)}(\cdot,t)-\omega^{(m)}(\cdot,t) \big\|_{L^{2}(\real^3)}  \: {\leq} \:  C \: \big\| \omega_0^{(n)}- \omega_0^{(m)} \big\|_{L^{2}(\real^3)},
\end{equation}
where 
\begin{equation*}
	C{=}C \Big( \: T, \: \big\|\omega_0^{(n)} \big\|_{2}, \: \big\|\omega_0^{(m)} \big\|_{2} \Big),
\end{equation*}
and $C{\neq}0$ when $\|\omega_0^{(n)}\|_{2}{=}\|\omega_0^{(m)}\|_{2}$.

Let $\omega_0 \in L^{2}(\real^3)$ and $\omega_0^{(n)} \rightarrow \omega_0$ in $L^{2}(\real^3)$, where $\omega_0^{(n)} \in C_c^{\infty}(\real^3)$. In view of (\ref{eq:vort-alpha-Y-bound}), (\ref{eq:vel-alpha-Y-bound}), (\ref{eq:vort-n-m-l2-bound}) and the well-posedness of the operator ${\mathbf{S}}$, the sequence of solutions
$\{ \omega^{(n)} = {\mathbf{S}} (t) \omega_0^{(n)} \}_{n=1}^{\infty}$
converges in $C((0,T], L^{2}(\real^3))$ to a function $\omega(t) \in C((0,T], L^{2}(\real^3))$. For every $t$ in $0 < t \leq T^*$, it follows that, by interpolating (\ref{eq:vort-l2-linfty-bound}) and (\ref{eq:vort-n-m-l2-bound}), the sequence $\{\omega^{(n)}(\cdot,t)\}_{n=1}^{\infty}$ converges in space $Y$.
In other words, 
\begin{equation*}
	\omega^{(n)}(\cdot,t) \rightarrow \omega(\cdot,t)\;\;\; \in C((0,T],Y).
\end{equation*}
It is not difficult to establish the regularity. Since ${\mathbf{S}}$ is regular, for every multiple-index $\alpha$ and every integer $\beta{\geq}0$, the sequence 
$\{ \partial_t^{\beta} \partial_x^{\alpha} \omega^{(n)}(\cdot,t) \}_{n=1}^{\infty}$
converges in $C((0,T], Y)$. It follows that $\omega(x,t)$ is smooth in $\real^3{\times}(0,T]$. 
In view of (\ref{eq:nabla-p-lp-bound}) and (\ref{eq:mom-ie}), the velocity can be shown to be smooth as well. The vorticity and the velocity constitute a classical solution of the vorticity equation. 
\subsection*{Uniqueness and compatibility}
Since the mappings (\ref{eq:t-mapping}) are equibounded and equicontinuous, bound (\ref{eq:vort-l2-linfty-bound}) can be extended to 
\begin{equation} \label{eq:vort-op-l2-linfty-bound}
 \big\| \: {\mathbf{S}} \omega_0(t) \: \big\|_p \; {\leq} \; \vartheta(\nu t;{\mathscr K} ) \: (\nu t)^{-3/4+{3}/{(2p)}}, \;\;\; \omega_0 \in {\mathscr K}, \;\;\; 2<p\leq\infty,
 \end{equation}
where ${\mathscr K} \subseteq L^{2}(\real^3)$ is precompact. Consider a second solution of the vorticity equation $\varphi(x,t)$. Suppose that the initial data $\varphi_0$ are essentially bounded.  
Preceding analyses show that $\varphi(x,t)$ and the associated velocity $v(x,t)$ satisfy the vorticity integral equation (\ref{eq:vort-ieqn}) with $\omega, \omega_0, u$ replaced by $\varphi, \varphi_0, v$. We then repeat the analysis steps leading up to the bound (\ref{eq:vort-l2-linfty-bound}) for $\varphi(x,t)$. In effect, we obtain the bounds,
\begin{equation} \label{eq:nabla-vel-linfty-bound-K}
	\big\|\varphi(\cdot,t) \big\|_{\infty} \; {\leq} \; \vartheta(\nu t;{\mathscr K}) \: (\nu t)^{-3/4}, \;\;\; t>0,
\end{equation}
and
\begin{equation} \label{eq:vel-linfty-bound-K}
	\big\|v(\cdot,t) \big\|_{\infty} \; {\leq} \; \vartheta(\nu t;{\mathscr K}) \big( C_1 + C_2 \vartheta(\nu t;{\mathscr K}) \big) \: (\nu t)^{-1/4},\;\;\;t>0.
\end{equation}
Writing $\tilde{\varphi}{=}{\mathbf{S}} \varphi_0(t)$, we deduce that the difference in the two vorticity vectors is given by
\begin{equation}
	\big\|\tilde{\varphi}(\cdot,t) - \varphi(\cdot,t) \big\|_p \: {\leq} \: C(\vartheta) \int_0^t ( \nu t {-} \nu s )^{-1/2} (\nu s)^{-1/4} \big\|\tilde{\varphi}(\cdot,s) - \varphi(\cdot,s) \big\|_p \:  \rd s
\end{equation}
for $2 < p < 3$. Hence $\tilde{\varphi}(\cdot,t)$ coincides with $\varphi(\cdot,t)$ over the interval $0< t < T^*$. The uniqueness is shown to be true for $\varphi_0 \in L^{2}(\real^3) \cap L^{\infty}(\real^3)$. Next consider the initial data $\varphi_0 \in L^{2}(\real^3)$. In view of (\ref{eq:nabla-vel-linfty-bound-K}), $\varphi(\cdot,s)$ is essentially bounded for any $s{>}0$. Setting $\varphi(\cdot,s)$ as the initial data, the preceding argument indicates that
\begin{equation} \label{eq:time-extension}
	{\mathbf{S}} \varphi(\cdot,s)(t) = \varphi(\cdot,t+s),\;\;\;s >0,\;\;\;t\geq0.
\end{equation}
Because $\varphi(\cdot,t) \in C((0,T], L^2(\real^3))$, we have the precompact set ${\mathscr K} = \{ {\varphi(\cdot,s),}$ ${0 < s \leq 1} \}$ $\subseteq L^{2}(\real^3)$. It follows that bounds (\ref{eq:vort-op-l2-linfty-bound}) and (\ref{eq:time-extension}) imply
\begin{equation*}
	\big\|\varphi(\cdot,s+t) \big\|_p \; {\leq} \; \vartheta(\nu t;{\mathscr K}) \: (\nu t)^{-3/4+{3}/{(2p)}} ,\;\;\;0<s<1,\;\;\;2<p\leq\infty.
\end{equation*}
As $s {\rightarrow} 0$ at fixed $t$, this inequality reduces to 
\begin{equation}
	\big\|\varphi(\cdot,t) \big\|_p \; {\leq} \; \vartheta(\nu t;{\mathscr K}) \: (\nu t)^{-3/4+{3}/{(2p)}}, \;\;\; 2<p\leq\infty,\;\;\;0<t<T^*.
\end{equation}
In particular, we recover the bounds in (\ref{eq:nabla-vel-linfty-bound-K}) and (\ref{eq:vel-linfty-bound-K}). Since $\vartheta(\nu t;{\mathscr K}) {\rightarrow} 0$ as $t {\rightarrow} 0$ and is continuous, we obtain 
\begin{equation}
	\lim_{t \rightarrow 0} \: (\nu t)^{3/4-{3}/{(2p)}} \: \big\|\varphi(\cdot,t) \big\|_p = 0,\;\;\; 2<p\leq\infty.
\end{equation}
This compatibility condition necessarily holds by every strong solution. It is obvious how to establish $\varphi(\cdot,t) = \tilde{\varphi}(\cdot,t) = {\mathbf{S}} \varphi_0(t) \; \forall t {>}0$.       
The method of prolongation can be used to extend the present results to any finite time $T$.

We summarize the conclusion for the well-posedness of the vorticity equation for $\omega_0 \in L^2(\real^3)$:
{\itshape The operator ${\mathbf{S}}$, defined for $\omega_0 \in C_c^{\infty}(\real^3)$, can be extended as a continuous operator as follows:
\begin{equation}
{\mathbf{S}}:\; L^{2}(\real^3) \rightarrow C((0,T], L^{2}(\real^3)) \cap C((0,T], H^{1}(\real^3) \cap L^{\infty}(\real^3)).
\end{equation}
For every $\omega_0 \in L^{2}(\real^3)$ and $\forall t \in (0,T]$, the function $\omega(t) {=} ({\mathbf{S}} \omega_0)(t)$ and the associated velocity $u$ satisfy the vorticity equation weakly in $\real^3{\times}(0,T]$. In addition, there exist
constants $C_i {>}0, \; i{=}0,1,2$ such that for all $t > 0$,
\begin{align}
& \big\| \omega(\cdot,t) \big\|_{L^{\infty}(\real^3)} \: {\leq} \: C_0 \: (\nu t)^{-3/4} \: \big\| \omega_0 \big\|_{L^{2}(\real^3)}, \\ 
& \big\| u(\cdot,t) \big\|_{L^{\infty}(\real^3)} \: {\leq} \: C_1 \: (\nu t)^{-1/4} \: \big\| \omega_0 \big\|_{L^{2}(\real^3)} \: \big( 1 + C_2 \big\| \omega_0 \big\|_{L^{2}(\real^3)}\big).
\end{align}

Let $\tilde{\omega}$, $\tilde{u}(x){=}{\cal K} {*} \tilde{\omega}(x)$ be a weak solution in $\real^3{\times}(0,T]$ of
\begin{equation}
	{\partial \tilde{\omega}}/{\partial t} - \nu \Delta \tilde{\omega} = (\tilde{\omega} . \nabla )\tilde{u}  - (\tilde{u} . \nabla ) \tilde{\omega} = \nabla \times (\tilde{u} \times \tilde{\omega} ),
\end{equation}
where
\begin{equation}
	\tilde{\omega} \in C((0,T], L^{2}(\real^3) \cap L^{\infty}(\real^3)),
\end{equation}
and
\begin{equation}
	\tilde{\omega}(x,0) = \omega_0(x) \in L^{2}(\real^3). 
\end{equation}
Then $\tilde{\omega}(\cdot,t){=}({\mathbf{S}} \omega_0)(t) \: \forall t \in (0,T]$.

For every $\omega_0 \in L^{2}(\real^3)$, the function ${\omega}(\cdot,t){=}({\mathbf{S}} \omega_0)(t)$ is in $C^{\infty}(\real^3{\times}(0,T])$. The vorticity equation is satisfied classically by $\omega$ and $u$. 
Moreover, for every multi-index $\alpha$ and every integer $\beta{\geq}0$, the maps:
\begin{equation*} 
\partial^{\beta}_t \partial_x^{\alpha} {\mathbf{S}}: \: L^{2}(\real^3) \rightarrow C((0,T], L^{2}(\real^3) \cap L^{\infty}(\real^3)), 
\end{equation*}
are continuous. Hence the vorticity is regular. On the basis of (\ref{eq:nabla-p-linf-bound}) and (\ref{eq:mom-ie}), the velocity $u$ is also regular. The pressure gradient can be evaluated from (\ref{eq:dt-ddp}). Hence
\begin{equation*} 
\big| \partial^{\beta}_t \partial_x^{\alpha} u \big| < \infty, \;\;\; \big| \partial^{\beta}_t \partial_x^{\alpha} \nabla p \big| < \infty. 
\end{equation*}
Consequently, $u, \nabla p \;(\mbox{or}\; p) \; \in C^{\infty } (\real^3{\times}(0,T])$}. 

The weak solutions obtained satisfy the energy equality
\begin{equation*} 
	\frac{1}{2}\big\|u(t)\big\|^2_{L^2 (\real^3)} + \nu \int_0^t \!\!\! \big\|\omega (t) \big\|^2_{L^2 (\real^3)} \rd t  \; {=} \; \frac{1}{2}\big\|u_0 \big\|^2_{L^2 (\real^3)}.
\end{equation*}
By comparing with bound (\ref{eq:u2-by-mom-ie}), we see that our interpolation approach grossly over-estimates the kinetic energy when $\|\omega_0\|_2$ is considerably large.

So far we have been explicit in expressing our solutions in the parabolic cylinder which excludes $t=0$. We do not know, in general, what extra smoothness bounds at start of motion for the initial velocity need to be given in order to achieve smoothness back to the initial instant, as the Navier-Stokes system may be altered into an over-specified problem. The matter becomes more delicate in the presence of a solid boundary. The reasons for the difficulties as well as their remedies have been reappraised by Heywood (2007). In virtue of the Leray-Hopf local in-time solutions in $0 \leq t \leq T_L$ and in view of the uniqueness of our solutions, we assert that, by an alternative approach, such as the one used by Heywood, our solutions can be shown to assume the initial data continuously up to the starting instant $t=0$ and are globally regular in the cylinder $\real^3 \times [0,T]$. In this respect, we may view our conclusion as an extension of the classical local in-time solutions to arbitrary finite time interval $T_L < T < \infty$.
\subsection*{Extension to initial velocity data other than $H^1_0$}
Although we have dealt with the initial data $\omega_0 {\in} L^2$, our analysis is clearly applicable to the scale-invariant case $\omega_0 {\in} L^{3/2}$.  Consider the critical {\itshape a priori} bounds,
\begin{equation*}
	\nabla u \in L^{\infty}((0,T],L^{3/2}(\real^3))\;\;\; \mbox{and} \;\;\; u \in L^{\infty}((0,T],L^3(\real^3)).
\end{equation*}
According to the well-developed theory of parabolic differential equation of second order, the vorticity equation is well-posed.  It is straightforward to verify that, by H\"{o}lder's inequality, the integrals,
\begin{equation*}
	\sup_{(x,t) {\in} \real^3{\times}[0,T]} \int_{{t-\varepsilon}}^t \! \int {\mathbf Z}(x{-}y,t{-}s) \big|\nabla u(y,s) \big| \rd y \rd s \:{\leq}\:  C \: \big\| \nabla u \big\|_{L^{3/2}(\real^3)}  \log \big( t/(t{-}\varepsilon) \big),
\end{equation*}
and
\begin{equation*}
	\sup_{(x,t) {\in} \real^3{\times}[0,T]} \int_{t-\varepsilon}^t \! \int {\mathbf Z}(x{-}y,t{-}s) \big|u(y,s) \big| s^{-1/2} \: \rd y \rd s 
	\:{\leq}\:  C \: \big\| u \big\|_{L^{3}(\real^3)}  \log \big( t/(t{-}\varepsilon) \big),
\end{equation*}
vanish as ${\varepsilon}{\rightarrow}0$ so that functions, $u$ and $\nabla u$, belong to certain parabolic class (see, for example, Zhang 1995; Liskevich \& Semenov 2000). It follows that there exists a weak fundamental solution for the vorticity equation. The fundamental solution can be constructed by the parametrix method and is unique by an adjoint analysis. 
For $\omega_0 {\in} {\mathscr K} {\subseteq} C_c^{\infty}(\real^3)$, being precompact in $L^{3/2}(\real^3)$, the analogous bounds to (\ref{eq:vort-l2-linfty-bound}) and (\ref{eq:vel-linfty-bound-strong}) are
\begin{equation*} 
	\big\| \omega(\cdot,t) \big\|_{L^p(\real^3)} \: {\leq} \: C(T) \: \vartheta(\nu t;{\mathscr K}) \: (\nu t)^{-1+{3}/{(2p)}}, \;\;\; \forall p \in (1,\infty],
\end{equation*}
and
\begin{equation*} 
\big\| u(\cdot,t) \big\|_{L^{\infty}(\real^3)} \:{\leq}\: (\nu t)^{-1/2} \: \vartheta(\nu t;{\mathscr K}) \big( C_1(T) + C_2(T) \: \vartheta(\nu t;{\mathscr K}) \big)
\end{equation*}
respectively. Now the vorticity is characterized by the function $\vartheta$:
\begin{equation*}
	\big\| \omega(\cdot,t) \big\|_p \leq \vartheta(\nu t;{\mathscr K}) (\nu t)^{-1+{3}/{(2p)}} + C \int_0^t \! (\nu t{-} \nu s)^{-{3}/{2}+{3}/{(2r)}} \big\| \omega(\cdot,t)\big\|^2_p \: \rd s,
\end{equation*}
where $1/p{+}1/r{=}4/3$. The rest of proof is akin to that for $L^2$ initial data. We shall refrain from going into the technical detail. For the Navier-Stokes equations with the initial velocity $u_0 \in L^3(\real^3)$, Escauriaza {\em et al} (2003) show that the solutions are smooth. Their work is based on an analysis of an adjoint heat equation and Carleman's inequalities. 

For the sub-critical cases ($\omega_0 {\in} L^{r}(\real^3), 1{\leq}r{<}3/2$), the present approach does not seem to be adequate. In particular, for $\omega_0 {\in} L^1(\real^3)$, integral inequality (\ref{eq:vort-lp-int}) becomes
\begin{equation*}
\big\|\omega(\cdot,t) \big\|_p \:{\leq}\: C (\nu t)^{-3/2(1-1/p)} \big\|\omega_0 \big\|_1 + C \int_0^t  (\nu t{-} \nu s)^{-2+3/(2r)} \big\| \omega(\cdot,s) \big\|^2_p \rd s .
\end{equation*}
This inequality does not admit any regular bound for large $\nu t$. A possible reason for the failure is that the interpolation techniques used in our current theory may have underestimated the viscous smoothing effect on the non-linearity. A vorticity perturbation, no matter how small it may be, from the strongly singular initial data may be an ill-defined concept even in the locally Lebesgue-integrable topology. 

The Cauchy problem for fluid dynamics becomes mathematically tractable once coercive {\itshape a priori} bounds, such as the bounds (\ref{eq:du-time-bound-y}) and (\ref{eq:du-time-bound}), have been established. For initial velocity $u_0 {\in} L^2(\real^3)$, as specified in the Leray-Hopf weak formulation, there exist alternative methods to deduce global regularity in suitable function spaces (see, for example, Temam 1977; Doering \& Gibbon 1995; Heywood 2007).
\section{Vorticity equation as an integral equation}
Our mathematical solutions obtained so far are adequate in establishing the well-posedness of the Navier-Stokes equations. However, they do not appear to reveal any ubiquitous nature of turbulence, which is the most general form of fluid motions. Effectively, the results of our interpolation theory determine some upper bounds on the dynamic solutions without providing any details on the actual flow evolution. The state of affairs is not entirely satisfactory. In the next two sections, we introduce our second method of solution. In essence, we attempt to explore the possibility of solving the vorticity equation by construction. 

The analytic form on the left-hand side of vorticity equation (\ref{eq:vorticity}) is nothing more than a diffusion operator. It is then anticipated that the smoothing effects of viscosity are essential during the initial phase of the vorticity evolution no matter how large the initial data may be.
By Duhamel's principle and in view of the {\itshape a priori} bounds derived in the preceding sections, the vorticity equations can be converted into integral equations:
\begin{equation} \label{eq:duhamel-comp}
	\omega_i(x,t) = \varpi_i(x,t) + \int_0^{t} \int Z(x,t,y,s) \Big( \: (\omega . \nabla) u_{i}  - (u . \nabla )\omega_{i} \: \Big)(y,s)  \rd y \rd s,
\end{equation}
where $\varpi$ is called the caloric mollified initial vorticity. It is given by
\begin{equation} \label{eq:mollified-vort-ic}
	\varpi(x,t) = \int \mathbf{Z}(x{-}y,t) \: \omega_0(y) \rd y.
\end{equation}
The function $\varpi$ is the solution of the pure initial value problem of heat equation
\begin{equation*}
	\partial_t \varpi - \nu \Delta \varpi =0,\;\;\; \varpi(x,t{=}0)=\omega_0(x).
\end{equation*}
The initial value is postulated to be localized according to (\ref{eq:vort-ic}) or (\ref{eq:ic-localization}). Thus the mollified vorticity is unique, and 
\begin{equation*}
	\varpi(x,t) \; \in \; C^{\infty}\;\;\;\mbox{in}\;\;\;\real^3 \times [0,t\leq T].
\end{equation*}
Moreover, $\varpi$ is continuous in $\real^3 {\times} [0,t{\leq}T]$; $\varpi(x,t)=\omega_0(x)$ as $t\rightarrow0$ if $\omega_0$ is continuous. If $\omega_0 \in L^p(\real^3)$ for $1 {\leq} p {\leq} \infty$, then 
\begin{equation*}
	\big\| \varpi(x,t) - \omega_0(x) \big\|_{L^p} \: \rightarrow 0 \; \;\;\;\mbox{as}\;\;\; t \rightarrow 0.
\end{equation*}
For $t>0$, the following time-wise bound is well-established:
\begin{equation*}
	\big\| \varpi(\cdot,t) \big\|_{L^q(\real^3)}  \leq  C \: (\nu t)^{-3/2\:(1/r-1/q)} \; \big\| \omega_0 \big\|_{L^r(\real^3)}, 
\end{equation*}
where $1{\leq}r{\leq}q{\leq}{\infty}$ except the case $r{=}1, q{=}\infty$, and the constant $C{=}C(r,q,T)$.

The gradient operators $\nabla$ in the integrand of (\ref{eq:duhamel-comp}) can be smoothed out via integration by parts:
\begin{equation} \label{eq:duhamel-ie}
	\omega_i(x,t) = \varpi_i(x,t) - \int_0^{t} \int \Big( \: (\omega . \nabla Z) u_i  - (u . \nabla Z) \omega_i \: \Big)(y,s)  \rd y \rd s,\;\;\; i=1,2,3.
\end{equation}
The boundary term arising from the integration,
\begin{equation*}
	\int_0^t \Big| Z (u \omega_i - \omega u_i) \Big|_{|y|{\rightarrow}\infty} \rd s,
\end{equation*}
vanishes in view of the {\itshape a priori} $L^p$-bounds on $\omega$, and of the decay properties of $Z$ at infinity, for any $0 < s < t \leq T $. 
\subsection{Transformation of vorticity integral equation}
For given time $T$, we divide the time interval $[0,T]$ into $n{+}1$ equal sub-intervals such that
\begin{equation} \label{eq:time-intervals}
	0 \;{<}\; t_0 \;{<}\; t_1\; {<}\; t_2 \;{<}\; \; {\cdots} \; \;{<}\; t_k \;{<}\; \; {\cdots} \; t_{n-1} \;{<}\; t_n = t \leq T.
\end{equation}
We have chosen the equal intervals for the sake of convenience. It is equally admissible to use unequal ones. For any $0{\leq}s{<}t_k, \: k{=}0,1,2,{\cdots},n$, equations (\ref{eq:duhamel-ie}) hold, namely,
\begin{equation} \label{eq:vort-k00}
	\omega_i(x,t_k) =  \varpi_i(x,t_k) + 
	\int_0^{t_k} \!\!\! \int \Big( (\omega_i u - u_i\omega).\nabla_y Z(x,t_k,y,s) \Big)\: \rd y \rd s.
\end{equation}
Making use of the Biot-Savart relation (\ref{eq:biot-savart}), these equations can be rewritten as
\begin{equation} \label{eq:vort-k}
	\omega_i(x,t_k) = \varpi_i(x,t_k) +  \int_0^{t_k} \!\! \int \!\! \int   \sum_{j=1}^3 G_{ij}(x,t_k,y,s,z,s) \omega_j(z,s)\omega(y,s)  \rd z \rd y \rd s.
\end{equation}
For every vorticity component, we use notation $G_{ij}$ to denote the ${1 {\times} 3}$ row matrix, where
\begin{equation*}
	G_{ij}=\big( \; \alpha_{ij} \;\;\; \beta_{ij} \;\;\; \gamma_{ij} \; \big).
\end{equation*}
The elements are the products of heat kernel's derivatives and the derivatives of Newtonian potential embed in the Biot-Savart formula. 
In particular, only the vorticity with the independent variable $z$ is related to the velocity. Thanks to a symmetry in expression $\omega_i u  -  u_i\omega$, there are cancellations among the nine terms in the sum in (\ref{eq:vort-k}). For each vorticity, one term associated with the vorticity convection neutralizes the identical term related to the vorticity stretching. The cancellations are a result of the kinematics in the vorticity field; they express the fact that component $\omega_i$ cannot induce any velocity on component $u_i$. 
\subsection*{Structure matrix $G$}
It is convenient to introduce some abbreviations:
\begin{equation} \label{eq:abbrev}
	Z_i = Z_i(x,\tau,y,s) = \frac{\partial Z(x,\tau,y,s)}{\partial y_i}, \;\;\; N_i = N_i(y,z) = \frac{\partial {\cal N}(y{-}z)}{\partial y_i},\;\;i=1,2,3,
\end{equation} 
where ${\cal N} (x){=} {(4 \pi)^{-1}}{|{x}|^{-1}}$ is the Newtonian potential, the spatial differentiations on ${\cal N}$ are taken at every instant of time $s$.
In terms of these notations, the elements of $G_{ij}$ are given by
\begin{equation} \label{eq:gij}
\left. \begin{aligned} 
	{\alpha}_{11} & = Z_2 N_3{-}Z_3N_2 & {\alpha}_{12} & = 0, & {\alpha}_{13} & = 0, \\
	{\beta}_{11}  & = Z_3N_1, & {\beta}_{12} & = Z_2N_3, & {\beta}_{13}    & =Z_3N_3, \\
	{\gamma}_{11} & = -Z_2N_1,  & {\gamma}_{12}& = -Z_2N_2,  & {\gamma}_{13}   &= -Z_3N_2, \\
	& & & & & \\
	{\alpha}_{21} & = -Z_1N_3  & {\alpha}_{22}& = -Z_3N_2,  & {\alpha}_{23} & = -Z_3N_3, \\
	{\beta}_{21}  & = 0,       & {\beta}_{22} & = Z_3N_1{-}Z_1N_3, & {\beta}_{23}  & = 0, \\
	{\gamma}_{21} & = Z_1N_1,  & {\gamma}_{22}& = Z_1N_2,  & {\gamma}_{23} &= Z_3N_1, \\ 
	& & & & & \\
	{\alpha}_{31} & = Z_1N_2  & {\alpha}_{32}& = Z_2N_2,  & {\alpha}_{33} & = Z_2N_3, \\
	{\beta}_{31}  & = -Z_1N_1,  & {\beta}_{32} & = -Z_2N_1,   & {\beta}_{33}  & = -Z_1N_3, \\
	{\gamma}_{31} & = 0,       & {\gamma}_{32}& = 0,  & {\gamma}_{33} &= Z_1N_2{-}Z_2N_1. 
 \end{aligned}
 \right \} 
\end{equation}
They are characterized by the geometry of the problem under consideration. Both the stretching as well as the convection contribute to the synthesis of the matrix.

Consider the case $k{=}0$, for $0 {\leq} s {<} t_0$. The integral equations read
\begin{equation} \label{eq:vort-k0}
 \begin{split}
	\omega_i(x,t_0) & =  \varpi_i(x,t_0) + \int_0^{t_0} \!\! \int \!\! \int  \sum_{j=1}^3 G_{ij}(x,t_0,y,s,z,s) \omega_j(z,s)\omega(y,s)  \rd z \rd y \rd s \\
	\quad & =  \varpi_i(x,t_0) + g^{(0)}_i(x,t_0).
 \end{split}
\end{equation}
At time $t_1> t_0$, for $i{=}1, j{=}1$, we pre-multiply equation (\ref{eq:vort-k0}) by 
\begin{equation} \label{eq:pre-factor}
	G_{1j}(x_1,t_1,y_1,t_0,x,t_0)
\end{equation}
and post-multiply the resulting product by
\begin{equation} \label{eq:post-factor}
	\omega(y_1,t_0).
\end{equation}
We carry out similar multiplications for $j{=}2,3$. Adding the resulting three products and integrating the sum over space, and over time from $0$ to $t_1$, we obtain the integral identity,
\begin{equation*} 
 \begin{split}
 \int_0^{t_1} & \!\! \int \!\! \int   \sum_{j=1}^3 G_{1j}(x_1,t_1,y_1,t_0,x,t_0) \omega_j(x,t_0)\omega(y_1,t_0)  \rd x \rd y_1 \rd t_0 \\
	= &\int_0^{t_1} \!\! \int \!\! \int  \sum_{j=1}^3 G_{1j} (x_1,t_1,y_1,t_0,x,t_0) 
	\Big( \varpi_j(x,t_0)+g_j^{(0)}(x,t_0) \Big)  \omega(y_1,t_0) \rd x \rd y_1 \rd t_0.
	\end{split}
\end{equation*}
Renaming the independent variables $x {\rightarrow} z$, $y_1 {\rightarrow} y$, $x_1 {\rightarrow} x$ and $t_0 {\rightarrow} s$ $(0{<}s{<}t_1)$, the last displayed equation reduces to
\begin{equation*} 
 \begin{split}
	\omega_1(x,t_1) = \varpi_1(x,t_1) + \int_0^{t_1} \!\! \int \!\!  \int \sum_{j=1}^3 G_{1j} &  (x,t_1,y,s,z,s) \\
	& \Big( \varpi_j(z,s)+g_j^{(0)}(z,s) \Big)  \omega(y,s) \rd z \rd y \rd s. 
	\end{split}
\end{equation*}
It has been implicitly assumed that the time independent variable in $g^{(0)}$ has also been renamed accordingly. 
\subsection*{Capacity matrix $K$}
The three terms involving $G_{1j}$, $\varpi_j$ and $\omega$ in the integrand can be combined into one term as follows:
\begin{equation}
 \begin{split}
	(\alpha_{11} \varpi_1  + \alpha_{12} \varpi_2 + \alpha_{13} \varpi_3) \omega_1 \:+\: &
	(\beta_{11} \varpi_1    + \beta_{12} \varpi_2 + \beta_{13} \varpi_3) \omega_2 \:+\: \\
	&(\gamma_{11} \varpi_1   + \gamma_{12} \varpi_2 + \gamma_{13} \varpi_3) \omega_3 
	= \sum_{j=1}^3 K_{1j} \omega_j.
 \end{split}	
\end{equation}
The new kernel $K_{1j}$ is a function of the initial vorticity and has the form
\begin{equation}
	K_{1j}(x,t_1,y,s)= \int G_{1j}(x,t_1,y,s,z) \varpi(z) \rd z.
\end{equation}
We repeat the procedures of the multiplications and the integration by replacing $G_{1j}$ in (\ref{eq:pre-factor}) by $G_{2j}$ and $G_{3j}$ in turn, while keeping the vorticity factor of (\ref{eq:post-factor}) at the same space-time location. We obtain the first transformation for the vorticity components, namely,
\begin{equation} \label{eq:vort-k1}
	\omega_i(x,t_1) = \varpi_i(x,t_1) + \int_0^{t_1} \!\! \int  \sum_{j=1}^3 K_{ij} (x,t_1,y,s) \omega_j(y,s) \rd y \rd s + g^{(1)}_i(x,t_1),
\end{equation}
where 
\begin{equation*}
	g^{(1)}_i(x,t_1)=\int_0^{t_1} \!\! \int \!\! \int  \sum_{j=1}^3  G_{ij}(x,t_1,y,s,z,s) g^{(0)}_j(z,s) \omega(y,s) \rd z \rd y \rd s.
\end{equation*}
For convenience, the integral terms in (\ref{eq:vort-k1}) are denoted by
\begin{equation*} 
	h^{(1)}_i(x,t_1) = \int_0^{t_1} \!\! \int  \sum_{j=1}^3 K_{ij} (x,t_1,y,s) \omega_j(y,s) \rd y \rd s.
\end{equation*}

Every element of $K_{ij}$ is proportional to the product of two factors,
\begin{equation*} 
	a(x,\tau,y,s) \;\;\; \mbox{and} \;\;\; b(y;s).
\end{equation*}
The first factor $a$ is related to the derivatives of the heat kernel
\begin{equation*} 
	\frac{x_i-y_i}{(\tau-s)^{5/2}}\: \exp \Big(-\frac{|x{-}y|^2}{4{\nu}(\tau{-}s)} \: \Big).
\end{equation*}
The second is the double layer potential due to the moment of the mollified initial vorticity and has the representative form   
\begin{equation*} 
	b(y) = C \int \frac{y_j-z_j}{|y-z|^3} \: \varpi (z) \rd z
\end{equation*}
at every instant $s < \tau$. In fact, the kernel $K$ is a $3 {\times} 3$ matrix whose elements are evaluated in terms of $G_{ij}$:
\begin{equation*}
	\int \left( \begin{array}{ccc}
G_{11}\varpi_1 & G_{12}\varpi_2 & G_{13}\varpi_3 \\
 G_{21}\varpi_1 & G_{22}\varpi_2 & G_{23}\varpi_3 \\
G_{31}\varpi_1 & G_{32}\varpi_2 & G_{33}\varpi_3
\end{array} 
\right) \rd z,
\end{equation*} 
where $G_{ij}$'s are given by (\ref{eq:gij}). 

Let $\psi$ be the Newtonian potential associated with the mollified vorticity $\varpi$. It satisfies the second order elliptic equation $\Delta \psi(y) {=} {-}\varpi(y)$. The solution is related to the harmonic function $\cal N$ 
\begin{equation} \label{eq:sol-poisson-by-f}
	\psi(y) = - \int {\cal N}(y{-}z) \varpi(z) \rd z.
\end{equation}
The spatial derivatives of the solution are the contributions to the induced velocity due to the vorticity. We extend the use of abbreviations (\ref{eq:abbrev}) to the velocity 
\begin{equation} \label{eq:ui-def}
	U_i(z) = -\int \frac{\partial {\cal N}(y{-}z)}{\partial y_i} \varpi_1(z) \rd z, \;\;\; i=1,2,3.
\end{equation}
Similarly we introduce notations
\begin{equation} \label{eq:vi-wi-def}
	V_i(y) = -\int \frac{\partial {\cal N}(y{-}z)}{\partial y_i} \varpi_2(z) \rd z \;\;\; \mbox{and} \;\;\;
	W_i(y) = -\int \frac{\partial {\cal N}(y{-}z)}{\partial y_i} \varpi_3(z) \rd z.
\end{equation}
Now the matrix kernel $K{=}K(x,t,y,s)$ is given by
\begin{equation} \label{eq:k-elements}
	\left( \begin{array}{ccc}
Z_2 U_3 {-} Z_3 U_2 \!&\! Z_3V_1{+}Z_2V_3{+}Z_3V_3 \!&\! {-}Z_2W_1{-}Z_2W_2{-}Z_3W_2 \\
 & & \\
{-}Z_1U_3{-}Z_3U_2{-}Z_3U_3 \!&\! Z_3V_1{-}Z_1V_3 \!&\! Z_1W_1{+}Z_1W_2{+}Z_3W_1 \\
& & \\
Z_1U_2{+}Z_2U_2{+}Z_2U_3 \!&\! {-}Z_1V_1{-}Z_2V_1{-}Z_1V_3 \!&\! Z_1W_2{-}Z_2W_1 
\end{array} 
\right).
\end{equation} 
For structure matrix $G$, the elements are completely determined once the initial vorticity is specified. It is essential to note that even if the initial vorticity is specified on a lower dimension, the kernel $K$ always transforms a vorticity vector into three-dimensional space. 
Familiarly, the velocity calculated from solution (\ref{eq:sol-poisson-by-f}) can be expressed as the Biot-Savart relation 
\begin{equation*}
	\nabla {\times} \psi(y) = - \int \nabla_y {\cal N}(y{-}z) {\times} \varpi(z) \rd z = \bar{U}(y).
\end{equation*}
Explicitly, we denote the components of $\bar{U}$ by $(\bar{u},\bar{v},\bar{w})$. They are linked to the shorthand notations in (\ref{eq:ui-def}) and in (\ref{eq:vi-wi-def}) by
\begin{equation*}
	\bar{u} = W_2 - V_3,\;\;\;\bar{v} = U_3 - W_1, \;\;\;\bar{w} = V_1 - U_2.
\end{equation*}
We give some properties of the matrix $K$. The trace of $K$ is found to be
\begin{equation*}
	tr(K)= Z_1 \bar{u} + Z_2 \bar{v} + Z_3 \bar{w}.
\end{equation*}
The sums of the rows of $K$ are written in the row matrix
\begin{equation*}
	S_r=\big( \;Z_2(\bar{v}{-}\bar{u})+Z_3(\bar{w}{-}\bar{u}), \; Z_3(\bar{w}{-}\bar{v})+Z_1(\bar{u}{-}\bar{v}), \; Z_1(\bar{u}{-}\bar{w})+Z_3(\bar{v}{-}\bar{w})\; \big).
\end{equation*}
Similarly, the sums of the columns are
\begin{equation*}
	\begin{split}
	S_c= \big(\;  (Z_1{+}Z_2{-}2Z_3)U_2 &+ (2Z_2{-}Z_1{-}Z_3)U_3, \\
	\quad  (Z_2{+}Z_3{-}2Z_1)V_3  &+ (2Z_3{-}Z_1{-}Z_2)V_1, \\
	\quad  (Z_3{+}Z_1{-}2Z_2)W_1  &+ (2Z_1{-}Z_3{-}Z_2)W_2 \; \big).
	\end{split}
\end{equation*} 
\subsection*{Complete reduction of non-linearity}
For $t_1{<}t_2$, for $i,j{=}1,2,3$ in turn, we repeat the process of the transformation on (\ref{eq:vort-k1}) using pre-factor
\begin{equation*}
	G_{ij}(x_1,t_2,y_1,t_1,x,t_1)
\end{equation*}
and post-factor 
\begin{equation*}
	\omega(y_1,t_1).
\end{equation*}
Collecting the resulting terms, we get
\begin{equation} \label{eq:vort-k2}
  \begin{split}
	\omega_i(x,t_2) = \varpi_i(x,t_2) + \int_0^{t_2} \!\! \int  & \sum_{j=1}^3  K_{ij} (x,t_2,y,s) \omega_j(y,s) \rd y \rd s \\ 
	& + g^{(2)}_i(x,t_2) + h^{(2)}_i(x,t_2),
	\end{split}
\end{equation}
where
\begin{equation*}
	g^{(2)}_i(x,t_2)=\int_0^{t_2} \!\! \int \!\! \int  \sum_{j=1}^3   G_{ij}(x,t_2,y,s,z,s) g^{(1)}_j(z,s) \omega(y,s)  \rd z \rd y \rd s,
\end{equation*}
and 
\begin{equation} \label{eq:vort-h2}
	h^{(2)}_i(x,t_2)=\int_0^{t_2} \!\! \int \!\! \int   \sum_{j=1}^3  G_{ij}(x,t_2,y,s,z,s) h^{(1)}_j(z,s) \omega(y,s)  \rd z \rd y \rd s.
\end{equation}
Furthermore, we perform the transformation for $t_2{<}t_3$ on (\ref{eq:vort-k2}) once more, we obtain
\begin{equation} \label{eq:vort-k3}
  \begin{split}
	\omega_i(x,t_3) = \varpi_i(x,t_3) + \int_0^{t_3} \!\! \int & \sum_{j=1}^3 K_{ij}(x,t_3,y,s) \omega_j(y,s) \rd y \rd s \\ 
	& + g^{(3)}_i(x,t_3) + h^{(3)}_i(x,t_3) + h^{(2)}_i(x,t_3),
	\end{split}
\end{equation}
where
\begin{equation*}
	g^{(3)}_i(x,t_3)=\int_0^{t_3} \!\! \int \!\! \int  \sum_{j=1}^3   G_{ij}(x,t_3,y,s,z,s) g^{(2)}_j(z,s) \omega(y,s)  \rd z \rd y \rd s,
\end{equation*}
and
\begin{equation*} 
	h^{(3)}_i(x,t_3)=\int_0^{t_3} \!\! \int \!\! \int   \sum_{j=1}^3  G_{ij}(x,t_3,y,s,z,s) h^{(2)}_j(z,s) \omega(y,s)  \rd z \rd y \rd s.
\end{equation*}
Suppose that after $k$ times of the  transformation, the vorticity is given by
\begin{equation} \label{eq:vort-kk}
  \begin{split}
	\omega_i(x,t_k) = \varpi_i(x,t_k) + \int_0^{t_k} \!\! \int & \sum_{j=1}^3 K_{ij}(x,t_k,y,s) \omega_j(y,s) \rd y \rd s \\ 
	& + g^{(k)}_i(x,t_k) + \sum_{m=2}^k h^{(m)}_i(x,t_k),
	\end{split}
\end{equation}
where the new non-linear terms are
\begin{equation*}
	g^{(k)}_i(x,t_k)=\int_0^{t_k} \!\! \int \!\! \int  \sum_{j=1}^3  G_{ij}(x,t_k,y,s,z,s) g^{(k-1)}_j(z,s) \omega(y,s)  \rd z \rd y \rd s,
\end{equation*}
and
\begin{equation} \label{eq:vort-hk}
	h^{(k)}_i(x,t_k)=\int_0^{t_k} \!\! \int \!\! \int   \sum_{j=1}^3  G_{ij}(x,t_k,y,s,z,s) h^{({k-1})}_j(z,s) \omega(y,s)  \rd z \rd y \rd s.
\end{equation}
We carry out a further transformation on (\ref{eq:vort-kk}). The result on the last term becomes
\begin{equation*}
\int_0^{t_{k+1}} \!\! \int \!\! \int  \sum_{j=1}^3 G_{ij}(x,t_{k+1},y,s,z,s) \sum_{m=2}^k h^{(m)}_i(z,s) \omega(y,s) \rd z \rd y \rd s {=} \sum_{m=3}^{k+1} h^{(m)}_i(x,t_{k+1}).
\end{equation*}
The second term is redefined as
\begin{equation*}
  \int_0^{t_k} \!\! \int \sum_{j=1}^3 K_{ij}(x,t_k,y,s) \omega_j(y,s) \rd y \rd s = h_i^{(1)} (x,t_{k}).
\end{equation*}
This integral equality is then converted into $h_i^{(2)}(x,t_{k+1})$. Hence we obtain
\begin{equation*} 
  \begin{split}
	\omega_i(x,t_{k+1}) = \varpi_i(x,t_{k+1}) + \int_0^{t_{k+1}} \!\! \int & \sum_{j=1}^3 K_{ij}(x,t_{k+1},y,s) \omega_j(y,s) \rd y \rd s \\ 
	& + g^{(k+1)}_i(x,t_{k+1}) + \sum_{m=2}^{k+1} h^{(m)}_i(x,t_{k+1}),
	\end{split}
\end{equation*}
where
\begin{equation*}
	g^{(k+1)}_i(x,t_{k+1})=\int_0^{t_{k+1}} \!\! \int \!\! \int  \sum_{j=1}^3   G_{ij}(x,t_{k+1},y,s,z,s) g^{(k)}_j(z,s) \omega(y,s)  \rd z \rd y \rd s,
\end{equation*}

\begin{equation*} 
	h^{(k+1)}_i(x,t_{k+1})=\int_0^{t_{k+1}} \!\! \int \!\! \int   \sum_{j=1}^3  G_{ij}(x,t_{k+1},y,s,z,s) h^{({k})}_j(z,s) \omega(y,s)  \rd z \rd y \rd s.
\end{equation*}

After $n$ similarity transformations ($n{\geq}2$), equations (\ref{eq:vort-k0}) have been converted into
\begin{equation} \label{eq:vort-vie-comp}
	\omega_i(x,t) = \varpi_i(x,t) + \int_0^t \!\! \int \sum_{j=1}^3 K_{ij}(x,t,y,s) \omega_j(y,s) \rd y \rd s + q_i(x,t),
\end{equation}
where
\begin{equation} \label{eq:vort-vie-comp-q}
	q_i(x,t) = g^{(n)}_i(x,t) +  \sum_{k=2}^{n} \;  h^{(k)}_i(x,t).
\end{equation}
To simplify notations in our exposition, we write (\ref{eq:vort-vie-comp}) in vector form
\begin{equation} \label{eq:vort-VIE}
	\omega(x,t) = \varpi(x,t) + \int_0^t \!\! \int K(x,t,y,s) \omega(y,s) \rd y \rd s + q(x,t).
\end{equation}
In non-linear term $q(x,t)$, we have implicitly used $G{=}G(x,t,y,s,z,s)$ for the matrix $G_{ij}$.
\subsection{Properties of the integral kernels}
The mathematical structure of (\ref{eq:vort-vie-comp}) suggests that we can keep track of the complete evolution of any fluid motion by piecing together a sequence of flow development resulting over an incremental time, which can be as small as convenient.  As the time intervals in (\ref{eq:time-intervals}) can be divided into arbitrarily small steps, we wish to determine the limit of this time divisibility as $n{\rightarrow} {\infty}$. Therefore, we must first establish some properties of the capacity matrix $K$ and the structure kernel $G$.

Recall that the vorticity is {\itshape a priori} bounded. Let
\begin{equation} \label{eq:vort0-y-norm}
	A_0 = \sup_{x \in \real^3} \big(\: \|\omega_0\|_{L^1(\real^3)}+\|\omega_0\|_{L^\infty(\real^3)} \: \big).
\end{equation}

We note that real-valued function $\beta^\alpha \re^{-\beta}$ reaches its maximum $\alpha^\alpha \re^{-\alpha}$ at $\alpha{=}\beta$ for $0{\leq}\beta{<}\infty$ and $\alpha{>}0$. Let $\beta{=}|x{-}y|^2/[4{\nu}(\tau{-}s)]$ for $\tau{>}s$. We readily see that
\begin{equation} \label{eq:kernel-a-bound}
	\frac{|x-y|}{(\tau-s)^{5/2}}\: \re^{-\beta} =\frac{(4 \nu)^{\kappa-5/2}}{(\tau-s)^{\kappa}\;|x-y|^{4-2\kappa}} \beta^{5/2-\kappa} \re^{-\beta}\;{<}\;\frac{C}{(\tau-s)^{\kappa}\;|x-y|^{4-2\kappa}},
\end{equation}
for some constant $C$, and a parameter $\kappa < 1$. In view of the embedding $C^{\infty} {\subset} C^{0,1}$, we consider the case that the mollified initial vorticity $\varpi$ is Lipschitz continuous
\begin{equation*}
	\varpi \in C^{0,1}(\real^3).
\end{equation*}
Such a choice is contrived to simplify our analysis. The derivative of the heat kernel is bounded by
\begin{equation} \label{eq:del-heat-bound}
	\big| \nabla Z (x,t,y,s) \big| \leq C (\nu t {-} \nu s)^{-1/2} |x{-}y|^{-3} \exp\Big( {-} \nu^* \frac{|x{-}y|^2}{4 \nu (t{-}s)}\Big),
\end{equation}
where $C$ is constant, and $0{<}\nu^*{<}1$ (see, for example, Friedman 1964; Ladyzhenskaya 
{\itshape et al} 1968; Edel'man 1969).

Let $\psi_1(x)$ be the Newtonian volume potential of the mollified vorticity $\varpi$. It satisfies Poisson's equation $\Delta \psi_1(x) {=} {-} \varpi(x)$
for every fixed value of $s {\in} [0,T]$ in the light of the incompressibility hypothesis.
It follows that every component of the velocity vector function, $b(x;s){=} {-}{\partial \psi_1}/{\partial x}$, can be calculated from the derivatives of the potential, namely,
\begin{equation} \label{eq:kernel-b}
	b_i(x) = - \int \frac{\partial}{\partial x_i} {\cal N}(x-y) \varpi(y) \rd y.
\end{equation}
By virtue of Schauder's estimates (see, for example, Gilbarg \& Trudinger 1998; Evans 2008), it is straightforward to  establish that the velocity functions in (\ref{eq:ui-def}) and (\ref{eq:vi-wi-def}) are all continuous
\begin{equation*} 
U_i, \; V_i,\; W_i \; \in C^{1,1}(\real^3),\;\;\; i=1,2,3.
\end{equation*}
Moreover, by the partition of $\real^3$, we deduce that 
\begin{equation} \label{eq:b-bound}
	\| \:b(x)\: \|_{L^\infty(\real^3)} \leq C A_0,
\end{equation}
where $C$ denotes a constant. In view of (\ref{eq:del-heat-bound}) and (\ref{eq:b-bound}), there exists some constant $C$ such that 
\begin{equation} \label{eq:k-int-bound}	
\big| K(x,t,y,s) \big| \leq C A_0 (\nu t - \nu s)^{-1/2}|x-y|^{-3} \exp\Big( - \nu' \frac{|x{-}y|^2}{4 \nu (t{-}s)}\Big),
\end{equation}
where $0{<}\nu'{<}1$. It follows that the singularity of $K$ is integrable in space and in time. Consequently, for any integrable function $f(x,t)$ which is essentially bounded, we have
\begin{equation*}
	\Big| \int K(x,t,y,s) f(y,s) \rd y \Big| \leq C A_0 (\nu t - \nu s )^{-1/2} \: \big\| f \big\|_{L^{\infty}},
\end{equation*}
$C$ is a constant. More generally, we verify that
\begin{equation*} 
\big| K(x,t,y,s) \big| \leq A_0 \: (\nu t - \nu s)^{-\mu'}\;|x-y|^{2\mu'-4} \exp\Big( - \nu' \frac{|x{-}y|^2}{4 \nu (t{-}s)}\Big),
\end{equation*}
where $0{\leq}\mu'{\leq}2$, and $0{<}\nu'{<}1$. Let $H$ denote the resolvent kernel of $K$. It satisfies
\begin{equation} \label{eq:resolvent-int}
\begin{split}
	H(x,t,y,s)-K(x,t,y,s) & = \int_{s}^{t} \!\! \int H(x,t,z,\tau) K(z,\tau,y,s) \rd z \rd \tau \\
	\quad & = \int_{s}^{t} \!\! \int K(x,t,z,\tau) H(z,\tau,y,s) \rd z \rd \tau 
	\end{split}
\end{equation}
(see, example, Volterra 1930; Tricomi 1957; Courant \& Hilbert 1966; Miller 1971).
The explicit form of the resolvent kernel is given by 
\begin{equation} \label{eq:resolvent}
	H(x,t,y,s) = \sum_{i=0}^{\infty} K_{i+1}(x,t,y,s)\;\;\; \mbox{a.e.}
\end{equation}
The iterated kernels (denoted by $K_i$) are related to the kernel $K$ by
\begin{equation} \label{eq:resolvent-sum}
\begin{split}
K_{i+1}(x,t,y,s) & = \int_{s}^{t} \!\! \int K(x,t,z,\tau) K_{i}(z,\tau,y,s) \rd z \rd \tau, \\
\quad & = \int_{s}^{t} \!\! \int K_{j}(x,t,z,\tau) K_{l}(z,\tau,y,s) \rd z \rd \tau, \\
K_1(x,t,y,s)& =K(x,t,y,s), \;\;\;\;\;\; j{=}1,2,{\cdots},i,\;l{=}i{+}1{-}j.
\end{split}
\end{equation}
In general, the resolvent kernel is bounded by
\begin{equation} \label{eq:h-bound}	
\big| H(x,t,y,s)\big| \leq H_0 \: (\nu t - \nu s)^{-\mu^*}\;|x-y|^{2\mu^*-4}\exp\Big( {-} \lambda^* \frac{|x{-}y|^2}{4 \nu (t{-}s)}\Big), 
\end{equation}
where the constant,
\begin{equation*}
	H_0=H_0(A_0),
\end{equation*}
for given $T$, and $0{<}\lambda^*{<}1$. Choosing $\mu^*{=}1$, we obtain
\begin{equation} \label{eq:h-int-bound}	
	\Big| \int H(x,t,y,s) f(y,s) \rd y \Big| \leq C H_0 (\nu t - \nu s)^{-1/2} \: \big\| f \big\|_{L^{\infty}},
\end{equation}
where $C$ is a constant.

There exists a volume potential $\psi_2$ in $\real^3$ which is also governed by elliptic equation $	\Delta \psi_2 (x; s) {=} {-} f(x; s)$
at every instant of time. We then introduce a vector function $d$ such that its components are evaluated according to
\begin{equation*} 
d_i(x;s)=  - \int \frac{\partial}{\partial x_i} {\cal N}(x-y) f(y;s) \rd y.
\end{equation*}
Thus we obtain the bound, $ \|d\|_{L^\infty} \leq C ( \| f \|_{L^1} + \| f \|_{L^\infty} )$,
where $C$ is a constant. Since
\begin{equation*}
	\int G(x,t,y,s,z,s) f(z,s) \rd z = \nabla_y Z(x,t,y,s) d(y,s),
\end{equation*}
any integral convolution involving the structure kernel $G$ can be calculated:
\begin{equation} \label{eq:g-int-bound}
\begin{split}
	\Big| \int \!\! \int G(x,t,y,s,z,s) f(z,s) \rd z \rd y \Big|  & \leq  \Big| \int \nabla_y Z(x,t,y,s) d(y,s) \rd y \Big| \\
	& \leq C (\nu t - \nu s)^{-1/2} \big( \| f \|_{L^1(\real^3)} + \| f \|_{L^\infty(\real^3)} \big).
	\end{split}
\end{equation}

To determine the behaviour of the function $q$ in the limit $n{\rightarrow}\infty$, we verify that
\begin{equation*}
	\big| g^{(0)}(x,t) \big| \: {\leq} \: C A_0^2 \int_0^t \!\! \int \nabla Z(x,t,y,s) \rd y \rd s \leq C A_0^2 \: t^{1/2},
\end{equation*}
where $C$ is constant. The factor $A^2_0$ bears out the quadratic non-linearity in vorticity. From the definition of $g^{(1)}$, we compute the bound
\begin{equation*}
	\big| g^{(1)}(x,t) \big| \: {\leq} \: C A_0^3 \: t \: B(1,1/2),
\end{equation*}
where $B(p,q)$ denotes the Euler Beta function, and $C$ denotes a constant. The Beta function has the property
\begin{equation*}
	B(p,q)=\Gamma(p)\Gamma(q)/\Gamma(p+q). 
\end{equation*}
By induction, it is easy to show that
\begin{equation*}
	\big| g^{(k)}(x,t) \big| \: {\leq} \: C A_0^{k+2} \: t^{(k+1)/2} \: \big(\Gamma(1/2)\big)^{k}\: \big( \Gamma((k+2)/2) \big)^{-1}
\end{equation*}
for any $k{>}1$. Hence we deduce that
\begin{equation} \label{eq:iter-tail}
	g^{(k)} \: {\rightarrow} \: 0 \;\;\; \mbox{a.e.} \;\;\; \mbox{as} \;\;\; k \: {\rightarrow} \: \infty.
\end{equation}
As $n \rightarrow {\infty}$, equation (\ref{eq:vort-vie-comp-q}) reduces to
\begin{equation} \label{eq:iterated-q}
q(x,t)= \lim_{n {\rightarrow} \infty} \sum_{k=2}^{n} h^{(k)}(x,t).
\end{equation}
Consequently, the equation governing the vorticity dynamics has been transformed into an equivalent integral equation. 
The non-linear functions, $h^{(k)}$, are defined recursively in terms of the vorticity:
\begin{equation}
 \begin{split}
	h^{(k)}(x,t)& =\int_0^{t} \!\! \int \!\! \int  G(x,t,y,s_k,z,s_k) h^{(k-1)}(z,s_k) \omega(y,s_k)  \rd z \rd y \rd s_k, \\
	h^{(k-1)}(x,s_k)& =\int_0^{s_k} \!\! \int \!\! \int  G(x,s_k,y,s_{k-1},z,s_{k-1}) \\ 
	\quad & {\hspace{3cm}} h^{(k-1)}(z,s_{k-1}) \omega(y,s_{k-1})  \rd z \rd y \rd s_{k-1}, \\
	\quad & \cdots \\
	h^{(2)}(x,s_2)& =\int_0^{s_2} \!\! \int \!\! \int  G(x,s_2,y,s_1,z,s_1) h^{(1)}(z,s_1) \omega(y,s_1)  \rd z \rd y \rd s_1, \\
	h^{(1)}(x,s_1)& =\int_0^{s_1} \!\! \int K(x,s_1,y,s) \omega(y,s) \rd y \rd s,
	\end{split}
\end{equation}
where $s_k$'s satisfy $0{<}s{<}s_1{<}s_2 \cdots {<}s_{k-1}{<}s_k{<}t$. 
The non-linear term $q$ consists of space-time convoluted vorticity integrals of arbitrarily large order if $n \rightarrow \infty$. In essence, equation (\ref{eq:vort-VIE}) demonstrates the fact that the complete fluid motion is solely determined by the non-linearity in the vorticity equation. The entire flow evolution can be calculated by considering the superposition of those highly interactive and diffusive vorticity convolutions. Clearly, the actual magnitudes of these individual vorticity terms depend strongly on the initial vorticity matrix, particularly those small values of $k$ for the initial phase of the evolution. However, equation (\ref{eq:vort-VIE}) is not yet in a form for further analytic treatment. 
\subsection*{First-order approximation}
Instructively, we view equation (\ref{eq:vort-VIE}) as a Volterra-Fredholm {\itshape linear} integral equation of the second kind. If the term $q$ in (\ref{eq:vort-VIE}) is small enough so that its effect may be neglected, we obtain a first-order approximation $(0 \leq s < t)$ for fluid motions:
\begin{equation} \label{eq:vort-first-order-VIE}
	\omega(x,t) = \varpi(x,t) + \int_0^t \!\! \int K(x,t,y,s) \omega(y,s) \rd y \rd s,
\end{equation}
this is simply a linear Volterra-Fredholm integral equation in $\omega$. The solution is explicitly given by
\begin{equation} \label{eq:vort-sol-first-order-VIE}
	\omega(x,t) = \varpi(x,t) + \int_0^t \!\! \int H(x,t,y,s) \varpi(y,s) \rd y \rd s.
\end{equation}
Since the eigenvalue spectrum of any linear Volterra integral equation is always empty, it follows that there exist no bifurcations in the first-order approximation of the Navier-Stokes Cauchy problem. Attention should be paid to the fact that the capacity kernel $K$ is interlinked with the caloric mollified vorticity $\varpi$ which represents the commencement stage of the motion.
We reiterate the use of the description, ``first-order approximation'', in order to emphasize the fact that equation (\ref{eq:vort-first-order-VIE}) {\itshape is not} a result of linearization of the equations of motion. The solution (\ref{eq:vort-sol-first-order-VIE}) describes the dynamic evolution of the linearly diffused initial vorticity over a short time interval from the starting of the motion. We shall address the issue of linearization in a later section.

As the last operation in our transformation process, let us return to (\ref{eq:vort-VIE}). Consider the linear integral equation
\begin{equation} \label{eq:voltIC-VIE}
	\gamma(x,t) = \varpi(x,t) + \int_0^{t} \!\! \int K(x,t,y,s) \gamma(y,s) \rd y \rd s.
\end{equation}
Its solution is given by, $(0{\leq}s{<}t)$,
\begin{equation} \label{eq:solIC-VIE}
	\gamma(x,t) = \varpi(x,t) + \int_0^{t} \!\! \int H(x,t,y,s) \varpi(y,s) \rd y \rd s. 
\end{equation}
In view of (\ref{eq:h-int-bound}), we are able to estimate the right-hand side. We find that
\begin{equation*}
	\big\| \gamma(\cdot,t) \big\|_{L^\infty} \leq \big\| \varpi \big\|_{L^\infty} + \big\| \varpi \big\|_{L^\infty} \int_0^t \!\! \int \big| H(x,s,y) \big| \rd y \rd s \leq C A_0 H_0 (\nu t)^{1/2}.
\end{equation*}
We assume that $(\nu t)^{1/2}H_0 {>} 1$ for $t{>}0$ as we are mainly interested in problems with large initial data. Similarly, we verify that, by Young's inequality for convolution, 
\begin{equation*}
	\big\| \gamma(\cdot,t) \big\|_{L^1} \leq \big\| \varpi \big\|_{L^1} + \big\| \varpi \big\|_{L^1} H_0 \int_0^t  (\nu t- \nu s)^{-1/2}  \rd s \leq C A_0  H_0 (\nu t)^{1/2}, 
\end{equation*}
where $C$ is a constant. The $\| \gamma \|_{L^q}$ bound ($1{<}q{<}\infty$) may be derived by interpolation. As the existence of bifurcations has been ruled out, the transformed vorticity (\ref{eq:vort-VIE}) can be expressed as 
\begin{equation} \label{eq:vort-nonlinear-VIE}
 \begin{split}
	\omega(x,t) & =  \gamma(x,t) + q(x,t) + \int_0^t \!\! \int H(x,t,y,s) q(y,s) \rd y \rd s \\
	& =  \gamma(x,t) + q(x,t) + \bar{q}(x,t).
	\end{split}
\end{equation}
The non-linear terms in the vorticity equation have been thoroughly transformed into the last two integral terms, each characterized by superposition of convoluted vorticity. The last term $\bar{q}$ is called the Volterra-Fredholm filtered non-linearity.  

{\itshape We have showed how the non-linearity in the vorticity integral equation can be converted into two infinite series by means of similarity transformation (or non-linearity reduction). Each of the series is a sum of spatio-temporal convolutions of the vorticity to arbitrarily large orders. The kernel of every convolution is regulated by the caloric mollified initial vorticity. Consequently, to solve the vorticity equation amounts to the determination of the convergence of these series when successive approximations are applied.}
\subsection*{Remarks}
The way we derive the solution of integral equation (\ref{eq:voltIC-VIE}) follows an orthodox and mundane approach which can be found in many standard textbooks on the theory of integral equations. To illustrate the motivation behind our similarity transformations, let us consider the linear equation
\begin{equation*} 
	f(t) = g(t) + \int_0^t k(t,s) f(s) \rd s,
\end{equation*}
where the kernel $k(t,s)$ is a Volterra kernel.
Intuitively, we solve this equation in the following manner: We multiply the equation by the kernel $k$ and integrate over time (represented by $\int'$) to obtain,
\begin{equation*} 
	f(t) = g(t) +  \int' k(t,s) g(s) \rd s + \int' k(t,s) \int' k(s,r) f(r) \rd r \rd s.
\end{equation*}
Continuing this process of reduction $n$ times, we write the result in the symbolic form,
\begin{equation*} 
	f(t) = g(t) +  \int' k g +  \int' k \int' k g + \int' k \int' k \int' k g +  \cdots \cdots + \phi_n(f). 
\end{equation*}
The tail of the summation $\big| \phi_n\big| \:{\leq}\: \max(f) \;k^n_0\; t^n / n!$,
where $k_0 {=} \sup | k |$, and $f$ is assumed bounded {\itshape a priori}. Evidently, the tail vanishes as $n \rightarrow \infty$.
The sum of the integral terms is nothing more than the convolutions of the kernel and the initial data $g$. If a proper definition of the iterated kernels (or integrals) is introduced, the convolution sum coincides with the resolvent of $k$. In the theory of linear integral equations, it is well-known that the resolvent kernel of a Volterra equation has a majorant which has an infinite radius of convergence while the resolvent of a Fredholm kernel converges only when its eigenvalue is less than the inverse of the norm of the kernel. 

The vanishing tail of the summation is a result of the boundedness of kernel $k$. In contrast, the limit of $g^{(n)} \rightarrow 0$ as $n \rightarrow \infty$ in (\ref{eq:iter-tail}) is a consequence of the {\itshape diffusion gradients}. As far as the equations of motion are concerned, the characters of flow evolution are determined by the effectuation of the gradients $\nabla Z$ and the non-linearity $(u.\nabla)u$. 

Liapunov (1906) attempted to solve a non-linear integral equation in order to establish the equilibrium figures of rotating homogeneous fluid in a sphere. An iterated integral, resembling those in (\ref{eq:vort-nonlinear-VIE}), is known as an integro-power series of infinite order (Liapunov 1906; Schmidt, 1908). Schmidt investigated a class of non-linear integral equation over a fixed domain so as to demonstrate the phenomena of bifurcations or branchings. Schmidt's integral equation is related to certain Dirichlet or Neumann {\itshape boundary value} problems, though he has not mentioned any particular application to physical problems. In particular, he showed that bifurcations do occur if linearized problems admit eigenvalues. The determination of bifurcation in integral equations is known as the Liapunov-Schmidt method.
The key difference between Schmidt's equation and (\ref{eq:vort-VIE}) lies in the presence of the time variable -- a consequence of {\itshape initial value} problems. Volterra (1930) referred the series of the iterated integrals, 
\begin{equation*}
\begin{split}
	G[f(x)] = k_0 +  \int_0^x \!\! & k_1(x,y) f(y) \rd y + \int_0^x \!\!\! \int_0^x \!\! k_2(x,y_1,y_2) f(y_1) f(y_2) \rd y_1  \rd y_2 +   {\cdots \cdots}	\\
\quad & + \int_0^x \!\!\! {\cdots} \!\! \int_0^x \!\! k_i(x,y_1,{\cdots},y_i) f(y_1) {\cdots} f(y_i) \rd y_1 {\cdots} \rd y_i + {\cdots \cdots},
\end{split}
\end{equation*}
as a functional power series. Every term in the series represents a regular homogeneous functional of degree $i$. The leading term $k_0$ is constant. All the kernels are considered to be symmetric with respect to the independent variables. Set $k_0{=}0$, a non-linear integral equation can be defined in the abbreviated form
\begin{equation*}
	f(x) + G[f(x)] = g(x),
\end{equation*}
where the right-hand side is given, and the functions $f$ and $g$ are continuous functions. He mentioned one application of this equation in elasticity (Volterra 1912). If the upper limits in all the functionals are fixed, the equation becomes the Schmidt non-linear integral equation with symmetric kernels.
\section{Construction of solution}
Although integral equation (\ref{eq:vort-nonlinear-VIE}) has a rather complex analytic character,  it provides us a recipe for the construction of the vorticity solution by successive approximation. Every convolution term in the series consists in a vortical flow of a particular size. We are mainly interested in those cases where the magnitude of the initial vorticity is of a moderate to large size. For $t<T$, we approximate every vorticity $\omega(x,t)$ by a series expansion in terms of the mollified initial data $\gamma(x,t)$. Then we examine the convergence of the resulting series in such an approximation scheme.  
\subsection{The formal solution}
The following expression is called an integro-power term of degree $m$ with respect to function $\phi$:
\begin{equation} \label{eq:ip-term}
  \begin{split}
\idotsint	{\mathbf K}(x,t,x_i,t_i,{\cdots},x_2,t_2,x_1,t_1) \phi^{k_1}(x_1,t_1)\phi^{k_2}&(x_2,t_2) {\cdots} \phi^{k_i}(x_i,t_i) \\
	\quad & \rd x_1 \rd t_1 \rd x_2 \rd t_2 {\cdots} \rd x_i \rd t_i,
	\end{split}
\end{equation}
where $k_1, k_2, {\cdots}$ are non-negative integers, and they satisfy the Diophantine equation
\begin{equation} \label{eq:diophantine}
 k_1+k_2+\cdots+k_i=m.
\end{equation}
The domains of the integral are understood to cover $\real^3$ in space and over the time interval $[0, t]$. The integral kernel ${\mathbf K}$ in (\ref{eq:ip-term}) is in the form of consecutive convolutions. The symbol $V[x,t;\phi]^m$ or simply $V[\phi]^m$ will be used to represent a generic integro-power {\itshape term} of degree $m$ with respect to the functional argument $\phi=\phi(x,t)$. The integro-power terms which have the same power $m$ but differ by their integral kernels are of the identical type. The number of different types of integro-power terms of degree $m$ is identical with the number of solutions to the Diophantine equation. It is more convenient to denote the integro-power terms using different letters or subscripts:
\begin{equation*}
	U[\phi]^m, \;\; V[\phi]^m,\;\; W[\phi]^m,\;\;V_2[\phi]^m,\;\;V_3[\phi]^m, \;\; \cdots.
\end{equation*}
Thus the rules of operation for the integro-power terms can be expressed in terms of these notations
\begin{equation} \label{eq:ip-term-op}
	U[\phi]^m\: V[\phi]^n = P[\phi]^{m+n},\;\;\;U[V[\phi]^m]^n=Q[\phi]^{mn}.
\end{equation}
The sum of a {\itshape finite} number of integro-power terms of degree $m$ of different types is called an integro-power {\itshape form} of degree $m$ with respect to $\phi$. It is denoted by
\begin{equation} \label{eq:ip-form}
	\sum_{k=1}^{S_m}\: V_k[\phi]^m = S_m \: V[\phi]^m = W_m\left( \begin{array}{c}
x,t  \\
\phi  
\end{array} 
\right), 
\end{equation}
where the integer $S_m$ is the number of different types of the integro-power terms of degree $m$. An integro-power series refers to the expression
\begin{equation} \label{eq:ip-series}
	\sum_{m=1}^{\infty}W_m\left( \begin{array}{c}
x,t  \\
\phi  
\end{array} 
\right).
\end{equation}
The concepts of integro-power type and form can be generalized to integro-power terms with more than one functional argument. For instance, we call
\begin{equation} \label{eq:ip-term-2arg}
	R_{mn}\left( \begin{array}{c}
x,t  \\
\phi,  \psi
\end{array} 
\right) 
\end{equation}
an integro-power form of degree $m$ with respect to argument $\phi{=}\phi(x,t)$ and of degree $n$ with respect to $\psi{=}\psi(x,t)$. An integro-power series for the integro-power form with more than two arguments can be defined accordingly. 

In terms of integro-powers, equation (\ref{eq:vort-nonlinear-VIE}) can be written as
\begin{equation} \label{eq:VIE-sum}
	\omega(x,t) = \gamma(x,t) + \sum_{m{\geq}2} W_m 
\left( \begin{array}{c}
x,t  \\
\omega
\end{array} 
\right).
\end{equation}
Let us set the leading term ($S_1=1$)
\begin{equation} \label{eq:sm-v1}
	V_1
	\left( \begin{array}{c}
x,t  \\
\omega  
\end{array} 
\right)=\gamma(x,t).
\end{equation}
In (\ref{eq:VIE-sum}), we substitute $\omega$ by $V_1$ on the right-hand side and collect the integro-power terms of second degree. Effectively we are deriving the integro-power form of degree $2$ which is just the sum
\begin{equation*}
	\sum_{k=1}^{S_2}\: V_k[\gamma]^2 = S_2 \:V[\gamma]^2.
\end{equation*}
Repeating this substitution procedure for 
\begin{equation*}
	V_j
	\left( \begin{array}{c}
x,t  \\
\omega  
\end{array} 
\right),\;\;\;j=2,3,{\cdots},
\end{equation*}
we replace every $\omega$ in (\ref{eq:VIE-sum}) by the sum
\begin{equation*}
	\sum_{j=1}^{S_m - 1} V_j[\gamma]^{m-1},
\end{equation*}
we collect the integro-power terms of degree $m$. The sum of these terms gives us the integro-power form
\begin{equation*} 
	V_m
	\left( \begin{array}{c}
x,t  \\
\gamma
\end{array} 
\right)=S_m\; V[\gamma]^m.
\end{equation*}
This substitution process enables us to obtain the ${\mathit formal}$ solution for $\omega$ in terms of $\gamma$ as an integro-power series,
\begin{equation} \label{eq:vort-formal-sol}
\omega(x,t) =	\sum_{m=1}^{\infty} V_m
	\left( \begin{array}{c}
x,t  \\
\gamma
\end{array} 
\right).
\end{equation}

Every integro-power form in the series is the sum of finite number of integro-power terms of different types which differ from each other in general. They are closely related to vorticity scales in fluid motions. It must be revealing to know the exact number of such terms in each integro-power form. The substitution procedures and the structure of (\ref{eq:vort-nonlinear-VIE}) suggest that $S_m$ can be identified as the coefficient of $z^{m+1}$ in the expansion 
\begin{equation} \label{eq:sm-sub}
\sum_{j=1}^m \; \Big\{ \;\Big(\; \sum_{k=1}^{m} \; 2 \; a_k \;z^k \;\Big)^{j+1} \; \Big\}
\end{equation}
with $a_1{=}1$. The factor $2$ refers to the fact that the non-linearity in the equations of motion has been reduced to two infinite convolution sums, $q$ and $\bar{q}$, in the transformed vorticity (\ref{eq:vort-nonlinear-VIE}). 

\begin{figure}[ht] \centering
  \includegraphics[width=3.5cm]{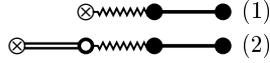}
  \vspace{3mm}
  \caption{The tree representations for the vorticity integral power terms $V_k[\gamma]^2, \; k{=}1,2$. A tree defines an integro-power term. The two trees represent $q$ and $\bar{q}$ in (\ref{eq:vort-nonlinear-VIE}). Every tree is the convoluted integrals starting from the root (the right-hand solid circle) which is the inner most integration. Every symbol $\otimes$ indicates the top of a tree. Every solid circle $\bullet$ stands for one $\gamma$. The open circle $\circ$ connects two integral convolutions. The lines linking the symbols are related to the integral convolutions. Every solid line stands for the space-time convolution with the kernel $K$ and a double line for $H$. A zigzag line denotes the space-time convolution for the space integral of $G$.}\label{fig:V2} 
\end{figure}

Interestingly, the numbers $S_m$ form an integer sequence. In the present context, we give a combinatoric interpretation of them in terms of allowable integral convolutions. Specifically, the second integro-power form is simply
\begin{equation*}
V_2\left( \begin{array}{c}
x,t  \\
\gamma
\end{array} 
\right)= \sum_{k=1}^2\: V_k[\gamma]^2.
\end{equation*}
Every term is an integral convolution of a binary product in $\gamma$. The two terms in the above expression take different numerical values at fixed $(x,t)$ because the integral kernels are different in each case. In figure~\ref{fig:V2}, we give the tree structure which shows the relation between $\gamma$ and the integral convolutions.

\begin{figure}[ht] \centering
  \includegraphics[width=6.268cm]{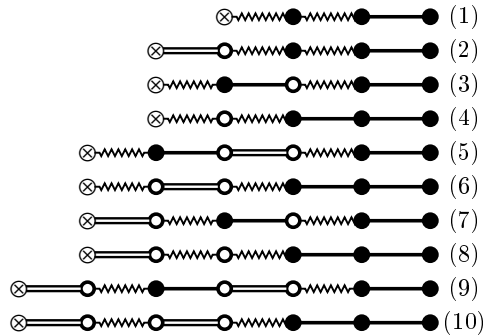}
  \vspace{3mm}
  \caption{The tree representations for the vorticity integral power terms $V_k[\gamma]^3,\; k{=}1,2,\cdots,10$ or $10 \: V[\gamma]^3$.} \label{fig:V3} 
\end{figure}

The third integro-power form consists of a total of 10 integral convolutions. We use notation $\{\gamma\}$ to represent a space-time integral convolution in $\gamma$. Two convolutions are the integrals of the ternary products $\{\gamma\}\{\gamma\}\{\gamma\}$, and $8$ convolutions of the binary product $\{\gamma\}\{\gamma\}^2$. Thus the third integro-power form is found to be 
\begin{equation*}
V_3\left( \begin{array}{c}
x,t  \\
\gamma
\end{array} 
\right)	= \sum_{k=1}^{10}\: V_k[\gamma]^3.
\end{equation*}
In figure~\ref{fig:V3}, we list the complete tree presentations.

\begin{figure}[ht] \centering
  \includegraphics[width=7.986cm]{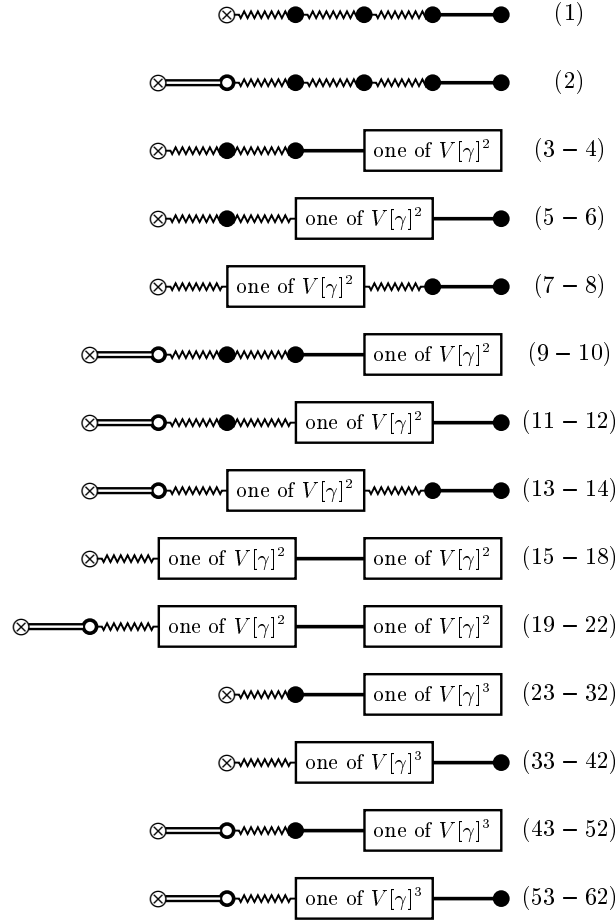}
  \vspace{3mm}
  \caption{The complete tree representations for the vorticity integro-power terms $V_k[\gamma]^4,\;k{=}1,2,\cdots,62$. In each tree, the relation to the structure of lower-order forms is evident. Every tree has its own unique structure; it is formed by the distinct combinations of the circles and the lines. Given the initial condition, every integro-power term in general takes distinct numerical values at different spatio-temporal locations $(x,t)$. }\label{fig:V4} 
\end{figure}

In the evaluation of the next form of $m=4$, the sum in recurrence (\ref{eq:sm-sub}) has $7$ terms for the non-linearity $q$ (cf. (\ref{eq:vort-nonlinear-VIE})). It in turn gives a number of integral convolutions, $1$ quaternary product $\{\gamma\}\{\gamma\}\{\gamma\}\{\gamma\}$; $6$ ternary products $\{\gamma\}\{\gamma\}\{\gamma\}^2$; $4$ binary products $\{\gamma\}^2\{\gamma\}^2$; and $20$ binary products $\{\gamma\}\{\gamma\}^3$. The total number of the integral convolutions doubles due to the presence of the filtered non-linear term $\bar{q}$.
The complete vorticity tree structure is given in figure~\ref{fig:V4}. 

By the rules of operations on the integro-power terms, we readily verify that 
the full expansion (\ref{eq:vort-formal-sol}) can be expressed as
\begin{equation} \label{eq:vort-series-sol}
 \begin{split}
	\omega(x,t) = \gamma(x,t) \: & + \sum_{m{\geq}2}\: \Big( \sum_{k=1}^{S_m} \: V_k[\gamma]^m\Big) = \gamma(x,t) \: + \sum_{m{\geq}2}\: {S_m} \: V[\gamma]^m  \\
	\quad & \\
	= \gamma(x,t) \: & + 2 \: V[\gamma]^2 + 10 \: V[\gamma]^3 + 62 \: V[\gamma]^4 + 430 \: V[\gamma]^5 + 3194 \: V[\gamma]^6  \\ 
	 \quad & \\
	\quad & + 24850 \: V[\gamma]^7 + 199910 \: V[\gamma]^8 + 1649350 \: V[\gamma]^9 \\
	 \quad & \\
	\quad & + 13879538 \: V[\gamma]^{10}  + 118669210 \: V[\gamma]^{11} \\
	\quad & \\
	\quad & + 1027945934 \: V[\gamma]^{12} + \cdots \cdots \: . 
 \end{split}
\end{equation}

The generating function for the integer sequence is given by 
\begin{equation} \label{eq:sm-gf}
	\sum_{m=1}^{\infty}S_m \: g^m = \frac{1}{6} \Big( 1+g - \sqrt{1-10\:g + g^2 } \; \Big) = w.
\end{equation}
As there are two convolution sums in (\ref{eq:vort-nonlinear-VIE}), this fact suggests that we seek the following expansion: 
\begin{equation*}
	g + 2 w^2 + 2 w^3 + 2 w^4 + 2 w^5 + {\cdots} = g + \frac{2 w^2}{1-w} = w.
\end{equation*}
The last equality defines a quadratic equation in $w$
\begin{equation*}
	3w^2 -(1+g)w + g =0.
\end{equation*}
Hence the generating function equals to one of the roots of the equation.
Differentiating $w$ with respect to $g$, we obtain
\begin{equation*}
	\frac{\rd w}{\rd g} = \frac{w-1}{6w-g-1}=\frac{(g-5)w-g+1}{g^2-10g+1}.
\end{equation*}
This relation is nothing more than an ordinary differential equation, namely,
\begin{equation}
	(g^2-10g+1)({\rd w}/{\rd g})-(g-5)w+g-1=0.
\end{equation}
The differential equation can be solved by the method of series expansion. More specifically, the relation for $S_m$ can be established from the coefficients of $g^{m-1}$. We find that   
the numbers $S_m$ satisfy the remarkably simple recurrence relation:
\begin{equation} \label{eq:recurrence}
  \begin{split}
	S_{m} & = \frac{1}{m} \Big( \: 5\: (2m -3)S_{m-1} - (m-3)S_{m-2} \: \Big), \;\;\; m{\geq}3,\\
	S_2 & = 2, \;\;\;\;\;\; S_1 = 1.
	\end{split}
\end{equation}
The magnitude of the integer members in the sequence increases very rapidly with $m$. We notice that ${S_{16} \sim  10^{12}}$ and ${S_{27} \sim 10^{23}}$. When $m\sim O(100)$, $S_m$ is a googol. Good order-of-magnitude estimates can be obtained from the formula 
\begin{equation*}
	S_m \sim 10^{m-4},\;\;\; m < 200.
\end{equation*}
The theoretical asymptotic approximation can be found by the method of singularity analysis (see, for example, Flajolet \& Sedgewick 2009). We find that 
\begin{equation*}
	S_m = \sqrt{\frac{12+5\sqrt{6}}{36 \pi}} \; \frac{\big( 5+2\sqrt{6} \big)^m} {m^{3/2}} \; \Big( 1 + O\big(\frac{1}{m}\big) \Big) \;\;\; \mbox{as} \;\;\; m \rightarrow \infty,
\end{equation*}
where the quantity, $5{+}2\sqrt{6}$, is one of the two reciprocal roots of the quadratic equation implied in (\ref{eq:sm-gf}). The power scaling, $m^{-3/2}$, reflects the nature of the square-root singularity. 
In the sequel, we need the knowledge of the ratio of the consecutive terms in the sequence. From the recurrence relation, we verify that
\begin{equation} \label{eq:sm-ratio}
	\frac{S_{m+1}}{S_{m}} < 10.
\end{equation}
In \ref{app:a}, we list the complete $S_m$ up to $m{=}51$.

Since there are a large number of terms in every $S_m$, it may be the case that only some of the vorticity integro-powers are required to describe a vorticity solution. We shall use notation $(S_m)$ to indicate a partial sum. 
\subsection{Convergence of the series solution}
In our subsequent analysis, we frequently employ the identities,
\begin{equation}
	2 \int_0^{\infty} r^m \exp(-\kappa r^2)\rd r = \frac{\Gamma(p)} {\kappa^p},\;\;\; p=\frac{m+1}{2},\;\;\;\kappa,m\;{>}\:0,
\end{equation}
where $\Gamma(p)$ denotes the gamma function, and
\begin{equation}
	\int_0^t (t-s)^{p} s^q \rd s = t^{p+q+1} \int_0^1 (1-x)^p x^q \rd x = t^{p+q+1} B(p+1, q+1),\;\;\;p,q\;{>}\:-1.
\end{equation}
Particular values of the gamma function are $\Gamma(1/2){=}\sqrt{\pi}$, and $\Gamma(n{+}1){=}n!$.

Given the analytic structure of series (\ref{eq:vort-series-sol}), its convergence can be established by the method of majorant for given $\omega_0$ (cf. (\ref{eq:vort0-y-norm})).
We are mainly interested in long-time flow development for large initial data. The norm in (\ref{eq:vort0-y-norm}) exists as long as the initial vorticity $\omega_0$ satisfies the regularity and the localization requirements (\ref{eq:ns-ic}) and (\ref{eq:ic-localization}). The regulating property of heat equation guarantees the existence of the smooth function $\varpi$ (cf. (\ref{eq:mollified-vort-ic})). According to the well-established theory of linear Volterra integral equation, function $\gamma$, the filtered initial data, is well-behaved.

Consider the integro-power form of degree $2$. In the leading integro-power term, 
\begin{equation}
 \begin{split}
	V_1[\gamma]^2=
\int_0^{t} \int \!\! \int G(x,t,& y,s,x_0,s) \gamma(y,s) \; \\
\Big( &\int_0^{s} \!\! \int K(x_0,s,z,r) \gamma(z,r) \rd z \rd r \Big) \rd x_0 \rd y \rd s,
 \end{split}
\end{equation}
($r{<}s{<}t$), we let $\eta$ stand for the expression for the inner space-time integral. In view of (\ref{eq:kernel-b}), $\eta$  is in fact the spatial derivative of the caloric volume potential of $b(x,s) \gamma(x,s)$:
\begin{equation*} 
	\eta(x_0,s)= \int_0^s \!\! \int K \gamma \rd y \rd \tau = \int_0^s \!\! \int \nabla Z(x_0,s,y,\tau) (b \gamma)(y,\tau) \rd y \rd \tau.
\end{equation*}
The function $b$ is given by (\ref{eq:kernel-b}). By direct calculation and by virtue of (\ref{eq:k-int-bound}), it follows that 
\begin{equation} \label{eq:v2-inner}
	\| \eta \|_{L^{\infty}} \leq  C \:A_0^2 H_0 (\nu s) \: B\big({3}/{2},{1}/{2} \big),
\end{equation}
where $C$ denotes a constant. (Similarly, $\|\eta\|_{L^1}$ can also be bounded by the right-hand side of (\ref{eq:v2-inner}) in view of H\"{o}lder's inequality and Young's inequality for convolution with a different constant $C$.) Next we verify that
\begin{equation} \label{eq:v2-outer}
\begin{split}
\int_0^{t} \Big| \int \Big( \int & G(x,t,y,s,z,s) \eta(z,s) \rd z \Big)  \gamma(y,s) \rd y \Big| \rd s \\
&  \leq C A_0^3 H_0^2 B\big({3}/{2},{1}/{2} \big) \int_0^t (\nu t - \nu s )^{-1/2} (\nu s)^{3/2} \rd s.
\end{split}
\end{equation}
Combining results (\ref{eq:v2-inner}) and (\ref{eq:v2-outer}), we have
\begin{equation}
	\Big| V_1[\gamma]^2 \Big| \: {\leq} \: C \: B\big({3}/{2},{1}/{2} \big) B\big({5}/{2},{1}/{2} \big) \: A_0^3 H_0^2 \: (\nu t)^2.
\end{equation}
The second integro-power term $V_2[\gamma]^2$ is given by
\begin{equation}
	V_2[x,t,\gamma]^2 = \int_0^{t} \!\! \int H(x,t,y,s) \: V_1[y,s, \gamma]^2
 \rd y \rd s.
\end{equation}
In view of the property of the resolvent kernel $H$ (cf. (\ref{eq:h-int-bound})), we obtain
\begin{equation}
  \begin{split}
	\Big| V_2[\gamma]^2 \Big| & \: {\leq}\: C A_0^3 H_0^3 \: B\big({3}/{2},{1}/{2} \big) B\big({5}/{2},{1}/{2} \big) \int_0^t (\nu t {-} \nu s)^{-1/2} (\nu s)^2 \rd s \\
	& \: {\leq}\: C \: B\big({3}/{2},{1}/{2} \big) B\big({5}/{2},{1}/{2} \big) B\big({6}/{2},{1}/{2} \big) \: A_0^3 H_0^3 \: (\nu t)^{2+1/2}, 
	\end{split}
\end{equation}
where $C$ is a constant. The term $V_2[\gamma]^2$ is an order of magnitude larger than $V_1[\gamma]^2$. So it is dominant in the integro-power form. Therefore we assert that
\begin{equation} \label{eq:v2-bound}
	\Big | V_2\left( \begin{array}{c}
x,t  \\
\gamma
\end{array} 
\right) \Big | \: {\leq}\: C_2 S_2 L_2 \: A_0^3 H_0^3 \: (\nu t)^{2{+}1/2} \frac{\big(\Gamma(1/2)\big)^{3}}{\Gamma(3{+}1/2)}
\end{equation}
for some constant $C_2$ which is independent of $A_0$, $H_0$ and $t$. The constant 
\begin{equation*}
	L_2 = \frac{\Gamma(2{+}1/2)}{\Gamma(2)}=\Big(\frac{1}{2}\Big) \Big(\frac{3}{2}\Big) \sqrt{\pi} = \lambda_2 \Big(\frac{3}{2}\Big) \sqrt{\pi}.
\end{equation*}
Alternatively, we confirm the bound (\ref{eq:v2-bound}) heuristically by considering the tree structure in figure~\ref{fig:V2} where, in the case of the second tree, the magnitude of the kernels $K$ and $H$ can easily be estimated.

In parallel to (\ref{eq:v2-bound}), the dominant terms in $V_k[\gamma]^3$ come from those terms being proportional to
\begin{equation*}
 \int^* \!\! H \int^* \!\! \int \!\! G \gamma  \int^* \!\! K \: V_2[\gamma]^2,
\end{equation*}
which is in the order of $A_0^5 H_0^5$. (The integral signs $\int^*$ stand for space-time integral.) They are one order of magnitude  (i.e., $O(A_0 H_0 t)$) larger than the term of the form
\begin{equation*}
\int^* \!\! H \int^* \!\! \int \!\! G \gamma \int^* \!\! \int \!\! G \gamma  \int^*  \!\! K \gamma. 
\end{equation*}
Hence we infer, from the analysis for the bound on $V_2$, that
\begin{equation} \label{eq:v3-bound}
	\Big | V_3\left( \begin{array}{c}
x,t  \\
\gamma
\end{array} 
\right) \Big | \: {\leq}\: C_3 S_3 (L_2L_3) \: A_0^5 H_0^5 \: (\nu t)^{4+1/2} \frac{\big(\Gamma(1/2)\big)^{6}}{\Gamma(5{+}1/2)}
\end{equation}
for some constant $C_3$, and
\begin{equation*}
	L_3 = \frac{\Gamma(4{+}1/2)}{\Gamma(4)}=\Big(\frac{1}{2}\Big) \Big(\frac{3}{4}\Big)\Big(\frac{5}{6}\Big)\Big(\frac{7}{2}\Big) \sqrt{\pi} < \lambda_3^3 \Big(\frac{7}{2}\Big) \sqrt{\pi},
\end{equation*}
where $\lambda_3{=}5/6$. The tree structures in figure~\ref{fig:V3} suggest that the dominant term is the tree labelled $(9)$ in the list.
Furthermore, we estimate the integro-power form $V[\gamma]^4$ which is bounded by
\begin{equation*}
	C_4 S_4 (L_2 L_3 L_4) A_0^7 H_0^7 \: (\nu t)^{6{+}1/2} \frac{\big(\Gamma(1/2)\big)^{9}}{\Gamma(7{+}1/2)}
\end{equation*}
for some constant $C_4$, and
\begin{equation*}
	L_4 = \frac{\Gamma(6{+}1/2)}{\Gamma(6)}=\Big(\frac{1}{2}\Big) \Big(\frac{3}{4}\Big)\Big(\frac{5}{6}\Big)\Big(\frac{7}{8}\Big)\Big(\frac{9}{10}\Big)\Big(\frac{11}{2}\Big) \sqrt{\pi} < \lambda_4^5 \Big(\frac{11}{2}\Big) \sqrt{\pi},
\end{equation*}
where $\lambda_4{=}9/10$.
The ideas of establishing the bounds such as (\ref{eq:v2-bound}) and (\ref{eq:v3-bound}) can be generalized to the integro-power form of $V[\gamma]^m$ with arbitrary value $m$. Suppose that 
\begin{equation} \label{eq:vk-bound}
\begin{split}
	\Big | V_m\left( \begin{array}{c}
x,t  \\
\gamma
\end{array} 
\right) \Big | & \: {\leq}\: C_m S_m \Big( \prod_{j=2}^{m}L_j \Big)\: A_0^{2m-1}H_0^{2m-1} \: (\nu t)^{2m-3/2} \frac{\big(\Gamma(1/2)\big)^{3(m-1)}}{\Gamma(2m{-}1/2)} \\
& = C_m S_m f_m \: (\nu t)^{2m-3/2} 
\end{split}
\end{equation}
for some constant $C_m$, and 
\begin{equation*}
	L_m = \frac{\Gamma(2m{-}2{+}1/2)}{\Gamma(2m{-}2)}=\Big(\frac{1}{2}\Big) \Big(\frac{3}{4}\Big){\cdots} \Big(\frac{4m{-}7}{4m{-}6}\Big) \Big(\frac{4m{-}5}{2}\Big) \sqrt{\pi} < \lambda_m^{2m-3} \Big(\frac{4m{-}5}{2}\Big) \sqrt{\pi},
\end{equation*}
where 
\begin{equation} \label{eq:lambda-ratio}
	\lambda_m=\frac{4m-7}{4m-6}, \;\;\; \lambda_{m-1}<\lambda_m, \;\;\;  \mbox{and} \;\;\; m\geq2.
\end{equation}
Among the possible solutions for the Diophantine equation, 
\begin{equation*}
 l_1+l_2+\cdots+l_p=m+1
\end{equation*}
for positive integers $l_i$, the integro-power term
\begin{equation*}
\underbrace{\int^* \!\! H  \int^* \int \!\! G \gamma \cdots \int^* \!\! \int \!\! G \gamma  \int^*  \!\! K \gamma}_{m+2 \;\; \mbox{integral convolutions}} 
\end{equation*}
represents the lowest possible powers in $A_0 H_0$. The size of convolution increases substantially in
\begin{equation*}
\int^* \!\! H  \int^* \!\! \int G \: V_i[\gamma]^l \int^*  \!\! K \: V_j[\gamma]^{2m-3-l}\;\;\; \mbox{or} \;\;\; \int^* \!\! H  \int^* \!\! \int G \: V_i[\gamma]^{2m-3-l} \int^*  \!\! K  \: V_j[\gamma]^l,
\end{equation*}
where the positive powers in $[\gamma]$ are restricted by $1{\leq}i{\leq}S_l$, $1{\leq}j{\leq}S_{2m{-}3{-}l}$, and $2{\leq}l{\leq}2m{-}4$. Either term is proportional to $ (A_0H_0)^{2m-1}$. The procedures can be generalized to any convolutions with more than just two middle terms. Generally, the case of $l_p$ has the expression 
\begin{equation*}
\int^* \!\! H  \int^* \!\! G \: V_i[\gamma]^{l_1} \int^* \!\! G \: V_j[\gamma]^{l_2} \cdots  \int^*  \!\! K \: V_p[\gamma]^{2m-3- (l_1 + l_2 +\cdots +l_p)},
\end{equation*}
which is in the order of $(A_0H_0)^{2m-1}$.
The dominant term is 
\begin{equation} \label{eq:sm1-term}
 \begin{split}
\int_0^t \!\! \int H(x,t,y,s) & \int_0^s \!\! \int \!\! \int  G(y,s,z,r,x_0,r) \gamma(x_0;r) \\ 
 & \int_0^r \!\! \int K(z,r,x',t') V_i[x',t',\gamma]^{2m-3} \rd x' \rd t' \rd x_0 \rd z \rd r \rd y \rd s
 \end{split}
\end{equation}
for $r{<}s{<}t$, and $1{\leq}i{\leq}S_m$. We let 
\begin{equation*}
	I(z,r)=\int^* K V_{i}[\gamma]^{2m-3},
\end{equation*}
and we calculate
\begin{equation*} 
\begin{split}
	\big| I(z,r)\big| & \leq C f_m A_0 \int_0^r \!\! (\nu r {-} \nu \tau )^{-1/2} (\nu \tau)^{2m-3/2} \rd \tau \\
	& \leq C f_m A_0 (\nu r)^{2m-1} B\big(2m{-}1/2,{1}/{2} \big) = g_m A_0 (\nu r)^{2m-1}. 
	\end{split}
\end{equation*}
Similarly, we let
\begin{equation*}
	J(y,s)=\int^* \!\! \Big( \int G I \: \Big) \gamma,
\end{equation*}
and we find
\begin{equation*} 
\begin{split}
	\big| J(y,s)\big| & \leq C g_m A_0^2 H_0 \int_0^s \!\! (\nu s {-} \nu \tau )^{-1/2} (\nu \tau)^{2m-1/2} \rd \tau \\
	& \leq C g_m A_0^2 H_0 (\nu s)^{2m} B\big(2m{+}1/2,{1}/{2} \big) = h_m A_0^2 H_0 (\nu s)^{2m}. 
	\end{split}
\end{equation*}
Finally, we evaluate the outer integral in (\ref{eq:sm1-term}) and bound it by
\begin{equation*}
	 C h_m A_0^2 H_0 \int_0^t \!\! (\nu t {-} \nu \tau )^{-1/2} (\nu \tau)^{2m} \rd \tau \leq C h_m A_0^2 H_0^2 (\nu t)^{2m+1/2} B\big(2m{+}1,{1}/{2} \big).
\end{equation*}
Combining these bounds, we obtain the estimate for $V_k[\gamma]^{m{+}1}$,
\begin{equation} \label{eq:vmplus1-bound}
	\Big |  \sum_{k=1}^{S_{m+1}} V_k[\gamma]^{m+1} \Big| \:{\leq} \: C_{m+1} S_{m+1} \Big( \prod_{j=2}^{m+1}L_j \Big)\: (A_0H_0)^{2m+1} \: (\nu t)^{2m+1/2} \frac{\big(\Gamma(1/2)\big)^{3m}}{\Gamma(2m{+}3/2)},
\end{equation}
where $C_{m+1}$ is constant, and
\begin{equation*}
	L_{m+1} = \frac{\Gamma(2m{+}1/2)}{\Gamma(2m)}=\Big(\frac{1}{2}\Big) \Big(\frac{3}{4}\Big){\cdots} \Big(\frac{4m{-}3}{4m{-}2}\Big) \Big(\frac{4m{-}1}{2}\Big) \sqrt{\pi}.
\end{equation*}

We conclude that series (\ref{eq:vort-formal-sol}) or (\ref{eq:vort-series-sol}) is absolutely convergent since either can be majorized by the following series, for any $t \leq T$,
\begin{equation}
	\sum_{m=2}^{\infty} C_{m}S_m \Big(\prod_{j=2}^m L_j \Big)\: A_0^{2m-1} H_0^{2m-1}\: (\nu T)^{(2m-3/2)} \frac{\Gamma^{3m-3}(1/2)}{\Gamma(2m{-}1/2)}. 
\end{equation}
In view of D'Alembert test and by virtue of inequalities (\ref{eq:sm-ratio}) and (\ref{eq:lambda-ratio}), the ratio of the consecutive terms in integro-power series (\ref{eq:vort-formal-sol}) is proportional to
\begin{equation}
	 (A_0 H_0 \nu T)^2\:\lambda_{m+1}^{2m-1}\:(4m{-}1)\frac{\Gamma(2m{-}1/2)}{\Gamma(2m{+}3/2)}\: \sim \:(A_0 H_0 \nu T)^2 / m \rightarrow 0 \;\; \mbox{as}\;\; m \rightarrow \infty,
\end{equation}
where the proportionality constant is omitted, as it is independent of $A_0$, $H_0$ and~$T$.
The radius of the convergence of this series is infinity for $0 < T < \infty$. 
\subsection*{Regularity of the solution}
Every term in the series expansion (\ref{eq:vort-series-sol}) constitutes a space-time convoluted vorticity having the dimensions of vorticity $[\mbox{Time}^{-1}]$. Each constituent is referred to as an eddy. Evidently, the initial vorticity comprises the core of every eddy and hence must be an integral part of every fluid motion. Moreover, it is evident that the entire vorticity field is populated by the vorticity integro-power terms of all the degrees. 
 
For any smooth and localized initial vorticity $\omega_0$, we see that
\begin{equation} \label{eq:f-smooth-x-t}
	\varpi \in C^{\infty} \;\;\; \mbox{for}\;\;\; x\in \real^3, \;\;t>0,
\end{equation}
thanks to the smoothing property of the heat kernel ($\nu{>}0$). It follows that
the kernel $K$ and hence the resolvent kernel $H$ are also smooth. Thus we infer that
\begin{equation} \label{eq:gamma-smooth-x-t}
	\gamma \in C^{\infty} \;\;\; \mbox{for}\;\;\; x\in \real^3, \;\;t>0.
\end{equation}
Since every term $V_m$ in series (\ref{eq:vort-formal-sol}) consists of smooth integrals, this fact in turn implies that the regularity of every term $V_m$ inherits the regularity of $\gamma$. It follows that
\begin{equation} \label{eq:omega-smooth-x-t}
	\omega \in C^{\infty} \;\;\; \mbox{for}\;\;\; x\in \real^3, \;\;t>0.
\end{equation} 
In addition, we have $\omega \rightarrow \gamma \rightarrow \varpi \rightarrow \omega_0$ as $t\rightarrow0$. Hence $\omega$ is continuous in $\real^3$ and for $t \geq 0$. The regularity behaviour of the vorticity at the beginning of the motion resembles the development of heat in the pure initial value problem in $\real^3$.  The continuity of vorticity at $t = 0$ is therefore a well-known problem and we do not need to repeat the proof (see, for example, \S 7 of John 1982). If the initial velocity $u_0$ is smooth, then the regularity of vorticity can be upgraded to include the initial instant of time,
\begin{equation} \label{eq:omega-total-smooth-x-t}
	\omega \in C^{\infty} \;\;\; \mbox{for} \;\;\; (x,t) \in \real^3 {\times}[0,T].
\end{equation}
We can determine the complete regularity for the triplet $(u,p,\omega)$ from the vorticity and momentum equations as we have done for the iteration systems (\ref{eq:iterative-vorticity}) and (\ref{eq:iterative-momentum}). 

Let $\omega_m$ be the approximate vorticity solution of $m$ terms in (\ref{eq:vort-series-sol}), the sequence of vorticity $\{ \omega_m\}$ is a Cauchy sequence. For any given $\varepsilon{>}0$, there exists an integer $N$ such that $|\omega_i - \omega_k|\;{<}\;\varepsilon$ for any $i{>}N$ and $k{>}N$ at any $(x,t)$. It follows that the solution of the non-linear vorticity integral equation is unique as well as stable. In \ref{app:b}, we outline two additional methods for uniqueness.

{\itshape The series solution (\ref{eq:vort-series-sol}) represents a spatio-temporal flow structure which forms a basis for the general state of fluid motion: turbulence. It is found that the non-linear term in the equations of motion has an intrinsic capability to proliferate vorticity eddies in huge quantity. The mechanism of the vorticity production
is a combined effect of vorticity stretching and convection. Nevertheless, the non-linearity alone cannot produce the hierarchy of the eddies in the complete absence of viscosity. }
\subsection{Navier-Stokes solution as $t{\rightarrow}\infty$}
The solution given in (\ref{eq:vort-series-sol}) provides a starting point to estimate the long-time decay of the Navier-Stokes equations. All the terms $V_m[\gamma]^m$ are interlaced with the heat kernel $Z$, hence they decay at large time at every fixed spatial location. It is plausible to expect the asymptotic approximation, 
\begin{equation} \label{eq:vort-decay}
\omega(x,t) \sim \gamma(x,t) \sim \varpi(x,t),	
\end{equation}
holds as $t \rightarrow \infty$. 
The last relation shows that the decays of the Navier-Stokes equation is identical to those of the initial value problem of heat equation, namely,
\begin{equation*}
	\omega(x,t) \; {\sim} \; \int {\mathbf Z}(x{-}y,t) \omega_0(y) \rd y \;\;\;\;\;\; \mbox{as}\;\;\; t \rightarrow \infty.
\end{equation*}
By direct calculation, we obtain the bounds (cf. (\ref{eq:vort-lq-lp-bound}))
\begin{equation*} 
 \begin{split}
	\big\| \omega \big\|_{L^q(\real^3)} & \; {\leq} \; C \; (\nu t)^{-3/2 \: (1/p-1/q)} \; \big\| \omega_0 \big\|_{L^p(\real^3)}, \\
		\big\| \nabla \omega \big\|_{L^q(\real^3)} & \; {\leq} \; C \; (\nu t)^{-3/2 \:(1/p-1/q)-1/2} \; \big\| \omega_0 \big\|_{L^p(\real^3)},
 \end{split}
\end{equation*}
where $1{\leq}p{\leq}q{\leq}{\infty}$ except the case $q{=}\infty, p{=}1$, and $C{=}C(p,q)$ is constant. 
Alternatively, since
\begin{equation*}
	\int {\mathbf Z}(x{-}y,t) \omega_0(y) \rd y = - \int \nabla_y {\times} {\mathbf Z}(x{-}y,t) u_0(y) \rd y,
\end{equation*}
it is more convenient to express the decay bound in terms of the initial velocity norm,
\begin{equation*} 
	\big\| \omega \big\|_{L^q(\real^3)} \; {\leq} \; C \; (\nu t)^{-3/2\:(1/p-1/q)-1/2} \; \big\| u_0 \big\|_{L^p(\real^3)}.
\end{equation*}

From the preceding analysis, we can extract some long-time decay estimates:
\begin{equation} \label{eq:vort-linfty-bound1}
	\frac{\|\omega(\cdot,t) \|_{L^\infty(\real^3)}} {\| u_0 \|_{L^2(\real^3)}} \:\; {\sim} \:\;  (\nu t)^{-5/4} \;\;\;\;\;\; \mbox{as}\;\;\; t \rightarrow \infty,
\end{equation}

\begin{equation} \label{eq:vort-l2-bound}
	 \frac{\|\omega(\cdot,t) \|_{L^2(\real^3)}}{\| u_0 \|_{L^2(\real^3)}} \;\; {\sim} \;\; (\nu t)^{-1/2} \;\;\;\;\;\; \mbox{as}\;\;\; t \rightarrow \infty,
\end{equation}
and
\begin{equation} \label{eq:dvort-linfty-bound}
	 \frac{\|\nabla \omega(\cdot,t) \|_{L^{\infty}(\real^3)}}{\| \omega_0 \|_{L^2(\real^3)}} \;\; {\sim} \;\; (\nu t)^{-5/4} \;\;\;\;\;\; \mbox{as}\;\;\; t \rightarrow \infty.
\end{equation}
Naturally, the asymptotic relation on the enstrophy is a temporal cumulation of the energy dissipation. This point-wise decay confirms the belief that the energy of fluid motion must tend to zero after a sufficiently long time.

In a similar fashion, the corresponding velocity decays can be estimated from these vorticity decays. From the early study on homogeneous turbulence, the asymptotic behaviour of the momentum equation has been linked to the equation of diffusion (see, for example, \S 5.4 of Batchelor 1953). Thus the momentum equation can be approximated, to a good degree of validity, by
\begin{equation} \label{eq:ns-u-decay}
	\partial_t u - \nu \Delta u = 0
\end{equation}
as $t \rightarrow \infty$. In other words, during the final period of decay, the complete flow field evolved according to linear diffusion. The underlying assumption was that the Navier-Stokes equations are globally regular. Hence we can treat any solution pair $(u,p)$ as well-behaved functions. To derive the decay bounds, we make use of the approximation
\begin{equation*}
	u(x,t) \; {\sim} \; \int {\mathbf Z}(x{-}y,t) u_0(y) \rd y \;\;\;\;\;\; \mbox{as}\;\;\; t \rightarrow \infty.
\end{equation*}
By direct calculation, we have the decay bound
\begin{equation} \label{eq:vel-linfty-bound1}
	 \frac{\| u(\cdot,t) \|_{L^\infty(\real^3)}} {\| u_0 \|_{L^2(\real^3)}} \:\; {\sim} \:\;  (\nu t)^{-3/4} \;\;\;\;\;\; \mbox{as}\;\;\; t \rightarrow \infty.
\end{equation}
It is not difficult to deduce that
\begin{equation} \label{eq:p-linfty-bound}
	 \frac{\| \nabla p(\cdot,t)/\rho \|_{L^\infty(\real^3)}} {\| u_0 \|^2_{L^2(\real^3)}} \:\; {\sim} \:\;  (\nu t)^{-2} \;\;\;\;\;\; \mbox{as}\;\;\; t \rightarrow \infty.
\end{equation}
Within the framework of the present analysis, the decay bound on $\|u\|_{L^2}$ appears to be best described by
\begin{equation} \label{eq:vel-l2-bound}
	 \frac{\| u(\cdot,t) \|_{L^2(\real^3)}} {\| \omega_0 \|_{L^1(\real^3)}} \:\; {\sim} \:\;  (\nu t)^{-1/4} \;\;\;\;\;\; \mbox{as}\;\;\; t \rightarrow \infty.
\end{equation}
This bound may be sharpened to  
\begin{equation} \label{eq:vel-l2-l1-bound}
	 \frac{\| u(\cdot,t) \|_{L^2(\real^3)}} {\| u_0 \|_{L^1(\real^3)}} \:\; {\sim} \:\;  (\nu t)^{-3/4} \;\;\;\;\;\; \mbox{as}\;\;\; t \rightarrow \infty.
\end{equation}

The proportionality constants in asymptotic relations, (\ref{eq:vort-linfty-bound1}) to (\ref{eq:dvort-linfty-bound}), and (\ref{eq:vel-linfty-bound1}) to (\ref{eq:vel-l2-l1-bound}), are known and all dimensionless.

{\itshape There exists an instant of time after which every fluid motion evolves into a state which can be described by the leading term in the solution (\ref{eq:solIC-VIE}). All the intricate vorticity characters of the motion eventually disappear due to viscous dissipation. As $t \rightarrow \infty$, the general solution of the Navier-Stokes equations decays and the motion ultimately restores to its equilibrium vorticity-free state}.
\section{Kinetic theory of gases and its relation to turbulence}
The Maxwell-Boltzmann kinetic theory of dilute gases (Maxwell 1867;  Boltzmann 1905) is a beautiful and sophisticated subject. It is well-known that the Navier-Stokes equations can be derived, as an approximation, from the theory (see, for example, Chapman \& Cowling 1970; Sommerfeld 1956; Grad 1958; Cercignani 1969; Liboff 2003). 
To vindicate the cause of observed macroscopic randomness in turbulent fluid motions, we must investigate the motion beyond the continuum. In reality, evolution of fluid motions is irreversible owning to viscous dissipation. The dissipative process converts kinetic energy of flows into fluids' internal energy. When we solve the Navier-Stokes equations, we have made no direct relevance to any property of the energy sink. The consequence of the dissipation is roughly referred to as heat. In fact, no definitive knowledge on the process of the heat production is available. For fluid dynamics on the continuum this lack of knowledge is not particularly severe since energy conservation is valid. We shall review the derivation of the equations of fluid dynamics from the kinetic theory. In this respect, we still work within the constraints imposed by the constitutive relations in hydrodynamics because we have to close the set of the derived equations. Intuition suggests that the apparent randomness cannot be a property of the Navier-Stokes equations as fluid motions are modelled on deterministic configuration space. Therefore we have to study the behaviour of fluids' constituents (simply referred as molecules or particles) and to examine fluid motions from particles' phase space. By considering Maxwellian molecules with cut-off as a model, we show that the density function of the molecules has a structure of turbulence. At least in a qualitative sense, a link between molecular fluctuations and the commonly-observed flow randomness on the continuum can be demonstrated. 
\subsection{Basic ideas in kinetic theory}
Consider a fluid occupying a finite region in $\real^3$. We model the occupying volume to consist of a swamp of identical monatomic molecules, which are under random motion. The total number of the particles ${\mathrm N}$ is huge. For instance, there are about $10^{23}$ water molecules in $18$ grams of pure water.  For each particle, we consider its kinetic energy and neglect its rotational and vibrational energy. The occupying volume is divided into small hypercubes of size $(\Delta x)^3 {\times} (\Delta \xi)^3 $ (phase space). We label these cubes by $j{=}1,2,3,{\cdots}$. Obviously, the total number of the cubes cannot exceed ${\mathrm N}$. The particles occupy the cubes in a random manner. Let ${\mathrm N}_j$ denote the number of particles in the $j$-th cube. Then the energy is ${\mathrm E}_j{=}{\mathrm E}_j(x_j,{\xi}_j){=}m {\xi}_j^2/2$, where ${\xi}_j$ is particle velocity, and $m$ its mass. Hence every cube is characterized by its energy content ${\mathrm N}_j m {\xi}_j^2/2$ which is called a microstate. The conservation of the number of particles and the conservation of energy hold at all time, namely,
\begin{equation} \label{eq:particles-conserves}
	\sum_j {\mathrm N}_j = {\mathrm N} \;\;\;\mbox{and}\;\;\; \sum_j {\mathrm N}_j {\mathrm E}_j = {\mathrm E}.
\end{equation}
Particles may take any of their velocity values in the range ${-} \infty {<} {\xi} {<} {+} \infty$. We assume that all accessible microstates are equally likely. The number of microstates is given by the binomial coefficient,
\begin{equation} \label{eq:boltz-w}
	{\mathrm W} = \frac{{\mathrm N}!}{\prod_{j} \: {\mathrm N}_j!}.
\end{equation}
The probability for observing the energy state ${\mathrm E}_j$ is simply ${\mathrm P}_j{=}{\mathrm N}_j/{\mathrm N}$. To facilitate our calculations, we first take the (natural) logarithm of (\ref{eq:boltz-w}) and make use of Stirling's approximation $\log f {\approx} f \log f {-} f $. We obtain
\begin{equation} \label{eq:log-w}
	\log {\mathrm W}  = {\mathrm N} \log {\mathrm N} - \sum_j {\mathrm N}_j \log {\mathrm N}_j.
\end{equation}
We are interested in the most probable distribution (denoted by ${\mathrm N}^*_j$) which can be found by solving a variational problem of (\ref{eq:log-w}) subject to constraints (\ref{eq:particles-conserves}). Now
\begin{equation*}
	\delta \log {\mathrm W}  - \delta( \alpha {\mathrm N} ) - \delta ( \beta {\mathrm E} ) = 0,
\end{equation*}
where $\alpha$ and $\beta$ are Lagrangian multipliers. Performing the variations and simplifying the result, we obtain the following expression: $\log {\mathrm N} {-} \log {\mathrm N}^*_j = (\alpha {+} \beta {\mathrm E}_j)$ or
\begin{equation*}
	{\mathrm N}^*_j = {\mathrm N} \exp \big(- (\alpha + \beta {\mathrm E}_j) \big).
\end{equation*}
The total ${\mathrm N}^*_j$ is just ${\mathrm N}$ so that $\alpha$ can be determined. Hence the probability for the most probable distribution to occur is given by
\begin{equation} \label{eq:boltz-distri}
	{\mathrm P}^*_j=\frac{{\mathrm N}^*_j}{{\mathrm N}}=\frac{ \exp \big( - \beta {\mathrm E}_j \big)} {\sum_{i} \: \exp \big( - \beta {\mathrm E}_i  \big)}=\frac{ \exp \big( - \beta {\mathrm E}_j \big)} {{\mathrm Z}}. 
\end{equation}
This is known as the Boltzmann distribution. The denominator $\mathrm Z$ is just the partition function which measures the total number of the energy states. Explicitly, we get
\begin{equation*}
	\alpha = \log {\mathrm Z}.
\end{equation*}
Substituting the expression ${\mathrm N}^*_j$ into (\ref{eq:log-w}) and in view of Boltzmann's formula for entropy, we deduce an expression that shows how the total entropy is related to the Lagrange multipliers,
\begin{equation} \label{eq:boltz-entropy}
	{\mathrm S} = k_B \log {\mathrm W} = k_B (\alpha {\mathrm N} + \beta {\mathrm E}),
\end{equation}
where Boltzmann's constant $k_B$ has the numerical value 
$	k_B {=} 1.381{\times}10^{-23}$J/K.

As temperature is related to the entropy by ${\mathrm T}^{-1} {=} {\partial{\mathrm S}}/ {\partial {\mathrm E}}$, we evaluate the rate of change in the entropy. The definition of temperature contains two important aspects of thermodynamics; in the energy transfer due to a temperature difference, energy flows from one part in a gas with high temperature to one at low temperature via heat. The entropy in the gas system and its surroundings never decreases. 
Specifically, we verify that
\begin{equation*} 
	\beta = {(k_B {\mathrm T})}^{-1}.
\end{equation*}

If the gas as whole is set into motion from a particular instant and is left to evolve, experience tells us that the kinetic energy of the flow decays in time due to viscous dissipation. The energy is converted into the internal energy of the particles and there must be a temperature rise in the flow. The exponential decay in the Boltzmann distribution (\ref{eq:boltz-distri}) puts a strong limitation on the likelihood of the most probable state. If the temperature increases slightly, the particle populations migrate from lower energy states to higher energy states {\itshape without} drastic changes in the internal structure of the gas (unless the gas is close to its phase transition). 
Second, an observed value of a gas property ${\mathrm Q}$ (a macrostate) is the ensemble average over a great number of microstates,
\begin{equation} \label{eq:mstates}
\left\langle {\mathrm Q} \right\rangle = 	\sum_{j} {\mathrm P}_j {\mathrm Q}_j.
\end{equation}
Even though the particles fluctuate randomly, the average is well-defined. Third, the internal energy of the gas is given by
\begin{equation} \label{eq:prob-ie}
	\left\langle {\mathrm E} \right\rangle = - \frac{\rd \alpha}{\rd \beta} = \frac{1}{\mathrm Z} \sum_{j} {\mathrm E}_j \exp( - \beta {\mathrm E}_j).
\end{equation}
As ${\mathrm N} \rightarrow \infty$, we then treat the two summations as integrals. We readily obtain 
\begin{equation} \label{eq:ie-mean-v}
	\left\langle {\mathrm E} \right\rangle = \frac{3}{2} \: k_B {\mathrm T} =  \frac{m}{2}\:\overline{{\xi}^2},
\end{equation}
where the last quantity stands for the mean kinetic energy of the gas. The most probable velocity (${\xi}^*$) can be calculated from the extreme value of the most probable distribution (\ref{eq:prob-ie}). It is easy to find that
\begin{equation} \label{eq:ie-mprob-v}
	\frac{1}{2}m{{\xi}^*}^2 = k_B {\mathrm T}. 
\end{equation}
Hence the root mean square velocity is given by
\begin{equation*}
	{\xi}_{rms} = \sqrt{\:\overline{{\xi}^2}\:} = 1.2247 \: {\xi}^*.
\end{equation*}
The mean velocity or the ensemble average velocity is
\begin{equation*}
	\left\langle {\xi} \right\rangle = 	\sum_{j} {\mathrm P}_j {\xi}_j = 1.1284 \: {\xi}^*.
\end{equation*}
As an approximation, the most probable velocities for nitrogen and oxygen at $25^{\circ}$C and under atmospheric pressure are estimated to be $420$m/s and $390$m/s respectively. It is quite remarkable that we can establish so many basic facts about the swamp of randomly agitating particles from a probability consideration. 
\subsection{The Boltzmann equation and conservation laws in hydrodynamics}
In this subsection, we follow a standard exposition of the subject. The purpose is to review the relevant concepts which form the basis of our subsequent discussion. We consider a density function $F(x,\xi,t)$ for a single particle in the phase space $(x,\xi) {\in} \real^3{\times}\real^3$ having velocity $\xi{=}(\xi_1, \xi_2, \xi_3)$. As the particles are moving in space-time in random manner, we can also interpret $F$ as the probability of finding a particle having velocity $\xi$ at location $(x,t)$. 
We consider a gas which is neither too dense so that only binary elastic collisions are considered, nor too dilute so that the continuum properties of the gas do not vary over the length scales comparable to the mean free path. This assumption does not hold for all gases and liquids in general. It does not even hold for monatomic gases under extreme pressure and temperature conditions. The justification is that the equations of motion derived from the Boltzmann equation work for many real fluids under practical conditions. Moreover, the velocity of gas particles is independent of the position.
This is Boltzmann's hypothesis of complete molecular chaos. 

We denote $\xi$ and $\xi_*$ the velocities of two particles before a collision, $\xi'$ and $\xi'_*$ the velocities after the collision. We are interested in collisions which conserve momentum and energy, namely,
\begin{equation*}
	\begin{split}
	\xi + \xi_* & = \xi' + \xi'_*, \\
	|\xi|^2 + |\xi_*|^2 & = |\xi'|^2 + |\xi'_*|^2.
	\end{split}
\end{equation*}
We use the unit vector $\vec{\alpha}$ to present the angle of the colliding particles. The vector is parallel to the line joining the mass centres. During the collision, the components of particles' velocities along $\vec{\alpha}$ are exchanged while the components normal to the vector are conserved.  
Thus the following relations hold:
\begin{equation} \label{eq:collision-relations}
	\xi' = \xi - (\vec{\alpha} \cdot (\xi - \xi_*)) \vec{\alpha}, \;\;\; \xi'_* = \xi_* + (\vec{\alpha} \cdot (\xi - \xi_*)) \vec{\alpha}.
\end{equation}
It is convenient to use notation
\begin{equation*}
	{\mathrm V} = |\xi - \xi_*| = |\xi' - \xi'_*|
\end{equation*}
for the magnitude of the relative velocity of the colliding particles. The Boltzmann equation for the density function reads
\begin{equation} \label{eq:boltzmann}
	\frac{\partial F}{\partial t} + \xi_i \frac{\partial F}{\partial x_i} + {X_i}\frac{\partial F}{\partial \xi_i} = {\cal Q}(F,F),
\end{equation}
where $X(x){=}(X_1,X_2,X_3)(x)$ is an external force per unit mass. For binary collisions, the collision kernel ${\cal Q}$ is given by
\begin{equation}
  {\cal Q}(F,F) = \int_{\real^3} \int_{\mathbb{S}^2} \big|\vec{\alpha} \cdot (\xi{-}\xi_*)\big| \; \big(\: F'F'_* - F F_*\:\big) \; \rd \xi_* \rd \vec{\alpha},
\end{equation}
where
\begin{equation*}
F_*=F(x,\xi_*,t), \;\;\; F' = F(x,\xi',t), \;\;\; F'_*=F(x,\xi'_*,t),
\end{equation*}
and the meaning for these notations is clear in view of (\ref{eq:collision-relations}). Equation (\ref{eq:boltzmann}) must be solved subject to initial data 
\begin{equation} \label{eq:boltzmann-ic}
	F_0=F(x,\xi,0)\;{>}\;0, \;\;\;(x,\xi) \in \real^3{\times}\real^3.
\end{equation}
Although there are no particular restrictions on the initial data, the conservation laws and the second law of thermodynamics require that
\begin{equation} \label{eq:boltz-ic-bound}
	\int_{\real^3} \Big( 1 + {|\xi|^2}/{2} \Big) F_0 \rd \xi \;{<}\; \infty,\;\;\; \int_{\real^3} \big( F_0 \log F_0 \big) \rd \xi \;{<}\; \infty.
\end{equation}
Additional requirements may be supplemented in specific applications. 

We introduce a spherical co-ordinates system $(z,\theta,\varepsilon)$ being attached to the particle moving with $\xi$. The $z-$axis orients in the direction of $\xi_*{-}\xi$, the angular-axes $\theta$ and $\varepsilon$ are the polar and azimuthal angles of the particle moving with $\xi'$. The angle $\theta$ lies between the unit vector $\vec{\alpha}$ and the vector $\xi_*{-}\xi$.
For repulsive intermolecular potentials with an inverse power law
\begin{equation*} 
	U(r)\:{\propto}\:r^{-(p-1)}, \;\;\; p>2,
\end{equation*}
where $r$ is the distance between the colliding particles, Maxwell (1867) showed that the kernel $|\vec{\alpha} \cdot (\xi{-}\xi_*)|$ can be evaluated,
\begin{equation} \label{eq:max-mols}
 |\vec{\alpha} \cdot (\xi{-}\xi_*)| =
	B \big( {\mathrm V},\theta \big) \propto {\mathrm V}^{{(p-5)}/{(p-1)}} \; \Theta(\theta),
\end{equation}
where $0{\leq}\theta{\leq}\pi/2$. The precise form of function $\Theta$ is not required here. It is sufficient to know several special cases. If $\theta{=}\pi/2$, it is a head-on collision. If $\theta{=}0$, the case is singular and it is called a grazing collision. 
Maxwellian molecules are the molecules that have the potential of an inverse fourth power ($p{=}5$). The other cases, $2{<}p{<}5$ and $p{>}5$, correspond to the soft and hard potentials respectively. 

From the dynamics of collisions (see, for example, Landau \& Lifshitz 1981; Liboff 2002), the function $B$ is known to be nonnegative and can be related to the impact parameter $r_i$ which is defined as the perpendicular distance from $\xi$ to the asymptotic orbit of $\xi_*$. 
The total cross-section $\sigma{=}\sigma({\mathrm V},\theta)$ of collision is given by
\begin{equation} \label{eq:boltz-cutoff}
	{\mathrm V} \sigma = \int_{\mathbb{S}^{2}} B({\mathrm V},\theta) \rd \theta \rd \varepsilon = {\mathrm V}  \int_{\mathbb{S}^{2}} r_i \rd r_i \rd \varepsilon,
\end{equation}
where $\mathbb{S}^{2}$ covers the support of $B$.
If the collisions are due to inter-molecular forces of infinite space range ($r_i{\rightarrow}\infty$), a mathematical artifact, called a cut-off, is required to make the value of the cross-section finite. If the forces are of finite range, no cut-off is necessary. The cut-off is also known as an angular cut-off which imposes a limitation on $\theta$ such that $\theta{>}0$, thus avoiding the singularity caused by the grazing collision. 

Boltzmann's entropy ${\mathrm H}$ is defined in terms of the distribution function by
\begin{equation*}
	{\mathrm H}(x,t)= - k_B \int F \log F \: \rd \xi.
\end{equation*}
This is a local entropy. Significantly, its time variation is
\begin{equation*}
	\frac{\partial {\mathrm H}}{\partial t} = - k_B \int \big( 1 + \log F \big) \frac{\partial F}{\partial t} \rd \xi.
\end{equation*}
We substitute the time variation $\partial F/\partial t$ from the Boltzmann equation (\ref{eq:boltzmann}). One simplification can be made; the integral over the term involving the force $X_i$ is zero in view of Gauss' divergence theorem. Expressing the result in the spherical co-ordinates, we obtain
\begin{equation} \label{eq:boltz-h}
	\frac{\partial {\mathrm H}}{\partial t} + \nabla . {\cal H} = - k_B \int \int \int_{\mathbb{S}^2}  B\big( {\mathrm V},\theta \big) \big( 1 + \log F \big) \big(F'F'_* - F F_*\big)  \rd \theta \rd \varepsilon  \rd \xi \rd \xi_*,
\end{equation}
where $\cal H$ is the transformed term relating to $\xi_i$,
\begin{equation*}
	{\cal H} = - k_B \int \xi F \log F \rd \xi.
\end{equation*}
The integrand in (\ref{eq:boltz-h}) is symmetric in the four functions of $F$. The symmetry remains in collisions. It is not difficult to verify that
\begin{equation} \label{eq:boltz-h-sym}
	\frac{\partial {\mathrm H}}{\partial t} + \nabla . {\cal H} = - \frac{k_B}{2} \int \int \int_{\mathbb{S}^2}  B\big( {\mathrm V},\theta \big)  \Big( \log \frac{F' F_*'}{F F_*} \Big) \big(F'F'_* - F F_*\big) \rd \theta \rd \varepsilon  \rd \xi \rd \xi_*.
\end{equation}
Since the integrand cannot be negative as the last two factors (containing $F$'s) always have the same signs, the integral expression on the right cannot be less than zero. To obtain the total entropy, we carry out an integration over space. The resulting integral over $\nabla.{\cal H}$ vanishes by Gauss' divergence theorem. In view of the non-negativity constraint for the integrand in (\ref{eq:boltz-h-sym}), we assert that the total entropy satisfies 
\begin{equation} \label{eq:boltz-h-theorem}
	\frac{\rd}{\rd t} {\mathrm S} = \frac{\rd}{\rd t} \int  {\mathrm H} \: \rd x = - k_B \frac{\rd}{\rd t} \Big( \int \int \big(F \log F\big) \rd \xi \rd x \Big) \geq 0.
\end{equation}
This is the second law of thermodynamics or the Boltzmann ${\mathrm H}$-theorem. For a gas motion slightly out of equilibrium, the entropy strictly increases. 

The equality sign holds for equilibrium Maxwellian state. It is well-known that the Boltzmann equation admits the local Maxwellian as a solution (with $X{=}0$), namely, ${\cal Q}(M^{\star},M^{\star}){=}0$ in (\ref{eq:boltzmann}). Then we have the symmetry condition 
$F'F'_* {=} FF_*$. Let $\log \chi {=} F $, the symmetry equality translates into
\begin{equation*}
	\chi'+\chi'_* = \chi + \chi_*
\end{equation*}
which is known as the collision invariant. The conservation laws of mass, momentum and energy define five invariants:
\begin{equation} \label{eq:boltz-invariant}
	\chi_1=1, \;\;\; \chi_2=\xi_1, \;\;\; \chi_3=\xi_2, \;\;\; \chi_4=\xi_3 \;\;\; \mbox{and} \;\;\;  \chi_5=\xi^2/2. 
\end{equation}
The most general linear combination of the invariants has the form
\begin{equation*}
	\chi = a_1  + a_2  \xi + a_3  \xi^2, 
\end{equation*}
where $a_1$ to $a_3$ are constant. Since $\chi$ is the inverse logarithm of the density function, we obtain $\log F = \log \alpha_0  {-} \beta_0  m(\xi {-} \bar{\xi})^2/2$.
Thus the Maxwellian distribution is given by
\begin{equation} \label{eq:maxwellian}
M^{\star}(x,\xi,t) = \alpha_0 \; \exp \Big(- \beta_0 \: {m(\xi - \bar{\xi})^2/2} \:  \Big),
\end{equation}
where $\alpha_0, \beta_0$, and $\bar{\xi}$ are constants to be determined. As we are mainly interested in particle's space-time properties, we only need to sum all the velocities of the particle. For instance, the particle density is the integral of the distribution,
\begin{equation} \label{eq:maxwellian-rho}
	\varrho(x,t)=\int_{\real^3} F(x,\xi,t) \rd \xi = \int_{\real^3} M^{\star}(x,\xi,t) \rd \xi = \alpha_0 \big( 
	2\pi/(m\beta_0) \big)^{2/3}.
\end{equation}
The mass density on macroscopic description is related to the particle density by $\rho = \varrho m$.
The mean velocity due to particle's random fluctuations is calculated from
\begin{equation} \label{eq:maxwellian-u}
	\frac{1}{\varrho} \int_{\real^3} \xi F(x,\xi,t) \rd \xi = \frac{1}{\varrho} \int_{\real^3} \xi M^{\star}(x,\xi,t) \rd \xi = \bar{\xi}(x,t) \equiv \left\langle {\xi} \right\rangle(x,t).
\end{equation}
This mean value is interpreted as the macroscopic velocity. The mean kinetic energy of the particles moving with the mean velocity is found to be
\begin{equation} \label{eq:maxwellian-u2}
	\int_{\real^3} (\xi - \bar{\xi})^2 F(x,\xi,t) \rd \xi = \int_{\real^3} (\xi - \bar{\xi})^2 M^{\star}(x,\xi,t) \rd \xi = \frac{3}{2}\: \frac{\varrho}{\beta_0}.
\end{equation}
Then $\beta_0{=}\beta$ by comparing with (\ref{eq:ie-mean-v}). As explicitly stated, equations (\ref{eq:maxwellian-rho}) to (\ref{eq:maxwellian-u2}) are satisfied by function $F$. It is expected that an irreversible fluid motion characterized by $F$ quickly restores to equilibrium over Maxwell's relaxation time, the distribution $F$ adjusts to a local Maxwellian so that $F{=}M^{\star}$. Then the entropy attains a maximum; the motion is now in equilibrium. If the motion undergoes a further relaxation, the values of $\varrho,u$ and $\mathrm T$ are independent of $(x,t)$. The final equilibrium stage is called the absolute Maxwellian. The fluid consisting of the swamp of the particles is either in a state of rest or in a uniform translational motion. The dynamics of fluid motions in the equilibrium Maxwellian stages is governed by the Navier-Stokes equations (see, for example, Liboff 2003).

Multiplying (\ref{eq:boltzmann}) by $\chi_{\alpha}$ and integrating the result over all $\xi$, we obtain
\begin{equation*}
\int	\chi_{\alpha}(\xi) \Big( \frac{\partial F}{\partial t} + \xi_i \frac{\partial F}{\partial x_i} + {X_i}\frac{\partial F}{\partial \xi_i}\Big) \rd \xi = \int \chi_{\alpha}(\xi)  {\cal Q}(F,F) \rd \xi.
\end{equation*}
Symmetry in the integrand on the right suggests that $\chi_{\alpha}$ can be replaced by the invariant
\begin{equation*}
	\chi_{\alpha}(\xi) \rightarrow \frac{1}{4} \; \Big( \chi_{\alpha}(\xi) + \chi_{\alpha}(\xi_*) - \chi_{\alpha}(\xi') - \chi_{\alpha}(\xi'_*) \Big).
\end{equation*}
The conservation laws can be expressed as the collision invariants,
\begin{equation} \label{eq:boltz-invariant-int}
	\int_{\real^3} {\chi}_{\alpha} {\cal Q}(F,F) \rd \xi = 0,\;\;\;\alpha=1,2,3,4,5.
\end{equation}
As explained by Maxwell, the particle velocity $\xi$ is decomposed into a sum of a mean velocity $u$ and a random velocity $u'$:
\begin{equation} \label{eq:vel-decomp-ke}
	\xi = u + u'.
\end{equation}
The mean velocity is the velocity perceived in macroscopic description. The fluctuation $u'$ is particle's relative velocity with respect to the mean. 

A macroscopic quantity, denoted by $\left\langle {\mathrm Q} \right\rangle$, is defined by
\begin{equation*}
	\left\langle {\mathrm Q} \right\rangle = \frac{1}{\varrho}\int_{\real^3} {\mathrm Q} F (x,\xi,t) \rd \xi.
\end{equation*}
This is the integral form of (\ref{eq:mstates}). Multiplying both sides of the Boltzmann equation by the collision invariant $\chi_{\alpha}$, integrating over particle velocity $\xi$, and taking the average, we obtain the following formula for the derivation of fluid dynamics equations:
\begin{equation} \label{eq:boltz-invariants-fd}
	\frac{\partial}{\partial t} \Big( \rho \left\langle \chi_{\alpha} \right\rangle \Big) + \frac{\partial}{\partial x_i} \Big( \rho \left\langle \xi_i \chi_{\alpha} \right\rangle \Big) + {\rho} X_i \left\langle \frac{\partial \chi_{\alpha}}{\partial \xi_i} \right\rangle  = 0. 
\end{equation}
To relate particle's density function to the macroscopic fluid quantities, we introduce the following definitions:
\begin{equation} \label{eq:macro-fluiddyn}
\begin{aligned}
	u_i(x,t)&= \frac{1}{\varrho} \int_{\real^3} \xi_i F(x,\xi,t)  \rd \xi  & \mbox{(velocity)},\\
  u_i u_j(x,t)&= \frac{1}{\varrho} \int_{\real^3} \xi_i \xi_j F(x,\xi,t) \rd \xi & \mbox{(momentum)},\\
  	p_{ij}(x,t)&= \int_{\real^3} u'_i u'_j  F(x,\xi,t) \rd \xi &\mbox{(stress tensor)}, \\
  {\mathscr E}(x,t)&= \frac{1}{2 \varrho}\int_{\real^3}  |u'|^2  F(x,\xi,t) \rd \xi & \mbox{(internal energy/mass)}, \\ 
 Q_i(x,t)&= \frac{1}{2}\int_{\real^3}  u'_i |u'|^2  F(x,\xi,t) \rd \xi & \mbox{(heat flux)}.
\end{aligned}
\end{equation}
The product of the velocities is calculated according to 
\begin{equation*} 
\left\langle \xi_i \xi_j \right\rangle = \left\langle (u_i + u'_i)(u_j+u'_j) \right\rangle = u_i u_j + \left\langle u'_i u'_j \right\rangle	
\end{equation*}
because $\left\langle u'_i \right\rangle {=} 0$. 
Substitution of the invariants in (\ref{eq:boltz-invariant}) into equation (\ref{eq:boltz-invariants-fd}) in turn gives rise to the equations of fluid dynamics:
\begin{equation} \label{eq:fluiddyn}
 \begin{aligned}
	\frac{\partial \rho}{\partial t} + \frac{\partial}{\partial x_i} \big( \rho u_i \big)  & = 0 & \mbox{(continuity)}, \\
	\frac{\partial}{\partial t} \big(\rho u_i \big) + \frac{\partial}{\partial x_j} \big( \rho u_i u_j + p_{ij} \big)  & = {\rho} X_i & \mbox{(momentum)}, \\
	\frac{\partial}{\partial t} \Big( \frac{1}{2} \rho u^2 + \rho {\mathscr E}  \Big) + \frac{\partial}{\partial x_i} \Big( \frac{1}{2} \rho u^2 u_i + \rho {\mathscr E} u_i + p_{ij} u_j + Q_i  \Big)  & = {\rho} u_i X_i & \mbox{(energy)}.
	\end{aligned}
\end{equation}
There are actually $13$ variables: $\rho$, ${\mathscr E}$, two vectors ($u$ and $Q$), and five $p_{ij}$'s. System (\ref{eq:fluiddyn}) is a set of five equations and evidently it does not form a closed system for the variables. For Newtonian viscous fluids, we postulate the constitutive relations
\begin{equation} \label{eq:fluiddyn-constitute}
 \begin{split}
	p_{ij}& = p \delta_{ij} - \mu S_{ij} - \Big( \eta - \frac{2}{3}\mu \Big) (\nabla. u) \delta_{ij}, \\
	Q_i & = - {\mathrm k} \partial {\mathrm T} / \partial x_i,
	\end{split}
\end{equation}
where $\mu$ and $\eta$ are the coefficients of viscosity, and ${\mathrm k}$ the thermal conductivity or the heat conduction coefficient. The minus sign in the Fourier law implies the fact that the energy flux flows from a region of higher temperature to one with lower temperature. Thus we see that the energy density consists of two parts,
\begin{equation*}
	\frac{1}{2 \varrho} \int \xi^2 F(x,\xi,t) \rd \xi = \frac{1}{2} u^2 + \frac{1}{2} |u'|^2 = \frac{1}{2} u^2 +  {\mathscr E}.
\end{equation*}
In addition, the macroscopic pressure is simply 
\begin{equation} \label{eq:macro-p}
	p = - \frac{1}{3}p_{ii} = \frac{m}{3} \int |u'|^2 F \rd \xi = \frac{2}{3} \rho {\mathscr E} = \varrho k_B {\mathrm T}.
\end{equation}
This is Bernoulli's concept of pressure which states that pressure is the rate of change of momentum of particles in the direction normal a surface.
Specifically, the equations of Navier-Stokes dynamics are recovered from our microscopic description (\ref{eq:fluiddyn}) and (\ref{eq:fluiddyn-constitute}). For fluids with vanishing transport coefficients ($\mu=\eta={\mathrm k}=0$), the dynamics is governed by the Euler equations (Euler 1755). 
\subsection{Maxwellian molecules as a generic microscopic turbulence}
For Maxwellian molecules ($p{=}5$) in (\ref{eq:max-mols}), the collision function $B$ is independent of the relative velocity ${\mathrm V}$. Consider spatially homogeneous density function $F{=}F(\xi,t)$. For $X_i{=}0, t{>}0$, the Boltzmann equation (\ref{eq:boltzmann}) reduces to the special case
\begin{equation} \label{eq:mm-boltz}
	{\partial F}/{\partial t} + \vartheta(F) F = \int_{\real^3} \int_{\mathbb{S}^{2}} B(\theta) F'(\xi') F'_*(\xi'_*) \rd \theta \rd \varepsilon \rd \xi_* = {\cal Q}_s(F,F).
\end{equation}
It can be shown that the particle density is an invariant of time: $\partial \varrho/\partial t {=} 0$ (see, for instance, Wild 1951). Then the mean collision frequency $\vartheta$ can be simplified: 
\begin{equation*}
	\vartheta(F) = \int_{\real^3} \int_{\mathbb{S}^{2}} B(\theta) F(\xi_*) \rd \theta \rd \varepsilon \rd \xi_* = {\mathrm V}  \sigma \varrho =  {\mathrm V}  \sigma  \varrho_0, 
\end{equation*}
where $\varrho_0$ is the initial particle density. The cross-section $\sigma$ can be expressed explicitly as
${\mathrm V} \sigma = \int_{\mathbb{S}^{2}} B(\theta) \rd \theta \rd \varepsilon$ so that the frequency is independent of ${\mathrm V}$ as well. We consider the solution of (\ref{eq:mm-boltz}) in the Banach space $L^1(\real^3)$. The operator ${\cal Q}_s$ is a bounded map; ${\cal Q}_s: L^1(\real^3){\times}L^1(\real^3){\rightarrow}L^1(\real^3)$. It is easy to establish that $\|{\cal Q}_s\|_{L^1}{=}{\mathrm V} \sigma \varrho^2$. 

Equation (\ref{eq:mm-boltz}) is a linear first-order differential equation with constant coefficients. It can be converted into an integral equation if we integrate it along its characteristics. The result is
\begin{equation} \label{eq:mm-boltz-ie}
	F(\xi,t) = F_0(\xi) \re^{- \vartheta t} + \int_0^t 
	\re^{- \vartheta(t-s)} \vartheta \big\{ F \circ F \big\}(\xi,s) \rd s,\;\;\; t>s,
\end{equation}
where notation $\{F \circ F \}$ denotes the convolution of two distributions. Explicitly, it~is defined by
\begin{equation} \label{eq:wilds-convol}
	\vartheta(\xi) \big\{ f \circ g \big\} (\xi) = \int_{\mathbb{S}^{2}} \int f(\xi') g(\xi_*') \: {\mathrm V}  \sigma \: B(\theta)\: \rd \xi_* \rd \theta \rd \varepsilon.
\end{equation}

The local in-time solution of integral equation (\ref{eq:mm-boltz-ie}) for initial value of arbitrary magnitude was established by Wild (1951) who solved the Boltzmann equation by considering the iterates,
\begin{equation*} 
	F_{k+1}(\xi,t) = \re^{- \vartheta t} F_0(\xi) + \int_0^t 
	\re^{- \vartheta(t-s)} \vartheta \big\{ F_k \circ F_k \big\} (\xi,s)\rd s,\;\;\;k=1,2,\cdots,
\end{equation*}
with $F_1{=}0$. Wild's proof is essentially based on the Picard iteration scheme or the Banach contraction principle for non-linear equations. As expected, the principle works for the {\itshape local} existence for initial data of arbitrary size.
By an indirect approach (working with particle density rather than distribution function), Morgenstern (1954) showed that Wild's solution is global. The solution can be represented as a combination of the collision operators ${\cal Q}_s(\exp({- \vartheta t}) F_0(\xi))$. For interpretation of the solution and its combinatorial properties, see the review by Villani (2002).

Let us denote the boundedness of density function by $\sup_{(t, \xi)} |F| < A$.
Now we demonstrate how to construct the solution of the Boltzmann equation which shows characters of turbulence. The essential idea is to transform the integral in (\ref{eq:mm-boltz-ie}) into an infinite series as we have done for the vorticity equation. In parallel to vorticity non-linearity, the non-linear term in the present context is also quadratic. 
Let $t_k, k{=}0,1,2,\cdots,n{+}1$, be equal time steps in interval $[0,T]$ (cf. (\ref{eq:time-intervals})). It follows that equation (\ref{eq:mm-boltz-ie}) holds for every $t_k$,
\begin{equation} \label{eq:mm-boltz-ie-tk}
	F(\xi,t_k) = F_0(\xi) \re^{- \vartheta t_k} + \int_0^{t_k} 
	\re^{- \vartheta(t_k-s)} \vartheta \big\{ F \circ F \big\}(\xi,s) \rd s,\;\;\; t_k>s.
\end{equation}
The dual to this integral equation is
\begin{equation*} 
	F(\xi',t_k) = F_0(\xi') \re^{- \vartheta t_k} + \int_0^{t_k} 
	\re^{- \vartheta(t_k-s)} \vartheta \big\{F \circ F \big\}(\xi',s) \rd s,\;\;\; t_k>s,
\end{equation*}
where the dual convolution to (\ref{eq:wilds-convol}) is given by
\begin{equation} \label{eq:wilds-convol-dual}
	\vartheta(\xi') \big\{ f \circ g \big\} (\xi') = \int_{\mathbb{S}^{2}} \int f(\xi) g(\xi_*) \: {\mathrm V}  \sigma\: B(\theta)\: \rd \xi_*' \rd \theta \rd \varepsilon.
\end{equation}
Equation (\ref{eq:mm-boltz-ie-tk}) at a short time from the start of motion ($t=t_0$) is written as 
\begin{equation*} 
	F(\xi,t_0) = F_0(\xi) \re^{- \vartheta t_0}  + J_0(\xi,t_0). 
\end{equation*}
To follow the evolution, we multiply this equation by 
\begin{equation*}
	\re^{- \vartheta(t_1-t_0)} {\mathrm V} \sigma B(\theta) F(\xi_*,t_0)
\end{equation*}
and integrate the result over
\begin{equation*}
	\int_0^{t_1} \int \int_{\mathbb{S}^{2}}\;\; \big( \cdots \big) \;\; \rd \theta \rd \varepsilon \rd \xi_*' \rd t_0.
\end{equation*}
The distribution develops into the next instant $t=t_1$ is given by
\begin{equation} \label{eq:mm-boltz-sim-red-1}
\begin{split}
	F(\xi',t_1) & = F_0(\xi') \re^{-\vartheta t_1}  + \int_0^{t_1}  \re^{- \vartheta t_1} \vartheta \big\{ F_0 \circ F \big\} (\xi',s) \rd s + J_1(\xi',t_1) \\
	& = F_0(\xi') \re^{-\vartheta t_1} + I_1(\xi',t_1)+ J_1(\xi',t_1),
\end{split}
\end{equation}
where	
\begin{equation*} 
	J_1(\xi',t_1) = \int_0^{t_1} \re^{- \vartheta (t_1-s) } \vartheta \big\{J_0 \circ F \big\} (\xi',s) \rd s
\end{equation*}
in view of the duals.
Multiply equation (\ref{eq:mm-boltz-sim-red-1}) by 
\begin{equation*}
	\re^{- \vartheta(t_2-t_1)} {\mathrm V}  \sigma B(\theta) F(\xi_*',t_1)
\end{equation*}
and integrate the result over
\begin{equation*}
\int_0^{t_2} \int \int_{\mathbb{S}^{2}}\;\; \big( \cdots \big) \;\; \rd \theta \rd \varepsilon \rd \xi_*' \rd t_1,
\end{equation*}
we obtain the distribution up to $t=t_2$,
\begin{equation} \label{eq:mm-boltz-sim-red-2}
	F(\xi,t_2) = F_0(\xi) \re^{-\vartheta t_2} + \int_0^{t_2}  \re^{- \vartheta t_2} \vartheta \big\{ F_0 \circ F \big\} (\xi,s) \rd s + I_2(\xi,t_2)+J_2(\xi,t_2),
\end{equation}
where	
\begin{equation*}
	I_2(\xi,t_2) = \int_0^{t_2} \re^{- \vartheta (t_2-s) } \vartheta \big\{I_1 \circ F \big\} (\xi,s) \rd s,
\end{equation*}
and
\begin{equation*} 
	J_2(\xi,t_2) = \int_0^{t_2} \re^{- \vartheta (t_2-s) } \vartheta \big\{J_1 \circ F \big\} (\xi,s) \rd s.
\end{equation*}
Obviously we can carry out our similarity reduction on (\ref{eq:mm-boltz-sim-red-2}). The process can be repeated as many times as we wish. Let us concentrate on the distribution of the first particle after collisions. After $k(>2)$ reductions, we arrive at
\begin{equation} \label{eq:mm-boltz-sim-red-k}
	F(\xi',t_k) = F_0(\xi') \re^{-\vartheta t_k} + \int_0^{t_k}  \re^{- \vartheta t_k} \vartheta \big\{ F_0 \circ F \big\} (\xi',s) \rd s + \sum_{j=2}^{k-1} I_j(\xi',t_k) + J_k(\xi',t_k),
\end{equation}
where	
\begin{equation*} 
	I_j(\xi',t_k) = \int_0^{t_k} \re^{- \vartheta (t_k-s) } \vartheta \big\{ I_{j-1} \circ F \big\} (\xi',s) \rd s,
\end{equation*}
and
\begin{equation*} 
	J_k(\xi',t_k) = \int_0^{t_k} \re^{- \vartheta (t_k-s) } \vartheta \big\{ J_{k-1} \circ F \big\} (\xi',s) \rd s.
\end{equation*}
In reference to (\ref{eq:boltz-cutoff}), we assume that a cut-off has been introduced such that the cross-section $\sigma$ is finite. Equivalently, we require
\begin{equation*}
	\int \int_{\mathbb{S}^{2}} {\mathrm V}  \sigma B(\theta) \rd \theta \rd \varepsilon  \rd \xi \leq \kappa^* < \infty.
\end{equation*}
Then it is elementary to verify that 
\begin{equation*}
	\big| J_0 \big| < \kappa^* A^2 \Big( 1- \re^{\vartheta t_0}\Big),\;\;\;
	\big| J_1 \big| < {\kappa^*}^2 A^3 \Big( 1 - \re^{\vartheta t_1} - \vartheta t_1 \re^{\vartheta t_1} \Big),\;\;\; \cdots .
\end{equation*}
If we choose to divide the time interval $[0,T]$ into infinitely small increments, then $t_k \rightarrow 0$ as $k \rightarrow \infty$. Hence $|J_k| \rightarrow 0$. Consequently, equation (\ref{eq:mm-boltz-sim-red-k}) reduces to
\begin{equation*}
	F(\xi',t) = F_0(\xi') \re^{-\vartheta t} + \int_0^{t}  \re^{- \vartheta t} \vartheta \big\{ F_0 \circ F \big\} (\xi',s) \rd s + \sum_{j=2}^{\infty} I_j(\xi',t).
\end{equation*}
This integral equation can be rewritten as
\begin{equation} \label{eq:mm-boltz-sim-red-ie}
	F(\xi',t) = F_0(\xi') \re^{-\vartheta t} + \int_0^{t}  \int K_s(\xi',t,\xi_*,s) F (\xi_*,s) \rd \xi_* \rd s + q_s(\xi',t),
\end{equation}
where
\begin{equation*}
	K_s(\xi',t,\xi_*,s) =  \re^{- \vartheta t } F_0(\xi') \Big(\int_{\mathbb{S}^{2}}  {\mathrm V}  \sigma B(\theta) \rd \theta \rd \varepsilon\Big).
\end{equation*}
This equation is viewed as a Volterra-Fredholm equation, and its solution can be expressed as
\begin{equation} \label{eq:mm-boltz-red}
\begin{split}
	F(\xi',t) = F_0(\xi') \re^{-\vartheta t} & + \int_0^{t}  \int H_s(\xi',t,\xi_*,s) F_0 \re^{-\vartheta s} \rd \xi_* \rd s \\
	& + q_s(\xi',t) + \int_0^{t}  \int H_s(\xi',t,\xi_*,s) q_s(\xi_*,s) \rd \xi_* \rd s,
\end{split}
\end{equation}
where $H_s$ is the resolvent of kernel $K_s$. We notice that the transformed Boltzmann equation (\ref{eq:mm-boltz-red}) has the analogous mathematical structure to (\ref{eq:vort-nonlinear-VIE}). Without repeating the procedures of successive approximations, we assert that the solution is given by
\begin{equation} \label{eq:mm-boltz-series-sol}
	F(\xi',t) = \tilde{F_0}(\xi',t) + \sum_{m{\geq}2} S_m \:\tilde{V}\: [\:\tilde{F_0}(\xi',t)\:]^m,
\end{equation}
where
\begin{equation*}
	\tilde{F_0}(\xi',t) = F_0(\xi') \re^{-\vartheta t} + \int_0^{t}  \int H_s(\xi',t,\xi_*,s) F_0(\xi_*) \re^{-\vartheta s} \rd \xi_* \rd s.
\end{equation*}

Since the particle density for Maxwellian molecules with cut-off is calculated from the distribution function $\upvarrho (t) {=} \int F(\xi,t) \rd \xi$,
we infer that the density $\upvarrho$ must also have a series expansion for the sake of consistency. 
Integrating equation (\ref{eq:mm-boltz}) over $\real^3$ and transforming the result into an integral equation by method of characteristics, we obtain
\begin{equation} \label{eq:mm-boltz-rho-ie}
	\upvarrho(t) = \re^{- \vartheta t} {\upvarrho_0} + \int_0^t 
	\re^{- \vartheta(t-s)} {\mathrm V}  \sigma \big( \upvarrho(s) \big)^2 \rd s,\;\;\;t>s.
\end{equation}
This equation holds for every $t{=}t_k$. Now the multiplicative factor in our similarity reductions reads
\begin{equation*}
	{\mathrm V}  \sigma \: \upvarrho \: \re^{- \vartheta(t_k - s)}.
\end{equation*}
Without going into technical detail, we write the result as follows:
\begin{equation} \label{eq:mm-boltz-nie}
	\upvarrho(t) = \re^{- \vartheta t} \upvarrho_0 + \int_0^t g(t,s) \upvarrho(s) \rd s + {\mathrm q} (t,\upvarrho),\;\;\;t>s.
\end{equation}
The integral kernel $g$ has a very simple analytic form $g(t,s) =  {\mathrm V}  \sigma \upvarrho_0 \exp(- \vartheta t) > 0$, where the initial density $\upvarrho_0$ acts like a mollifier (in space variable) on the collision cross-section. The non-linear term ${\mathrm q}$ has an analogous form to $q_s$ in (\ref{eq:mm-boltz-sim-red-ie}). Hence the density $\upvarrho$ does have an analogous series expansion to (\ref{eq:mm-boltz-series-sol}).
Interestingly, we notice that $\upvarrho(t)=\upvarrho(0), \forall t\geq 0$. It follows that the density expansion must possess this invariant property as well.

{\itshape Let Maxwellian molecules in a dilute gas undergo binary collisions according to the spatially homogeneous Boltzmann equation. If the number of molecules at initial instant is given, then the Cauchy problem has a unique solution for all time. In addition, if the size of the initial data exceeds a certain magnitude such that a large number of terms in (\ref{eq:mm-boltz-series-sol}) are needed to represent the converged solution to prescribed accuracy, then there is a time interval over which the density function of the molecules has a phase-space distribution resembling the continuum incompressible turbulence. The distribution is strongly dependent on the initial data. }

We remark that the results we just obtained can be generalized to the spatially inhomogeneous Boltzmann equation (in the absence of an external force),
\begin{equation*} 
	{\partial F}/{\partial t} + (\xi. \nabla_x) F = {\cal Q}(F,F),
\end{equation*}
because we can always write $F(x{+}\xi t, \xi,t)=F^{\sharp}(x,\xi,t)$, as there are no space boundaries (Arkeryd 1972). The function $F^{\sharp}$ is the periodic continuation in the space variable $x$ with period $L$ (say) of the density function $F$ in (\ref{eq:boltzmann}). The period $L$ is supposed to be bounded as the molecular velocity $\xi$ can be considered as finite at all time on the ground of physics. Thus the Boltzmann equation can be rewritten as 
\begin{equation*} 
	{\partial F^{\sharp}}/{\partial t} = {\cal Q}(F^{\sharp},F^{\sharp}).
\end{equation*}
This equation can be solved with initial data (\ref{eq:boltzmann-ic}). 

If we are free to specify an additional requirement on the initial data, the need for a cut-off can be relaxed once and for all. The cut-off is in reality a space-related quantity. To be consistent with the space-homogeneity, it is plausible to stipulate a {\itshape space-wise} constraint on the initial data. Suppose that the total number of the molecules at $t{=}0$ is bounded, namely,
\begin{equation*}
	\int_{\real^3{\times}\real^3} F(x,\xi,0) \rd \xi \rd x \;{<}\; \infty.
\end{equation*}
Then the product of the impact parameter $r_i$ and the density $\varrho_0$ in the neighbourhood of infinity must render the collision frequency $\vartheta$ finite. In physics, there are simply no molecules at large distances and there exist no grazing collisions.
\subsection{Origin of macroscopic randomness}
Because the solution of the Navier-Stokes Cauchy problem is unique, any flow field can be reproduced from an identical initial condition. The spatio-temporal structure of the solution implies that the velocity field evolves orderly and regularly in space and in time for any large initial data of finite energy. The occurrence of a local stagnation state in the vorticity field has to be an instantaneous and temporary event except in the final stage of the decay. In theory, the evolution records at any particular $(x,t)$ can be predicted and repeated. Therefore the equations of fluid dynamics do not govern any stochastic aspect in flow development. This analytical property appears to be in contradiction to experience. For example, the turbulent flow in Reynolds' pipe experiments (Reynolds 1883) contains a generic character: The velocity measurements at fixed space-time locations for nominally identical initial conditions are irregular and do not have a repeatability property. What are the random processes which underlie the fluctuations in fluid motions? 

In the continuum description of fluid motions, the dynamic viscosity $\mu$ and the thermal conductivity ${\mathrm k}$ are defined by Newton's viscosity law and Fourier's heat law respectively, 
\begin{equation*}
	\tau = \mu \frac{\partial u}{\partial y},\;\;\;\;\;\; {\mathrm Q}= -{\mathrm k} \frac{\partial {\mathrm T}}{\partial y}.
\end{equation*}
The existence of the gradients matters. One of the most important results of the kinetic theory is that the transport coefficients are all due to the molecular agitations (see, for example, Chapman \& Cowling 1970; de Groot \& Mazur 1984; Chandler 1987). For real dilute gases, the viscosity has the expression,
\begin{equation*}
\mu \: {\propto} \: \big(m k_B {\mathrm T}\big)^{1/2}/{\bar{d}},
\end{equation*}
where $\bar{d}$ denotes the mean diameter of the molecule.
A similar relation holds for the thermal conductivity, 
\begin{equation*}
	{\mathrm k} \propto {k_B}^{3/2}{\mathrm T}^{1/2}/{\big({m^{1/2}} \: \bar{d} \big)}.
\end{equation*}
For non-reactive monatomic gases, these two transport coefficients are related by ${\mathrm k}/\mu{=}c_V 5/2$, where $c_V$ is the specific heat at constant volume. We also observe that the last three terms in the macroscopic quantities of fluid dynamics (\ref{eq:macro-fluiddyn}), $p_{ij}$, $\rho {\mathscr E}$ and $Q_i$, are all defined in terms of {\itshape fluctuations} of particle velocity $\xi$. In statistical physics, the coefficients of monatomic gases can be estimated according to the fluctuation-dissipation theorem. Significantly, the net effect of the velocity and temperature gradients leads to transport phenomena; the viscosity is a result of momentum transport, the conductivity energy transport. The presence of the fluctuations prompts these transport processes.

The solution we obtained for Maxwellian molecules with cut-off (\ref{eq:mm-boltz-series-sol}) is expedient and provides some evidence to vindicate the turbulence structure implied in the Navier-Stokes dynamics. Taking the inner product with $u$ on momentum equation (\ref{eq:ns}), integrating over space-time and integrating by parts, we get the law of conservation of energy,
\begin{equation} \label{eq:energy-equality}
	\frac{1}{2}\int u^2(x,t) \rd x  + \nu \int_0^{t} \int \omega^2(x,\tau) \rd x \rd \tau = \frac{1}{2} \int u^2_0(x) \rd x.
\end{equation}
To illustrate the connection between viscous dissipation and increase in the internal energy, we integrate the last equation in (\ref{eq:fluiddyn}) over space, assuming $X_i{=}0$ and $Q_i{=}0$. In view of Gauss' divergence theorem, the term involving spatial derivative drops out. Thus we express the law of energy conservation in the invariance form,
\begin{equation} \label{eq:energy-balance}
	\frac{\rd}{\rd t } \int \Big( \frac{1}{2}\rho u^2 + \rho {\mathscr E} \Big)(x,t) \rd x = 0.
\end{equation} 
Furthermore the increment in the internal energy up to time $t$ is related to the enstrophy by
\begin{equation} \label{eq:internal-energy-increase}
 \begin{split}
	\rho \! \int {\mathscr E} (x,t) \rd x - \rho \! \int {\mathscr E} (x,0) \rd x &  = 
	\frac{3}{2} k_B \Big( \int \varrho \:{\mathrm T}(x,t) \rd x  - \int \varrho \: {\mathrm T}(x,0) \rd x  \Big) \\
	\quad & = \mu \int_0^t  \! \int \omega^2(x,\tau) \rd x \rd \tau 
	\end{split}
\end{equation}
in view of the last relation in (\ref{eq:macro-p}). As a result of the dissipation, the loss of the kinetic energy is accompanied by a temperature rise; the internal energy of fluid's molecules is the heat sink. 

Consider an initial data and the subsequent motion given by the series solution  (\ref{eq:vort-series-sol}). The last integrand in (\ref{eq:internal-energy-increase}) is in fact an expansion in the vorticity scales in the series. The dissipation must be operative on every term in the expansion; the dissipation of energy gives effect to all the vorticity scales wherever there are non-vanishing shears. Since the eddies are subject to {\itshape manifold diffusions} as implied in the convolutions, the energy of the small eddies is being trimmed off in successive and iterative viscous processes during the flow evolution. Roughly speaking, the sizes of the smallest eddies are estimated to be in the order of $\nu$ and are larger than the mean free path of fluid's molecules by a few orders of magnitude, say $\sim O(10^3)$, in incompressible turbulence (see, for example, Bradshaw 1971; Tennekes \& Lumley 1972; Leslie 1973; Davidson 2004). In the sea of the vorticity eddies, it is plausible that the dissipation must be most effective at the smallest-scales where, as expected, the viscous forces are comparable to eddies' inertia forces. The viscous diffusive effects of the smallest vortices ought to be sensitive to molecular concentrations which vary due to the thermal fluctuations. For the eddies of diffusive sizes, the local velocity gradients per unit area are much stronger, hence the momentum transports across the surface areas of these eddies have to be much more efficient. Recall that the transports are the aggregates of the molecules transversing adjacent layers of a fluid. Thus, the momentum transports on the dissipative eddies must depend on the local fluctuations of the molecules involved. In the present context, the internal thermal energy serves as an energy reservoir whose local specifications are under constant irregular variations. Because the viscous dissipation {\itshape interfaces} the thermal reservoir, the process of annihilating shears must be non-uniform among the dissipative eddies as soon as there exist sufficiently numerous small-scale motions because, on the dissipative scales, both the rate of the momentum transports and local molecules' concentrations are subject to random fluctuations. At a particular moment during the motion when the non-linearity becomes strong, this dissipative irregularity initiates localized embryonic randomness due to local eddy decimation, which most likely populates at the smallest scales. The initial randomness is relayed to the entire flow by the surviving eddies according to the Biot-Savart law. As the flow evolves, birth of the embryonic fluctuations in large quantity is incessant and the randomness character persists. The higher the initial energy content the more intensive the fluctuations. (For high Reynolds-number flows such as sustained jets or wakes, the number of the small vortices is abundant, so it is entirely plausible that there exists a well-defined mean or ensemble fluctuation.) Once a large portion of the small-scale eddies dies off in the dissipation process, the randomness ceases to be pronounced and perceptible. 
Soon a flow field with mild unsteadiness reinstates and it evolves predominantly by diffusion. After any remaining residual shears have been smoothed out, the flow reverses to equilibrium vorticity-free state (and fluid's molecules become the absolute Maxwellian). 

{\itshape In turbulent incompressible flows, the macroscopic velocity appears to be irregular in space and in time. Being the only irreversible process, viscous dissipation on the copious eddies of small scales expedites entropy production in order to restore the flows to the equilibrium of shear-free state. The dissipation on individual small-scale eddies is a random process in space and in time as fluid's thermal energy in every place fluctuates constantly. In brief, fluid motions on the continuum inhere the microscopic randomness through the dissipation.}
\section{Implications for turbulence}
In the classical theories of turbulence, statistical approaches are used to capture the complexity of the velocity field (see, for example, Richardson 1922; Taylor 1935; Kolmogorov 1941$a, b$; Heisenberg 1948; von K\'{a}rm\'{a}n 1948; Batchelor 1953; Kraichnan 1967). The vorticity solution in (\ref{eq:vort-series-sol}) comprises basic ingredients for turbulence. In comparison with observed features of turbulence, it becomes necessary to clarify a number of conceptual issues on the nature of turbulence. Conventionally, the origin of turbulence has been imputed to instability in fluid motions. By a back-to-back analysis of the initial value problems involved, we show that linearization of the equations of motion consists in a prognostic scheme which is too simplistic to describe the flow evolutions beyond a small time interval from the start (depending on the initial conditions). In particular, none of the flow structure resembling turbulence exists over this short interval in the linearized flows. We then study the rate of production of vorticity during flow evolutions; we explain why small perturbative differences in initial conditions cannot be regarded as the cause of randomness in turbulent motions. In terms of the vorticity solution, we next provide an account for the transition in Reynolds' pipe flow experiments. We show that the changeover from a laminar flow to turbulence, in a graduate or abrupt manner, is an inevitable consequence of the non-linearity in fluid dynamics. In a nutshell, the fallacies of instability theories on fluid motions are discussed. Similarly, we find that the properties of the series solution are valuable to avoid confusions on energy hypotheses in turbulence. Given the fact that the statistics methods are based on the {\itshape assumption} that the instantaneous velocity in turbulence decomposes into a mean motion and a fluctuation, an inference of the decomposition is given in the light of the Navier-Stokes determinism and the kinetic theory of gases.
\subsection{Consequence of linearization}
Every fluid motion is an initial value problem of the Navier-Stokes dynamic equations. We have shown that the general solution of the equations is unique and exists for all finite time. Before this fact has been established, a host of approximation methods has been proposed to analyze the dynamics. One of the well-known proposals is to linearize the dynamic equations. On the basis of the linearized equations, criteria are derived to determine if a flow becomes stable or not when it is subject to disturbances. Consider a mean fluid motion, denoted by the triplet $(\bar{U}(x,t), \bar{P}(x,t), \bar{\Omega}(x,t))$ that is a solution of the Navier-Stokes equations. At a subsequent moment ($t_s$), a perturbation $u'_0(x,t_s)$ is introduced into the established mean motion. The superposition of the mean flow and the perturbation {\itshape is assumed} to satisfy the Navier-Stokes equations; the aim is to figure out the development of the perturbation. Furthermore, the perturbation flow ($u'(x,t), p'(x,t), \nabla{\times}u'(x,t))$) is supposed to remain infinitesimal at all subsequent time $t{>}t_s$: 
\begin{equation} \label{eq:small-pert}
	|u'| \;{\ll}\; |\bar{U}|,\;\;\; |\omega'| \;{\ll}\; |\bar{\Omega}|  \;\;\; \forall (x,t) \in \real^3{\times}[t_s,T],
\end{equation}
so that the modifications due to the disturbance on the mean motion are small and thus negligible.
In practice, the size of any perturbation quantity is taken as a few percentage of the mean flow quantity. 

In order to analyze the most general cases, we treat both the mean flow and the perturbation as unsteady velocity fields. Let
\begin{equation*}
	u(x,t)=\bar{U}(x,t)+u'(x,t),\;\;\;p(x,t)=\bar{P}(x,t)+p'(x,t).
\end{equation*}
Substituting these expressions into the Navier-Stokes equations ((\ref{eq:continuity}) and (\ref{eq:ns})) and simplifying, we get
\begin{equation} \label{eq:linear-ns}
	\partial_t u' - \nu \Delta u' = -(u'.\nabla) \bar{U} - (\bar{U} . \nabla) u' -(u'.\nabla) u' - \nabla p' / \rho,\;\;\; \nabla.u'=0.
\end{equation}
The initial condition is given by
\begin{equation} \label{eq:linear-ns-ic}
	u'(x,t_s)=u'_0(x),
\end{equation}
which is a smooth, localized function in $\real^3$. We only consider flows of finite energy i.e. $\|u'_0\|_{L^2(\real^3)}{<}\infty$. Linearized momentum (\ref{eq:linear-ns}) can be rewritten as
\begin{equation} \label{eq:ns-pert}
	\partial_t u' - \nu \Delta u' = u'{\times}\omega' + u'{\times}\bar{\Omega} + \bar{U}{\times}\omega' + \nabla \chi',
\end{equation}
where the expression, $\chi'{=}{-}(p'/\rho+\nabla(u'.u')/2{+}u'.\bar{U})$, denotes the modified pressure. Then we take the curl of equation (\ref{eq:ns-pert}) to obtain the vorticity equations for the perturbation components ($i{=}1,2,3$),
\begin{equation} \label{eq:pert-vort}
 \begin{split}
 \partial_t \omega'_i - \nu \Delta \omega'_i =  (\omega'.\nabla)u'_i - (u'.\nabla)\omega'_i &+ (\bar{\Omega}.\nabla)u'_i - (u'.\nabla)\bar{\Omega}_i \\
 \quad & \hspace{0mm}  + (\omega'.\nabla)\bar{U}_i - (\bar{U}.\nabla)\omega'_i = {\mathscr R}'_i (x,t).
 \end{split}
\end{equation}
The initial perturbation vorticity is given by
\begin{equation} \label{eq:pert-vort-ic}
	\omega'(x,t_s)=\omega'_0(x) \; \in C_c^{\infty}.
\end{equation}

Our objective is to examine the perturbed vorticity equations (\ref{eq:pert-vort}) when a linearization is made. Only in this way can the solutions be compared with the results of the full dynamic equations without ambiguity. It is clear that 
\begin{equation*}
	\int_{t_s}^T \!\! \int \sum_{i=1}^3 {\mathscr R}'_i (x,t) \; \rd x \rd t =0
\end{equation*}
because the incompressibility hypothesis holds for both the mean velocity and the perturbation. Thus the total perturbed vorticity is an invariant of the motion:
\begin{equation} \label{eq:pert-vort-inv}
	\frac{\rd }{\rd t }\int \partial_t^{\beta} \partial_x^{\alpha} (\omega_1'+\omega_2'+\omega_3')(x,t) \rd x = 0.
\end{equation}
From this invariant principle and its generalization, we establish that
\begin{equation} \label{eq:pert-vort-bound}
	\omega'(x,t) \in C^{\infty},\;\;\; x \in \real^3,\;\;\; t>t_s.
\end{equation}
Thus we will work in the space of smooth functions.
To recover the perturbation velocity from the vorticity, we make use of the Biot-Savart law,
\begin{equation} \label{eq:pert-biot-savart}
u'(x;t) = \frac{1}{4 \pi} \int \frac{(x{-}y)}{|x{-}y|^3} {\times} \omega'(y;t) \rd y.	
\end{equation}

In view of the assumptions for linearization, the perturbation of the non-linearity in (\ref{eq:ns-pert}), or any terms involving  the product of $u'$ and $\omega'$, may be neglected. Thus equations (\ref{eq:pert-vort}) are simplified as
\begin{equation} \label{eq:linear-ns-vort}
 \partial_t \omega'_i - \nu \Delta \omega'_i = (\bar{\Omega}.\nabla)u'_i - (u'.\nabla)\bar{\Omega}_i + (\omega'.\nabla)\bar{U}_i - (\bar{U}.\nabla)\omega'_i.
\end{equation}
The left-hand side describes the classical linear diffusion process. None of the terms on the right contains the mechanism of vorticity stretching. All the terms are product of the mean flow and the perturbation; all the non-linear effects in fluid dynamics have been weakened. Mathematically, equations (\ref{eq:linear-ns-vort}) form a system of {\itshape linear} second-order parabolic equations. By virtue of invariance principle (\ref{eq:pert-vort-inv}), linearized equations (\ref{eq:linear-ns-vort}) can be converted into
\begin{equation*}
 \begin{split}
	\omega'_i(x,t) = & \int Z(x,t{-}t_s,y) (\omega'_0)_i(y) \rd y \\
	\quad & + \int_{t_s}^t \!\! \int \Big( (\nabla_y Z.u')\bar{\Omega}_i - (\nabla_y Z.\bar{\Omega})u'_i 
	+ (\nabla_y Z.\bar{U})\omega'_i - (\nabla_y Z.\omega')\Bar{U}'_i \Big) \rd y \rd s. 
	 \end{split}
\end{equation*}
In view of the Biot-Savart law (\ref{eq:pert-biot-savart}), we find that the perturbed vorticity components satisfy the following linear Volterra-Fredholm integral equations:
\begin{equation} \label{eq:pert-vort-ie}
	\omega'_i(x,t) = \varpi'_i(x,t) + \int_{t_s}^t \!\! \int  \sum_{j=1}^3 k'_{ij}(x,t,y,s) \omega'_j(y,s) \rd y \rd s,\;\;\; i=1,2,3,
\end{equation}
where the kernel, $k'_{ij}{=}k'_{ij}(Z,\bar{U}, \bar{\Omega})$, is a $3{\times}3$ matrix, and the function $\varpi'$ is the counterpart of the mollified initial vorticity (\ref{eq:mollified-vort-ic}),
\begin{equation} \label{eq:pert-vort-ic-mod}
	\varpi'(x,t) = \int {\mathbf Z}(x,t{-}t_s,y) \omega'_0(y) \rd y.
\end{equation}
Equations (\ref{eq:pert-vort-ie}) have integrable Volterra kernels for $(x,t) {\in} \real^3{\times}[t_s,T]$. The solution has the explicit form 
\begin{equation} \label{eq:pert-vort-sol}
	\omega'_i(x,t) = \varpi'_i(x,t) + \int_{t_s}^t \!\! \int \sum_{j=1}^3 h'_{ij} (x,t,y,s) \varpi'_j(y,s) \rd y \rd s.
\end{equation}
The matrix $h_{ij}'$ stands for the resolvent kernel of $k_{ij}'$ which can be found according to (\ref{eq:resolvent}). The eigenvalue spectrum of (\ref{eq:pert-vort-ie}) is known to be empty in view of the well-established theory of Volterra integral equations. Hence there can be no bifurcations in the perturbation vorticity. Since the linearization restricts the size of the perturbation $u'$, the summation series in (\ref{eq:pert-vort-sol}) are expected to converge fairly quickly to give a good approximation to perturbation's development. 

Instead of the problem defined in (\ref{eq:pert-vort}) and (\ref{eq:pert-vort-ic}), the mathematical problem posed by the perturbation $\omega'$ on the mean flow $\bar{\Omega}$ must be formulated as follows: To solve the full vorticity equation,
\begin{equation} \label{eq:ns-pert-vort}
 \partial_t \omega^* - \nu \Delta \omega^* =  (\omega^*.\nabla)u^* - (u^*.\nabla)\omega^*,
\end{equation}
subject to the initial condition 
\begin{equation} \label{eq:ns-pert-vort-ic}
	\omega^*(x,t_s)=\omega^*_0(x)=\bar{\Omega}(x,t_s)+\omega'_0(x), \;\;\; x\in \real^3.
\end{equation}
The perturbation is a solenoidal velocity field throughout the motion $\nabla.u^*{=}0$ and $\nabla.u_0'{=}0$. We use the superscript $^*$ for the flow quantities of the problem. The Cauchy problem defined by (\ref{eq:ns-pert-vort}) and (\ref{eq:ns-pert-vort-ic}) is known as the problem of receptivity which intends to model the effect of forced excitations on the mean flow. The mathematical functions of the excitations are arbitrary as long as they are finite and localized. In practice, they are often waves output from wave-generators. In our theory, we consider the allowable functions as smooth functions or the functions well-approximated by the Dirac impulse functions. In comparison to equations (\ref{eq:vorticity}) and (\ref{eq:vort-ic}), we assert that the nature of the solution of this Cauchy problem is essentially identical to that of the Navier-Stokes Cauchy problem. Compared with (\ref{eq:pert-vort-ic-mod}) the mollified initial vorticity has two terms,
\begin{equation} \label{eq:ns-pert-vort-ic-mod}
	\varpi^*(x,t) = \int {\mathbf Z}(x,t{-}t_s,y) \Big( \omega'_0(y) + \bar{\Omega}(x,t{-}t_s) \Big) \rd y,\;\;\; t{-}t_s >0.
\end{equation}
Then the Volterra-Fredholm filtered initial data are given by
\begin{equation*}
	\gamma^*(x,t) = \varpi^*(x,t) + \int_{t_s}^t \!\! \int H(x,t,y,s) \varpi^*(y,s) \rd y \rd s.
\end{equation*}
Thus the vorticity solution of the receptivity problem has the following analogous series expansion:
\begin{equation} \label{eq:vort-series-sol-pert}
	\omega^*(x,t) = \gamma^*(x,t) + \sum_{m{\geq}2} S_m V^*[\gamma^*(x,t)]^m,
\end{equation}
where $V^*$ is analogous to $V$ of (\ref{eq:vort-series-sol}). This is a general solution in the sense that no restriction has been imposed on the size of the initial data $\omega'_0$. In fact, we see that the perturbation factor in $\varpi^*$ must be capable of instigating large modifications not only in $\varpi^*(x,t)$ but also in every integro-power term in the sum (\ref{eq:vort-series-sol-pert}).

{\itshape In brief, the solution of the receptivity problem consists in a fluid motion of turbulence as soon as the initial perturbation vorticity superimposed on the mean motion is strong enough, so that numerous terms in (\ref{eq:vort-series-sol-pert}) are required for convergence. Moreover, if the mean motion itself is a turbulent flow, then the effect of introducing forced perturbations is to modify the original turbulent motion into another motion, possibly a turbulent one.} 

Given the initial data (\ref{eq:ns-pert-vort-ic-mod}) and (\ref{eq:pert-vort-ic-mod}), we are in a position to appraise the effect of linearization by comparing (\ref{eq:vort-series-sol-pert}) with (\ref{eq:pert-vort-sol}). Attention is drawn to the fact that these solutions can be evaluated to an arbitrary degree of accuracy for any identical initial condition and any mean flow, within the fundamental laws of continuum physics. It is evident that there is a short time interval $t_{\epsilon}$ over which the two solutions agree; the function,
\begin{equation} \label{eq:zeroth-order-heat}
	\int {\mathbf Z}(x,t_1,y) \omega'_0(y) \rd y, \;\;\; t_1-t_s \approx t_{\epsilon} \rightarrow 0^+,
\end{equation}
is identical in these solutions. We call function (\ref{eq:zeroth-order-heat}) over $t_{\epsilon}$ as the zeroth-order solution since the mean motion does not yet play a role. It is apparent that solution (\ref{eq:zeroth-order-heat}) portrays the {\itshape linear diffusion} of the initial perturbation vorticity up to the time $t_1$. In other words, good comparison can only be maintained in the designated time interval immediately after the introduction of the perturbation. The diffusion has a generic character; it is independent of the mean flow. In view of the properties of the heat kernel $Z$, the zeroth-order flow must closely resemble the initial perturbation vorticity,
\begin{equation*}
	\omega^*(x,t) \sim \omega'(x,t) \sim \omega_0(x),\;\;\; t \in (t_s, t_1],
\end{equation*}
particularly for small $\nu {\rightarrow} 0$. Beyond the interval $t_{\epsilon}$, the solution in (\ref{eq:pert-vort-sol}) from the linearized equations cannot faithfully describe perturbation's development any longer because the function $\gamma^*$ must be in substantial disparity with the second term in (\ref{eq:pert-vort-sol}). In fact, the first-order approximation for $\gamma^*$ (cf. (\ref{eq:vort-sol-first-order-VIE})) does not even agree with the zeroth-order solution for $t{>}t_{\epsilon}$. Furthermore, both solutions (\ref{eq:pert-vort-sol}) and (\ref{eq:vort-series-sol-pert}) depend on space as well as on time as every motion of fluid dynamics is an initial value problem governed by the complete Navier-Stokes equations. Thus the temporal character is indispensable. 

There is another issue related to the receptivity problem which is of conceptual importance. Turbulence ensues so long as the initial data (or the unit Reynolds number) are large enough even in the complete absence of perturbation; turbulence {\itshape is not} a result of the forced excitations by disturbances. Pragmatically, it is extremely difficult to design disturbance-free test facilities for laminar flows of purposeful Reynolds number, for example, a wind tunnel of zero free-stream turbulence and negligible acoustic interference. Although noisy environments have often disguised the very nature of turbulence, it is important to quantify the effects of disturbances. Should it be desirable to study the influence of particular {\itshape external} perturbations, we must solve the generalized initial value problem of receptivity (for fixed boundary conditions for the sake of illustrating the principle): Given a mean flow with vorticity $\omega^*(x,t_s)$. Let $(\omega_0'(x_j))_k$ be introduced into the flow at time $t_k, k=1,2,\cdots,k$, where $t_s < t_1 < t_2 < \cdots < t_k$, and at the locations $x_j,j=1,2,\cdots,l$. Generally, we must solve the Cauchy receptivity problem (\ref{eq:ns-pert-vort}) $k$-times. For every value of $k$, the initial data must be assigned as
\begin{equation} \label{eq:ns-pert-vort-gen-ic}
	\omega^*_0(x,t_k)= \omega^*(x,t_k) + \sum_{j=1}^l\big(\omega_0'(x_j) \big)_k,
\end{equation}
where $\omega^*(x,t_k)$ stands for the solution of the Navier-Stokes equations.

{\itshape We have shown how the linearization of the Navier-Stokes equations degrades the essence of the non-linearity in fluid dynamics; the linearized equations are unable to describe the leading behaviour of the Volterra-Fredholm filtered non-linearity $\gamma^*$, and thus they do not offer a reliable model for initiating turbulence.}  
\subsection{Rate of vorticity production}
For any function $g(x) \in C_c^1$, the following identity holds:
\begin{equation} \label{eq:total-vort-integral}
	\int g(x) \sum_{i=1}^3 \Big( (\omega.\nabla) u_i - (u.\nabla)\omega_i \Big) \rd x =0,
\end{equation}
because of the solenoidal conditions $\nabla.u {=} 0$ and $\nabla.\omega {=}0$. This is a generalized integrability condition of (\ref{eq:vort-rhs-integrability}). Consider the evolution equation for the total vorticity,
\begin{equation} \label{eq:total-vort-ns}
	\partial_t \upomega - \nu \Delta \upomega = \sum_{i=1}^3 \Big( (\omega.\nabla) u_i - (u.\nabla)\omega_i \Big).
\end{equation}
We denote the initial data by $\upomega_0$. To simplify our analysis, we take into account the decay that $\upomega \rightarrow 0$ sufficiently fast as $|x| \rightarrow {\infty}$. Multiplying the equation by $\upomega$ and integrating over $\real^3$, we arrive at the bound
\begin{equation} \label{eq:vort-enstrophy}
	\frac{\rd }{\rd t} \int \upomega^2 \rd x = - 2\; \nu \int (\nabla \upomega)^2 \rd x \leq 0,
\end{equation}
where the last equality sign holds only when the viscosity $\nu$ vanishes. {\itshape Hence the rate of change of the total enstrophy in the Navier-Stokes system decreases monotonically over the entire course of flow evolution}. The result is plausible as there exists no source for production of vorticity (such as a solid surface). Similarly, we derive the following inequality for the rate of change of the enstrophy gradient:
\begin{equation} \label{eq:d-vort-enstrophy}
	\frac{\rd }{\rd t} \int (\nabla \upomega)^2 \rd x = - 2\; \nu \int (\Delta \upomega)^2 \rd x \leq 0.
\end{equation}
The generalization of (\ref{eq:d-vort-enstrophy}) to higher derivatives is straightforward and hence their derivations are omitted. Evolution equation (\ref{eq:total-vort-ns}) for the total vorticity satisfies a quasi-linear parabolic equation; the total vorticity behaves like heat flow in linear diffusion, where
it is known that the rate of entropy production never increases. This is known as the Glansdorff-Prigogine evolution criterion for linear diffusion of irreversible systems (Glansdorff \& Prigogine 1971). 
 
For the initial value problem of freely-evolving flows in $\real^3$,  the fluid motion forms a closed thermodynamic system in which there is an exchange of energy between the fluid motion and the internal energy of fluid's molecules. Evidently, the process of the evolution is an irreversible process because the kinetic energy of the fluid dissipates into heat by viscosity. The enstrophy is just a measure of the bulk entropy of the motion (see, for example, Kondepudi \& Prigogine 1998). For the sake of convenience, the change of the entropy $\rd {\mathrm S}$ of the system is regarded to consist of two parts:
\begin{equation*}
	\rd {\mathrm S} = \rd_e {\mathrm S} + \rd_i {\mathrm S}.
\end{equation*}
It is clear that $\rd_e {\mathrm S} = \rd Q/{\mathrm T} > 0$. The second part $\rd_i {\mathrm S} (\geq 0 )$ is the entropy change due to the irreversible process of the dissipation.
The second law of thermodynamics says $\rd {\mathrm S} \geq 0$.
As $t \rightarrow \infty$, the fluid system restores to its equilibrium vorticity-free state which can be characterized either by zero entropy production or by the fact that the entropy reaches a maximum.

As a variational problem, the energy Lagrangian at each instant of time can be expressed as
\begin{equation*}
\mathscr{L}	= \frac{1}{2} \big| \nabla \uppsi \big|^2 - \upomega \uppsi 
\end{equation*}
for some smooth function $\uppsi{=}\uppsi(x)$. To minimize the energy functional
\begin{equation*}
\int \mathscr{L}(x) \rd x ,
\end{equation*}
we make use of the Euler-Lagrange equation (see, for example, Courant \& Hilbert 1966; Byron \& Fuller 1969). The minimization yields Poisson's equation
\begin{equation*}
	\Delta \uppsi (x) = - \upomega(x),
\end{equation*}
if we impose the decay condition $\uppsi \rightarrow 0$ sufficiently fast as $|x| \rightarrow \infty$.
Now $\uppsi$ is identified as the scalar stream function for the total vorticity (cf. (\ref{eq:stream-func-poisson})). Standard elliptic theory assures us of the existence of the stream function. In other words, {\itshape the Navier-Stokes dynamics defines a motion with a minimum kinetic energy among all admissible finite-energy motions}. 

Consider a second motion with total vorticity $\bar{\upomega}$ which satisfies (\ref{eq:total-vort-ns}). We regard this motion as a perturbation with initial data $\bar{\upomega}_0$. We are particularly interested in motions due to fluctuations in the initial conditions. Denote the perturbed solenoidal velocity and thereby the associated vorticity by $\bar{u}$ and $\bar{\omega}$. We also postulate that $\bar{u}$ and $\bar{\omega}$ decay at infinity. 
Evidently, the excess vorticity ($\upomega - \bar{\upomega}$) is governed by
\begin{equation*} 
	(\partial_t - \nu \Delta )(\upomega - \bar{\upomega}) = \sum_{i=1}^3 \Big( (\omega.\nabla) u_i - (u.\nabla)\omega_i  - (\bar{\omega}.\nabla) \bar{u}_i + (\bar{u}.\nabla)\bar{\omega}_i \Big).
\end{equation*}
By the analogous procedures of deriving (\ref{eq:vort-enstrophy}) and (\ref{eq:d-vort-enstrophy}), we assert that
\begin{equation} \label{eq:delta-vort-enstrophy}
	\int (\upomega - \bar{\upomega})^2 (x,t) \rd x \:{\leq}\: \int (\upomega_0 - \bar{\upomega}_0)^2(x) \rd x,
\end{equation}
and 
\begin{equation} \label{eq:d-delta-vort-enstrophy}
\int \big( \nabla (\upomega - \bar{\upomega}) \big) ^2 (x,t) \rd x \;{\leq}\; \int \big( \nabla (\upomega_0 - \bar{\upomega}_0) \big)^2(x) \rd x.
\end{equation}
Both the inequalities are independent of the viscosity.
In essence, the fluctuation in the initial conditions {\itshape is not} amplified over the flow evolution. If the two smooth initial conditions are close to each other, the entropy difference of the motions must be small. This is the interpretation of the Navier-Stokes uniqueness according to the concept of entropy. In practice, local observation of instantaneous bounds (\ref{eq:delta-vort-enstrophy}) and (\ref{eq:d-delta-vort-enstrophy}) must have been aberrantly misleading because of the presence of macroscopic randomness. The global character of the velocity and of the pressure further obscures a meaningful interpretation of the observables. However, the ensemble averages of measurements in turbulence flows are perfectly repeatable, as well-known. This important fact can be reiterated in the present context: The randomness has an origin which must be independent of the differences in the initial conditions.

Suggestions have been put forward to explain turbulence by means of dynamic systems, particularly by chaos theory which has been developed as a diagnostic tool into many complex phenomena in science and engineering. The best known {\itshape model} emphasizes on a system's sensitivity to initial conditions and is characterized by the presence of a strange attractor. Imperatively, the ansatz of chaos has not been derived from the full Navier-Stokes equations and hence the limitation of the theory is apparent. First, a common route leading to chaos or deterministic chaos is due to the appearance of a sequence of subharmonic bifurcations which has double-orbit portrays in systems' phase diagram. The non-unique character contradicts the well-posedness property of the Navier-Stokes dynamics. Second, the phase dimensions of a chaos system are far too low compared to those of turbulence. This in turn implies that the ensemble averages of chaos solutions from nominally identical initial conditions may not be regarded as repeatable quantities in the sense of statistics. Generally, the chaos solutions of contiguous initial conditions are apparently random in solutions' analytical characters; the solutions grow exponentially in time as a result of the presence of positive Liapunov exponents. The symptom of solutions' exponential divergence implies that the solutions soon lose the dependence of the past history, and hence the prediction of system's evolution can only be made with confidence not far from the initial time. In contrast, the solutions of the Navier-Stokes equations assert that the fluid dynamics equations do not contain elements of randomness. {\itshape The estimates (\ref{eq:delta-vort-enstrophy}) and (\ref{eq:d-delta-vort-enstrophy}) suggest that the Navier-Stokes system is not sensitive to small differences in the initial conditions. In particular,
the initial data are explicitly embodied in every term of the solution series (\ref{eq:vort-series-sol}). The spatio-temporal complexities exhibited in the solution differ essentially from the typical, erratic characters of chaos systems}. 
\subsection{Transition in Reynolds' pipe flow}
The evolution of the Hagen-Poiseuille flow in a pipe was first observed by Reynolds when the speed of the water was varied in a controlled manner (Reynolds 1883).  
Specifically, he found that the water in the glass pipe underwent three distinct stages; a steady direct motion (known as laminar flow), a transitional phase and a sinuous eddying motion (turbulent flow), depending on the Reynolds number
\begin{equation*}
	Re=\frac{U_m d}{\nu},
\end{equation*}
where $U_m$ is the mean speed of the flow, and $d$ is the pipe diameter. 

We next show that the principal causes for the flow evolution are intimately related to the non-linearity in the equations of motion. To see the important relationship between the flow development and the Reynolds number, we let
\begin{equation*}
	t_*=t \:U_m/d,\;\;x_*=x/d,\;\;,u_*=u/U_m\;\;\; \mbox{and}\;\;\;p_*=p/(\rho U_m^2).
\end{equation*}
%
%
The equations of motion (\ref{eq:continuity}) and (\ref{eq:ns}) can be written in the dimensionless form,
\begin{equation} \label{eq:ns-re}
	\nabla . u_* =0 \;\;\; \mbox{and} \;\;\;  {\partial_{t_*} u_*} - \frac{1}{Re} \Delta u_* = - (u_* . \nabla) u_*  - \nabla p_*,
\end{equation}
where all the spatial derivatives are with respect to $x_*$. 
The non-dimensional vorticity equation reads
\begin{equation} \label{eq:vorticity-re}
	{\partial_{t_*} \omega_*} - \frac{1}{Re} \Delta \omega_* = (\omega_* . \nabla) u_*  - (u_* . \nabla )\omega_*.
\end{equation}
To specify the initial condition, we denote a cross-section in the pipe by $\Upomega_i$. It can be the inlet cross-section or it can be any cross-section at some distance just downstream of the inlet where measurements of the local velocity may be more practical.  
Then the Cauchy problem for Reynolds' experiments is to solve (\ref{eq:ns-re}) or (\ref{eq:vorticity-re}) subject to the initial condition,
\begin{equation} \label{eq:ns-re-ic}
 u_*(x_*,0) = u_s(x_*)\;\;\; \mbox{or} \;\;\; \omega_* (x_*,0) = \nabla {\times} u_s(x_*), \;\;\; x_* \in \Upomega_i,
\end{equation} 
and to the boundary condition
\begin{equation} \label{eq:ns-re-bc}
 u_*(x_*) = 0, \;\;\; x_* \in \Upomega_s,
\end{equation} 
where $\Upomega_s$ denotes the internal surface of the pipe. 
The initial velocity $u_s$ is assumed to be a known function. Mathematically, the Cauchy problem should be formulated in a cylindrical co-ordinates system rather than the Cartesian system which is chosen so that we are able to make direct comparison to our results of the Navier-Stokes equations.

Reynolds showed experimentally that there is a critical value of the Reynolds number ($Re_c$) which discriminates the laminar and the turbulent motions. (More precisely, there is {\itshape a range} of critical Reynolds' number.) The critical Reynolds number is known to be a function of the geometry at the inlet, the pipe surface roughness, background disturbance and test environment. In theory, we hypothesize that all these experiment-related conditions are well specified in $u_s$, as done in the receptivity problem. In analogy to (\ref{eq:vort-series-sol}), the solution of the Cauchy problem (\ref{eq:vorticity-re}) to (\ref{eq:ns-re-bc}) is given by
\begin{equation} \label{eq:pipe-vort-series-sol}
	\omega_*(x_*,t_*) = \gamma_*(x_*,t_*) + \sum_{m\geq2} S_m \:V_*[\gamma_*(x_*,t_*)]^m.
\end{equation}
It is assumed that the presence of the pipe wall has been fully taken into account in the analogous structure matrix $G_*$. The effect of the wall must be included in the heat kernel and the Biot-Savart relation. Accordingly, we expect some definite modifications in the integral kernels of the convolutions.  
In addition, the precise value of $Re_c$ does not concern us as long as the Reynolds number $Re$ is high enough such that the flow field needs to be approximated by many terms of (\ref{eq:pipe-vort-series-sol}). 

Take a cross-section sufficiently downstream of the inlet $\Upomega_i$ where the local velocity has a parabolic profile. For a flow at $Re \ll Re_c$, the vorticity distribution is given by the leading term $\varpi_*$ or $\gamma_*$, and the flow is said to be laminar. The dissipation of the kinetic energy is dominated by diffusion. As $Re$ increases but is still below the critical value, the complete vorticity is considered to consist of two-term expansion,
\begin{equation*}
	\omega_*(x_*,t_*) = \gamma_* + 2\: V_*[\gamma_*]^2.
\end{equation*}
The flow can still be regarded as laminar. However, some increased degree of unsteadiness is detectable in the flow field due to the interaction among the eddies. The observed unsteady wavy structures in the axial direction, also known as turbulent puffs, are expected to be larger than the pipe diameter. This is because the space-time convolutions in $\gamma_*$ are still comparable to the scale of the initial profile. If the Reynolds number is further increased until $Re \approx Re_c$, the instantaneous vorticity field is best described by
\begin{equation*}
	\omega_*(x_*,t_*) = \gamma_* + 2\: V_*[\gamma_*]^2 + (10) \: V_*[\gamma_*]^3,
\end{equation*}
or possibly, by 
\begin{equation*}
	\omega_*(x_*,t_*) = \gamma_* + 2\: V_*[\gamma_*]^2 + 10\: V_*[\gamma_*]^3 + (62)\: V_*[\gamma_*]^4.
\end{equation*}
The water in the pipe now consists of many eddies of multiple scales which bring about pronounced fluctuations in the flow. In practice, Reynolds did not observe the vorticity field but the velocity field. In view of the induction effect of the Biot-Savart law, the velocity field must show a space-time intermittent character due to the appearance of the localized vorticity patches which are related to the interaction of the interwoven $\gamma_*$ and the shears close to the wall. Reynolds referred the regions of the velocity induced by the vorticity concentrations as flashes (now commonly known as turbulent bursts). The whole flow field evolves into a structure which looks like the sinuous motion. As $Re$ is increased above the critical value, the number of eddies in the flow grows rapidly so that 
\begin{equation*}
	\omega_*(x_*,t_*) = \gamma_* + 2 \: V_*[\gamma_*]^2 + 10\: V_*[\gamma_*]^3 + 62\: V_*[\gamma_*]^4 + (430)\: V_*[\gamma_*]^5.
\end{equation*}
The vorticity field must appear to be chaotic and irregular due to the strong interactions among the vorticity eddies. For $Re \gg Re_c$, as there are numerous small-scale eddies, the viscous dissipation intensifies. As identified in the previous section, the randomness mechanism becomes marked and effective. The whole flow soon becomes chaotic as well as random as it contains a multitude of eddies in distinct scales. It is the fully developed turbulent flow which is strongly nonlinear everywhere. The mean velocity profile has to be substantially different from the parabolic distribution.

{\itshape Transition is an intrinsic, three-dimensional, non-linear process during any flow evolution. As the Reynolds number increases, the orderly streamlined structure of initial laminarity is successively modified by the proliferated eddies of progressively smaller size until the entire structure is dissolved into a composition of fluctuating shears.  The critical Reynolds number, as a function of the initial data (with fixed boundary conditions), marks the onset of the space-time flow configuration, in which the non-linearity first becomes conspicuous. }

The properties of the Navier-Stokes non-linearity show that the presence of a disturbance with infinitesimal or finite strength is not a prerequisite to initiate turbulence. In the conventional theories of transition, the introduction of the disturbance is absolutely necessary. The transition has been attributed to the eigen-states of linearized equations; the development of the disturbance is alluded as exponential growth, as transient amplification of damped modes, as pattern coalescence due to eigenvalue degeneracy or as non-linear interactions among non-Hermitian modes. On the other hand, when controlled perturbations are introduced into the flow, our explanation for the transition process needs some modifications because we must solve the generalized initial value problem of receptivity with the appropriate initial and boundary conditions (cf. (\ref{eq:ns-pert-vort-gen-ic})). Then the ``mean'' features in the complexity of the transition process observed in experiments can be fully accounted for (see, for example, Willis {\itshape et al} 2008; Monin \& Yaglom 1999; Avila {\itshape et al} 2011; Mullin 2011).

From application points of view, an alternative criterion for the critical Reynolds number may be chosen. For instance, we may define a threshold Reynolds number corresponding to a minimum drag value.
In general, the transition process can be continuous or intermittent in space as well as in time; it depends on the precise specification of the initial vorticity distribution and on the geometric configuration under consideration. For given geometry, the Reynolds-number dependence may well be super-sensitive and nurtures an abrupt transition; a small increase from a Reynolds number to the critical value ought to involve several $S_m$ terms in order to account for changes in the flow topology.
\subsection{Notion of instability and Navier-Stokes dynamics}
To elaborate our discussion of the present section, we examine an important property of the Hagen-Poiseuille flow in a circular pipe. The pipe flow is known as an exact solution of the Navier-Stokes equations, and it has a parabolic velocity distribution. In particular, the flow has been considered, or more appropriately accepted, to exist for {\itshape any} Reynolds number (Liepmann 1979). To appreciate the physical and theoretical implications, two separate derivations of the result can be found in standard fluid dynamics books (see, for example, Tritton 1988; Schlichting 1979; Batchelor 1973). 
In one derivation, the velocity distribution is derived by momentum balance in a control volume without reference to the equations of motion. In another, a simplified form of the equations is solved. No matter which method is used, the solution of the parabolic profile is in fact obtained under two assumptions: (1) The Navier-Stokes equations are solved as a steady boundary-value problem; (2) The flow is idealized in a two-dimensional flow field. The underlying hypotheses effectively define an over-determined system of differential equations for the pipe flow. In a fundamental sense, it is definitely more appropriate to analyze Reynolds' pipe experiments as an {\itshape initial-boundary} value problem. For convenience, we can specify the flow condition at the pipe inlet. The initial solenoidal velocity must satisfy the no-slip boundary condition on the pipe wall. The simplest initial value is a flow with non-zero vorticity component in the axial direction (denoted by $x_1$); the initial condition may be specified as a two-dimensional flow. So let the vorticity $\omega_0(x) {=}\big(f(x_2,x_3),0,0\big)$, where $f$ is a given suitable function. In experiments, it is plausible that there exist well-behaved initial conditions which give rise to a parabolic velocity profile somewhere downstream of the inlet. From (\ref{eq:k-elements}), we see that the three elements in the first column of the capacity matrix $K$ do not vanish in general. It follows that, as long as the solution goes beyond the leading term in the first-order approximation (cf. (\ref{eq:vort-sol-first-order-VIE})), the solutions for the full Navier-Stokes are wholly {\itshape three-dimensional} on all occasions. As a consequence, the Hagen-Poiseuille flow is best considered as the transient solution of a suitable initial flow. In practice, the parabolic velocity profile ceases to exist when the non-linear, three-dimensional effects of the dynamic equations become significant. 

By analogy, we conclude that, in the boundary layer approximations, Blasius' profile on a flat plate does not exist for any arbitrary Reynolds number (based on the free-stream velocity and the displacement thickness). Once the Reynolds number is sufficiently large, the laminar flow intrinsically evolves into a three-dimensional turbulent boundary layer. In low free-stream turbulence wind tunnels, the well-defined laminar boundary layers are best identified with the one-term expansion of the full Navier-Stokes solutions. The detailed processes of the transition may be overwhelmingly intricate to study because the plate surface not only acts as a source of vorticity but also complicates non-linear interactions among shears. In dedicated laboratory experiments (see, for example, Schubauer \& Skramstad 1947), the so-called naturally excited oscillations have been found. For a measuring probe fixed in the boundary layer, the oscillations were recorded and interpreted as amplified instability waves that travelled downstream, having a broad range of frequencies. In fact, these detected oscillations can be reinterpreted in the current context: They must be ensemble-averaged time-series records of a single vorticity eddy or a group of eddies of various size transversing over the probe. Moreover, controlled artificial disturbances of selective frequency were also generated in the experiments by wave-makers or loudspeakers so as to excite the laminar boundary layer. The resulting flow over the plate defines a proximate fluid dynamics problem of modified initial and boundary conditions which has to be formulated according to (\ref{eq:ns-pert-vort-gen-ic}). Thus the observed evolutionary characters, commonly known as the Tollmien-Schlichting waves, primarily recount the reverberations of the diffused artificial disturbances (cf. (\ref{eq:ns-pert-vort-ic-mod}) and (\ref{eq:vort-series-sol-pert})) whose presence reshapes the otherwise Navier-Stokes {\itshape natural transition}.  

Above all, {\itshape the dynamics of the Navier-Stokes equations shows that, given initial conditions, instability is not a genuine mechanism for triggering the laminar-turbulent transition whose occurrence does not demand the presence of infinitesimal or finite disturbances as a precondition. Every fluid motion, no matter how simple its geometry may appear, must be analyzed as an initial-boundary value problem of the complete Navier-Stokes equations in three space dimensions.} 
\subsection{Liapunov stability analysis in fluid motion}
The main application of the linearized equations of motion is to establish stability of uni-directional time-independent mean flow, $\bar U(x,t)=\bar U(x_3)$, which is subject to a disturbance having a {\itshape wave} form,
\begin{equation*}
	(u',p')(x,t) = (\tilde{u}, \tilde{p})(x_3)\exp\Big(\: \ri\:\big(\tilde{\alpha}_1 x_1 + \tilde{\alpha}_2 x_2 - \tilde{\beta} t\big)\Big).
\end{equation*}
A general disturbance can be described by summing all the Fourier components. Use of the wave-form renders the linearized partial differential equations separable, and thus equations (\ref{eq:linear-ns-vort}) become a system of ordinary differential equations governing the development of the disturbance. With appropriate boundary conditions, specified wave-numbers and characteristic Reynolds number, a dispersion relation of the system can be derived:
\begin{equation} \label{eq:dispersion}
 {\mathscr D}(\beta,\alpha_1,\alpha_2,Re)=0	
\end{equation}
which determines admissible (complex) eigenvalues $\beta$ and, as a result, the amplification or attenuation of the disturbances as time $t {\rightarrow} \infty$. 

In particular, the linear theory anticipates that the first instability is always associated with two-dimensional waves (Squire 1933); we need consider only two-dimensional disturbances ($\alpha_2{=}0$, say) in order to obtain the minimum critical Reynolds number. Evidently, the prediction can only be true over the minute interval $t_{\epsilon}$ in which the linear diffusion prevails (cf. (\ref{eq:zeroth-order-heat})). Furthermore, as a problem of receptivity, suppose that a wave-like disturbance $v'$ is introduced into a mean motion at start $t{=}0$. Then the initial total vorticity is given by
\begin{equation*}
	\int \sum \Big(\: \nabla{\times}\bar{U}(x) + {\Real}\big( \nabla{\times}v'(x,0) \big) \: \Big) \rd x 
\end{equation*}
which is generally a non-zero constant. The summation sign denotes the sum of the vorticity components. Thus law (\ref{eq:vorticity-invariance}) imposes severe limitations on permissible solutions in (\ref{eq:dispersion}). Strictly, we require
\begin{equation*}
{\Real}	\Big(\: \int \sum \big( \nabla{\times}v'(x,0) \big)  \rd x \: \Big) = {\Real} \Big(\: \int \sum \big( \nabla{\times}u'(x,t) \big)  \rd x \: \Big), \;\;\;t>0.
\end{equation*}
For attenuated small-amplitude waves described by stable modes over a short time $t {\rightarrow} 0^+$, and for weakly sheared initial disturbances, it may be argued, against a possible violation of the invariance principle, that the requirement is approximately satisfied. Generally, {\itshape a large majority, if not all, of the eigen-modes defined by the dispersion relation do not accommodate the principle of vorticity invariance}. 

The idea of the stability analysis has been adopted from other branches of physics. To stand a ball-point pen on its point can hardly be achieved over a meaningfully long period of time. We say that the pen standing on its point is an unstable configuration in noisy environment. But the pen lying on its side consists in a stable equilibrium. In rigid body dynamics, the {\itshape identity form} of a system (i.e. pen's shape and geometry) is fixed in different stability circumstances. Also we pay little attention to the change of system's internal energy so that we concentrate on the dynamic change of its kinetic and potential energy (as a Hamiltonian system). The system is classified as stable, unstable or asymptotically stable with respect to external disturbances (Liapunov 1892). We ask ourselves: What happens to the configurations if the pen is made of fluid? A fundamental property of fluids, as opposed to solids, is the ability to sustain applied load by {\itshape deformation} (cf. figure \ref{fig:deft}). As demonstrated in experiments, a fluid flow of low viscosity readily disintegrates into smaller fluid parts owning to vorticity convection and stretching, within the law of mass conservation (Liouville's theorem), if a force or a disturbance is applied. The rate of deformation are quantitatively measured by the strains (cf. (\ref{eq:strain-tensors})). Thus the identity form of a fluid pen in motion cannot be devoid of the instantaneous dynamics. Any change in fluid's internal energy must not be neglected as viscous dissipation plays a key role in energy conservation. Conceptually, application of a Liapunov stability analysis in fluid dynamics is unlikely to bring forth reliable outcomes at large time, specifically for amplified disturbances. The superposition of a mean and a minute perturbation is not a well-justified assumption as these motions inevitably qualify each other's identity form due to the deformation propensity. 

On the basis of the mathematical inconsistency just discussed, it is natural to cast doubt on the credibility of the transition prediction algorithms, which are implemented according to the stability theory. As a matter of experience, the predictive capability is empirical in nature and thus may best be regarded as ``operational'' in preliminary design cycles, depending on the level of confidence in data filtering and calibration. The uncertainty is further compounded by the frustration that there has been a lack of alternative methods for independent verifications. Any apparent success of these tools ought to be a result of excessive degree of the empiricism. It must be admitted that the theoretical difficulties encountered in the stability analysis as a whole are deep-rooted. Hence use of the linear theory for transition prognosis in its conventional formulations cannot possibly be justified. {\itshape On the ground of rigorousness, any predictive tools based on the linearized theory for the onset of turbulence have virtually no practical values in high Reynolds-number applications.}
\subsection{Energy distribution and dissipation, spatio-temporal intermittency}
In the idea of an energy cascade envisaged by Richardson (1922), the production mechanism of the eddies in various scales was attributed to dynamical instability or to random vortex stretching. In high Reynolds-number flows, it has been since assumed that turbulent energy is transfered from the large energy-containing eddies to somewhat smaller eddies which become unstable and break up into weak eddies of even smaller sizes. In particular, the smallest eddies in the cascade dissipate the energy passed to them. Now the cascade process may only be possible under a condition of equilibrium; the rate of energy transfer from the largest eddies to the smallest ones equals to the rate of the dissipation. For many rapidly-changing turbulent flows, the consumption equilibrium can hardly be maintained, since the cascade is not an instantaneous process (Bradshaw 1994). 

On the other hand, as implied in the vorticity series solution (\ref{eq:vort-series-sol}), the non-linear term in the momentum equation defines a hierarchy of eddies and, in reality, as it manifests a dynamic distribution of the energy across the whole spectrum of the eddies in accordance with the law of energy conservation. The distribution of the initial energy among the eddies is clearly a function of the initial conditions, particularly for the eddies of larger sizes. As explained earlier in the present paper, the dynamic flow structures abide by the law of vorticity invariance and cannot be a consequence of instability or bifurcation. The quantities of the successive smaller-scale vortices proliferated by the non-linearity are significantly larger than those resulting from the consecutive binary fissions, as illustrated in Frisch (1995). 

Turbulent flow-fields are commonly decomposed into a hierarchy of mutually orthogonal Fourier modes (waves). Our treatment of the transition and turbulence has been carried out in the real space (the physical space) as opposed to the Fourier space which is particularly essential for statistical theories of turbulence. In an insightful discussion, Davidson has questioned the suitability of the Fourier decomposition in defining turbulence scales and energy cascade (see \S 6 of Davidson 2004) -- a justified and shared view. Nevertheless, our solutions of the hydrodynamic equations in the physical space (\ref{eq:vort-series-sol}) and in the phase space (\ref{eq:mm-boltz-series-sol}) precisely specify the number of degrees of freedom for the flow states at fixed space-time location prior to dissipation. It is then interesting to compare the theoretically values with the estimates from non-linear dynamics systems (see, for example, Grossmann 2000 and the references therein). As reported by Grossmann, there were insurmountable difficulties in evaluating the Fourier-transformed non-linearity. By means of simplifications, the phase-space dimension was calculated in terms of the Liapunov exponents of systems' spectrum. The dimension for transition at $Re{=}2000$ was quoted as $O(10^5)$. This estimate clearly overshoots. For fully developed turbulence at $Re{=}5{\times}10^6$, the quoted value was $O(10^{13})$ which is certainly an underestimate as a result of mitigating the non-linearity.

The knowledge of the quantitative nature of small-scale vortices is crucial in high Reynolds-number flows. The dynamics of the small eddies, however, is responsible for an enhanced diffusive capability and an increased local momentum transfer commonly detected in turbulence (see, for example, Taylor 1935; Batchelor 1953; Townsend 1976). The smaller eddies or the highly localized shears are more easily transported and entrained to any other parts of the flow fields, prior to stochastic viscous dissipation, by the impetus of the other remaining eddies, notably the larger ones.

Furthermore, every term in the series solution is a result of the repeated convolutions in space and in time; the heat or diffusion filter,
\begin{equation*}
	\exp\Big( - \frac{|x|^2}{4\nu t}\Big),
\end{equation*}
is present in every convolution. To be specific, we assume that the initial localized data are described by weighted polynomials decaying at infinity. The diffusion is particularly effective on the flows characterized by these functions. For large initial data or high Reynolds number, the convolutions have two major impacts. The vorticity terms comprising few convolutions, such as those in $S_2$ to $S_6$, still contain large amounts of kinetic energy. Thus the enstrophy distribution is in general spatially intermittent for the flows in $\real^3$. The other consequence of the repeated convolutions is that the small-scale eddies tend to lose their preference of spatial orientation, and to become isotropic even though the initial data may be given in a state of strong anisotropy. The vorticity eddies of the small scales rearrange themselves among the flow fields, mainly under the influence of the large eddies according to the Biot-Savart law. At least for a large class of the realizable initial data at high Reynolds number, all the small-scale fluid motions appear to be orientation-independent. Broadly speaking, Kolmogorov's hypotheses (Kolmogorov 1941$a, b$) that high Reynolds-number small-scale turbulent motions are statistically isotropic can thus be justified in view of vorticity dynamics, barring the effect of the presence of vorticity source such as a solid wall.  

It has been conceived that, according to Kolmogorov's similarity supposition (Kolmogorov 1941$a, b$), the mean dissipation rate per unit mass, $\varepsilon = \rd (\left\langle {u} \right\rangle /2) / \rd t$, remains finite in the limit of Reynolds number $Re \rightarrow \infty$. Generally, $Re=UL/\nu$, where $U$ and $L$ are the velocity and length scales respectively. Since the finiteness of viscosity $\mu > 0$ holds on the ground of physics, an infinity Reynolds number can only be approximated in incompressible motions whose length dimension $L \rightarrow \infty$. As long as the initial data possess finite energy, the Navier-Stokes non-linearity counter-balances the energy of the motion by generating sufficient quantities of small-scale eddies so that the smallest scales are bounded below. The existence of the lower bounds is consistent with the kinetic theory of dilute gases. Effectively, the mean viscous dissipation always out-performs the rate of possible local accumulation in the internal energy. Thus the scalings (\ref{eq:ns-scaling}) must have very limited physical validity. Technically, to scale any fluid motion of finite-energy by an infinite or arbitrarily large Reynolds number purely on the ground of dimensional analysis is tantamount to introducing an artificial singularity.

On the basis of the Navier-Stokes solutions, we are able to supplement the experimentally interpreted knowledge of the spatio-temporal intermittency in turbulence (see, for example, Batchelor 1953; Townsend 1976; Tritton 1988; McComb 1990). Consider a fixed finite portion in a free shear flow, such as a sustained self-preserving jet or wake developed in stationary surrounding (see \S 21.2 and \S21.3 of Tritton 1988). The intermittency factor is defined as the fraction of the time that vorticity fluctuations are occurring at a spatial location in the region. A collection of the factors throughout the region is a measure of the intermittency in the turbulent flow field. As the flow evolves downstream, it decays, spreads as well as entrains at interface with the ambient weakly rotational flow which is under constant influence from the main shears. The self-similarity of the velocity profiles preserves -- a direct consequence of the law of invariant total vorticity. There is no {\itshape a priori} reason that the instantaneous dissipation ($\propto \nu S_{ij} S_{ij}$) should be uniform in turbulence (one exception could be the idealized case of homogeneous isotropic turbulence). Thus the vorticity eddies of the smallest scales are bound to distribute unevenly throughout the flow unless the initial vorticity is strictly symmetric and uniform in all the three space dimensions -- a practical impossibility. Moreover, the viscous dissipation itself has been contemplated as a random process in space and in time. Plausibly, annihilation eddies below the dissipative scales must be more effective in the neighbourhood of the interface; local relaminarization of certain degree ought to occur temporarily, albeit as a random event. The detected spatio-temporal intermittency can be attributed to the competition between the vorticity proliferation and the decimation due to action of viscosity. Although the two processes have different origins in the dynamic equations, the proliferation by the non-linearity $(u.\nabla) u$, while the dissipation by the diffusion Laplacian $\nu \Delta u$, it is the inherent randomness of matter's internal energy which precipitates an imbalance in the evolution of these processes. At locations sufficiently far downstream, all the intermittency factors must fall below the peak unity to lower values, and ultimately to nil; the dissipation is dominant in the long run since the initial energy of the flow is finite. However, precise quantitative evaluation of the intermittent characters must be a daunting task since the various vorticity scales in the flow are intimately related, and the individual velocity fields are coupled to one another.
\subsection{Instantaneous characters of Navier-Stokes turbulence}
In nearly all experiments for turbulent flows, the test results are analyzed according to statistics rules (see, for example, Batchelor 1953; Bradshaw 1971; Tennekes \& Lumley 1972). The standard practice is that the instantaneous velocity is decomposed into a mean motion and a fluctuation:
\begin{equation} \label{eq:vel-decomp}
	V(x,t) = v(x,t) + v'(x,t),
\end{equation}
assuming that turbulence already exists. This is known as Reynolds' decomposition (Reynolds 1895; Lamb 1975). The mean motion has the following meaning in statistics. In view of ergodic hypothesis (see, for example, Monin \& Yaglom 1999), an ensemble average is used to describe the flow, namely,
\begin{equation*} 
	\left\langle {\mathrm Q} \right\rangle (x,t) = \frac{1}{\mathrm N} \: \sum_{i=1}^{\mathrm N} {\mathrm Q}^{(i)} (x,t),
\end{equation*}
where $\mathrm N$ is the total number of repeats of an experiment. The $i$th realization of the series of the nominally similar experiments are denoted by ${\mathrm Q}^{(i)}$. We assume that $\mathrm N$ is large in view of the strong law of large numbers so that the average is insensitive to minute variations in $\mathrm N$. According to the definition of the mean, the fluctuation has a zero-average,
\begin{equation} \label{eq:zero-mean}
	\left\langle {v'} \right\rangle = 0.
\end{equation} 
In other words, the mean and the fluctuation must be completely uncorrelated. In Reynolds' original derivation, $\left\langle {v'} \right\rangle \approx 0$ as Reynolds did not rely on the mean used in the current sense.

Substituting the decomposition (\ref{eq:vel-decomp}) into the equations of motion, 
we take the mean term by term. The operation of differentiating with respect to time (or space) commutes with the operation of the averaging. In particular, we notice that
\begin{equation} \label{eq:v-uncorr}
	\left\langle {(v. \nabla ) \: v'} \right\rangle = (\left\langle {v} \right\rangle . \nabla) \: \left\langle {v'} \right\rangle = 0, \;\;\; \left\langle {(v' . \nabla ) \: v} \right\rangle = (\left\langle {v'} \right\rangle . \nabla) \: \left\langle {v} \right\rangle = 0.
\end{equation}
Thus the Reynolds equations for the mean motion can be derived
\begin{equation} \label{eq:reynolds-mean-flow}
	\frac{\partial \left\langle {v_i} \right\rangle} {\partial t}  - \nu \frac{\partial^2 \left\langle {v_i} \right\rangle }{\partial x_i^2 }  = - \left\langle {v_j} \right\rangle \frac{\partial \left\langle {v_i} \right\rangle}{\partial x_j}  - \frac{1}{\rho} \frac{\partial \left\langle {\pi} \right\rangle}{\partial x_i} + \frac{\partial}{\partial x_j}\Big( 2 \nu \: \left\langle {S_{ij}} \right\rangle - \rho \: \left\langle {v'_i v'_j} \right\rangle \Big),  
\end{equation}
where $\left\langle {S_{ij}} \right\rangle{=}S_{ij}$ (cf. (\ref{eq:strain-tensors})). The {\itshape ad hoc} symbol $\pi$ denotes the mean pressure. Furthermore, the averaging procedure reduces the continuity equation to
\begin{equation} \label{eq:reynolds-continuity}
	\nabla . \left\langle{v} \right\rangle = 0,\;\;\;\;\; \nabla.v' = 0
\end{equation}
because the mass conservation is a linear integral constraint.

The decomposition (\ref{eq:vel-decomp}) is an analogy to Maxwell's decomposition (\ref{eq:vel-decomp-ke}). Reynolds referred the motion $v$ as the mean-mean motion in reference to Maxwell's mean motion $u$, and the fluctuation $v'$ as the relative mean-motion. Maxwell called $u'$ the relative motion with reference to the motion $u$. It is instructive to notice that Reynolds' velocity decomposition is made on the basis of fluid motion as a continuum while Maxwell's decomposition is based on particles' microscopic description where randomness is an intrinsic property. In the kinetic theory of gases, the mean velocity is well-defined because fluid's particles or molecules are numerous in {\itshape any} fluid element where hydrodynamics equations of motion are presumed to hold. It is physically plausible that the fluctuation of an individual molecule $u'$ is completely independent of the macroscopic velocity $u$. Alternatively, the zero-average fluctuation $\left\langle {u'} \right\rangle {=}0$ is a consequence of the Maxwellian distribution, which is Gaussian with a zero-mean (cf. (\ref{eq:maxwellian})).
 
However, the analogous ``molecules'' in turbulence were thought to be vorticity eddies. Richardson realized the analogy of the two decompositions require clarification (see \S 4/8/0 of Richardson 1922). He wrote: (Reynolds' decomposition) ``was true however large was the interval of time or the volume over which the mean was taken, with the limitation that the eddies within this time and space must be sufficiently numerous.'' He clarified Reynolds' decomposition by pointing out that the mean value must be taken over a sphere with a radius large enough to include a considerable number of eddies. Then the centre of the sphere moves with the mean-mean motion.

When the decomposition (\ref{eq:vel-decomp}) is substituted into the Navier-Stokes equations, an underlying assumption is being made that any infinitesimal fluid element on the continuum contains a large number of eddies. The general solution of the Navier-Stokes equations in $\real^3$ contradicts the supposition. Many experimental measurements show that turbulent flows consist of an intermittent vorticity field; the actual numbers of vortices at a particular space-time location depend on the unit Reynolds, and necessarily they have not to be abundant enough for a turbulence-continuum to exist. Thus there exists no {\itshape a priori} reason that a turbulent flow, no matter how large the Reynolds number may be, has a uniform distribution of the vorticity eddies of all scales through its occupying space. Specifically, the numbers of dissipative eddies vary dynamically across a typical turbulent flow field; the phenomena of the spatio-temporal intermittency in turbulence are well-founded. The lack of spatial eddy homogeneity mirrors the complaint made by Liepmann on the basis of experiments that there exists a nearly bimodal probability distribution for the coefficient of the second velocity correlation in a turbulent flow (Liepmann 1979). 
Furthermore, we have explained how the transition occurs; turbulence {\itshape cannot} be in a state of continuous instability or of bifurcation. It is then logical to ask: What are the ``microstates'' which characterize the fluctuation $v'$ and engender the mean $ \left\langle{v} \right\rangle$?

The Navier-Stokes solutions show that velocity $u(x,t)$ oscillates vigorously in the turbulent flow once the transition process has mostly completed. In experiments, randomness in $v(x,t)$ becomes progressively intensive with increasing Reynolds number because of the dissipation of the small-scale eddies. During transition process, a laminar flow and the ``new-born'' turbulent flow depend strongly on the initial data. The mechanism of the Navier-Stokes transition establishes the fact that there is no essential difference between a laminar flow and a turbulent flow in the sense that both are governed by the Navier-Stokes equations of motion. It follows that the dynamic stipulations,
\begin{equation*}  
	u = \left\langle {v} \right\rangle \;\;\; \mbox{and} \;\;\; p = \left\langle {\pi} \right\rangle,
\end{equation*}
must be satisfied at any space-time location so that the laws of conservation of momentum and energy hold. In fact, the continuity for the fluctuation in 
(\ref{eq:reynolds-continuity}) is a redundant constraint as the zero-average condition (\ref{eq:zero-mean}) implies the fluctuation continuity. However the converse is not true.

To understand the implications, we consider the possible microscopic cause that relates to the fluctuation in Reynolds' decomposition. By combining Reynolds' decomposition (\ref{eq:vel-decomp}) with Maxwell's kinetic theory (\ref{eq:vel-decomp-ke}), we have
\begin{equation} \label{eq:sum-decomp}
	\xi(x,t) = v(x,t) + \big( v'(x,t) + u'(x,t) \big). 
\end{equation} 
It is known that, for shear flows in $\real^3$, the percentage increment in temperature in (\ref{eq:internal-energy-increase}) is estimated to be
\begin{equation} \label{eq:temp-increase}
	\frac{\Delta {\mathrm T}} {\mathrm T} \propto \frac{M^2_a}{4(\gamma_a-1)},
\end{equation}
where $M_a$ is the Mach number based on the maximum velocity, and $\gamma_a$ is the ratio of specific heats for an ideal gas. For instance, at $25^{\circ}$C and atmospheric pressure, the average velocity of air molecules is roughly $470$m/s. The Boltzmann distribution (\ref{eq:boltz-distri}) then gains a slightly longer tail toward high velocity due to the small increase in the temperature. Let us imagine that the whole initial energy is instantaneously converted into the internal energy with perfect efficiency. Even in this extreme situation, the viscous dissipation may increase the average velocity of the molecules by half of the amount in (\ref{eq:temp-increase}). In view of the incompressibility hypothesis ($M_a {\leq} 1/3$ say), the kinetic energy content in a typical turbulence encountered in nature and in laboratory under standard temperature and pressure cannot substantially modify the material structure. Nor can such a turbulence change the state of the fluid. It follows that $v'$ must be designated to represent the identical physical effects as $u'$; the velocity $v'$ must refer to the fluctuation in one of the molecules. In essence, decomposition (\ref{eq:sum-decomp}) is identical to Maxwell's decomposition; $\left\langle {v} \right\rangle$ represents the macroscopic velocity which is the ensemble average of the microstates. Pragmatically, incompressible turbulent flows can be regarded as in the state of non-equilibrium processes, which slightly deviate from their equilibrium Maxwellian (\ref{eq:maxwellian}). The implication is that the second law of thermodynamics must hold for turbulent flows (cf. (\ref{eq:boltz-h-theorem})). Consequently, we have two alternatives for consistency of fluid dynamics. The first one is primarily important for computation of turbulence: {\itshape ab initio} $v'{\equiv}0$ and the last term in (\ref{eq:reynolds-mean-flow}) must be identically zero in any flow evolution determined by the Navier-Stokes dynamics.
  
{\itshape Under the hypotheses of the continuum and incompressibility, turbulent flows can be uniquely determined by solving the Cauchy problem posed by the Navier-Stokes dynamics with given initial data. The general solutions represent the instantaneous flow topology; the overall flow manifests as an intricate vorticity field with a broad spectrum of spatio-temporal scales. In principle, the Navier-Stokes equations do not predict any random character in the flow field. Accordingly, these solutions have to be interpreted as the descriptions of the mean flow quantities. The presence of the numerous vorticity eddies induces dynamic interactions among themselves which contribute to the enhanced diffusivity and increased shear stresses commonly perceived in turbulence. }  

The second alternative is motivated by the theory of Langevin (1908). If we are adamant to compute complete fluid quantities instead of the mean values, we must somehow introduce a reformulation of the dynamics equations, albeit an {\itshape ad hoc} approach. Essentially, we need to modify momentum equation (\ref{eq:ns}) by adding a stochastic function in order to explicate the random fluctuations. The statistical properties of the additional term have to be either postulated or empirically derived.
\section{Extended analysis of fluid motion}
\subsection*{Generalized vorticity}
So far we have taken for granted that vorticity is non-zero throughout flow field. For fluid motions with zero vorticity, see \ref{app:c}. 

The idea of vorticity can be generalized by repeatedly taking the curl of vorticity equation (\ref{eq:vorticity}). The result is written as
\begin{equation} \label{eq:k-fold-vort}
 \begin{split}
	\partial {\omega^{[k]}} / \partial t - \nu \Delta \omega^{[k]} & = \underbrace{ \nabla {\times} \nabla {\times} {\cdots} \; \nabla {\times} }_{k \;\mbox{fold} } \; \{\nabla {\times} (u {\times} \omega)\},\\
	\nabla.\omega^{[k]}& =0, \\ 
	\omega^{[k]}_0(x) & =\underbrace{ \nabla {\times} \nabla \times {\cdots} \; \nabla {\times} }_{k+1 \;\mbox{fold} }u_0(x),
	\end{split}
\end{equation}
where the integer $k{\geq}0$, and $\omega^{[k]}$ stands for the $k$-fold curl of $\omega$. The lowest order vorticity is $\omega^{[0]}{\equiv}\omega$. By virtue of the identity $\nabla.(\nabla {\times} \omega){=}0$, the solenoidal condition remains unaltered every time we take the curl. The coefficient functions in (\ref{eq:k-fold-vort}) are defined in the following elliptic equation:
\begin{equation*} 
\Delta \omega^{[k]}(x;t)  = - \omega^{[k+2]}(x;t) .
\end{equation*}
The Laplacian can always be inverted. In particular, the equation governing divorticity $\omega^{[1]}$ reads 
\begin{equation*}
	\partial {\omega^{[1]}} / \partial t - \nu \Delta \omega^{[1]} = \nabla (\nabla . (u {\times} \omega)) - \Delta (u {\times} \omega).
\end{equation*}
Although we do not need the full quantitative knowledge of this complicated dynamic equation, we notice, given the regularity of $u$, that the right-hand side can be expressed in term of certain regular integral operators involving only the unknown. 
\subsection*{Finite-time singularity in the Euler equations}
An answer to the question whether singularities may form in finite time in the solutions of fluid dynamics equations would enable us to identify a possible cause of turbulence. It has been contemplated that breakdown of the continuum assumption, specifically in the limit of vanishing viscosity $\nu \rightarrow 0$, would demonstrate not only the existence of turbulence but also the inadequacy of the Navier-Stokes equations (at least in their present formulation). 
Physically, the presence of viscosity, no matter how minute it may be, amounts to the smoothing property of fluids. This property is natural and inherent, and serves to maintain the initially smooth motion over the time span of the flow evolution. Theoretically, viscosity is critically essential in ruling out any finite-time singularities in the flow field as implied in the global regularity of the Navier-Stokes equations. 

It is a well-known fact that the conservation laws used to derive the equations of motion can be expressed in an integro-differential form which contains the spatial derivatives of only the first order (see, for example, \S6 \& \S7 of Oseen 1927). Thus it is not necessary for the velocity Laplacian to be bounded and continuous. In initial value problems, the implications suggest that the initial vorticity needs only to be essentially bounded. A plausible elaboration is that, as the initial data are not smooth, the damping effect of viscosity may not be strong enough to regulate large velocity gradients in the flow evolution. Consequently, flow fields may develop singularities in finite time. It has been conjectured (Leray 1934$b$) that should a solution behave like $u \sim (t-t^*)^{-\sigma}, \sigma > 0 $, the equations of motion break down on the continuum, and turbulence ensues beyond the singular time $t^*$. But the existence of the self-similar singular behaviour has been ruled out by the work of Nec\v{a}s {\itshape et al} (1996) and Tsai (1998). However, algebraic singularities of general nature, conceived as a result of intensive vortex stretching against diffusion and ineffectual strain attenuation, might well exist. To a large extent, such views on possible blow-ups are often indoctrinated by the scalings (\ref{eq:ns-scaling}) for the equations of motion. Alternative arguments are found by considering the Leray-Hopf {\itshape weak} solutions which are known to hold only for an energy {\itshape inequality}:
\begin{equation*} 
	\frac{1}{2}\big\|u(t)\big\|^2_{L^2 (\real^3)} + \nu \int_0^T \!\!\! \big\|\omega (t) \big\|^2_{L^2 (\real^3)} \rd t  \; {\leq} \; \frac{1}{2}\big\|u_0 \big\|^2_{L^2 (\real^3)}.
\end{equation*}
Thus the law of energy conservation is apparently violated; a certain amount of energy is speculated to percolate into fluid's material structure, giving rise to a hallmark state of small-scale fluid motion beyond the continuum. Nevertheless, turbulence is ubiquitous. Our common experience and dedicated high Reynolds-number laboratory experiments of various designs have been unable to identify a generic flow motion showing features of finite-time singularity and the energy leakage. There are no specific reasons, on the ground of physics, that those perceived anomalous flow phenomena should be proved so elusive to observe in practice. Even for the Euler equations, the endeavour to pin down a breakdown scenario, chiefly by means of numerical computations, has been a subject of intense controversy over the last few decades (Gibbon 2008). To close the knowledge gap, we have made a detour from dealing with real fluids to an exploit on the behaviour of fluid motion as $\mu \rightarrow 0$ (see \ref{app:d}). It is concluded that {\itshape the occurrence of finite-time singularity is out of the question in finite-energy flows having smooth localized initial data}.
\subsection{Presence of a prescribed force} \label{sect:force}
It is perhaps meaningless to consider the effects of ``an external force'' due to the absence of a boundary in $\real^3$. Hence we prefer to regard a force as prescribed; a force is given at specific spatial locations and time. We denote the force by ${\cal F}$ which is assumed to have the decomposition,
\begin{equation*}
	{\cal F}(x,t)= - \rho^{-1} \: \nabla \Phi(x,t) + {\mathrm F}(x,t), 
\end{equation*}
where $\Phi$ is the potential, and ${\mathrm F}$ is non-conservative. Then the Navier-Stokes momentum equation has the modified form,
\begin{equation} \label{eq:ns-force}
	{\partial u}/{\partial t} - \nu \Delta u = - (u . \nabla) u  - {\rho}^{-1} \nabla {\tilde p} + {\mathrm F}(x,t), 
\end{equation}
where ${\tilde p}$ stands for the modified pressure, and ${\tilde p} = p + \Phi$. The continuity equation remains unchanged. In order to avoid a radical rework of our theory, we impose certain localization constraints on ${\cal F}$:
\begin{equation*} 
	\Big| \Big(1 + | x | +t \Big)^{j} \: \partial_x^{\alpha} \: \partial_t^{\beta} \; {\cal F}  (x,t)  \Big| \;{<}\; \infty, \;\;\; \forall (x,t) \in \real^3 {\times} [0, T],
\end{equation*}
for any values of $j$, $\alpha$, and $\beta$. Therefore if the contribution from the non-conservative component vanishes (${\mathrm F}{=}0$), there is no essential change in our preceding theory.

Let ${\mathrm F}{\neq}0$, vorticity equation (\ref{eq:vorticity}) becomes
\begin{equation} \label{eq:vorticity-force}
	{\partial \omega}/{\partial t} - \nu \Delta \omega = (\omega . \nabla) u  - (u . \nabla )\omega + \nabla {\times} {\mathrm F}.
\end{equation}
Consider
\begin{equation*}
	{\mathrm F}(x,t),\; \partial_x {\mathrm F}(x,t) \in L^1(\real^3{\times}[0, T]).
\end{equation*}
It is evident that bounds (\ref{eq:vort-l1-norm}) and (\ref{eq:vort-infty-bound}) hold with the appropriate modifications that allow for the force. The vorticity integral equations (\ref{eq:vort-k00}) are modified as
\begin{equation*}
 \begin{split}
	\omega_i(x,t_k) =  {\tilde \varpi}_i(x,t_k) + 
	\int_0^{t_k} \int \Big( & \big(\omega_i u - u_i\omega \big).\nabla_y \Big) Z(x,t_k,y,s) \: \rd y \rd s,
 \end{split}
\end{equation*}
where
\begin{equation*}
	{\tilde \varpi}_i(x,t_k) = \varpi_i(x,t_k) + \int_0^{t_k} \int Z(x,t_k,y,s) (\nabla {\times} {\mathrm F})_i \rd y \rd s.
\end{equation*}
There is one essential aspect that calls for attention. We are still interested in the case 
\begin{equation*}
	{\tilde \varpi}_i \in C^{0,1} (\real^3),
\end{equation*}
which in turn imposes certain restrictions on the allowable ${\cal F}$. Thus the solution of the resulting integral equation has an analogous expression to (\ref{eq:vort-series-sol}). It follows that the global regularity of (\ref{eq:ns-force}) or (\ref{eq:vorticity-force}) can be established accordingly.
Although it appears that only minor modifications are necessary to include the effect of a force, it must be borne in mind that the presence of the non-conservative contribution may well dominate the evolution of the vorticity field. Specifically, consider an initially stationary flow and then suddenly set it into motion by a force ${\mathrm F}$ alone; the flow will certainly become turbulent if the force is sufficiently strong. 
\subsection{Generalization to initial data of finite Dirac measure}
The proceeding theory has been developed with the analytical requirements on the initial data (cf. (\ref{eq:ns-ic}) and (\ref{eq:ic-localization})). The theory also applies to $\omega_0 \in L^p, p \geq 1$ by virtue of the heat mollifier. However, it is frequent in practice to encounter fluid motions which are initiated by well-controlled point vortices.  A mathematical idealization of the vortices is to model them in terms of finite Dirac measure. Since a finite Radon measure can be expressed as a superposition of a continuous part and a discrete part (see, for example, Reed \& Simon 1972; Royden \& Fitzpatrick 2010): $\upmu = \upmu_c + \upmu_d$. The total variation is $\|\upmu\|_{var} = \|\upmu_c\|_{var} + \|\upmu_d\|_{var}$. Essentially, we only need to consider the evolution of the motion due to the discrete measure. Let the initial vorticity be a sum of $N (< \infty)$  isolated point masses,
\begin{equation} \label{eq:ic-dirac}
	\bar{\omega}_0(x) = \sum_{l=1}^{N} \upalpha_l \: \delta(x-y_l),\;\;\;(x,y_l)\;\in \; \real^3,
\end{equation}
where the supports of the Dirac measures are located at $x = y_l$. The supports, $\upalpha_l {\in} \real$, are assumed to be finite so that the initial vorticity has a finite total variation  $\|\bar{\omega}_0\|_{var} = \sum_l|\upalpha_l| < \infty$. Evidently, the initial velocity induced by $\bar{\omega}_0$ is a solenoidal field except at the singular points $y_l$. In addition, we assume, with suitable choice of $\alpha_l$ and $y_l$, that the initial vorticity satisfies the localization constraint,
\begin{equation*}
	\big\|(1+|x|)^j \bar{\omega}_0 \big\| < \infty
\end{equation*}
for any $j \geq 1$ and $\forall \; x \neq y_l$. The caloric mollified vorticity initial vorticity is the solution of the pure initial value problem of diffusion equation,
\begin{equation*}
	\partial_t \bar{\mathrm w} - \nu \Delta \bar{\mathrm w} =0,\;\;\; \bar{\mathrm w}_0(x)=\bar{\omega}_0(x).
\end{equation*}
Explicitly, the solution is given by
\begin{equation*}
	\bar{\mathrm w}(x,t) = \int_{\real^3}{\mathbf Z}(x,y,t) \bar{\omega}_0(y) \rd y = {(4 \pi \nu t )^{-3/2}} \sum_{l=1}^N \upalpha_l \: \exp \Big( - \frac{|x{-}y_l|^2}{4 \nu t} \Big),\;\;\;t > 0.
\end{equation*}
The physics is clear here: Viscosity acts as a damping agent to smooth out the singular vortex cores via diffusion. The global regularity of fluid motions with initial data (\ref{eq:ic-dirac}) follows from our general theory. The requirement that the initial point vortices have a finite total variation means that $\bar{\gamma}$, the analogous function to ${\gamma}$ in (\ref{eq:solIC-VIE}), is finite. Note that the mollified vorticity $\bar{\mathrm w}$ is a smooth function for $(x,t)$ in $\real^3 {\times} (0,T]$. At the beginning $t=0$, the smoothness property cannot be true. 
\subsection{Non-linear integral equation for momentum}
The advantages of using integral equations in solving problems of fluid dynamics have long been recognized in the early development of hydrodynamics (see, for example, Liapunov 1906; Lichtenstein 1925; Oseen 1927; Odqvist 1930; Leray 1934$b$; Lamb 1975).  Suppose that we carry out an analysis with {\itshape no reference to the concept of vorticity}. Technically, to convert the momentum equation into an integral equation is equivalent to inverting the operator,
\begin{equation} \label{eq:momentum-operator}
	\partial_t  - \nu \Delta  + \rho^{-1} \nabla. 
\end{equation}
To achieve this goal, we have to hypothesize that the rates of spatial decay for the velocity and pressure are sufficiently fast:
\begin{equation*}
	|u| \rightarrow 0, \;\;\; p \rightarrow p_0 \;\;\; \mbox{as} \;\;\; |x| \rightarrow \infty,
\end{equation*}
where $p_0$ is a constant. 

Consider the homothetic transformation, $t'{=}t$ and $x'{=} a x	$, where $a{=}l/l_0$, and $l$~is a length scale. The constant, $l_0$, is a reference length scale which can be fixed in relation to the initial condition. For instance, given $\delta_0{>}0$, $l_0$ is defined by 
\begin{equation*}
	\inf_{x {\in} \real^3} \Big( \int_{l_0}^{\infty} |u_0(x)|^2 \rd x \Big) < \delta_0.
\end{equation*}
Denote the scaled $(\hat{u}{-}\hat{p})$ pair by
\begin{equation*} 
	u(x,t)=\hat{u}(x',t')/a^{m+1},\;\;\; p(x,t)= \hat{p}(x',t') / a^m,
\end{equation*}
where $m$ is an integer, and $m{\geq}{-}1$. The momentum equation becomes
\begin{equation*}
a^{-2} \partial_{t'} \hat{u} - \nu \Delta \hat{u} = - a^{-(m+2)}  \:(\hat{u}.\nabla)\hat{u} - {\nabla \hat{p}}/\rho,
\end{equation*}
where all the spatial derivatives are with respect to $x'$.
Assuming $l{\gg}l_0$, the equations of motion reduce to the approximation, 
\begin{equation*} 
	\nabla . \hat{u}  = 0, \;\;\;
	\mu \Delta \hat{u} - \nabla \hat{p}  = 0.
\end{equation*}
Evidently, the pressure is harmonic near infinity. It follows that
\begin{equation*}
	p \rightarrow p_0 \;\;\; \mbox{as}\;\;\; |x| \rightarrow {\infty}
\end{equation*}
according to Liouville's theorem. The constant $p_0$ is the reference pressure and its value cannot be determined from the dynamics equations alone. However, the exact value is immaterial in our analysis.

Let ${\mathbf T}_{ij}$ and $P_j$ be the fundamental solutions of operator (\ref{eq:momentum-operator}). They can be determined by solving the following equations:
\begin{equation} \label{eq:oseen-tensor}
	\partial_t {\mathbf T}_{ij}  - \nu \Delta {\mathbf T}_{ij} + \rho^{-1} \partial_{x_i} P_j = \delta_{ij} \: \delta(t) \delta(x), \;\;\;
\sum_{i=1}^3 \frac{\partial {\mathbf T}_{ij}}{\partial x_i}=0.
\end{equation}
The function ${\mathbf T}$ is known as the Oseen tensor (see, for example, Oseen 1927; Odqvist 1930; Solonnikov 1964; Ladyzhenskaya 1969; Fabes {\em et al} 1972). Evoking Fourier transforms, the solutions are found to be
\begin{equation} \label{eq:riesz}
	{\mathbf T}_{ij}(x,t)= (\delta_{ij} + R_{ij} ) Z(x,t), \;\;\;  P_i(x{-}y;t) = \frac{\delta(t)}{4 \pi} \frac{(x_i-y_i)}{|x-y|^3},
\end{equation}
where $R_{ij}$ is known as the Riesz transform. The Riesz transform belongs to the special case of the more general class of the Calder\`{o}n-Zygmund singular integral operators (see, for example, Stein 1970; Adams \& Fournier 2003). Particularly $R_{ij}$ is related to the inverse of the Laplacian in $\real^3$ for any measurable function $f$ by 
\begin{equation*}
	R_{ij}f=R_i R_j f = - \frac{\partial^2}{\partial x_i \partial x_j} \Delta^{-1} f =  \frac{\partial^2}{\partial x_i \partial x_j} \Big(  \int {\cal N}(x-y) f(y) \rd y \Big).
\end{equation*}
By the same token, the pressure may be expressed in terms of the velocity derivatives,
\begin{equation} \label{eq:p-inverse}
	p(x)= \rho^{-1} \sum_{i,j=1}^3 R_{ij} (u_i u_j)(x) + p_0.
\end{equation}
It follows that, by direct evaluation, 
\begin{equation*} 
	\int \partial_{x_i} P_i (u.\nabla)u_i \rd y = 0.
\end{equation*}
We have implicitly {\itshape assumed} that the non-linearity satisfies
\begin{equation} \label{eq:vel-rhs}
	\int_0^t \int (u.\nabla)u \rd y \rd s\;{<}\; {\infty}.
\end{equation}
This boundedness condition holds as long as the velocity decays sufficiently fast at infinity.
The tensor ${\mathbf T}_{ij}$ in (\ref{eq:riesz}) can be further simplified as
\begin{equation} \label{eq:tensor-t}
	{\mathbf T}_{ij}(x,t)= \delta_{ij} Z(x,t) + (\nu t) \lim_{\sigma{\rightarrow}\infty}\!\int_1^{\sigma} \! \frac{\partial^2}{ {\partial x_i} {\partial x_j}} Z(x, t \tau) \rd \tau.
\end{equation}
By Duhamel's principle and in view of assumption (\ref{eq:vel-rhs}), the components of momentum equation (\ref{eq:ns}) may be converted into integral equations:
\begin{equation} \label{eq:vel-k}
	u_i(x,t_k) = \int \! Z(x,t_k,y) (u_0)_i(y) \rd y + \int_0^{t_k} \!\!\! \int \sum_{j=1}^3 \frac{\partial {\mathbf T}_{ij}} {\partial y_j} (x,t_k,y,s) u_j(y,s) u(y,s) \rd y \rd s.
\end{equation}
These equations hold for all $t_k$'s in the time interval (\ref{eq:time-intervals}) and, evidently, they are an analogy to the transformed vorticity in (\ref{eq:vort-k}). 

A well-known property of the kernel in (\ref{eq:vel-k}) is that there exist constants $0{<}\varepsilon_1,\varepsilon_2{<}{1}/{2}$ such that 
\begin{equation*}
	\big| \partial_x {\mathbf T}_{ij} \big| \: {<} \: C \: |x|^{-3+\varepsilon_1} \: t^{-{1}/{2}-\varepsilon_2} 
\end{equation*}
(Solonnikov 1964; Fabes {\em et al} 1972). Thus the integral equation can be transformed into a sum of velocity integro-powers of infinite order. We start from the case $k{=}1$ in (\ref{eq:vel-k}). Our similarity transformation proceeds as follows: Multiply the equations ($k{=}0$) by 
\begin{equation*}
	\frac{\partial {\mathbf T}_{mi}} {\partial x_i} (z,t_1,x,t_0) u(x,t_0),
\end{equation*}
sum over $m{=}1,2,3$, and integrate over $\real^3{\times}[0,t_1]$. We obtain
\begin{equation*}
	u_i(x,t_1) = \tilde{u}_i(x,t_1) + \int_0^{t_1} \!\!\! \int \sum_{j=1}^3 \Big( \frac{\partial {\mathbf T}_{ij}} {\partial y_j} \tilde{u}_j \Big)(x,t_1,y,s) u(y,s) \rd y \rd s + \tilde{q}^{(0)}(x,t,u),
\end{equation*}
where $\tilde{u}$ is the caloric mollified initial velocity, and $\tilde{q}^{(0)}(u)$ stands for the first non-linear term. Particularly, the kernels have been ``regulated'' by the initial velocity. The next reduction starts from
\begin{equation*}
	\frac{\partial {\mathbf T}_{m k}} {\partial x_k} (z,t_2,x,t_1) u(x,t_1).
\end{equation*}
The non-linear term $\tilde{q}^{(0)}(u)$ is updated to $\tilde{q}^{(1)}(u)$. 
Obviously we can repeat the reductions as many times as we wish. Equations (\ref{eq:vel-k}) can be reduced to a system of non-linear integral equations (cf. (\ref{eq:vort-VIE})). In view of the integrability of the kernel, the solution of the Navier-Stokes momentum equation can also be expressed as a convergent series. The main advantage is that this solution directly gives the velocity and its derivatives at any location $(x,t) \in \real^3{\times}[0,T]$. Then the pressure gradient is calculated from the tabulated data using formula (\ref{eq:p-inverse}). In contrast, to recover the velocity from the vorticity solutions in (\ref{eq:vort-series-sol}), we must first tabulate the velocity on $(x,t)$ using the Biot-Savart relation.
\subsection{Effect of thermal conduction and free convection}
\subsection*{Navier-Stokes-Fourier equations}
To account for heat transfer, we must consider the equation of energy (see, for example, Batchelor 1973; Landau \& Lifshitz 1976; Schlichting 1979). Let the heat flux density due to thermal conduction be $Q$. The density is taken to be a function of the temperature variation in the flow. Hence the Fourier law of heat conduction is expressed as 
\begin{equation*}
	Q = - {\mathrm k} \nabla {\mathrm T},
\end{equation*}
where ${\mathrm T}$ is the temperature measured in Kelvins (cf. (\ref{eq:fluiddyn-constitute})). The constant ${\mathrm k}$ is the thermal conductivity of the fluid. In general, ${\mathrm k}{=}{\mathrm k}({\mathrm T},p)$. By the hypothesis of incompressibility, the equation of heat transfer is expressed in terms of entropy $\mathrm S$,
\begin{equation} \label{eq:ns-energy}
	{\mathrm T} \: \Big( \frac{\partial {\mathrm S}} {\partial t} + (u . \nabla) {\mathrm S} \Big)  = {\mathrm k} \Delta {\mathrm T} + {2 \mu } S^2_{ij},
\end{equation}
where $S_{ij}$ is the symmetric tensor for strains (cf. \ref{eq:strain-tensors}). 
Making use of the thermodynamic relations, 
\begin{equation*}
	\frac{\partial {\mathrm S} }{ \partial t} = \Big(\frac{\partial {\mathrm S}} {\partial {\mathrm T}}\Big)_p \: \frac{\partial {\mathrm T}}{\partial t} , \;\;\; \nabla {\mathrm S} = \Big(\frac{\partial {\mathrm S}} {\partial {\mathrm T}}\Big)_p \: \nabla {\mathrm T},
\end{equation*}
we express the entropy in energy equation (\ref{eq:ns-energy}) in terms of the temperature. Thus the initial value problem for thermal conduction becomes
\begin{equation} \label{eq:nsf}
 \begin{split}
	{\partial u}/{\partial t} - \nu \Delta u & = - (u . \nabla) u  - {\rho}^{-1} \nabla p, \;\;\; \nabla.u =0, \\
	 {\partial {\mathrm T}}/{\partial t} - \kappa \Delta {\mathrm T} & = - (u . \nabla) {\mathrm T} + \frac{2\nu}{c_p}S^2_{ij}.
	 \end{split}
\end{equation}
These equations are to be solved subject to the initial velocity (\ref{eq:ns-ic}) and the initial (smooth) localized temperature,
\begin{equation*} 
 {\mathrm T}(x,0)={\mathrm T}_0(x),\;\;\; x \in \real^3. 
\end{equation*} 
To be specific, we assume that there exist positive constants $a_{{\mathrm T}}$ and $M_{{\mathrm T}}$ such that
\begin{equation*}
	\big| {\mathrm T}_0(x) \big| \:{<}\: M_{{\mathrm T}} \: \exp\big({a_{{\mathrm T}} |x|^2} \big).
\end{equation*}
The specific heat capacity at constant pressure and the thermal diffusivity of the fluid are denoted by
\begin{equation*}
	c_p = {\mathrm T} \big(\partial {\mathrm S}/ \partial {\mathrm T} \big)_p \;\;\; \mbox{and} \;\;\;
	\kappa = {\mathrm k}/\big( \rho \: c_p \big)
\end{equation*}
respectively. The combined motion of fluid and temperature (\ref{eq:nsf}) is known as the incompressible Navier-Stokes-Fourier system. Because of the temperature variations in the flow field, the incompressibility hypothesis is no longer self-consistent. Now the fluid density depends on the temperature and, strictly speaking, it cannot be regarded constant. In deriving (\ref{eq:nsf}), we have postulated that the variations in pressure are negligible. It follows that the density variations are approximately zero. In addition, the temperature differences are assumed to be small throughout the flow field. Note that the assumption of the small temperature variations does not impose any constraint on the spatio-temporal gradients of the temperature. Furthermore we neglect any thermal variation in $\mu$, ${\mathrm k}$ and $c_p$.

Let $Y$ be the analogous integral kernel to $Z$ (\ref{eq:heat-kernel}), 
\begin{equation*} 
	Y(x,t,y) = (4 \pi \kappa t)^{-{3}/{2}} \:  \exp \Big( - \frac{|x{-}y|^2}{4 \kappa t } \Big), \;\;\;{t>0}.
\end{equation*}
Since the Navier-Stokes equations are globally well-posed, then both $u$ and $S_{ij}$ are smooth and bounded. The vorticity of the combined motion is essentially bounded. The structure of the third equation in (\ref{eq:nsf}) suggests that the temperature ${\mathrm T}$ satisfies a maximum principle (see, for example, Friedman 1964; Protter \& Weinberger 1984; Evans 2008), 
\begin{equation*}
	\big\|{\mathrm T}(\cdot,t)\big\|_{L^{\infty}} \; {\leq} \; C \: T\: \big\|S^2_{ij}\big\|_{L^{\infty}}, \;\;\; 0 \leq t \leq T,
\end{equation*}
where constant $C$ is independent of $T$. Introducing short-hand notation
\begin{equation*}
	\tilde{\mathrm T}_0(x,t)=\int Y(x,t,y) {\mathrm T}_0(y) \rd y + \frac{2 \nu}{c_p} \int_0^t \!\!\! \int Y(x,t,z,s) S^2_{ij}(z,s) \rd z \rd s,
\end{equation*}
the differential equation for temperature is reduced to 
\begin{equation*}
	{\mathrm T} (x,t) = \tilde{\mathrm T}_0(x,t) - \int_0^t \!\! \int \Big( \nabla_z Y .u \Big) (x,t,z,s) {\mathrm T}(z,s) \rd z \rd s. 
\end{equation*}
This is a linear integral equation of Volterra-Fredholm type. It is completely solvable because its kernel is integrable for $(x,t) {\in} \real^3 {\times} [0,T]$. The solution for the temperature can be expressed by a Volterra series. It is interesting to notice that turbulence is embodied in the strain term $S_{ij}$ as well as in the kernel $(\nabla Y). u$. At any instant of time, turbulence has the strongest influence on the temperature field wherever the local strain is the highest. The transport term in the temperature equation does not appear in function $\tilde{\mathrm T}_0$ at all. The effect of heat transfer by momentum convection is only discerned in the second term (and beyond) in the Volterra series. 

{\itshape The Cauchy problem of the incompressible Navier-Stokes-Fourier system (\ref{eq:nsf}) is globally well-posed in $\real^3$ for localized initial smooth data of finite total energy. The vorticity field of fluid dynamics is responsible for turbulence characters. Turbulence is effectively transported into the thermal field by the strains of the velocity field.}
\subsection*{Oberbeck-Boussinesq equations}
Now we turn attention to the Cauchy problem of a flow motion with the effect of thermal convection. In the approximation of the thermal conduction, we have made an assumption that the density variation in the fluid is negligible. Now we are interested in the density variation due to a gravitational potential. Equivalently the variation defines a body force on the fluid:
\begin{equation*}
	F^* = - \rho \nabla \phi^* = \rho g,
\end{equation*}
where $\phi^*$ is the potential, and $g$ the (vector) gravitational acceleration.
Since the temperature is a function of the altitude in the gravitational field, the body force tends to restore the fluid motion back to mechanical equilibrium.

Consider the temperature variation in the form of ${\mathrm T}{=}{\mathrm T}_0{+}{\mathrm T}^*$,
where ${\mathrm T}_0$ is a constant reference temperature. We intend to investigate the fluid motion owning to the thermal variation ${\mathrm T}^*$ relative to this reference temperature. Furthermore, we assume that the variation is small in comparison to the reference value.
We write the density as $ \rho {=} \rho_0 {+} \delta \rho $,
where $\rho_0$ is a constant density. The density perturbation $\delta \rho \:{\ll}\: \rho_0$ so that
\begin{equation*}
	\delta \rho = (\partial \rho_0 / \partial {\mathrm T} )_p {\mathrm T}^* = - \rho_0 \beta {\mathrm T}^*,
\end{equation*}
where $\beta({=}{-}(\partial \rho / \partial {\mathrm T} ) / \rho)$ is the thermal expansion coefficient of the fluid. Consequently, the body force can be expressed in terms of the temperature change,
\begin{equation*}
	F^* = \rho_0 g - \rho_0 g \beta {\mathrm T}^*.
\end{equation*}
Similarly, we express the pressure as $p {=} p_0 {-} \rho_0 \phi^*$. Then the equations of motion read
\begin{equation} \label{eq:ns-ob}
 \begin{split}
 {\partial {\mathrm T}^*}/{\partial t} - \kappa \Delta {\mathrm T}^* & = - (u . \nabla) {\mathrm T}^*, \;\;\;\;\;\; \nabla.u =0, \\
	{\partial u}/{\partial t} - \nu \Delta u & = - (u . \nabla) u  - \nabla p_0/\rho_0 - g \beta {\mathrm T}^*. \\
	 \end{split}
\end{equation}
These equations are to be solved subject to initial velocity (\ref{eq:ns-ic}) and initial temperature,
\begin{equation*} 
 {\mathrm T}^*(x,0) = {\mathrm T}^*_0(x) \in C_c^{\infty}(\real^3). 
\end{equation*} 
The system (\ref{eq:ns-ob}) is known as the Oberbeck-Boussinesq equations governing free convection. The components of the thermal vorticity are found to be
\begin{equation} \label{eq:vort-thermal}
	{\partial \omega_i}/{\partial t} - \nu \Delta \omega_i = (\omega . \nabla) u_i  - (u. \nabla) \omega_i + \beta (g{\times} \nabla {\mathrm T}^*)_i,\;\;\; i=1,2,3.
\end{equation}	
The initial vorticity is specified as a smooth, localized function.

Let $u^{(0)}{=}0$. For $k=1,2, \cdots $, solve the following system of equations for $(x,t) \in \real^3{\times}[0,T],\:T < \infty$:
\begin{equation} \label{eq:ns-ob-iter}
 \begin{split}
 {\partial {{\mathrm T}^*}^{(k)}}/{\partial t} - \kappa \Delta {{\mathrm T}^*}^{(k)} & = - (u^{(k-1)} . \nabla) {{\mathrm T}^*}^{(k)}, \\
 {\partial \omega^{(k)}}/{\partial t} - \nu \Delta \omega^{(k)}  & =  (\omega^{(k)} . \nabla) u^{(k)}  - (u^{(k)}. \nabla) \omega^{(k)} + \beta \:g{\times} \nabla {{\mathrm T}^*}^{(k)},  \\
 u^{(k)} & =  {\cal K}* \omega^{(k)}, \\ 
	 \end{split}
\end{equation}
subject to the initial conditions
\begin{equation*}
 {{\mathrm T}^*}^{(k)}(x,0) = {\mathrm T}^*_0(x), \;\;\;
	\omega^{(k)}(x,0) = \nabla{\times}u_0(x). 
\end{equation*}
Let $k{=}1$, it is obvious that, at a moment immediately after the start of the motion $t{=}0^+$, linear diffusion is dominant while the drift effect (due to the drift velocity $u$) can be neglected. The temperature is described by
\begin{equation} \label{eq:t-k1}
	{{\mathrm T}^*}^{(1)}(x,t)=\int Y(x,t,y) {\mathrm T}^*_0(y) \rd y = \eta^*(x,t).
\end{equation}
The analytic structure of the equation for $\omega^{(1)}$ has the form of the Navier-Stokes vorticity in the presence of a prescribed force (cf. (\ref{eq:vorticity-force})). Thus we state that (\ref{eq:ns-ob-iter}) is well-posed for $k{=}1$.

Let $k{=}2$. The first equation in (\ref{eq:ns-ob-iter}) shows that the temperature must satisfy a maximum principle:
\begin{equation} \label{eq:temp-max-k2}
	\big\|{{\mathrm T}^*}^{(2)}(\cdot,t)\big\|_{L^{\infty}} \;{\leq}\; \big\|{\mathrm T}^*_0\big\|_{L^{\infty}},\;\;\; 0 \leq t \leq T.
\end{equation}
The solution of ${{\mathrm T}^*}^{(2)}$ can be obtained by solving the linear equation,
\begin{equation*}
	{{\mathrm T}^*}^{(2)}(x,t)=\int Y(x,t,y) {\mathrm T}^*_0(y) \rd y + \int_0^t \!\! \int (\nabla Y. u^{(1)}) {{\mathrm T}^*}^{(2)}(z,s) \rd z \rd s.
\end{equation*}
Thus the temperature ${{\mathrm T}^*}^{(2)}$ is completely determined. Moreover, 
${{\mathrm T}^*}^{(2)}$ decays at infinity. Hence vorticity invariance (\ref{eq:vorticity-invariance}), bounds (\ref{eq:du-space-lq-bound}) and (\ref{eq:u-space-bound}) hold for $\omega^{(2)}$ and $u^{(2)}$. It follows that system (\ref{eq:ns-ob-iter}) is well-posed for $k{=}2$. 

Suppose that it has been shown that (\ref{eq:ns-ob-iter}) is well-posed for $k$. In particular, $|u^{(k)}| < \infty$ for $(x,t)  \in \real^3{\times}[0,T]$. Then
\begin{equation*} 
	\big\|{{\mathrm T}^*}^{(k+1)}(\cdot,t)\big\|_{L^{\infty}} \leq  \big\|{\mathrm T}^*_0\big\|_{L^{\infty}},\;\;\; 0 \leq t \leq T,
\end{equation*}
and
\begin{equation} \label{eq:temp-k-vfie}
	{{\mathrm T}^*}^{(k+1)}(x,t)=\int Y(x,t,y) {\mathrm T}^*_0(y) \rd y + \int_0^t \!\! \int (\nabla Y. u^{(k)}) {{\mathrm T}^*}^{(k+1)}(z,s) \rd z \rd s.
\end{equation}
Since the vorticity ${\omega}^{(k+1)}$ is globally regular, we conclude that both sequences,
\begin{equation}
	{\mathrm T}^*(x,t)=\lim_{k{\rightarrow}\infty}{{\mathrm T}^*}^{(k)}(x,t) \;\;\; \mbox{and} \;\;\; \omega(x,t)=\lim_{k{\rightarrow}\infty}{\omega}^{(k)}(x,t),
\end{equation}
exist for $(x,t) {\in} \real^3{\times}[0,T]$ in view of the Arzel\`a-Ascoli theorem. Moreover ${\mathrm T}^*$ and~$\omega$ (and hence $u$) are smooth. The temperature is uniquely determined from the linear Volterra-Fredholm integral equation (\ref{eq:temp-k-vfie}). As the solution of the vorticity is unique, we assert that the solution of (\ref{eq:ns-ob-iter}) is also unique. 

The Cauchy problem of the coupled thermal-fluid equations of motion (\ref{eq:ns-ob}) for free convection has been shown to be globally well-posed. The temperature satisfies a maximum principle, 
\begin{equation*}
	\big\|{{\mathrm T}^*}(\cdot,t)\big\|_{L^{\infty}} \;{\leq}\; \big\|{\mathrm T}^*_0\big\|_{L^{\infty}}.
\end{equation*}
To construct the solution, we notice that system (\ref{eq:ns-ob}) can be written in the following form:
\begin{equation} \label{eq:ns-ob-ie}
 \begin{split}
 \omega(x,t) & = \int {\mathbf Z} \omega_0 (y) \rd y + \int_0^t \!\! \int \!\! \int G^*(x,t,y,s,z,s) \omega(y,s) \omega(z,s) \rd z \rd y \rd s \\
 \quad & {\hspace {4cm}}  + \int_0^t \!\! \int Q^*(x,t,y,s) {\mathrm T}^*(y,s) \rd y \rd s, \\
 {\mathrm T}^* (x,t) & = \int Y T^*_0(y) \rd y + \int_0^t \!\! \int \!\! \int R^*(x,t,y,s,z,s) \omega(z,s) {\mathrm T}^*(y,s) \rd z \rd y \rd s,
	\end{split}
\end{equation}
where the kernel functions $G^*, Q^*$ and $R^*$ are known functions of the initial data.

Our similarity transformation works as follows: We multiply the first equation in (\ref{eq:ns-ob-ie}) by
$G^* \omega$ and then integrate over space-time. Second we multiply the temperature equation by $Q^*$ and do the integration. Next we multiply the second equation in 
(\ref{eq:ns-ob-ie}) by $R^* \omega$ and integrate over space-time. Essentially, equations  (\ref{eq:ns-ob-ie}) can be reduced to
\begin{equation} \label{eq:ns-ob-series}
 \begin{split}
 \omega(x,t) & = \gamma(x,t) + \theta^*(x,t) + \sum_i A^*_i(\omega, {\mathrm T}^*) (x,t), \\
 {\mathrm T}^* (x,t) & = \eta^*(x,t) + \sum_i B^*_i(\omega, {\mathrm T}^*) (x,t),
	\end{split}
\end{equation}
where the functions, $A^*_i$ and $B^*_i$, are non-linear in $\omega$ and ${\mathrm T}^*$. Furthermore, 
\begin{equation*}
	\theta^*(x,t) = \int_0^t \!\! \int H^* \Big( \int_0^s \!\! \int Q^* \eta^* \rd z \rd \tau  \Big) \rd y\rd s + \int_0^t \!\! \int Q^* \eta^* \rd y \rd s,
\end{equation*}
where $H^*$ is the resolvent of $K^*$ (the analogous capacity kernel to $K$).
Thus the solutions of (\ref{eq:ns-ob-series}) can be expressed by series expansions in terms of $\gamma$ as well as $\eta^*$. 

{\itshape As opposed to the effect of the thermal conduction, the vorticity field and the convective thermal field interact with each other dynamically. Consequently, either has the potential to initiate turbulence}.
\section{Fluid dynamics in other space dimensions}
In the present section, we show that the equations of fluid motions in space dimensions other than $3$ are globally well-posed. Turbulence is an integral part in these motions. Practically, the results are obtained by analogy.
\subsection{Equations of motion in $\real^2$}
The global regularity of the Navier-Stokes equations in $2$ space dimensions has long been established by several theoretical approaches (see, for example, Leray 1934$a$; Ladyzhenskaya 1959; McGrath 1968). However, the outcomes of these investigations have not been linked to the well-known characters of turbulence. Our objective is to show that turbulence exists in two dimensional flows as well if the initial data are large. 

In 2 space dimensions, the equations of motion are
\begin{equation} \label{eq:2d-ns}
	{\partial w}/{\partial t} - \nu \Delta w = - w. \nabla w  - {\rho}^{-1} \nabla p, \;\;\; \mbox{and} \;\;\; \nabla.w = 0,
\end{equation}
where the velocity is denoted by $w=(w_1,w_2)$. We use notation $z=(x,y)$ for the space variable. The initial velocity is denoted by $w(z,0)=w_0(z)$,
which is assumed to be smooth (cf. (\ref{eq:ns-ic})) and to satisfy a localization requirement (cf. (\ref{eq:ic-localization})). Let the vorticity be $\zeta {=} \partial w_2 /\partial x {-} \partial w_1 /\partial y$.
The vorticity equation has a simple form 
\begin{equation} \label{eq:2d-vorticity}
 \partial_t \zeta  - \nu \Delta  \zeta = - (w.\nabla)\zeta,
\end{equation}
in which the mechanism of vorticity stretching is absent.
The initial vorticity data are considered to be smooth and localized,
\begin{equation*}
	\zeta(z,0)=\zeta_0(z),\;\;\; z \in \real^2.
\end{equation*}

It is well-known that the vorticity satisfies a maximum principle:
\begin{equation} \label{eq:max-vorticity}
	\max_{z \in \real^2, \; t \in [0,T] } \big|\: \zeta \: \big| \;{\leq}\; \max_{z \in \real^2 } \big| \: \zeta_0 \: \big|
\end{equation}
(see, for example, Friedman 1964; Protter \& Weinberger 1984; Evans 2008). Like its counterpart in three space dimensions, the velocity potential $\psi$ satisfies Poisson's equation $\Delta \psi(z;t) {=} - \zeta(z;t)$ for $t \in [0,T]$. In terms of the Newtonian logarithmic potential, we are able to invert the Laplacian to get the velocity components:
\begin{equation} \label{eq:2d-bs}
 \begin{split}
w_1(z) & = - \frac{\partial \psi}{\partial y} = -\frac{1}{2 \pi} \int_{\real^2} \frac{ y - \tilde{y} }{ |z-\tilde{z} |^2 } \zeta(\tilde{z}) \rd \tilde{z}, \\
w_2(z) & =  \frac{\partial \psi}{\partial x} = \frac{1}{2 \pi} \int_{\real^2} \frac{ x - \tilde{x} }{ |z-\tilde{z} |^2 } \zeta(\tilde{z}) \rd \tilde{z}.
 \end{split}
\end{equation}
The pressure can be found once the velocity derivatives are known. Denote the fundamental solution of the heat operator in two space dimensions by 
\begin{equation*}
	Y(z,t,\hat{z},s) = \big(4 \pi \nu (t{-}s)\big)^{-1} \exp \Big( - |z{-}\hat{z}|^2/ \big( 4 \nu (t{-}s) \big) \Big), \;\;\;{t>s}.
\end{equation*}
By virtue of (\ref{eq:max-vorticity}), we transform the vorticity equation in (\ref{eq:2d-vorticity}) into the following scalar integral equation:
\begin{equation} \label{eq:vort-2d-VIE}
	\zeta(z,t) =  \int_{\real^2} Y(z{-}\sigma,t) \zeta_0(\sigma) \rd \sigma + \int_0^t \int_{\real^2} Y(z,t,\sigma,s) (w.\nabla)\zeta (\sigma,s) \rd \sigma \rd s.
\end{equation}
Integrating over space, we obtain the invariance of the vorticity,
\begin{equation*}
	\frac{\rd}{\rd t } \int_{\real^2} \zeta (z,t) \rd z = 0.
\end{equation*}
The boundary term arising from the integration by parts vanishes in view of the maximum principle.
It is plain to generalize the invariance to space-time derivatives of arbitrary orders. The main application of the invariance theory is that the vorticity is {\itshape a priori} a smooth, bounded function. Substituting the velocity integrals in (\ref{eq:2d-bs}) into (\ref{eq:vort-2d-VIE}) and simplifying the result, we obtain
\begin{equation} \label{eq:vort-k0-2d}
	\zeta(z,t) =  \varpi_2(z,t) + \int_0^t \int_{\real^2} \int_{\real^2}  B(z,t,\sigma,s,\tilde{z},s)\zeta(\tilde{z},s) \zeta(\sigma,s)  \rd \tilde{z} \rd \sigma \rd s,
\end{equation}
where $B(z,t,\sigma,s,\tilde{z},s)$ denotes the product of the Biot-Savart law, and the derivative of heat kernel $Y$. The mollified initial data are denoted by $\varpi_2$. Following the transformation principles for the reduction of the $3D$ vorticity equation, we assert that
\begin{equation} \label{eq:vort-k-2d}
	\zeta(z,t) =  \varpi_2(z,t) + \int_0^t \int_{\real^2} A(z,t,z',s)\zeta(z',s) \rd z' \rd s + q_2(z,t, \zeta),
\end{equation} 
where the scalar function $q_2(z,t,\zeta)$ stands for the counterpart of the non-linear term $q$ in $\real^3$. The kernel is given by
\begin{equation*}
	A(z,t,z',t')= \int_{\real^2}B(z,t,z',t',\bar{z},t') \varpi_2(\bar{z},t') \rd \bar{z} 
	= \frac{\partial Y}{\partial x'} \; \frac{\partial \phi}{\partial y'} - \frac{\partial Y}{\partial y'} \; \frac{\partial \phi}{\partial x'},
\end{equation*}
where $Y{=}Y(z,t,z',t')$, and $\phi{=}\phi(z')$ is the potential satisfying Poisson's equation $\Delta \phi {=} {-}\varpi_2$.
For $\varpi_2 \in C^{0,1}(\real^2)$, standard theory suggests that there exists some constant $C$ so that the kernel $A$ is bounded,
\begin{equation*} 
	\big| A(z,t,z',t') \big|\:{<}\:\frac{C}{(t-t')^{\kappa}\:|z-z'|^{3-2\kappa}},\;\;\; 0 \leq \kappa \leq 3/2.
\end{equation*}
Since the singularity in $A$ is integrable, equation (\ref{eq:vort-k-2d}) can be solved by method of successive approximations. Thus we find that
\begin{equation} \label{eq:omega-2d-volterraIE}
	\zeta(z,t) =  \gamma_2(z,t) + q_2(z,t)+ \int_0^t \int_{\real^2} A_r(z,t,z',t') q_2(z',t') \rd z' \rd t',
\end{equation} 
where $A_r$ is the resolvent kernel of $A$. The function $\gamma_2$ is given by
\begin{equation*}
	\gamma_2(z,t) =  \varpi_2(z,t) + \int_0^t \int_{\real^2} A_r(z,t,z',t') \varpi_2(z',t') \rd z' \rd t'.
\end{equation*}
There are no essential differences in the mathematical structures between (\ref{eq:omega-2d-volterraIE}) and (\ref{eq:vort-nonlinear-VIE}). Hence we express the result as
\begin{equation} \label{eq:2d-vort-series-sol}
	\zeta(z,t) = \gamma_2(z,t) + \sum_{m{\geq}2}\: \Big( \sum_{i=1}^{S_m} \: U_i[\gamma_2]^m \Big).
\end{equation}
It follows that the existence, the uniqueness and the regularity can be established. 

The small-scale eddies in the vorticity field are generated purely by the non-linearity of the vorticity {\itshape convection}. The function $\gamma$ in three space dimensions is a matrix quantity while $\gamma_2$ is a scalar. Roughly speaking, the vorticity in $\real^3$ has more degrees of freedom to evolve. On the other hand, certain vorticity features, such as the coherent structures, are easier to maintain in planar flow fields. Hence it is not surprising to observe flows with an inverse energy cascade in $\real^2$ if the initial velocity is suitably given. Similarly, an enstrophy cascade from large to small scales may well be evident at a particular spatial location if the contributions of the integro-power terms in (\ref{eq:2d-vort-series-sol}) are temporarily increasing at the location.

{\itshape For freely-evolving flows in $\real^2$, the evolution of the vorticity field into turbulence is a strong function of the initial data. The convection mechanism in the vorticity equation is capable of mass-producing eddies of different scales and intensities. The properties of each flow have to be assessed on individual merits.}  
\subsection{Burgers' model for non-linearity}
It was contemplated by Burgers (1948) that a number of turbulence effects may be modelled by a set of equations which is simpler than the Navier-Stokes equations. In particular, he hoped that the question of the presence of the multiple fluid length scales (the spectrum in Fourier space) and the dissipation of energy can be elucidated by considering the Cauchy problem of the following one-dimensional equation: 
\begin{equation} \label{eq:burgers}
	{\partial v}/{\partial t} - \nu \Delta v + v \nabla v =0.
\end{equation}
It is to be solved subject to the initial data $v(x,0)=v_0(x), x \in \real^1$. The velocity is assumed to decay sufficiently fast at infinity,
\begin{equation*}
	v \rightarrow 0 \;\;\; \mbox{as} \;\;\; |x| \rightarrow \infty.
\end{equation*}
Equation (\ref{eq:burgers}) is known as Burgers' equation. From statistical descriptions of turbulence, the non-linearity was regarded as the interior mechanism of vorticity stretching which controls the transfer of the energy from the modes of high wave numbers to those of low wave numbers. The total number of these Fourier modes was in the order of the Reynolds number $\nu^{-1}$. In any particular flow, it was not necessary that every Fourier mode would be present. The geometry of the initial condition is very important. 

By virtue of the Cole-Hopf transformation, Burgers' equation can be simplified as a linear heat equation. Thus it can be solved in a closed form. Specifically, the velocity is found to be
\begin{equation} \label{eq:vel-1d-closed-sol}
	\frac{v(x,t)}{2 \nu} = {\int_{\real} \! \frac{\partial X(x{-}y,t)}{\partial y} \: V(y) \rd y } \; \Big/ {\int_{\real} \! X(x{-}y,t) \: V(y) \rd y},
\end{equation}
where
\begin{equation*}
	V(y) = \exp \Big(\: -\frac{1}{2 \nu} \int_0^y \! v_0(z) \: \rd z \Big),
\end{equation*}
and
\begin{equation*}
	X(x{-}y,t) = \big( 4 \pi \nu t \big)^{-1/2} \exp \Big( - \big( x{-}y \big)^2  \big( 4 \nu t \big)^{-1} \Big), \;\;\;{t>0}.
\end{equation*}
This solution has been extensively investigated as a model for the formation of shock waves in real high-speed flows (see, for example, Whitham 1974). However, a shock is a modest jump in {\itshape viscous} compressible flow field; there is a finite amount of energy in the shock structure and hence it is not a singularity in mathematical terms except in inviscid flows at extremely high Mach numbers. For instance, the flow visualization of an Ogive-cylinder at Mach 1.7 (Plate 261 of Van Dyke 1982) shows the contrast between shock wave and turbulence: The shock and the compressible turbulence in cylinder's wake and boundary layers have essentially different flow structures. Fundamentally, incompressible and compressible flows both have predominately identical turbulence structure (cf. Plate 151). 

If we are interested in the velocity solution that contains finite energy, then the law of energy conservation becomes
\begin{equation*} 
	\frac{1}{2} \big\| v \big\|_{L^2(\real)}^2 + \nu \int_0^t  \big\| \nabla v(\cdot,t) \big\|_{L^2(\real)}^2 \rd t \; {=} \; \frac{1}{2}\big\| v_0 \big\|_{L^2(\real)}^2.
\end{equation*}
In view of Duhamel's principle, (\ref{eq:burgers}) is converted into
\begin{equation} \label{eq:vel-VIE-1d}
	v(x,t)= r(x,t) + \frac{1}{2} \int_0^t \!\! \int_{\real} \frac{\partial X(x,t,y,s)}{\partial y} v^2(y,s) \rd y \rd s, 
\end{equation}
where $r(x,t)$ is the solution of the heat initial value problem for the initial data $v_0$. The integral kernel,
\begin{equation*}
	k(x,t,y,s) = \frac{1}{2} \frac{\partial X(x,t,y,s)}{\partial y} r(y,s),
\end{equation*}
is clearly integrable in $(y,s)$ because of the smoothness of the mollified function $r(x,t)$. Thus its integral resolvent exists and is denoted by $h(x,t,y,s)$. By means of similarity transformation, equation (\ref{eq:vel-VIE-1d}) can be converted into a non-linear integral equation for $v(x,t)$ involving the integral convolutions of $X(x,t,y,s)$ and $v_0(x)$. The resulting equation can be expressed as
\begin{equation} \label{eq:vel-VIE-1d-q}
	v(x,t)= \gamma_1(x,t) + q_1(x,t,v) + \int_0^t \!\! \int_{\real} h(x,t,y,s) q_1(y,s,v) \rd y \rd s,
\end{equation}
where $\gamma_1(x,t)$ is the solution of the linear Volterra-Fredholm integral equation with the kernel $k(x,t,y,s)$ (the counterpart of $\gamma(x,t)$). The function $q_1(x,t,v)$ is the analogous series to the non-linearity $q(x,t,\omega)$ in $\real^3$. The explicit solution is given by
\begin{equation} \label{eq:vel-1d-series-sol}
	v(x,t) = \gamma_1(x,t) \: + \sum_{m{\geq}2}\: \Big( \sum_{i=1}^{S_m} \: U_i[x,t,\gamma_1]^m\Big).
\end{equation}
The solutions expressed in (\ref{eq:vel-1d-series-sol}) and (\ref{eq:vel-1d-closed-sol}) must be two different representations of the same solution of (\ref{eq:burgers}). Consequently, we assert that ``turbulence" is inherently embodied in (\ref{eq:burgers}). 

The quintessential feature of turbulence in the Navier-Stokes dynamics is the presence of the non-linear term $(u.\nabla)u$. Indeed the Burgers equation contains this basic ingredient for turbulence. Analytically, the decay at infinity replaces the continuity hypothesis. With the assumption of the decay, it is straightforward to show that the velocity is in fact invariant of the motion,
\begin{equation*}
	\frac{\rd}{\rd t}\int_{\real} v(x,t) \rd x=0.
\end{equation*}
Nevertheless, there are several essential differences between the ``turbulent'' motion in (\ref{eq:vel-1d-series-sol}) and the Navier-Stokes turbulence in (\ref{eq:vort-series-sol}). In the latter, the Biot-Savart kernel has a dominant effect on the flow field, particularly on the vortices of small scales. The velocity at a particular point is a cumulative effect of all the vorticity in the entire flow field. In (\ref{eq:vel-1d-series-sol}), the evolution of the velocity field is mainly characterized by the spatio-temporal gradients of diffusion. It follows that Burgers' model has a weak diffusive capability compared to genuine turbulence. In fact, Burgers attempted to exemplify the effect of vortex stretching as the mechanism of energy cascade using Burgers' vortices. The stretching is necessarily a local process and its effectiveness is far less than the dynamics in which there is a global catalyst, such as the Biot-Savart induction. Although Burgers' equation is a plausible candidate to embellish the semblance of turbulence, its velocity field is radically dissimilar to that determined by the Navier-Stokes equations.

There are other non-linear equations other than the Burgers equation which contain  ``turbulence'' in their solutions, for given appropriate initial data. The best known example is the non-linear diffusion equation,
\begin{equation} \label{eq:diffusion-v2}
	{\partial \hat{v}}/{\partial t} - \hat{\mu} \: \Delta \hat{v} = \hat{v}^2,\;\;\;\hat{v}(x,0)=\hat{v}_0(x) \in C^{\infty}_c.
\end{equation}
Let us assume $\hat{v} \rightarrow 0$ sufficiently fast as $|x| \rightarrow \infty$. This differential equation can be reduced to the equivalent form,
\begin{equation*}
	\hat{v}(x,t) = \hat{r}(x,t) + \int_0^t \!\! \int_{\real} X(x,t,y,s) (\hat{v}(y,s))^2 \rd y \rd s,
\end{equation*}
where $\hat{r}$ stands for the caloric mollified initial data. Evidently, our method of similarity reduction works for this integral equation. Without going into the detail, we state the result of the transformations,
\begin{equation*}
	\hat{v}(x,t) = \hat{r}(x,t) + \int_0^t \!\! \int_{\real} \hat{k}(x,t,y,s) \hat{v}(y,s) \rd y \rd s + \hat{q}_1(x,t, {\hat v}).
\end{equation*}
After introducing the resolvent kernel $\hat{h}$ for the kernel $\hat{k}(=X \hat{r})$, we transform the equation into
\begin{equation*}
	\hat{v}(x,t) = \hat{r}(x,t) + \hat{q}_1(x,t,{\hat v}) + \int_0^t \!\! \int_{\real} \hat{h}(x,t,y,s) \big( \hat{r}(y,s) + \hat{q}_1(y,s,{\hat v}) \big) \rd y \rd s
\end{equation*}
(cf. (\ref{eq:vel-VIE-1d-q})). 
By analogy, the solution of the non-linear diffusion (\ref{eq:diffusion-v2}) has a series expansion similar to (\ref{eq:vel-1d-series-sol}). However, the integro-powers are dominated by diffusion rather than diffusion gradients. For data of sufficiently large size, the solution is of ``turbulence'' type.  
Differential equation (\ref{eq:diffusion-v2}) is in fact a prototype problem for the Landau-Coulomb equation which has a wide range of applications (Villani 2002, p$267$). We conjecture that ``turbulence'' is a general solution of the Landau-Coulomb equation, and ``blow-ups'' in finite time do not occur for $\hat{\mu} > 0$. 
\subsection{Navier-Stokes equations in $\real^n, n{>}3$}
The equations of fluid dynamics can be generalized to $n$ Euclidean space where the space dimensions $n$ is finite. The generalization has been considered in the work of Hopf (1951). In the book authored by  Arnold \& Khesin (1998), an extended section is devoted to the Euler equations. The local regular solution of the Navier-Stokes for the space dimensions $n{\leq}4$ is known; the analytic approach has been discussed in Serrin (1962, 1963) and in Temam (1977). In the present section, we present a succinct introduction of the global regularity for any space dimensions $n{>}3$. Apart from some analytic technicalities, the principles leading to the general theory do not differ substantially from those used for $n{=}3$. We will retain our notations $u,p \;(\mbox{or} \; \chi)$, and $\omega$ for the velocity, pressure and vorticity respectively. The co-ordinates $x$ reads 
\begin{equation*}
	x=(x_1,x_2, \cdots, x_n).
\end{equation*}
The equations of motion become
\begin{equation} \label{eq:ns-n}
	u' - \nu (\star \rd (\star \rd u)) = \star(u \wedge \omega) - \rd \chi,\;\;\; \star \rd (\star u)=0,
\end{equation}
where the non-linearity is written in terms of the Hodge star notation,  and of the exterior product. The differential operator $\rd$ denotes the exterior derivative and it has the well-known property $\rd^2=0$. The prime stands for time-wise differentiation. The initial data for $n{>}3$ have the generalized forms of (\ref{eq:ns-ic}) and (\ref{eq:vort-ic}) and they are assumed to satisfy a localization constraint (cf. (\ref{eq:ic-localization})). The pressure $p$ is governed by an elliptic equation and hence it represents a global quantity due to the assumption of incompressibility. Equations in (\ref{eq:ns-n}) may first be mollified in space as well as in time so that it is justifiable to perform differentiations on the equations (we prefer here to omit the use of the superficial notations resulting from the mollification approximations). 
Since the vorticity is given by $\omega = \star \rd u$, the vorticity equation reads
\begin{equation} \label{eq:vorticity-n}
	\omega' - \nu (\star \rd (\star \rd \omega)) = \star \rd \star (u \wedge \omega) = \big\{ \omega,u \big\},
\end{equation}  
where $\{\cdot,\cdot\}$ stands for the Poisson bracket. In view of the identity, $\rd (\rd \chi){=}0$, the pressure drops out of the vorticity equation. We note that $\star \rd( \star \rd (\star u)) {=} \star \rd (\star \omega){=}0$. Compared to the case in $\real^3$, the bracket shows an affine property for the vorticity; it is in fact a linear combination of the vorticity and its derivatives of all the components. The velocities play the role of the coefficients. In practice, it is more convenient to work in terms of bracket's components which have the form
\begin{equation*} 
	\big\{ \omega, u \big\}_i = - \big\{u, \omega \big\}_i = \sum_{j=1}^{n} \Big( \omega_j \partial_{x_j}u_i - u_j \partial_{x_j} \omega_i \Big).
\end{equation*}
Let $\widetilde Z(x,t)$ and $\widetilde N(x,y)$ be the fundamental solutions for heat equation and Laplace equation in $\real^n$ respectively. We have the well-known expressions
\begin{equation*}
 {\widetilde Z}(x,t) = \frac{1}{ (4 \pi \nu t)^{n/2}} \exp \Big( - \frac{|x|^2}{4 \nu t } \; \Big), \;\;\;{t>0},
\end{equation*}
and
\begin{equation*}
	{\widetilde N}(x,y) = \frac{1}{(2-n) s_n} \; \frac{1}{|x-y|^{n-2}},
\end{equation*}
where $s_n({=}2 {\pi}^{n/2} (\Gamma(n/2))^{-1})$ denotes the surface area of the unit hypersphere. 
It is instructive to verify that the Poisson bracket has a zero sum:
\begin{equation} \label{eq:pb-zero-mean}
	\int_0^T \!\! \int_{\real^n}  \sum_{i=1}^{n} \big\{ \omega, u \big\}_i  \; \rd x \rd t = 0.
\end{equation}
This result is due to the solenoidal conditions $\star \rd (\star u){=}0$ and $\star \rd (\star \omega){=}0$. Hence the Poisson bracket $\{\omega,u\} {\in} L_x^1 \: L_t^1\; \forall (x,t) {\in} \real^n {\times} [0,T]$. Denote $\widetilde \varpi{=}({\widetilde \varpi}_1, {\widetilde \varpi}_2, \cdots, {\widetilde \varpi}_n)$, as the solution of the pure initial value problem of heat equation in $\real^n$ with initial vorticity $\omega_0(x)$.
The time interval $[0,T]$ is divided into $m+1$ equal sub-intervals such that $0{<}t_0{<}t_1{<}{\cdots}{<}t_k{<}{\cdots}{<}t_m{=}T$. The vorticity can be expressed as
\begin{equation} \label{eq:duhamel-n}
	\omega_i(x,t) = {\widetilde \varpi}_i(x,t) + \int_0^t \!\! \int_{\real^n}  {\widetilde Z}(x{-}y,t{-}s) \sum_{j=1}^{n} \big\{ \omega,u \big\}_j(y,s) \rd y \rd s. 
\end{equation}
These integral equations hold for every time $t_k {\in} [0,T], k=0,1,2, \cdots, m$. In view of the zero sum property of the Poisson bracket (\ref{eq:pb-zero-mean}), we establish, by analogy, the integrability of the vorticity,
\begin{equation*}
	\omega(x) \in L^1 (\real^n),
\end{equation*}
since the heat kernel has the property 
\begin{equation*}
\int_{\real^n} \widetilde Z(x,t) \rd x = 1.	
\end{equation*}
Differentiating (\ref{eq:vorticity-n}) $\alpha({>}1)$ times with respect to the space variable, we obtain 
\begin{equation*} 
	(\partial^{\alpha}_x \omega)' - \nu (\star \rd (\star \rd (\partial^{\alpha}_x \omega))) = \partial^{\alpha}_x \big\{ \omega,u \big\}. 
\end{equation*}
Thus we establish 
\begin{equation*} 
	\int_0^T \!\! \int_{\real^n} \sum_{i=1}^{n} \Big( \partial^{\alpha}_x \big\{ \omega, u \big\}_i \Big) \; \rd x \rd t =0
\end{equation*}
because the Poisson bracket commutes with the differential operator $\partial^{\alpha}_x$. Therefore we deduce 
\begin{equation*}
	\omega(x) \in W^{n,1}(\real^n).
\end{equation*}
It follows that $\omega(x) \in L^{\infty}$
by virtue of the Sobolev embedding theorem. Naturally, we can improve the {\itshape a priori} regularity of the vorticity to
$W^{q,1}(\real^n)$ for arbitrary finite value $q{>}n$. Hence we assert that the diffusion by viscosity alone is effective to guarantee
\begin{equation*}
	\omega \in C^{\infty} (\real^n).
\end{equation*}
Alternatively, we may determine the smoothness by using the standard machinery of partial differential equations. The solenoidal conditions imply that there exists a vector function $\widetilde \Psi$ such that
\begin{equation*}
	\star \rd (\star \rd \widetilde \Psi)(x;t) = - \star \rd u (x;t).
\end{equation*}
The inversion of this equation gives the velocity, 
\begin{equation} \label{eq:u-poisson-n}
	u(x;t) = -\frac{\partial \widetilde \Psi}{\partial x}(x;t) = - \int_{\real^n} \frac{\partial \widetilde N}{\partial x}(x,y) \star \rd u (y;t) \rd y.
\end{equation}
This solution is kinematic in nature. Equation (\ref{eq:u-poisson-n}) is just the Biot-Savart law in $n$ dimensions. Let $\widetilde{\upomega}$ be the total vorticity (cf. (\ref{eq:total-vort})). The rate of change in the vorticity (analogous to (\ref{eq:vort-beta-time-deriv})) is given by
\begin{equation*} 
	(\partial_t^{\beta} \widetilde{\upomega})' - \nu \Delta (\partial_t^{\beta} \widetilde{\upomega}) = \partial_t^{\beta} \Big( \sum_{i=1}^n \big\{ \omega,u \big\}_i \Big)
\end{equation*}
for any $\beta{\geq}0$. Thus we obtain the generalized invariance
\begin{equation*} 
	\frac{\rd }{\rd t}\int_{\real^n} \partial_t^{\beta} \widetilde{\upomega}(x,t) \rd x = 0,\;\;\; t \in [0,T].
\end{equation*}
We have assumed that there exists a small time interval, depending on the initial data, in which the solution in $\real^n$ is smooth. The interval can be viewed as an analogy to the classical local in-time solution of the Navier-Stokes equations. It~follows that we are able to extend the smoothness bound in space to 
\begin{equation*} 
	\omega  \: \in \:  C^{\infty},\;\;\; (x,t) \in \real^n \times [0,T]. 
\end{equation*}
Consequently, we assert that the Cauchy problem for the motion of incompressible Newtonian fluids in $\real^n \;(n{>}3)$ is mathematically tractable. The global well-posedness can be established in accordance with the theory for systems of parabolic partial differential equations with bounded coefficients (see, for example, Eidel'man 1969; Ladyzhenskaya {\itshape et al} 1968; Friedman 1964). 

Integrating by parts and in view of (\ref{eq:u-poisson-n}), equations (\ref{eq:duhamel-n}) can be simplified as
\begin{equation*}
	\omega_i(x,t_0) = {\widetilde \varpi}_i(x,t_0) + \int_0^{t_0} \!\! \int_{\real^n}  \!\! \int_{\real^n} \sum_{j=1}^{n}  {\widetilde G}_{ij}(x,t_0,y,s,z)  \omega_j(z) \omega (y,s) \rd z \rd y \rd s. 
\end{equation*}
Evoking our similarity transformations, we obtain
\begin{equation*}
	\omega_i(x,t_1) = {\widetilde \varpi}_i(x,t_1) + \int_0^{t_1} \!\! \int_{\real^n}  \sum_{j=1}^{n} {\widetilde K}_{ij}(x,t_1,y,s) \omega_j (y,s) \rd y \rd s +  {\widetilde g}^{(1)}_i(x,t_1).
\end{equation*}
In the last two displaced equations, ${\widetilde G}_{ij}$ and ${\widetilde K}_{ij}$ are $n {\times} n $ matrices and they are the analogous matrices to $G_{ij}$ and $K_{ij}$ in $n{=}3$. Their elements are functions of the kernels ${\widetilde Z}$ and ${\widetilde N}$; they are uniquely defined for given initial data. By consecutive similarity transformations, the vorticity equation can be transformed into
the non-linear system:
\begin{equation*}
	\omega(x,t) = {\widetilde \varpi}(x,t) + \int_0^{t} \!\! \int_{\real^n} {\widetilde K}(x,t,y,s) \omega(y,s) \rd y \rd s +  {\widetilde q}(x,t,\omega),
\end{equation*}
where the series ${\widetilde q}$ is a non-linear function of vorticity ${\omega}$. If an initial vorticity data are given in a lower dimension ($<n$), the integral kernel ${\widetilde K}$ always transforms the initial vorticity vector into a full $n$-dimensional vorticity field. This non-linear Volterra-Fredholm system is first reduced to
\begin{equation*}
	\omega(x,t) = \widetilde \gamma(x,t) + {\widetilde q}(x,t) + \int_0^{t} \!\! \int_{\real^n} {\widetilde H}(x,t,y,s) {\widetilde q}(y,s) \rd y \rd s,  
\end{equation*}
where the tilde variables are exact the counterparts of those in $n{=}3$. The last two terms can be expanded in terms of $\widetilde \gamma$. The vorticity solution is given by 
\begin{equation*} 
	\omega(x,t) = \widetilde \gamma(x,t) \: + \sum_{m{\geq}2}\: \Big( \sum_{k=1}^{S_m} \: {\widetilde V_k}[x,t,\widetilde \gamma]^m\Big).
\end{equation*}
In parallel to the proof of convergence for $n{=}3$, we can show, by the method of majorant, that this series also converges for any given finite value $\widetilde \gamma(x,t)$. The global regularity of the velocity and the pressure can be established accordingly.
Fluid motions in $\real^n$ evolve in a manner analogous to that in $3$ space dimensions. This fact is the direct consequence of the affine property of the Poisson bracket. 
\section{Conclusion}
By considering the development of incompressible flows in $\real^3$, we have shown that the initial value problems of the Navier-Stokes equations, or more precisely, of the vorticity equation, are globally well-posed. The proof has been facilitated by the fact that the total vorticity is an invariant property in any fluid motion, thanks to a symmetry in the vorticity equation. If a fluid motion starts with its initial data which are bounded, localized and of considerable size, there exists a time interval in which the flow is known as laminar. Subsequently, the laminar flow undergoes the transition process and ultimately develops into turbulence. The whole flow field decays in time due to the viscous dissipation of its kinetic energy. After a sufficiently long period of time, the flow relaminarizes during the decay before the motion fully restores to its stationary thermodynamic equilibrium state in which any velocity gradients have been smoothed out. In particular, the law of energy conservation holds during the entire course of the flow evolution. 

In fluid dynamics, turbulence is the general solution of the vorticity equation which is characterized by a vorticity population of broad spatio-temporal scales and intensities. The non-linear terms in the equations of motion are solely responsible for giving rise to the intricate flow structure. Turbulence is a three-dimensional, intrinsic property of fluid motions. In observation, the operative interface of the viscous dissipation ramifies the microscopic fluctuations of fluid's molecular constituents so that turbulence shows randomness characters in space and in time. The Navier-Stokes equations expound turbulence as a continuum as well as a viscous being. In essence, turbulence can be analyzed in depth and computed with precision. As a result, turbulence is no longer a subject sheltered under phenomenology.

Fluid motions can be described by smooth functions; it is not necessary to rely on the idea of fractals to depict or predict their dynamics. Likewise, the uniqueness and the regularity of Navier-Stokes solutions imply that the velocity and its gradients do not bifurcate or become multi-valued in space-time. The finding is in direct contradiction to the instability hypotheses in which the growth of disturbances is suspected to precipitate eventual flow breakdown. Specifically, the concept of separating a fluid motion into a mean superimposed by wavy disturbances is fundamentally flawed because the eigenvalue relation entails spurious vorticity.

For smooth initial data with compact support, we have shown that the solutions of the Navier-Stokes equations, in the limit of vanishing viscosity (hence the Euler equations), cannot develop finite-time singularities -- a by-product of our {\itshape a~priori} analysis. In flows having finite energy, the absence of space-time blow-up is consistent with the classical vortex theorems of Helmholtz. In addition, Cauchy's invariance in the Lagrangian description implicitly asserts that the velocity field is an essentially bounded function in time.

The outcome of the present mathematical analysis provides us a practical means for computing turbulence. Any fluid motion, regardless of whether it is a laminar flow or a turbulent flow, has an identical microscopic origin, as formulated in the kinetic theory of gases. Hence the macroscopic flow quantities, such as velocity and pressure, must be the ensemble average of the microscopic fluctuations. On the continuum, the Navier-Stokes equations primarily govern the evolution of laminar flows {\itshape as well as} turbulent flows. The implication is that the mean flow quantities are well-defined by the equations which require no turbulence closure strategies. 
Nevertheless, it is recognized that reliable computations of complex flow fields involving turbulence may well become a routine task with advances in computer facilities and numerical algorithms. For certain applications, the effect of a solid boundary is critical and must be properly taken into account. 

On reflection, many theoretical problems in connection with turbulence have been either ill-defined or self-imposed, as contemplated by Saffman (1978), elaborated by Liepmann (1979), and expounded by Bradshaw (1994) from an experimental point of view. 
The investigation of pipe flows by Reynolds (1883) has set a scene for the study of turbulence. The fact that the transition process was never fully understood and satisfactorily explained in terms of varied instability-oriented notions is truly disappointing. On the other hand, turbulence measurements of tailor-made flows are mainly focused on understanding physical processes and, ambitiously, they have been used for prediction purposes in application. The difficulty in carrying out well-controlled experiments is notorious; we attempt to figure out the puzzling details of the vorticity jumble without guidance from a well-founded theory. At least, we have hoped, within a very limited scope, that certain theories of turbulence may be retrofitted by proper interpretation of experimental results. Evidently, incomplete knowledge of the non-linearity has been the key obstacle which bars us from achieving our objectives. The other reason is that the study of fluid motions is one discipline where the determinism of continuum Newtonian mechanics encounters the stochastic reality of molecular agitations, particularly on the dissipative scales. These essential difficulties have not been adequately addressed in methods of statistics and dimensional analysis. A quantitative analytic theory on the processes of energy dissipation is challenging but highly desirable though the subject is clearly outside the scope of the continuum fluid dynamics.

From the solution of the Boltzmann equation for Maxwellian molecules with cut-off, we conclude that the phase-space density distribution is consistent with the turbulence phenomena described by the continuum dynamics. It must be admitted that the molecular model used in our analysis is very limited, and it is tenably over-simplistic for real gases in application. Nevertheless, it is the consistency aspect between the continuum dynamics and the kinetic theory that is encouraging. For Newtonian fluids under extreme physical conditions or for non-Newtonian fluids in general, it has been known that modifications to the equations of motion are inevitable. In brief, it remains to be seen whether a conceptual framework derived from kinetic theories or quantum mechanics is necessary to generalize the principles of fluid dynamics beyond the continuum governed by the Navier-Stokes equations.
\vspace{10mm} 
\begin{acknowledgements}
\noindent 
The work on {\itshape the problems} of fluid dynamics started in the mid-nineties when the author was involved in design projects in the commercial aircraft industry in Bristol. Due to other job commitments, it has been put on hold for almost 15 years until 2008 when, in Bremen, I resumed background reading of technical materials. The work was further carried out in HKSAR and Cambridge, on a number of occasions. The entire work is self-funded. I would like to thank all my family members, particularly Jonathan, Dionne and Ling Ling, for their encouragement, support and understanding over the last few years. The author is indebted to Professor Tony Shing of The Chinese University of Hong Kong for his valuable help during the preparation of the present work. It has always been an enjoyable experience to discuss technical problems, for theoretical purposes and for practical applications, with my colleagues and friends; my sincere thanks go to Reginald Burrell, Stephen Chow, James Chu, Tony Davey, Axel Flaig, Shahid Mughal, Richard Sanderson, Stratos Saliveros, G\'eza Schrauf, Bruno Stefes, John Welch and Mr. D. Rozendal (NLR). It is a pleasure to acknowledge the Cambridge University libraries for access to their collections of scientific literature. It was my intention to present a preliminary version of the present work at the conference of ``Topological Fluid Dynamics (IUTAM Symposium)'' held during July 2012 at Isaac Newton Institute for Mathematical Sciences, Cambridge. I would like to thank the Institute for a financial support though, regrettably, my attendance has not been materialized due to unforeseeable reasons. 
\vspace{10mm}

\noindent 
17 September 2014

\noindent 
\texttt{f.lam11@yahoo.com}
\end{acknowledgements}
%
%
\newpage

\appendix{List of the coefficients $S_m$} \label{app:a}

The integer coefficients in the vorticity series solution are listed below \footnote{This is sequence {\ttfamily A107841} in The On-Line Encyclopedia of Integer Sequences on {\ttfamily www.oeis.org}}. 

\begin{center}
\begin{tabular}{cl} \hline \hline
$m$   & $S_m$ \hspace{25mm} Note: For $m{=}1$ to $11$, see (\ref{eq:vort-series-sol}). \\ \hline \hline
12  & 9002083870 \\
13  & 79568077034 \\
14  & 708911026210 \\
15  & 6359857112438 \\
16  & 57403123415350 \\
17  & 520895417047010 \\
18  & 4749381474135850 \\
19  & 43489017531266654 \\
20  & 399755692955359630 \\
21  & 3687437532852484442 \\
22  & 34121911117572911410 \\
23  & 316666408886000120582 \\
24  & 2946636961744936971430 \\
25  & 27486258777812752124114 \\
26  & 256973211157791603864250 \\
27  & 2407526588834623436461550 \\
28  & 22599603739290305631946750 \\
29  & 212529461593306742410678730 \\
30  & 2002045234365334208175990850 \\
31  & 18889388815475657436367735190 \\
32  & 178487761207390062612441116950 \\
33  & 1688910333503588969662234212290 \\
34  & 16002095810430452080950009867850 \\
35  & 151805250377580209220247213205630 \\
36  & 1441805206121252402768525955472750 \\
37  & 13709097801869521570731666349614650 \\
38  & 130487350981556551346787383248935250 \\
39  & 1243259837730752499259901772397608230 \\
40  & 11856807721130441176987397061795017350 \\
41  & 113179047461579269743065705735797031090 \\
42  & 1081279822970004169681227255764743423450 \\
43  & 10338717814007432319433088403746551742990 \\
44  & 98931744367299841862817375797502906320350 \\
45  & 947392638640132826939524118011471556252330 \\
46  & 9078950081470196664159989556598498515249250 \\
47  & 87064148315517405658794094968912199782490550 \\
48  & 835466096859655890640323175318974774795541750 \\
49  & 8022180840122075777891868351323277847721213858 \\
50  & 77076022415905177005994888656661014616310095210 \\ 
51  & 740967393439733320671359946584595610898043434654 \\ \hline \hline
\end{tabular}
\end{center}

\newpage
\appendix{Additional methods on uniqueness of vorticity} \label{app:b}

In the present appendix, two alternative approaches are given on the uniqueness of equation (\ref{eq:vort-nonlinear-VIE}). The first method is directly related to the vorticity integral equation and hence depends on the series solution. Second, it is well-known that if the initial velocity is identically zero, then the Navier-Stokes equations admit zero solution, $u(x,t){=}\nabla p(x,t){=}0$. It follows that if the initial vorticity vanishes, then $\gamma(x,t){=}0$ as $\varpi(x,t){=}0$. Hence $\omega(x,t){=}0$ is a solution of the vorticity equation. This observation suggests that we may generalize a method of Weierstrass for certain properties of the implicit functions defined by power series (\S 187 of Goursat 1904). The main feature of Weierstrass' method is its simplicity.

Let us consider two distinct series solutions to (\ref{eq:vort-VIE}), $w_1(x,t)$ and $w_2(x,t)$. Their difference, $\varphi=w_1{-}w_2$, satisfies 
\begin{equation} \label{eq:diff-w-VIE}
	\varphi(x,t)= \sum_{m{\geq}2}^{\infty} \Big\{
	W_m\left( \begin{array}{c}
x,t  \\
w_1
\end{array} 
\right)
-W_m\left( \begin{array}{c}
x,t  \\
w_2
\end{array} 
\right) \Big\} = \sum_{m+n{\geq}2}^{\infty} 
	Y_{mn}\left( \begin{array}{c}
x,t  \\
\varphi, w_2
\end{array} 
\right)
\end{equation}
(cf. (\ref{eq:VIE-sum}) and (\ref{eq:ip-term-2arg})). The sum on the right is obtained 
by substitution $w_2{+}\varphi$ for $w_1$ for every $W_m[x,t,w_1]$. Notation $Y_m$ stands for the integro-power form of {\itshape both} functional arguments $\omega_2$ and $\varphi$. We then replace all $\varphi$ {\itshape except one} in $Y_{mn}$ by $w_1{-}w_2$. The reduction gives rise to the result which is a sum of the integro-power forms in three functional arguments $\varphi, w_1,w_2$:
\begin{equation} \label{eq:um-sum}
	\sum_{m+n+l{\geq}2}^{\infty} \; U_{mnl}\left( \begin{array}{c}
x,t  \\
\varphi, w_1, w_2
\end{array} 
\right).
\end{equation}
Clearly $U_{mnl}$ takes more than one definite algebraic form. Nevertheless, every term in $U_{mnl}$ must be in one of the following three forms: (1) The function $\varphi$ is located in the inner most integral,
\begin{equation} \label{eq:inner-phi}
\int^* \!\!\! G w \int^* \!\!\! G w \cdots \int^* \!\!\! G w \int^* \!\!\! K \varphi  \;\;\; \mbox{or} \;\;\;  \int^* \!\!\! H \int^* \!\!\! G w  \cdots \int^* \!\!\! G w \int^* \!\!\! K \varphi;
\end{equation}
(2) the function $\varphi$ is involved in the outer most integral,
\begin{equation} \label{eq:outer-phi}
\int^* \!\!\! G \varphi \int^* \!\!\! G w \cdots \int^* \!\!\! G w \int^* \!\!\! K w  \;\;\; \mbox{or} \;\;\;  \int^* \!\!\! H \int^* \!\!\! G \varphi  \int^* \!\!\! G w \cdots \int^* \!\!\! G w \int^* \!\!\! K w;
\end{equation}
(3) the function $\varphi$ is interlocked somewhere in the middle of the integro-power form,
\begin{equation} \label{eq:middle-phi}
 \begin{split}
\int^* \!\!\! G w \cdots \int^* \!\!\! G w \int^* \!\!\! G \varphi \cdots & \int^* \!\!\! G w \int^* \!\!\! K w  \;\;\; \mbox{or} \;\;\; \\  
\int^* \!\!\! H & \int^* \!\!\! G w \cdots   \int^* \!\!\! G w \int^* \!\!\! G \varphi \cdots \int^* \!\!\! G w \int^* \!\!\! K w.
 \end{split}
\end{equation}
In (\ref{eq:inner-phi}) to (\ref{eq:middle-phi}), symbol $w$ stands for either $w_1$ or $w_2$. The integrals with $*$ are the shorthands for the space-time integrations.
All of these integrals are finite for $(x,t) \in \real^3 {\times} [0,t{<}T]$ as both $\omega_1$ and $\omega_2$ are solutions.

Either equation in (\ref{eq:inner-phi}) can be expressed as
\begin{equation} \label{eq:inner-phi-ie}
	\int_0^t \int \int_0^s \int F(x,t,y,s) K(y,s,z,r) \varphi(z,r) \rd z \rd r \rd y \rd s,
\end{equation}
where $F(x,t,y,s){=}F(x,t,\omega_1,\omega_2)$ is an integro-power form in arguments $\omega_1$ and $\omega_2$. 
The first expression in (\ref{eq:outer-phi}) can be reduced to
\begin{equation} \label{eq:outer-phi-ie}
	\int_0^t \int \tilde{G} (x,t,y,s) \varphi(y,s) E (y,s) \rd y \rd s,
\end{equation}
where the kernel, $E(y,s) {=} E(y,s,w_1,w_2)$, refers to the integral quantity,
\begin{equation*}
	\int^* \!\!\! G w \cdots  \int^* \!\!\! G w \int^* \!\!\! K w.
\end{equation*}
By analogy, the second expression in (\ref{eq:outer-phi}) has a similar reduction.
More explicitly, the first equation in (\ref{eq:middle-phi}) can be expressed as
\begin{equation} \label{eq:middle-phi-ie}
	\int_0^t \int \int_0^s \int D(x,t,y,s) {\tilde G}(y,s,z,r) \varphi(z,r) B(z,r) \rd z \rd r \rd y \rd s,
\end{equation}
where $D(x,t,y,s){=}D(x,t,\omega_1,\omega_2)$ is an integro-power form in arguments $\omega_1$ and $\omega_2$. The function, $D(z,r){=}D(z,r,w_1,w_2)$, is analogous to function $E(z,r)$.

Applying these reductions to every term in equation (\ref{eq:um-sum}) and in view of the integrals in  (\ref{eq:inner-phi-ie}) to (\ref{eq:middle-phi-ie}), we rewrite the integro-power forms in (\ref{eq:diff-w-VIE}) as
\begin{equation} \label{eq:phi-VIE}
	\varphi(x,t) = \int_0^t \int \bar{K}(x,t,y,s) \varphi(y,s) \rd y \rd s.
\end{equation}
The kernels $\bar{K}$ and $K$ contain the identical integrable singularity. Since $w_1$ and $w_2$ satisfy (\ref{eq:vort-formal-sol}), it follows that the kernel $\bar{K}$ is integrable.
Let $\bar{K}_{i}$ denote the $i$th iterated kernel of $\bar{K}$, we repeatedly multiply (\ref{eq:phi-VIE}) by $\bar{K}$ and integrate over space-time. The end result is that $\varphi$ is governed by the homogeneous Volterra-Fredholm equation,
\begin{equation*}
		\varphi(x,t) = \int_0^t \int \bar{K}_i(x,t,y,s) \varphi(y,s) \rd y \rd s.
\end{equation*}
Hence we deduce that
\begin{equation} \label{eq:phi-sol-VIE}
	0 = \int_0^t \int \Big(\: \bar{K}_i(x,t,y,s) - \bar{K}_j(x,t,y,s) \: \Big) \varphi(y,s) \rd y \rd s,
\end{equation}
where $i{\neq}j$. For arbitrary values of $i$ and $j$, the only bounded solution satisfies (\ref{eq:phi-sol-VIE}) is $\varphi(x,t){\equiv}0$ for $t{>}0$, $x {\in} \real^3$.
The uniqueness for the integro-power series of arbitrary order with symmetric kernels has been proved by Sabbatini (1925, cited by Volterra 1930).

The method of Weierstrass has been adapted in the following manner.
The substitution procedures show that the series solution  (\ref{eq:vort-series-sol}) satisfies equation (\ref{eq:vort-formal-sol}). We have shown that the series converges. For fixed $(x,t)$, introduce the abbreviation for the vorticity solution:
\begin{equation*} 
\bar{V} = \bar{V}(\gamma) = \sum_{m=1}^{\infty} V_m
	\left( \begin{array}{c}
x,t  \\
\gamma
\end{array} 
\right).
\end{equation*}
Rewrite (\ref{eq:VIE-sum}) as
\begin{equation} \label{eq:VIE-sum-decomp}
	F(\omega, \gamma) = \omega(x,t) - \Big\{ \gamma(x,t) + \sum_{m{\geq}2} W_m 
\left( \begin{array}{c}
x,t  \\
\omega
\end{array} 
\right) \Big\}=0.
\end{equation}
Let $V_{\varepsilon}$ be a perturbation quantity which is in general non-zero at a pair of $(x,t)$. Let the perturbed $\omega$ be
\begin{equation*}
	\omega=\bar{V} + V_{\varepsilon}.
\end{equation*}
Substituting the perturbed $\omega$ for $\omega$ in (\ref{eq:VIE-sum-decomp}), we obtain
\begin{equation*}
	\sum_{m{\geq}2} W_m 
\left( \begin{array}{c}
x,t  \\
\!\!\bar{V} {+} V_{\varepsilon}\!\!
\end{array} 
\right){=}
\sum_{m{\geq}2} \Big\{  W_m 
\left( \begin{array}{c}
x,t  \\
\bar{V}
\end{array} 
\right) + W_m 
\left( \begin{array}{c}
x,t  \\
V_{\varepsilon}
\end{array} 
\right)\Big\} + \sum_{m+n{\geq}2} U_{mn} 
\left( \begin{array}{c}
x,t  \\
\!\!\bar{V}, V_{\varepsilon}\!\!
\end{array} 
\right).
\end{equation*}
Every integro-power term in the rightmost integro-power form contains integral convolutions in both arguments $\bar{V}$ and $V_{\varepsilon}$. Moreover, the integro-power form can be expressed as
\begin{equation*}
	\sum_{m+n{\geq}2} U_{mn} 
\left( \begin{array}{c}
x,t  \\
\bar{V},  V_{\varepsilon}
\end{array} 
\right)=
\sum_{m{\geq}2} Q_m 
\left( \begin{array}{c}
x,t  \\
\gamma
\end{array} 
\right) P_n(V_{\varepsilon}),
\end{equation*}
where symbol $P_n$ stands for a polynomial in $V_{\varepsilon}$ with zero constant term, and $Q_m$~is another integro-power form. In effect, (\ref{eq:VIE-sum-decomp}) is reduced to
\begin{equation} \label{eq:vort-formal-sol-decomp}
F(\bar{V}+V_{\varepsilon},\gamma)  =	( \omega - \bar{V} ) \: \Big[ 1 - \sum_{m{\geq}2}^{\infty} \Big\{ Q_m
	\left( \begin{array}{c}
x,t \\
\gamma
\end{array} 
\right) \; \bar{P}_n(V_{\varepsilon}) + \bar{W}_m
	\left( \begin{array}{c}
x,t \\
V_{\varepsilon}
\end{array} 
\right)
\Big\} \Big]=0,
\end{equation}
where $\bar{P}_n$ is another polynomial in $V_{\varepsilon}$. This decomposition shows that the solution for $\omega$ given by (\ref{eq:vort-formal-sol}) is unique for $F(\omega,\gamma)=0$ since the numerical value of the term in the square brackets in (\ref{eq:vort-formal-sol-decomp}) is non-zero when $\gamma=0$. Obviously this argument is valid for any pair of $(x,t)$ as the perturbation $V_{\varepsilon}$ can be arbitrarily chosen.

\appendix{Solution for vanishing vorticity} \label{app:c}

The vorticity theory we have developed is based on the requirement of non-vanishing vorticity throughout $(x,t)$. Although it is extremely unlikely in practice that a fluid motion may be set in with vanishing vorticity, the mathematical problem for a fluid motion of zero vorticity remains valid and must be treated independently. Assume that the Navier-Stokes equations admit solutions for flows of zero vorticity ($\nabla {\times} u {\equiv} 0, (x,t) {\in} \real^3{\times}[0,T]$). We consider the velocity in the example of Serrin (1962)
\begin{equation*}
	u(x,t)= d(t) \nabla \vartheta(x),
\end{equation*}
where $d(t) \in L^1([0,T])$, and $\vartheta$ is harmonic. The initial condition is $u_0 = d_0 \nabla \vartheta(x)$, where $d_0$ is finite, and $|u_0| < \infty$. Moreover, we suppose that $\vartheta(x)$ is localized in order to avoid the trivial case. Because $\nabla.u{=}0$, we only need to consider the system,
\begin{equation} \label{eq:vort-free-vel}
	{\partial u} / {\partial t} - \nu \Delta u = - \nabla \chi_0, \;\;\; 
	\Delta \chi_0 =0,
\end{equation}
where $\chi_0 = p_0/\rho {+} u^2 /2$, and $p_0$ is the pressure.
By Liouville's theorem for harmonic functions in $\real^3$, we assert that $p_0 = c_0{-}\rho u^2/2$, where $c_0$ is an indeterminate constant. Thus the solution of (\ref{eq:vort-free-vel}) is given by
\begin{equation*}
	u(x,t) = d_0 \: \int {\mathbf Z}(x{-}y,t) \nabla \vartheta(y) \rd y.
\end{equation*}
If $\vartheta {\in} C^{\infty}({\Upomega}_0))$, where ${\Upomega}_0$ denotes a subset of $\real^3$, then $u$ is smooth in space as well as in time. Moreover, the velocity field has the point-wise decay
\begin{equation*}
	\big|\partial^{\alpha}_x  \:u(x,t) \big| \; {\sim} \; | x |^{-a} \; (\nu t)^{-b},
\end{equation*}
where $\alpha>0$, and $(a{-}\alpha)+ b=3$ as $|x| \rightarrow {\infty}$ or $\nu t \rightarrow {\infty}$. If $\vartheta$ is merely in $H^1_0(\Upomega_0)$, we have the estimates for the Stokes semi-groups $\exp(-t {\mathbf A}),\;\; {\forall} t > 0$ (cf. (\ref{eq:mom-semi-group})),
\begin{equation*}
 \begin{split}
	\big\| u \big\|_{L^q(\real^3)} & \; {\leq} \; C(q,T) \; (\nu t)^{-{3}/{2} \:({1}/{2}-{1}/{q})} \: \big\| u_0 \big\|_{L^2(\real^3)},\;\;\;2 \leq  q  \leq  \infty,\\
	\big\|\nabla u \big\|_{L^q(\real^3)} & \; {\leq} \; C(q,T) \; (\nu t)^{-{3}/{2} \:({1}/{2}-{1}/{q})-{1}/{2}} \: \big\| u_0 \big\|_{L^2(\real^3)},\;\;\; 2 \leq  q <  \infty.
	 \end{split}
\end{equation*}
Finally, it is easy to establish that
\begin{equation*}
	\big\|\partial_t u \big\|_{L^2(\real^3)} \; {\leq} \; C(T) \; (\nu t)^{-{1}/{2}} \: \big\| u_0 \big\|_{L^2(\real^3)}.
\end{equation*}
\appendix{Euler equations as limit of vanishing viscosity} \label{app:d}

In this appendix, we outline simple ideas which demonstrate that solutions of the Euler equations cannot develop finite-time singularities for localized initial data having finite energy. The objective is to provide further justifications to our Navier-Stokes theory of turbulence. It has been shown (Constantin 1986) that the spontaneous appearance of finite-time singularities in solutions of the Navier-Stokes equations cannot occur without blow-up of the corresponding solution of the Euler equations. 

Write the triplet $(u, p, \omega)$ as $(u^{(0)}, p^{(0)},\omega^{(0)})$ for the limit $\nu {\rightarrow} 0$.
For $\alpha, \beta \geq 0$, the following invariant relations for the total vorticity (cf. (\ref{eq:total-vort}), (\ref{eq:vort-rhs-integrability}) and (\ref{eq:vort-rhs})),   
\begin{equation} \label{eq:space-vort-int-euler}
\lim_{\nu \rightarrow 0} \frac{\rd }{\rd t} \int \partial_x^{\alpha} \upomega(x,t) \rd x = \frac{\rd }{\rd t} \int \partial_x^{\alpha} \upomega^{(0)}(x,t) \rd x = 0,
\end{equation}
and
\begin{equation} \label{eq:time-vort-int-euler}
\lim_{\nu \rightarrow 0} \frac{\rd }{\rd t} \int \partial_t^{\beta} \upomega(x,t) \rd x = \frac{\rd }{\rd t} \int \partial_t^{\beta} \upomega^{(0)}(x,t) \rd x = 0,\;\;\; t \in [0,T],
\end{equation}
are independent of viscosity, and they coincide with (\ref{eq:space-vort-int}) and (\ref{eq:time-vort-int}) respectively. In the second invariance, we have implicitly made use of the local in-time smooth solutions of the Euler equations (see, for example, Lichtenstein 1925; Swann 1971). In particular,  it is known that a unique solution to the Navier-Stokes equations exists in $\real^3$ for a small local time interval $[0,T_L]$ independent of viscosity. For initial data of finite energy, the smooth viscous solutions for varying viscosities are found to converge uniformly to a function that is a solution to the Euler equations for inviscid flow in $\real^3$.

For $(x,t) \in \real^3 \times [0,T< \infty]$, we verify {\itshape a priori} that every component of vorticity is essentially bounded in space and in time,
\begin{equation} \label{eq:vort-bound-euler}
\omega^{(0)} \in L_x^{\infty} L_t^{\infty},
\end{equation}
in view of Minkowski's inequality and the Sobolev embedding theorem.

To proceed further, we make the following {\itshape assumption}: There exists a fluid with zero viscosity and the equations of motion can be reduced from the Navier-Stokes equations by {\itshape formally} setting viscosity $\mu{=}0$. We recognize the fact that, for inviscid flows governed by the Euler equations,  
we will never be able to verify any ideal fluid motion in nature. The equations of motion read
\begin{equation} \label{eq:euler}
	 \partial_t u^{(0)}  + (u^{(0)} . \nabla) u^{(0)} = - {\rho}^{-1} \nabla p^{(0)},\;\;\; \nabla.u^{(0)}=0 
\end{equation}
(Euler 1755). We are interested in finite-energy initial value problems subject to the initial data (\ref{eq:ns-ic}) which are assumed to hold as $\mu{=}0$. The corresponding vorticity equations are given by
\begin{equation} \label{eq:vorticity-euler}
	\partial_t \omega^{(0)}_i   = (\omega^{(0)} . \nabla) u^{(0)}_i -  (u^{(0)} . \nabla )\omega^{(0)}_i.
\end{equation}
This is a set of hyperbolic equations. The initial data are specified in (\ref{eq:vort-ic}). Thus the total inviscid vorticity satisfies the dynamic equation
\begin{equation} \label{eq:total-vort-euler}
	{\partial \upomega^{(0)} } / {\partial t} = \sum_{i=1}^3{\mathscr R_i^{(0)}}(x,t,u^{(0)},\omega^{(0)}).
\end{equation}
By virtue of the space-time mollifications (cf. (\ref{eq:mollifiers})), the following equalities are easy to establish:
\begin{equation*} 
	{\partial \big( \partial^{\alpha}_x \upomega^{(0)}\big) } / {\partial t} = \partial^{\alpha}_x \sum_{i=1}^3{\mathscr R_i^{(0)}}(x,t,u^{(0)},\omega^{(0)}),
\end{equation*}
and
\begin{equation*} 
	{\partial \big( \partial^{\beta}_t \upomega^{(0)}\big) } / {\partial t} = \partial^{\beta}_t \sum_{i=1}^3{\mathscr R_i^{(0)}}(x,t,u^{(0)},\omega^{(0)}).
\end{equation*}
We then carry out an integration over space $\real^3$ and we recover the integral invariance relations in (\ref{eq:space-vort-int-euler}) and (\ref{eq:time-vort-int-euler}). {\itshape Thus a calculation in the limit of vanishing viscosity for the Navier-Stokes and a direct evaluation of the Euler equations produce the identical invariant integrals for the total vorticity.}

Given {\itshape a priori} bound (\ref{eq:vort-bound-euler}), the integral criterion (Beale {\itshape et al} 1984),
\begin{equation*}
	\int_0^T \big\|\:\omega(\cdot,t)\:\big\|_{L^{\infty}(\real^3)} \rd t < \infty,
\end{equation*}
is fulfilled by the Navier-Stokes solutions in the limit of vanishing viscosity, and by the Euler solutions in $0 \leq t \leq T < \infty$. To rule out a singularity, we assert that bound (\ref{eq:vort-bound-euler}) can be strengthened to
\begin{equation*}
\omega^{(0)}(x,t) \in C_B^{\infty} \;\;\; (x,t) \in \real^3 {\times} [0,T],
\end{equation*}
and that, for $\nu = 0$,
\begin{equation*}
	\frac{1}{2}\frac{\rd }{\rd t} \int u^2 \rd x = 0.
\end{equation*}
{\itshape The solutions of the Euler equations cannot develop any finite-time singularities from the smooth initial data (\ref{eq:ns-ic}).}

It is instructive to notice that, in the Lagrangian specification of fluid motion (\ref{eq:lagrangian-frame}), finite-time singularities can be completely ruled out in view of far-reaching Cauchy's invariant (Cauchy 1815) which takes the form,
\begin{equation} \label{eq:cauchy-invariant}
	\sum_{i=1}^3 \big( {\bar{\nabla}} u_i \times {\bar{\nabla}} x_i \big) = {\bar{\omega}}(a(x,t_0),t_0) \;(= {\bar{\omega}}_0),
\end{equation}
where ${\bar{\nabla}}$ denotes the gradient operation with respect to the Lagrangian co-ordinates $a=(a_1,a_2,a_3)$. The Jacobian of the Lagrangian map, $|{\bar{\nabla}}x|=1$. Given any smooth finite-energy initial flow having the vorticity ${\bar{\omega}}_0 = {\bar{\nabla}} {\times} u_0$, it is evident that {\itshape none of the quantities on the left-hand side becomes singular in finite time}. 

Cauchy's invariant is expressed in the {\itshape differential} form because the {\itshape Lagrangian} specification intends to follow the temporal trajectory of fluid elements having negligible geometric deformation. With hindsight, the {\itshape integral} of Cauchy's invariant turns out to be Kelvin's circulation theorem (Thomson 1869) in the {\itshape Eulerian} description, as first derived by Hankel in 1861 (for a historical perspective, see an interesting and informative article by Frisch \& Villone ({\ttfamily arXiv:1402.4957v1})). Finally, we make a remark on the issue of finite-time singularity according to the vortex theorems of Helmholtz (1858), which are stated below for reference: 
\begin{enumerate}
	\item Fluid flows remain irrotational if they are initially irrotational.
	\item A vortex line always consists of identical fluid particles at any time. 
	\item The circulation over any cross-sectional area of a vortex tube is an invariant of the motion. Thus vortex lines must either form closed curves or originate and terminate on the boundaries of the flow domain.
\end{enumerate}
As an initial-value problem in the whole space $\real^3$ (except flows with zero vorticity), the last two theorems cannot be true unless either the initial data possess an infinite amount of energy or all the initial vortex lines must be closed and localized having finite energy. In practice, it is a matter of obscurity how to generate singular fluid motions underlying boundless energy content. In the proof of his theorems, Helmholtz did not consider the initial value problem of the Euler equations. Since the uniqueness and the local regularity of the equations has been established; Helmholtz's theorems are indeed valid from a short time immediately after the commencement of the motion. Consequently, the circulation over any of the closed vortices,
\begin{equation*}
	\Gamma(t) = \oint_{C_v(t)} u \cdot \rd l = \int_{\partial S_v(t)} \omega \cdot \rd s, 
\end{equation*}
must remain {\itshape finite at all time} in any inviscid fluid motion of bounded energy. 
%
%
\addcontentsline{toc}{section}{\noindent{References}}

\label{lastpage}
\end{document}